\pdfoutput=1
\documentclass[oneside,notitlepage,11pt]{article}

\RequirePackage{ifpdf}
\RequirePackage{graphicx}

\ifpdf
\graphicspath{{./figures_PDF/}}
\else
\graphicspath{{./figures_EPS/}}
\fi

\RequirePackage[numbers,sort&compress]{natbib}   
\RequirePackage{amsthm}
\RequirePackage[newcommands]{ragged2e}
\RequirePackage{hyperref}
\RequirePackage[newcommands]{ragged2e} 
\RequirePackage{xspace}
\RequirePackage{layout}

\RequirePackage{fancyhdr}               % fancy headers, see fancyhdr.pdf
\RequirePackage[compact]{titlesec}      % customize sectioning commands; see 

\RequirePackage[centering,
          paperwidth = 8.5in,  %  top + bottom + textheight = paperheight
         paperheight = 11.00in,  %  top + bottom + textheight = paperheight
                 top = 0.9in,  %  top + bottom + textheight = paperheight
              bottom = 0.9in,  %
             headsep = 0.07in,  % separation between text and header
          headheight = 0.20in,  % header height (meaning unclear)
            footskip = 0.30in,  % bottom-of-footer to bottom-of-text
           textwidth = 6.00in % ,  % width of text
											]{geometry}

\RequirePackage{tocloft}  % needed for good-looking roman numerals in table of contents
\RequirePackage{color}    % for colored watermark       
  
\definecolor{QSEwatermark}{rgb}{0.85,0.85,0.88}
\definecolor{QSEwatermark}{rgb}{0.825,0.95,0.75}
\definecolor{QSEheadercolor}{rgb}{0,0,0}

\newlength{\mylen}        % need some extra space at end of number 
\cftsetindents{section}{0em}{3em}
\cftsetindents{subsection}{3em}{3em}
\cftsetindents{subsubsection}{5em}{4em}
                                        
\renewcommand{\thesection}{\arabic{section}} 
\renewcommand{\thesubsection}{\thesection.\arabic{subsection}}
\renewcommand{\thesubsubsection}{\thesubsection.\arabic{subsubsection}}

\titleformat{\section}%
{\vspace{0ex}\normalfont\sffamily\bfseries\large}% prior to \thesection
{\thesection.}{0.45em}%
{}% after the section text

\titlespacing*{\section}{0pt}{*1}{*1}

\titleformat{\subsection}%[runin]%
{\vspace{0ex}\normalfont\rmfamily\mdseries\itshape}% prior to \thesubsection
{\thesubsection.}{0.45em}%
{}% after the section text

\titlespacing*{\subsection}{0pt}{*1}{*1}

\titleformat{\subsubsection}[runin]%
{\vspace{0ex}\normalfont\rmfamily\mdseries\itshape}% prior to \thesubsubsection
{\thesubsubsection.}{0.45em}%
{}[\hspace*{0.5em}]% after the section text

\titlespacing*{\subsubsection}{0pt}{*1}{*1}

\newcommand{\subsubsubsection}[1]{\noindent\rule{0ex}{3ex}{%
\normalfont\sffamily\bfseries\upshape\small{#1}\hspace{0.05ex}:\hspace{0.75ex}}}

\input{QSE_environments}        
\input{QSE_macros}	 
\input{QSE_acronyms}
\newcommand{\fl}{}

\newcommand{\scriptpseudoinverse}{{\scriptstyle\text{\sffamily P}}}
\newcommand{\scriptscriptpseudoinverse}{{\scriptscriptstyle\text{\sffamily P}}}

\newcommand{\dimC}{{\dim \lcal{C}}}

\newcommand{\scriptscriptdimC}{{\scriptscriptstyle \dim \lcal{C}}}

\newcommand{\dimK}{{\dim \lcal{K}}}

\newcommand{\dimH}{{\dim \lcal{H}}}

\newcommand{\scriptscriptdimH}{{\scriptscriptstyle \dim \lcal{H}}}

\newcommand{\codim}{\operatorname{codim}}

\newcommand{\ioptext}[1]{{\text{\rmfamily\upshape#1}}}

\newcommand{\mycdot}{{\cdot}}

\DeclareMathOperator{\tr}{\ioptext{tr}}
\DeclareMathOperator{\Or}{\lcal{O}}
\DeclareMathOperator{\sgn}{\ioptext{sgn}}
\DeclareMathOperator{\csch}{\ioptext{csch}}

\newcommand{\myrho}[2][]{\rho_{#2}^\text{\upshape{#1}}}
\newcommand{\myqn}{q^{\text{\upshape n}}}
\newcommand{\myqm}{q^{\text{\upshape m}}}
\newcommand{\myfn}{f^{\,\text{\upshape n}}}
\newcommand{\myfext}{f^{\,\text{\upshape ext}}}

\newtheorem{mydefinition}{Definition}[section]
\newtheorem{mytheorem}{Theorem}[section]

\newtheorem{myDesignRule}[mytheorem]{Design Rule}

\newcommand{\myrenderasmathcharacter}[2][1]{%
\raisebox{-0.5\height}{\includegraphics[scale=#1]{#2}}}

\newlength{\myraise}
\newcommand{\myblockdiagramOne}{%
	\raisebox{\myraise}{%
	\raisebox{-0.03\height}{\myrenderasmathcharacter[1.00]{QSE_block_diagram_01}}}}

\newcommand{\myblockdiagramTwo}{%
	\raisebox{\myraise}{%
	\raisebox{0.09\height}{\myrenderasmathcharacter[1.00]{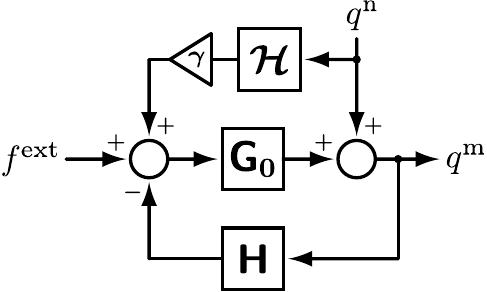}}}}

\newcommand{\myblockdiagramFour}{%
	\raisebox{\myraise}{%
	\raisebox{-0.0\height}{\myrenderasmathcharacter[1.05]{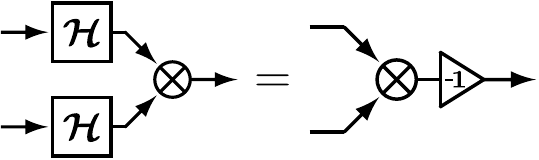}}}}

\newcommand{\myblockdiagramAddBlock}{%
	\raisebox{-0.85ex}[2.15ex][0.85ex]{%
	\includegraphics[height=3.0ex]{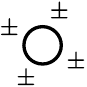}}}

\newcommand{\myblockdiagramMultBlock}{%
	\raisebox{-0.35ex}[2.15ex][0.85ex]{%
	\includegraphics[height=2.0ex]{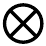}}}

\newcommand{\myblockdiagramGainBlock}{%
	\raisebox{-0.975ex}[2.15ex][0.85ex]{%
	\includegraphics[height=3.25ex]{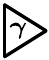}}}

\newcommand{\myblockdiagramBlankKernelBlock}{%
	\raisebox{-0.85ex}[2.15ex][0.85ex]{%
	\includegraphics[height=3.0ex]{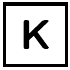}}}

\newcommand{\myblockdiagramDynamicKernelBlock}{%
	\raisebox{-1.0ex}[2.05ex][1.15ex]{%
	\includegraphics[height=3.00ex]{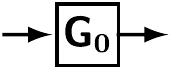}}}

\newcommand{\myblockdiagramFeedbackKernelBlock}{%
	\raisebox{-1.0ex}[2.05ex][1.15ex]{%
	\includegraphics[height=3.00ex]{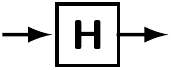}}}

\newcommand{\myblockdiagramGeneralKernelBlock}{%
	\raisebox{-1.0ex}[2.05ex][1.15ex]{%
	\includegraphics[height=3.00ex]{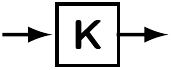}}}

\newcommand{\myblockdiagramHilbertKernelBlock}{%
	\raisebox{-1.0ex}[2.05ex][1.15ex]{%
	\includegraphics[height=3.00ex]{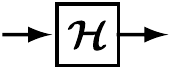}}}

\newcommand{\myblockdiagramFeedbackKernelSqrBlock}{%
	\raisebox{0.5ex}{%
	\raisebox{-0.5\height}[0ex][0ex]{%
	\scalebox{1.00}{\includegraphics{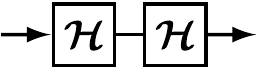}}}}}
	
\newcommand{\myblockdiagramInverterBlock}{%
	\raisebox{0.5ex}{%
	\raisebox{-0.5\height}[0ex][0ex]{%
	\scalebox{1.15}{\includegraphics{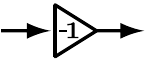}}}}}
	
\newcommand{\myref}[1]{(\ref{#1})}   % key shortcuts
\newcommand{\myeq}[1]{(\ref{#1})}
\newcommand{\myonehalf}{{\textstyle\frac{1}{2}}}

\newcommand{\myupsymbol}{{\scriptscriptstyle(\hspace{-0.15ex}{+}\hspace{-0.15ex})}}
\newcommand{\mydownsymbol}{{\scriptscriptstyle(\hspace{-0.17ex}{-}\hspace{-0.17ex})}}
\newcommand{\mytextupsymbol}{{\footnotesize(\hspace{-0.15ex}{+}\hspace{-0.15ex})}}
\newcommand{\mytextdownsymbol}{{\footnotesize(\hspace{-0.17ex}{-}\hspace{-0.17ex})}}

\newcommand{\mypartialtwo}{%
\sbar\partial{\otimes}\partial%
}

\newcommand{\mypartialone}{\sbar\partial}

\newcommand{\suba}{{\ensuremath{\scriptscriptstyle\text{\sffamily A}}}}
\newcommand{\subb}{{\ensuremath{\scriptscriptstyle\text{\sffamily B}}}}
\newcommand{\subab}{{\ensuremath{\scriptscriptstyle\text{\sffamily AB}}}}

\newcommand{\theObjective}{to enable the reader to design and implement practical quantum simulations, guided by an appreciation of the geometric, informatic, and algebraic principles that govern simulation accuracy, robustness, and efficiency}

\newcommand{\myversion}{}

\begin{document}

\title%
{\vspace*{-0.65in}\LARGE\bfseries\rmfamily%
Practical recipes for the model order reduction,\\[0.5ex]
dynamical simulation, and compressive sampling\\[0.5ex]
of large-scale open quantum systems}

\author{\normalsize%
J.~A.~Sidles\footnote{\scriptsize To whom correspondence should be addressed. %
$\sp{1}$Dept{.}\ of Orthop{\ae}dics and Sports Medicine, Box 356500,
School of Medicine, University of Washington, Seattle, Washington, USA; 
$\sp{2}$Dept{.}\ of Mechanical Engineering, University of Washington; 
$\sp{3}$Dept{.}\ of Physics, U.~S.~Military Academy, West Point; 
$\sp{4}$Dept{.}\ of Electrical Engineering, University of Michigan; 
$\sp{5}$Dept{.}\ of Bioengineering, University of Washington%
}~\,{$\sp{1}$}, %
J.~L.~Garbini$\sp{2}\!$, L.~E.~Harrell$\sp{3}\!$, A.~O.~Hero$\sp{4}\!$, \\[0.5ex]
\normalsize J.~P.~Jacky$\sp{1}\!$, %
J.~R.~Malcomb$\sp{2}\!$, A.~G.~Norman$\sp{5}\!$, A. M. Williamson$\sp{2}\!$}

\date{\rule{0pt}{3ex}\normalsize\today}

\maketitle

\thispagestyle{empty}

{\noindent%

\vspace{-0.25in}

\noindent This article presents practical numerical recipes for simulating high-temperature and non-equilibrium quantum spin systems that are continuously measured and controlled.  The notion of a ``spin system'' is broadly conceived, in order to encompass macroscopic test masses as the limiting case of large-$j$ spins.  The simulation technique has three stages: first the deliberate introduction of noise into the simulation, then the conversion of that noise into an informatically equivalent continuous measurement and control process, and finally, projection of the trajectory onto a \Kahlerian state-space manifold having reduced dimensionality and possessing a \Kahler potential of multilinear (\emph{i.e.}, product-sum) functional form.  These state-spaces can be regarded as ruled algebraic varieties upon which a projective quantum model order reduction (\QMOR) is performed. The Riemannian sectional curvature of ruled \Kahlerian varieties is analyzed, and proved to be non-positive upon all sections that contain a rule.   It is further shown that the class of ruled \Kahlerian state-spaces includes  the Slater determinant wave-functions of quantum chemistry as a special case, and that these Slater determinant manifolds have a Fubini-Study metric that is \Kahler-Einstein; hence they are solitons under Ricci flow. It is suggested that these negative sectional curvature properties geometrically account for the fidelity, efficiency, and robustness of projective trajectory simulation on ruled \Kahlerian state-spaces.   Some implications of trajectory compression for geometric quantum mechanics are discussed.  The resulting simulation formalism is used to construct a positive $P$-representation for the thermal density matrix and  to derive a quantum limit for force noise and measurement noise in monitoring both macroscopic and microscopic test-masses; this quantum noise limit is shown to be consistent with well-established quantum noise limits for linear amplifiers and for monitoring linear dynamical systems.  Single-spin detection by magnetic resonance force microscopy (\MRFM) is then simulated, and the data statistics are shown to be those of a random telegraph signal with additive white noise, to all orders, in excellent agreement with experimental results.  Then a larger-scale spin-dust model is simulated, having no spatial symmetry and no spatial ordering; the high-fidelity projection of numerically computed quantum trajectories onto low-dimensionality \Kahler state-space manifolds is demonstrated.  Finally, the high-fidelity reconstruction of quantum trajectories from sparse random projections is demonstrated, the onset of Donoho-Stodden breakdown at the Cand\`{e}s-Tao sparsity limit is observed, and methods for quantum state optimization by Dantzig selection are given. 
}

\newpage

\tableofcontents

\newpage 

\listoffigures

\listoftables

\newpage

\section{Introduction}
\label{sec: introduction}

This article describes practical recipes for the simulation of large-scale open quantum spin systems.  Our overall objective is \theObjective. 

\subsection{How does the Stern-Gerlach effect \emph{really} work?}
\label{sec: how does the S-G effect work}
This article had its origin in a question that Dan Rugar asked of us about five years ago: ``How does the Stern-Gerlach Effect \emph{really} work?''  The word ``\emph{really}'' is noteworthy because hundreds of articles and books on the Stern-Gerlach effect have been written since the original experiments in 1921 \cite{Gerlach:1921ys,Gerlach:1922fr,Gerlach:1922rt} {\ldots} including articles by the authors \cite{Sidles:92,Sidles:93} and by Dan Rugar \cite{Sidles:95} himself.  Yet we were unable to find, within this large literature, an answer that was satisfactory in the context in which the question was asked, that circumstance being the (ultimately successful) endeavor by Rugar's \IBM research group to detect the magnetic moment of a single electron spin by magnetic resonance force microscopy (\MRFM) \cite{Rugar:04}.

\subsubsection{Constraints upon the analysis}
\label{sec: Constraints upon the Analysis} 
Quantum theory has a reputation for mystery.  But as Peter Shor has remarked,  ``Interpretations of quantum mechanics, unlike Gods, are not jealous, and thus it is safe to believe in more than one at the same time.''  In particular, it is well known---and we will review the literature in this article---that advances in quantum information theory have provided Shor's principle with rigorous foundations.  

We will build upon these informatic foundations in answering Dan Rugar's question in accord with the following constraints: our analysis will be \emph{orthodox} in its respect for established principles of quantum physics.  It will be \emph{operational} in the sense that all its predictions are traceable to explicitly hardware-based measurement processes.  The analysis will be \emph{scalable} to accommodate large-dimension quantum systems (such as the spins in protein molecules that are the ultimate targets of \MRFM microscopy).  The analysis will be \emph{reductive} in the sense that the analysis will yield simple design rules that are in reasonable quantitative accord with the predictions of more accurate---but more complicated---large-scale numerical simulations.  The analysis will be \emph{synoptic} in the sense that when we are required to choose between equivalent analysis formalisms, a rationale for these choices will be provided, and the consequences of alternative choices noted.  And finally, the analysis will be \emph{extensible}---at least in principle---to the analysis and simulation of general quantum systems (such as spintronic devices, nanomechanical devices, and biomolecules).

There are of course strong practical motivations for seeking to analyze quantum systems by methods that are orthodox, operational, scalable, reductive, synoptic, and extensible: these same attributes are essential to practical methods for analyzing large-scale classical systems \cite{Johnson:05,Ramo:84}.  

\subsection{The feasibility of generic large-scale quantum simulation}
We did not begin our investigations with the idea that the numerical simulation of large-scale quantum spin systems was feasible.  Indeed, we were under the opposite impression, based upon the no-simulation arguments of Feynman \cite{Feynman:82} in the early 1980s.  These arguments have been widely---and usually uncritically---repeated in textbooks \cite[sec.~4.7]{Nielsen:00}.  But Feynman's arguments do not formally apply to noisy systems, and in the course of our analysis, it became apparent that this provides a loophole for developing efficient simulation algorithms. Furthermore, it became apparent that the class of noisy systems encompasses as a special case the low-temperature and strongly correlated systems that are studied in quantum chemistry and condensed matter physics.  In the concluding section of this article (Section~\ref{sec: sparse random projections}), we will develop the point of view that any quantum state that has been in contact with a thermal reservoir is an algorithmically compressible object.

This loophole helped us understand why---from an empirical point of view---simu\-lation capabilities in quantum chemistry and condensed matter physics have been improving exponentially in recent decades \cite{MCTCC:95,Friesner:05,Pauling:46}.  The analysis and simulation methods that we will present in this article broadly define a geometric and quantum informatic program for sustaining this progress.  

\subsubsection{The geometry of reduced-order state-spaces} 
This article's mathematical methods are novel mainly in their focus upon the geometry of reduced-order quantum state-spaces.  We will show that the quantum state-spaces that are most useful for large-scale simulation purposes generically have an algebraic structure that can be geometrically interpreted as a network of geodesic curves (rules) having nonpositive Gaussian curvature for all sections that contain a rule (Section~\ref{sec: sectional curvature}).  We~will see that these curvature properties are essential to the efficiency and robustness of model order reduction.  

\subsubsection{The central role of covert measurements} A~technique that is central to our simulation recipes is to simulate all noise processes (including thermal baths) as equivalent covert measurement and control processes (Section~\ref{sec: designing and implementing}).  From a quantum informatic point of view, covert quantum measurement processes act to quench high-order quantum correlations that otherwise would be infeasibly costly to compute and store (Section~\ref{sec: simulations}).  Thus the presence of noise can allow quantum simulations to evade the no-simulation arguments of Feynman \cite{Feynman:82}.

\subsubsection{Background assumed by the presentation}
No reader will be expert in all of the disciplines that our analysis and simulation recipes embody, which are (chiefly) quantum mechanics in both its physical and informatic aspects, the engineering theory of model order reduction (\MOR) and dynamical control, and the mathematical tools and theorems of algebraic and differential geometry.  Indeed, the writing this article has made the authors  acutely aware of their own considerable deficiencies in all of these areas.   Recognizing this, we will describe all aspects of our recipes at a level that is intended to be broadly comprehensible to nonspecialists.  

\subsubsection{Overview of the analysis and simulation recipes}
We begin by surveying our chain of reasoning in its entirety.  Figures~\ref{fig: QMOR formal}--\ref{fig: QMOR geometry} concisely summarize the simulation recipes and their geometric basis.  In a nutshell, the recipes embody the orthodox quantum formalism of Fig{.}~\ref{fig: QMOR formal}, as translated into the practical numerical algorithm of Fig{.}~\ref{fig: QMOR numerical}, which is based upon the algebraic structures of Fig{.}~\ref{fig: QMOR product sum}, whose functionality depends upon the fundamental geometric concepts of Fig.~\ref{fig: QMOR geometry}.  

The following overview summarizes those aspects of the simulation recipes that are fundamental, multidisciplinary, or novel, and it also seeks to describe the embedding of these recipes within the larger literature.

\subsection{Overview of the formal simulation algorithm}
 \label{subsubsec: operations} 
 \label{subsection: Overview} 
\begin{figure}[p]\centering
\includegraphics[scale=1.2]{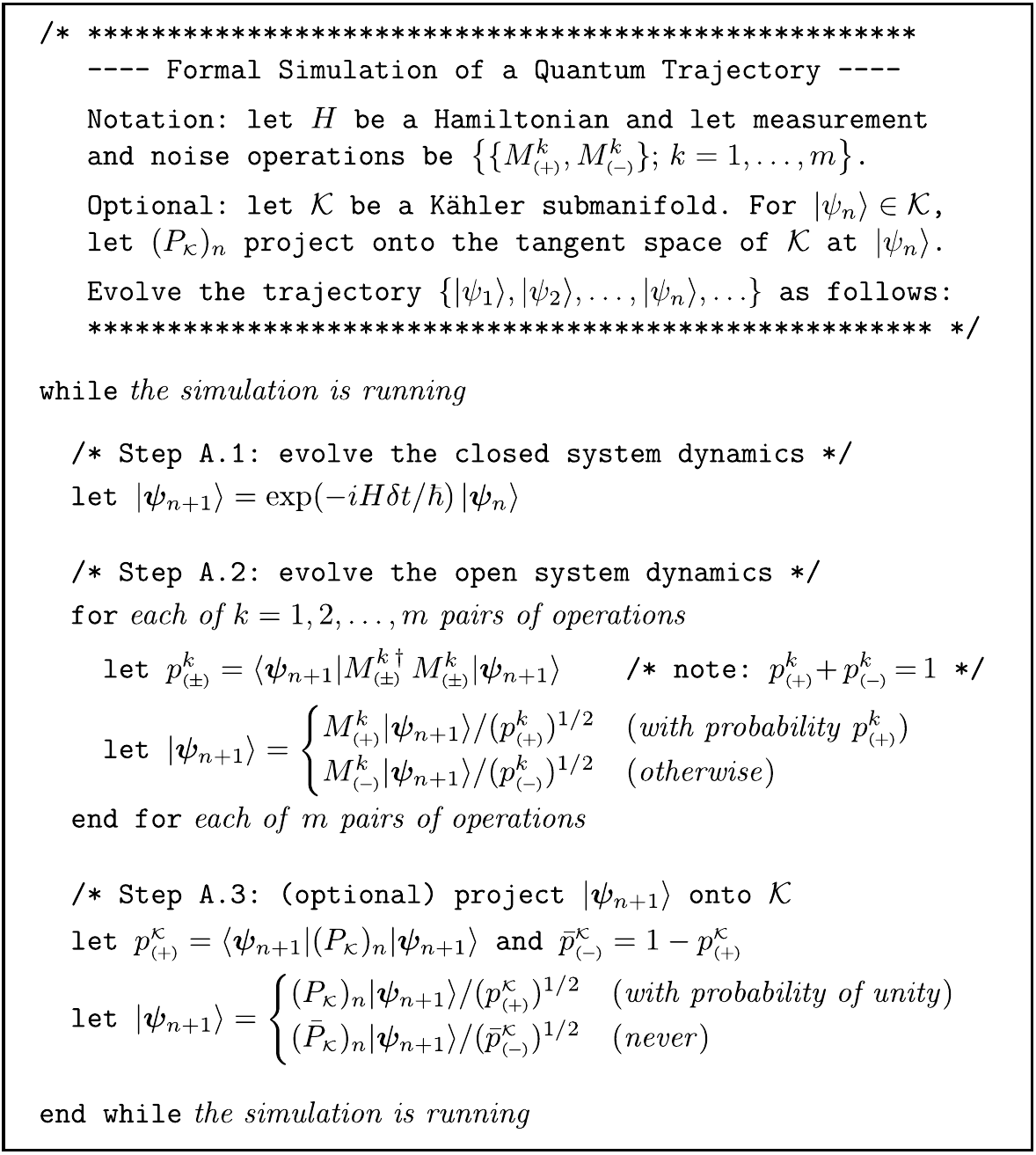}%
\caption[Formal algorithm for quantum model order reduction (\QMOR)]{%
\protect\justifying\label{fig: QMOR formal}%
Formal simulation by quantum model order reduction.\\[1ex]
Steps A.1--2 summarize the formal theory of the simulation of quantum systems (see, \emph{e.g.}, Nielsen and Chuang \cite[chs.~2 and~8]{Nielsen:00}).   Step~A.3 is a model order reduction of the Hilbert states $\ket{\psi_n}$ by projection onto a reduced-dimension \Kahler manifold~$\lcal{K}$ (see \emph{e.g.}  Rewie\'nski~\cite{Rewienski:04}).  Equivalently, Step~A.3 may be viewed as a variety of Dirac-Frenkel variational projection (see, \emph{e.g.}, \cite{Raab:00,Lubich:04}).}
\end{figure}

Formally, our simulation algorithms will be of the general quantum information-theoretic form that is summarized in Fig{.}~\ref{fig: QMOR formal}.  Steps A.1--2 of the algorithm are adopted, without essential change, from the axioms of Nielsen and Chaung \cite{Nielsen:00}, and our discussion will assume a background knowledge of quantum information theory at the levels of Chapters~2 and~8 of their text.  

Step~A.3 of the simulation algorithm---projection of the quantum trajectory onto a state-space of reduced dimensionality---will be familiar to system engineers as projective model order reduction (as we will review in Sec{.}~\ref{subsec: numerical algorithm}).  We will also establish that projective \MOR is formally identical to a method that is familiar to physicists and chemists as a variational order reduction of Dirac-Frenkel-McLachlan type \cite{Dirac:30,Frenkel:34,McLachlan:64} (see also the recent references \cite{Hairer:06,Raab:00,Lubich:04}).  

For purposes of exposition, we define \emph{quantum model order reduction} (\QMOR)  to~be simply classical \MOR extended to the complex state-space of quantum simulations.

\subsubsection{The operational approach to quantum simulation} 
Our simulation recipes will adopt a strictly operational approach to  measurement and control, in the sense that we will require that the \emph{only} information stream used for purposes of communication and control is the stream of binary stochastic outcomes of the measurement operations of Step A.2 of Fig.~\ref{fig: QMOR formal}.  Although it is not mathematically necessary, we will associate these binary outcomes with the classical ``clicks'' of physical measurement apparatuses, and we will develop a calibrated physical model of these clicks that will guide both our physical intuition and our simulation design.  

\subsubsection{The embrace of quantum orthodoxy} 
\label{sec: ultra-orthodox}
Because the binary ``clicks'' of measurement outcomes are \emph{all} that we seek to simulate, our analysis will regard the state trajectories $\{\gsb{\psi}_1,$ $\gsb{\psi}_2, \dots,$ $\gsb{\psi}_n, \dots\}$ as wholly inaccessible for all purposes associated with measurement and control, which is to say, as inaccessible for engineering purposes. We will analyze quantum state trajectories \emph{only} with the goal of tuning the simulation algorithms to compress the trajectories onto low-dimension manifolds.  In practice, this will mean that we mainly care about the geometric properties of quantum trajectories; this will be the organizing theme of our analysis.  

In the course of our analysis we will confirm---mainly to check our algebraic manipu\-lations---that several of the traditional quantum measurement short-cuts that deal directly with wave-functions (\emph{e.g.}, uncertainty principles, wave function collapse, quantum Zeno effects) yield the same results as our ``clicks-only'' reductive formalism.  But our simulations will not use these short-cuts, and in particular, we will never simulate quantum measurement processes in terms of von~Neumann-style projection operators.  

The resulting simulation formalism is wholly operational, and can be informally described as ``ultra-orthodox.''  The operational approach will require some extra mathematical work---mainly in the area of stochastic analysis---but it will also yield some novel mathematical results, including a closed-form positive $P$-rep\-resent\-ation \cite{Perelomov:86} of the thermal density matrix.  We will derive this  $P$-representation by methods that provably simulate finite-temper\-ature baths. Thus the gain in practical simulation power will be worth the effort of the extra mathematical analysis.

\subsubsection{The unitary invariance of quantum operations} 
\label{sec: quantum operations}
Our analysis will focus considerable attention upon the sole mathematical invariance of the simulation algorithm of Fig.~\ref{fig: QMOR formal}, which is a unitary invariance associated in the choice of the \emph{quantum operations} $M$ in Step~A.2.  Our main mathematical discussion of this invariance will be in Section~\ref{sec: QMOR respects the Theorema Dilectum}, our main discussion of its causal aspects will be in Section~\ref{sec: Causality and scattering}, our main review of the literature will be in Section~\ref{sec: dilectum literature}, and it will be central to the discussion of all the simulations that we present in Section~\ref{sec: simulations}.  

We will see that the short answer to the question ``What is this unitary invariance all about?'' is that (1)~it ensures that measured quantities respect physical causality, and (2)~it allows quantum simulations to be tuned for improved efficiency and fidelity.

In preparation, we caution readers that what we will call ``quantum operations'' are known by a great many other names too, including \emph{Kraus operators}, \emph{decomposition operators} and \emph{operation elements}.  These operations are discussed in textbooks by Nielsen and Chuang~\cite{Nielsen:00}, Alicki and Lendi~\cite{Alicki:87}, Carmichael~\cite{Carmichael:93,Carmichael:99}, Percival~\cite{Percival:98},  Breuer and Petruccione~\cite{Breuer:02}, and Peres~\cite{Peres:04}. These texts build upon the earlier work of the mathematicians Stinespring~\cite{Stinespring:55} and Choi~\cite{Choi:75} and the physicists Kraus~\cite{Kraus:71,Kraus:83}, Davies~\cite{Davies:76}, and Lindblad~\cite{Lindblad:76}.   Shorter, reasonably self-contained discussions of open quantum systems can be found in articles by Peres and Terna~\cite{Peres:04}, Adler~\cite{Adler:00}, Rigo and Gisin~\cite{Rigo:96}, and Garcia-Mata \emph{et al.}~\cite{Garcia-Mata:05}, and in on-line notes by Caves~\cite{Caves:00} and by Preskill~\cite{Preskill:06}.   

It is prudent for students to browse among these works to find congenial points of view, because two of the above references are alike in the significance that they ascribe to the unitary invariance of quantum operations.  This diversity arises because the invariance can be understood in multiple ways, including physically, algebraically, informatically, and geometrically.  Our analysis will touch upon all these aspects, but much more than any of the above references, our approach will be geometric.

\subsubsection{Naming and applying the \emph{Theorema Dilectum}}
\label{sec: Naming and applying}
It is vexing that no~short name for the unitary invariance associated with quantum operations has been generally adopted.  For example, this theorem is indexed by Nielsen and Chuang under the unwieldy phrase  ``theorem: unitary freedom in the operator-sum representation'' \cite[thm.~8.2, sec.~8.2]{Nielsen:00}.

Because we require a short descriptive name, we will call this invariance the \emph{Theorema Dilectum}, which means ``the theorem of choosing, picking out, or selecting'' (from the Latin \emph{deligo}).  As our discussions will demonstrate, this name is appropriate in both its literal sense and in its evocation of Gauss' \emph{Theorema Egregium}.  

In this article we will develop a geometric point of view in which the \emph{Theorema Dilectum} is mainly a theorem about trajectories in state-space, and that the central practical role of the theorem in quantum simulations is to enable noisy quantum trajectories to be algorithmically compressed, such that efficient large-scale quantum simulation is feasible.  To the best of our knowledge, no existing articles or textbooks have assigned to the \emph{Theorema Dilectum} the central geometric role that this article focusses upon.

The \emph{Theorema Dilectum} is the first of two main technical terms that we will introduce in this review.  To anticipate, the other is \emph{gabion}, which is the name that we will give to state-space manifolds that support a certain kind of affine algebraic structure (see Figs.~\ref{fig: QMOR product sum}--\ref{fig: QMOR geometry} and Section~\ref{sec:geometric ideas}).  When gabion manifolds are endowed with a \Kahler metric, we will call the result a \emph{gabion-\Kahler manifold} (\GK manifold%
\footnote{In a world in which every possible two-letter acronym is already in use, it is necessary to stipulate that this article's definition of \GK manifolds does not refer to the gyrokinetic (\GK) simulation codes of plasma physics \cite{Ethier:2008nx} nor to the generalized \Kahler (\GK) manifolds of quantum field theory \cite{Apostolov:2007kx}.}%
).  

\GK manifolds are the state-spaces onto which we will projectively compress the quantum trajectories of our simulations by exploiting the  \emph{Theorema Dilectum}.  When we further impose an  antisymmetry condition upon the state-space the result is a \emph{Grassmannian gabion-\Kahler manifold} (\GGK manifold), and we will identify these manifolds as being both the well-known Slater determinants of quantum chemistry, and the equally well-known Grassmannian varieties of algebraic geometry.

\subsubsection{Relation to geometric quantum mechanics}
\label{sec: intro to geometric}
Our recipes will embrace the strictly orthodox point of view that linear quantum mechanics is ``the truth'' to which  our reduced-order \Kahlerian state-spaces are merely a useful low-order approximation.  However, at several points our results will be relevant to a logically conjugate point of view, known as \emph{geometric quantum mechanics}, which is described by Ashtekar and Schilling as follows \cite{Ashtekar:99} (see also  \cite{Schilling:96,Benvegnu:04}):
\begin{QSEquote}
	{[}In geometric quantum mechanics{]} the linear structure which is at the forefront in text-book treatments of quantum mechanics is, primarily, only a technical convenience and the essential ingredients--Ñthe manifold of states, the symplectic structure and the Riemannian metricÑ--do not share this linearity.
\end{QSEquote}Thus in geometric quantum mechanics, \Kahlerian geometry is regarded as a fundamental aspect of nature, while in our quantum \MOR discussion, this same geometry is a matter of deliberate design, whose objective is optimizing simulation capability.  Because our main focus is upon quantum \MOR, we will comment only in passing upon those results that are relevant to geometric quantum mechanics (\emph{e.g.}, see the discussion in Section~\ref{sec: expectation}).

\subsection{Overview of the numerical simulation algorithm} 
\label{sec: central idea}  
\label{subsec: numerical algorithm}  
\begin{figure}[p]\centering
\includegraphics[scale=1.2]{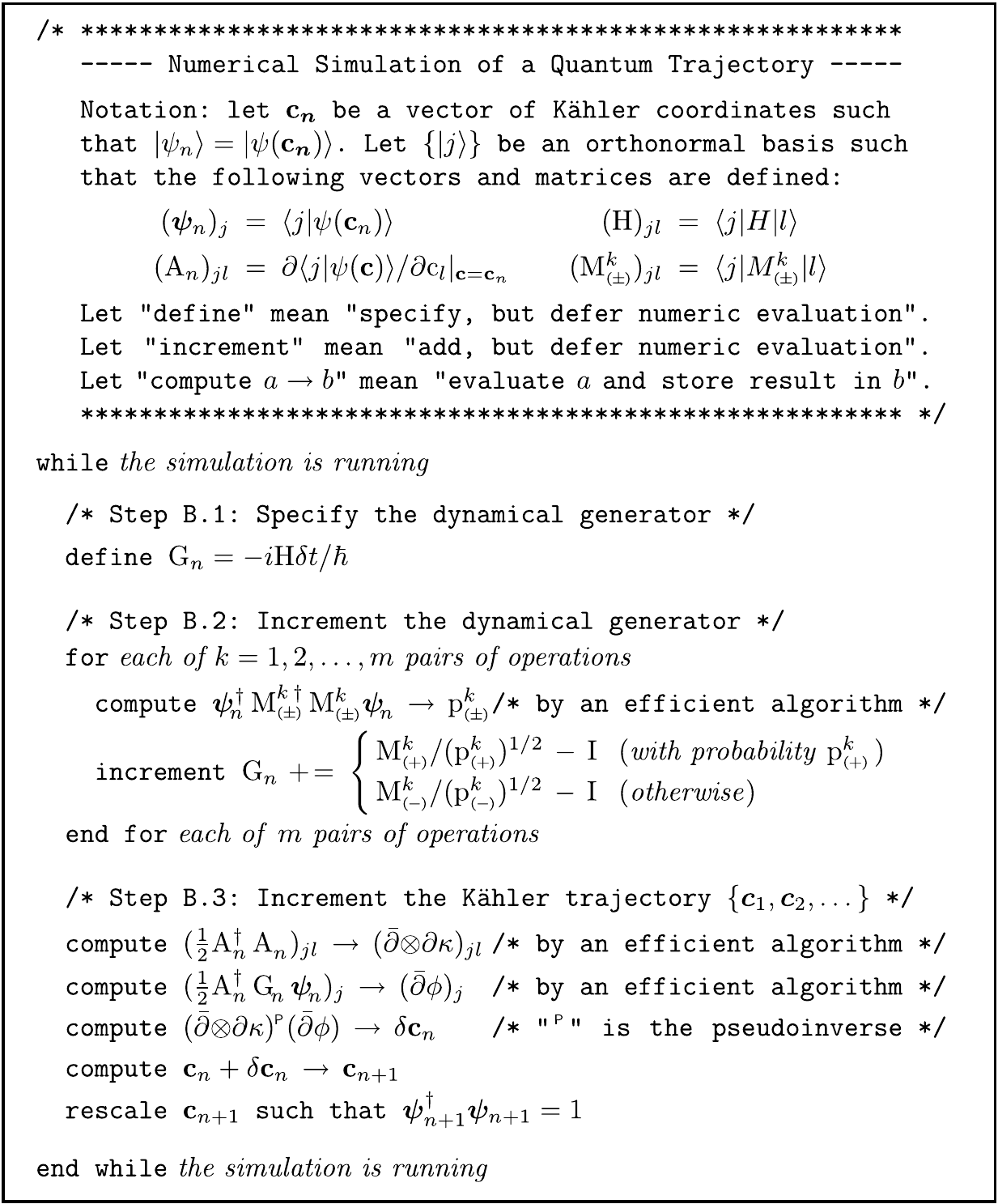}%
\caption[Numerical algorithm for quantum model order reduction (\QMOR)]{%
\protect\justifying\label{fig: QMOR numerical}%
Numerical algorithm for quantum model order reduction simulations.\\[1ex]
Steps B.1--3 are a numerical recipe that implements the simulation algorithm of Fig.~\ref{fig: QMOR formal}. The expressions $(\mypartialtwo\kappa)$ and $(\mypartialone\phi)$ that are introduced in Step~B.3 serve solely as variable names for the stored partial derivatives of the K\"ahler potential 
	\mbox{$\kappa(\gsb{\mathrm{\sbar{c}}},\gsb{\mathrm{c}}) \equiv 
		{\scriptscriptstyle\tfrac{1}{2}}\inner{\psi(\gsb{\mathrm{\sbar{c}}})}{\psi(\gsb{\mathrm{c}})}$}
and the dynamic potential 
  \mbox{$\phi(\gsb{\mathrm{\sbar{c}}},\gsb{\mathrm{c}}) \equiv 
     {\scriptscriptstyle\tfrac{1}{2}}\braket{\psi(\gsb{\mathrm{\sbar{c}}})}{\delta\mathrm{G}}{\psi(\gsb{\mathrm{c}})}$}; it is evident that these partial derivatives wholly determine the simulation's geometry and dynamics. %
}\end{figure}

The numerical simulation algorithm of Fig{.}~\ref{fig: QMOR numerical} is simply the formal algorithm of Fig{.}~\ref{fig: QMOR formal} expressed in a form suitable for efficient computation.  Note that Fig{.}~\ref{fig: QMOR numerical} adopts the \MATLAB-style engineering nomenclature of model order reduction, as contrasted with the physics-style bra-ket notation of Fig{.}~\ref{fig: QMOR formal}.  

The algorithm of Fig{.}~\ref{fig: QMOR numerical} is a fairly typical example of what engineers call \emph{model order reduction} (\MOR) \cite{Morris:77,Antoulas:05,Obinata:00,Noor:82,Noor:94}. Rewie\'nski's thesis is particularly recommended as a review of modern nonlinear \MOR~(\cite{Rewienski:04}, see also \cite{Rewienski:2006fy}).

\subsubsection{The main ideas of projective model order reduction}
We will now briefly summarize the main ideas of projective \MOR in a form that well-adapted to quantum simulation purposes.  We consider a generic \MOR problem defined by the linear equation $\delta\gsb{\psi} = \mathrm{G}\gsb{\psi}$.  Here $\gsb{\psi}$ is a state vector, $\delta\gsb{\psi}$ is a state vector increment, and $G$ is a (square) matrix.  For the present it is not relevant whether $\gsb{\psi}$ is real or complex.  It commonly happens that $\gsb{\psi}$ includes many degrees of freedom that are irrelevant to the practical interests that motivate a simulation.  

The central physical idea of \MOR is to adopt a reduced order representation $\gsb{\psi}(\lb{\mathrm{c}})$, where $\lb{\mathrm{c}}$ is a vector of model coordinates, having $\dim \lb{\mathrm{c}} \ll \dim \gsb{\psi}$.  The central mathematical problem of \MOR is to describe the large-dimension increment $\delta\gsb\psi$ by a reduced-order increment $\delta\gsb{\mathrm{c}}$.  It is convenient to organize the partial derivatives of $\gsb{\psi}(\lb{\mathrm{c}})$ as a non-square matrix $\mathrm{A}(\lb{\mathrm{c}})$ whose elements are $[\mathrm{A}(\lb{\mathrm{c}})]_{ij}\equiv \partial \gsb{\psi}_i/\partial c_j$.  The reduced-order increment having least mean-square error is obtained by the following sequence of matrix manipulations:
\begin{align}
\delta\gsb{\psi} = \mathrm{G}\,\gsb{\psi}
\ \ \rightarrow\ \ 
&\mathrm{A}\,\delta\gsb{\mathrm{c}} = \mathrm{G}\,\gsb{\psi}
\ \ \rightarrow\ \ 
\delta\gsb{\mathrm{c}} = \mathrm{A}^{\scriptpseudoinverse}\mathrm{G}\,\gsb{\psi}
\ \ \rightarrow\ \ 
\delta\gsb{\mathrm{c}} = (\mathrm{A}^\dagger \mathrm{A})^{\scriptpseudoinverse}(\mathrm{A}^\dagger\mathrm{G}\,\gsb{\psi})
\label{eq: Dirac-Frenkel}
.
\end{align}
Here ``$\sp{\scriptpseudoinverse}$'' is the Moore-Penrose pseudo-inverse that is ubiquitous in data-fitting and model order reduction problems \cite{Press:94}, ``$\sp{\dagger}$'' is Hermitian conjugation, and the final step relies upon the pseudo-inverse identity $\mathrm{X}^{\scriptpseudoinverse}=(\mathrm{X}^\dagger \mathrm{X})^{\scriptpseudoinverse}\,\mathrm{X}^\dagger$, which is exact for any matrix $\mathrm{X}$ \cite{Loan:96}.  This is the key step by which the master simulation equation is obtained that appears as Step~B.3 at the bottom of Fig{.}~\ref{fig: QMOR numerical}.  

The great virtue of (\ref{eq: Dirac-Frenkel}) for purposes of large-scale simulation is that $(\mathrm{A}^\dagger \mathrm{A})$ is a low-dimension matrix and $(\mathrm{A}^\dagger\mathrm{G}\,\gsb{\psi})$ is a low-dimension vector.  Provided that both $(\mathrm{A}^\dagger \mathrm{A})$ and $(\mathrm{A}^\dagger\mathrm{G}\,\gsb{\psi})$ can be evaluated efficiently, and provided also that $\gsb{\psi}(\lb{\mathrm{c}})$ represents the ``true'' $\gsb{\psi}$ with acceptable fidelity, substantial economies in simulation resources can be achieved.   We~will see that the required objectives of efficiency and fidelity both can be attained. 

\subsubsection{The natural emergence of \Kahlerian geometry} The simulation equations, when expressed in covariant form (Step~B.3 at the bottom of Fig{.}~\ref{fig: QMOR numerical}), provide a natural venue for asking fundamental geometric questions.  

For example, the low-dimension matrix $\mypartialtwo\kappa\equiv\tfrac{1}{2}\mathrm{A}^\dagger \mathrm{A}$ is obviously Hermitian (whether $\psi$ is real or complex).  Of what manifold is it the Hermitian metric tensor?  How does this manifold's geometry influence the simulation's efficiency, fidelity, and robustness?

To answer these questions, we will show that $\kappa$ is the \emph{\Kahler potential} of differential geometry, that the metric tensor $\mypartialtwo\kappa$ determines the Riemannian curvature of our reduced order state-space, and that the choice of an appropriate curvature for this state-space is vital to the simulation's efficiency, fidelity, and robustness.

In preparing this article, our search of the literature did not find a previous analysis of \MOR state-space geometry from this Riemannian/\Kahlerian point of view.  We did, however, find recent work in
communication theory by Cavalcante~\cite{Cavalcante:02} and coworkers~\cite{Cavalcante:05,Cavalcante:05b} that adopts a similarly geometric point of view in the design of digital signal codes.  Like us, these authors are unaware of previous similarly geometric work~\cite{Cavalcante:05b} ``To the best of our knowledge this [geometric] approach was not considered previously in the context of designing signal sets for digital communication systems.''  Like us, they recognize that ``[These state-spaces] have rich algebraic structures and geometric properties so far not fully explored.''  Also similar to us, they find~\cite{Cavalcante:05} ``The performance of a digital communication system depends on the sectional curvature of the manifold \ldots the best performance is achieved when the sectional curvature is constant and negative.''  

Our analysis will reach similar conclusions regarding the desirable properties of nonpositive sectional curvature in the context of quantum \MOR.  Because the mathematical basis of this apparent convergence of geometric ideas between \MOR theory and coding theory is not presently understood (by~us at least), we will not comment further upon it.  

It is entirely possible that related work exists of which we are not aware.  Model order reduction is ubiquitously practiced by essentially every discipline of mathematics, science, engineering, and business: the resulting literature is so vast, and the nomenclature so varied, that a comprehensive review is infeasible.  

It is fair to say, however, that the central role of Riemannian and \Kahlerian geometry in model order reduction is not \emph{widely} appreciated.  A major goal of our article, therefore, is to analyze the Riemannian/\Kahlerian aspects of \MOR, and especially, to link the \Kahlerian geometry of quantum \MOR to the fundamental quantum informatic invariance of the \emph{Theorema Dilectum}.

\begin{figure}[t]\centering
\vspace*{2ex}
\includegraphics[scale=1.2]{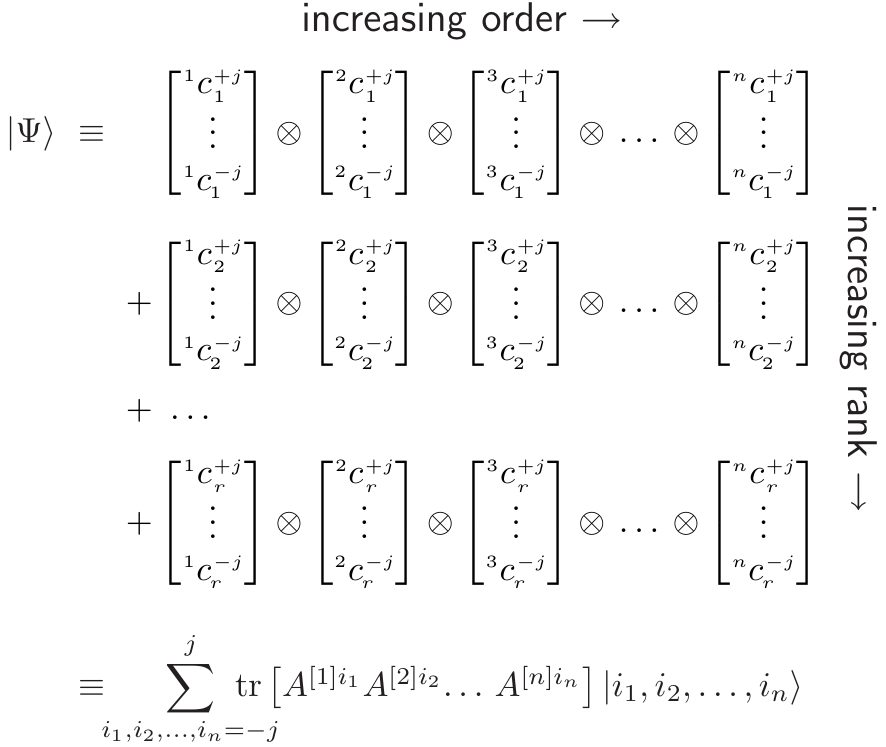}\\
\begin{minipage}{0.8\textwidth}
\caption[Algebraic definition of a gabion-\Kahler (\GK) state-space]{%
\protect\justifying\label{fig: QMOR product sum}%
Algebraic definition of a gabion-\Kahler (\GK) state-space.\\[1ex]
The algebraic definition of a gabion-\Kahler (\GK) state-space (top) expressed equivalently as a matrix product state (\MPS, bottom).  By definition, the \emph{order} of $\ket{\psi}$ is the number of elements (spins) in each row's outer product, the \emph{rank} of $\ket{\psi}$ is the number of rows.  The matrices $A^{[l]m}$ that appear at bottom are, by definition, \mbox{$r\times r$} matrices---hence rank $r$---having diagonal elements  \mbox{$(A^{[l]m})^{\hspace{1ex}}_{kk} \equiv {\sp{l\!}c^{m}_{k}}$} and vanishing off-diagonal elements.  Note that the matrix products are Abelian, such that the geometric properties of the state-space are invariant under permutation of the spins. Note also that when the above algebraic structure is antisymmetrized with respect to interchange of spins (equivalent to interchange of columns), the state becomes a sum of Slater determinants, or equivalently a join of Grasssmanian manifolds (a \GGK manifold).
}\end{minipage}
\end{figure}

\subsubsection{Preparing for a \Kahlerian geometric analysis} To prepare the way for our geometric analysis, at the bottom of Fig.~\ref{fig: QMOR numerical} the pseudo-code defines storage variables named ``$(\mypartialtwo\kappa)$'' and ``$(\mypartialone\phi)$.''  For coding purposes these names are of course purely conventional (an arbitrary string of characters would suffice), but these particular names are deliberately suggestive of partial derivatives of two scalar \mbox{functions:~$\kappa$}~and~$\phi$. 

To anticipate, $\kappa$ will turn out to be the \Kahler potential of complex differential geometry, which determines the differential geometry of the complex state-space, and $\phi$ will turn out to be a stochastic \emph{dynamical potential}, which determines the drift and diffusion of quantum trajectories on the \Kahlerian state-space.  

The link between geometry and simulation efficiency thus arises naturally because both the geometry and the physics of our quantum trajectory simulations are determined by the same two scalar functions.

\subsection{Overview of the unifying geometric ideas}
\label{sec:geometric ideas}
The main algebraic and geometric features of our state-space are summarized in Figs.~\ref{fig: QMOR product sum}--\ref{fig: QMOR geometry}. 

\subsubsection{The algebraic structure of the reduced-order state space} The state-space of all our simulations will have the algebraic structure shown in Fig.~\ref{fig: QMOR product sum}.  We will regard this algebraic structure as a geometric object that is embedded in a larger Hilbert space, and we will seek to understand its geometric properties, including especially its curvature, in relation to our central topic of quantum model order reduction.

In the language of algebraic geometry \cite{Cox:2007,Harris:1992}, the geometric objects we will study are the \emph{algebraic manifolds} that are associated with the \emph{projective algebraic varieties}  defined by the product-sums of Fig.~\ref{fig: QMOR product sum}.  Although the literature on algebraic varieties is vast (and it includes many engineering applications \cite{Cox:2007}) and the literature on Riemannian sectional curvature is similarly vast, the intersection of these two subjects has apparently been little studied from an engineering point of view.  This intersection, and especially its practical implications for quantum model order reduction, will be the main focus of our geometric investigations.

The general algebraic structure of Fig.~\ref{fig: QMOR product sum} is known by various names in various disciplines.  As noted in the caption to Fig.~\ref{fig: QMOR product sum}, these structures are known to physicists as a \emph{matrix product states} (often abbreviated \MPS) which are widely used in condensed matter physics and \emph{ab initio} quantum chemistry \cite{Schollwock:05,Daley:04,Klumper:93,Perezgarcia:2006,Vidal:04,Perez-Garcia:2007yu,Norbert-Schuch:2008qy}; these references provide entry to a rapidly growing body of \MPS-related literature. 

Quantum chemists have known the algebraic structures of Fig.~\ref{fig: QMOR product sum} as \emph{Hartree product states} \cite{Hartree:28a} since 1928.  Upon antisymmetrizing the outer products, we obtain the \emph{Slater determinants} \cite{Slater:29} that are the fundamental building-blocks of modern quantum chemistry; upon summing Slater determinants, and (optionally) imposing linear constraints upon these sums, we obtain \emph{post-Hartree-Fock quantum states}  \cite{Cramer:04}.   All of the theorems we derive will apply to Slater determinants and post-Hartree-Fock states as special cases (see Section~\ref{sec: Slater determinants}). We will comment later in this section, too, upon the intimate relation of these ideas to \emph{density functional theory} (\DFT).  Nuclear physicists embrace these same ideas under the name of \emph{wave function factorization} \cite{Papenbrock:2004ai}.  Beylkin and Mohlenkamp  \cite{Beylkin:05} note that statisticians call essentially the same mathematical objects \emph{canonical decompositions} and also \emph{parallel factors.}  

As Leggett and co-authors have remarked with regard to the similarly immense literature on two-state quantum systems: ``The topic of [this] paper is of course formally a problem in applied mathematics.  \ldots\ Ideas well known in one context have been discovered afresh in another, often in a language sufficiently different that it is not altogether trivial to make the connection. \ldots\ [In such circumstances] the primary purposes of citations are to help the reader understand the paper, and the references in the text are chosen with this in mind''  \cite{Leggett:1987pi}.  These same considerations will guide our discussion.

The general utility of affine algebraic structures for modeling purposes first came to our attention in a highly readable \emph{Mathematical Intelligencer} article by Mohlenkamp and Monz\'{o}n \cite{Mohlenkamp:05}; two further articles by Beylkin and Mohlenkamp \cite{Beylkin:02,Beylkin:05} are particularly recommended also.    Beylkin and Mohlenkamp call the algebraic structure of Fig.~\ref{fig: QMOR product sum} a \emph{separated representation}, and they have this to say about it \cite{Beylkin:02}:
\begin{QSEquote}
When an algorithm in dimension one is extended to dimension $d$, in nearly every case its computational cost is taken to the power $d$. This fundamental difficulty is the single greatest impediment to solving many important problems and has been dubbed the \emph{curse of dimensionality}. For numerical analysis in dimension $d$, we propose to use a representation for vectors and matrices that generalizes separation of variables while allowing controlled accuracy. 
\ldots 
The contribution of this paper is twofold. First, we present a computational paradigm. With hindsight it is very natural, but this perspective was the most difficult part to achieve, and it has far-reaching consequences. Second, we start the development of a theory that demonstrates that separation ranks are low for many problems of interest.
\end{QSEquote}In a subsequent article Beylkin and Mohlenkamp go on to say \cite{Beylkin:05} ``The representation seems rather simple and familiar, but it actually has a surprisingly rich structure and is not well understood.''  These remarks are remarkably similar in spirit to the coding theory observations of Cavalcante \emph{et al.} that were reviewed in Section~\ref{subsec: numerical algorithm}.  

For us, \Kahlerian algebraic geometry will provide a shared foundation for understanding the accelerating progress that all of the above large-scale computational disciplines have witnessed in recent decades.

\subsubsection{The medieval idea of a gabion, and its mathematical parallels} 
Deciding what to call the geometric state-space of quantum model order reduction is a vexing problem.  We have seen that various plausible names include  ``Hartree products,'' ``Slater deter\-min\-ants,'' ``separated representations,'' ``matrix product states,'' ``wave function factorizations,'' ``product-sum states,'' ``canonical decompositions,'' and ``parallel factors.''  

A shared disadvantage of the above names is that there is no precedent for associating them with the geometric properties that are the main focus of our investigations.  We therefore seek an encompassing name for these state-spaces viewed as \emph{geometric} entities.  

Finding no precedent in the literature, and desiring a short name having a long-established etymology, we will call them by the medieval name of \emph{gabions} \cite{LeDuc:1856}, or more formally, \emph{gabion manifolds} (this name arose spontaneously in the course of a seminar).

\begin{figure}[p]\centering
\includegraphics[width=0.90\textwidth]{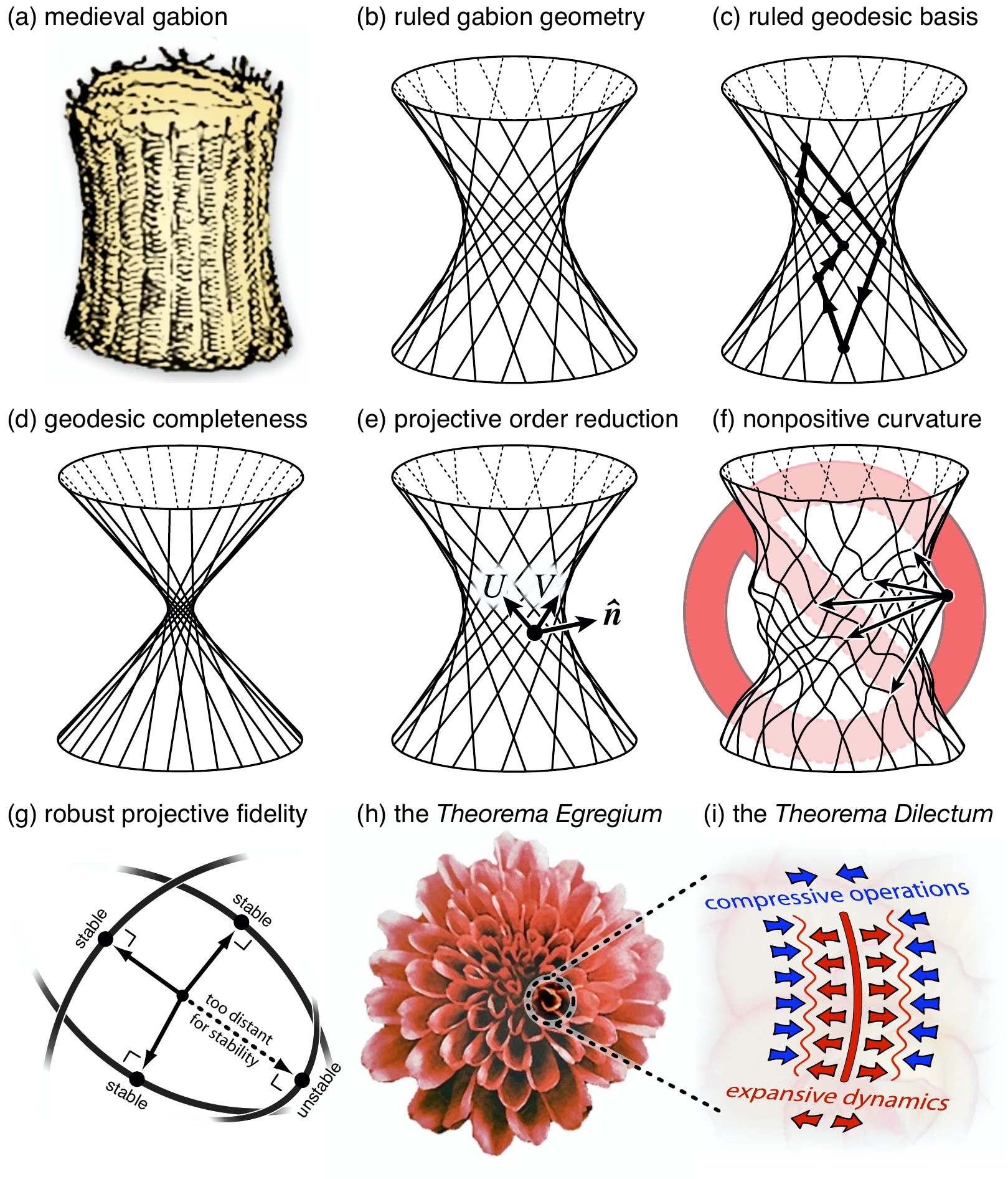}
\caption[Geometric principles of quantum model order reduction (\QMOR)]{%
\protect\justifying%
Geometric principles of quantum model order reduction (\QMOR).\\[1ex]
\label{fig: QMOR geometry}%
See Section~\ref{sec:geometric ideas} for a discussion of these principles.
}
\end{figure}

Most readers will have seen gabions numerous times, perhaps without recognizing that they have a well-established name.  ``Gabion'' is the generic engineering name for a mesh basket that is filled with a weighty but irregularly-shaped material such as rocks or lumber, then stacked for purposes of reinforcement, erosion control, and fortification.  In medieval times gabions were made of wicker or reed; Fig.~\ref{fig: QMOR geometry}(A)~shows a typical medieval gabion.    We will see that the defining geometry property of gabion manifolds is that they are possessed of a web of geodesic lines that constrain the curvature of the manifold, rather as the wicker reeds of a physical gabion constrain the curved rocks and boulders held inside.  Like physical gabions, gabion manifolds come in a wide variety of sizes and shapes that are suitable for numerous practical purposes.

We will postpone giving a formal---and necessarily rather abstract---mathemat\-ical definition of a gabion until Section~\ref{sec: formal definition}.  For the present our main objective is to informally describe the geometric properties that will motivate this formal definition.

\subsubsection{The geometric properties of gabion-\Kahler (\GK) manifolds} 
\label{sec: complex structure}
The main geometric properties of \GK manifolds that are relevant to quantum simulation are depicted in Fig.~\ref{fig: QMOR geometry}(B-H).   We will now survey these properties, and in doing so, we will introduce some of the nomenclature of \Kahlerian geometry.  
We begin our geometric overview by remarking that even though Hilbert space is a complex state-space, a common viewpoint among mathematicians is that a complex manifold is a real manifold that is endowed with an extra symmetry, called its \emph{complex structure} (see Sections~\ref{sec: real manifolds} and \ref{sec: Kahlerian gabions} for details).  For purposes of our geometric analysis, we will simply ignore this complex structure until we are ready to apply the quantum \emph{Theorema Dilectum} (Section~\ref{sec: Kahlerian gabions}).  Until then  we will treat gabion manifolds as real manifolds.

In particular, the state-space of quantum mechanics has a natural real-valued measure of length. Specifically, along a time-dependent quantum trajectory $\ket{\psi(t)}$ it is natural to define a real-valued velocity $v(t)$ whose formal expression can be written equivalently in several notations:
\begin{equation}
\label{eq: notation}
v(t)^2
=
g\big(\sdot\psi(t),\sdot\psi(t)\big)
=
\inner{\sdot\psi(t)}{\sdot\psi(t)}
=
\sdot{\gsbbar{\psi}}(t)\cdot\gsbdot{\psi}(t)
=
\sum_{i=1}^{{\scriptscriptdimH}/2} 
  \sdot{\sbar{\psi}}_i(t)\sdot\psi_i(t)\,.
\end{equation}
Here $\dot\psi(t) \equiv \partial \psi(t)/\partial t$, and we have used first the abstract notation of differential geometry (in which $g(\ldots)$ is a metric function), then the Dirac bra-ket notation of physics, then the matrix-vector notation of engineering and numerical computation, and finally the cumbersome but universal notation of components and sums over indices. We will assume an entry-level familiarity with all four notations, since this is a prerequisite for reading the literature. 

As a token of considerations to come, the factor of $\dim\lcal{H}/2$ in the index limit of (\ref{eq: notation}) above arises because we will regard a complex manifold like $\lfrak{C}^n$ as being a real manifold of dimension $2 n$.  Thus we will regard the complex plane $\lfrak{C}$ as a two-dimensional (real) manifold, and an spin-1/2 quantum state as a point in a Hilbert space $\lcal{H}$ having $\dim \lcal{H} = 4$ (real) dimensions.  This viewpoint leads to an ensemble of conventions that we will review in detail in Section~\ref{sec: Kahlerian gabions}. For now, we note that the arc length $s$ along a trajectory is \(
s = \int\!v(t)\,dt
\), so that geometric lengths in quantum state-spaces are dimensionless.  An equivalent differential definition is to assign a length increment ${d}s$ to a state increment $\ket{d\psi}$ via $({d}s)^2 = \inner{d\psi}{d\psi}$. 

Since we can now compute the real-valued length of an arbitrary curve on the gabion manifold, all of the usual techniques of differential geometry can be applied, without special regard for the fact that the state-space is complex.

\subsubsection{\GK manifolds are endowed with rule fields} Beginning a pictoral summary of our geometric results, we first note that state-space gabions resemble physical gabions in that they are naturally endowed with a geometric mesh, which is comprised of a network of lines called \emph{rules}, as depicted in Fig.~\ref{fig: QMOR geometry}(B).  More formally, they are equipped with vector fields having certain mathematical properties (see (\ref{eq: geodesics}) and Definition~\ref{def: gabion}) such that the integral curves of the rule fields have the depicted properties.    

Postponing a more rigorous and general definition of gabion manifolds until later (see Section~\ref{sec: formal definition}), we can informally define a \emph{gabion rule} to be the quantum trajectory associated with the variation of a single coordinate $\sp{n}c^k_r$ in the algebraic structure of Fig.~\ref{fig: QMOR product sum}, holding all the other coordinates fixed to some arbitrary set of initial values.  We see that gabion  rules are \emph{rays} (straight lines) in the embedding Hilbert space, and hence, the gabion rules are \emph{geodesics} (shortest paths) on the gabion manifold itself.   
As depicted in Fig.~\ref{fig: QMOR geometry}(C), the set of all gabion points that belong to a rule is (trivially) the set of all gabion points itself, which is the defining characteristic of a gabion being \emph{ruled}.  Furthermore, we will show that at any given point, the vectors tangent to the rules that pass through that point are a basis set.   

\subsubsection{\GK geometry has singularities}
\label{sec: singularities} 
Are gabion manifolds geometrically smooth, or do they have singularities?  As depicted in Fig.~\ref{fig: QMOR geometry}(D), we will show that gabion manifolds have pinch-like geometric singularities.  Algebraically these singularities appear whenever two or more rows of the product-sum in Fig.~\ref{fig: QMOR product sum} are equal.  Geometrically, we will show that the Riemann  curvature diverges in the neighborhood of these singular points.  However,  it will turn out that the continuity of the geodesic rules is respected even at the singular points, so that gabion manifolds are \emph{geodesically complete}. Pragmatically, this means that our numerical simulations will not become ``stuck'' at geometric singularities.  

\subsubsection{\GK projection yields compressed representations} As depicted in Fig.~\ref{fig: QMOR geometry}(E) model order reduction is achieved by the high-fidelity projection of an ``exact'' state $\ket\psi$ in the large-dimension Hilbert space onto a nearby point $\ket{\psi_{\lcal{K}}}$ of the small-dimension gabion.  It can be helpful to view this  projection as a data-compression process.  By analogy, the state $\ket\psi$ is like an image in \TIFF format; this format can store an arbitrary image with perfect fidelity, but consumes an inconveniently large amount of storage space.  The projected state $\ket{\psi_{\lcal{K}}}$ on the gabion is like an image in \JPEG format; lesser fidelity, but good enough for many practical purposes, and small enough for convenient storage and manipulation.  We thus appreciate that data compression can be regarded as a kind of model order reduction; the two processes are fundamentally the same. 

\subsubsection{\GK manifolds have negative sectional curvature} Practical quantum simulations require that the computation of order-reducing projections be efficient and robust, just as we require image compression programs to be efficient and robust.   As depicted in Fig.~\ref{fig: QMOR geometry}(F), order-reduction projection becomes ill-conditioned when the state-space manifold is ``bumpy'', in which case a numerical search for a high-fidelity projection can become stuck at local minima that yield poor fidelity.  We will prove that the presence of a ruled net guarantees that gabion manifolds are always smooth rather than bumpy. 

Resorting to slightly technical language to say exactly what we mean when we assert that gabion manifolds are not bumpy, in our Theorem~\ref{thm: one} we will prove that a gabion has nonpositive sectional curvature for all sections on its geodesic net.  This means that gabion manifolds can be envisioned as a net of surfaces that have the special property of being saddle-shaped everywhere (as contrasted with generic surfaces having dome-shaped ``bumps'').  As depicted in Fig.~\ref{fig: QMOR geometry}(G), the saddle-shaped curvature helps ensure that order-reducing projection onto gabion manifolds is a numerically well-conditioned operation.

\subsubsection{\GK manifolds have an efflorescing global geometry} 
\label{sec: efflorescing geometry}
As depicted in Fig.~\ref{fig: QMOR geometry}(H), when the number of state-space dimensions becomes very large, it becomes helpful to envision nonpositively curved manifolds as flower-shaped objects composed of a large number of locally Euclidean ``petals.''  This physical picture has been vividly conveyed by recent collaborative work between mathematicians and fabric artists~\cite{Belcastro:07}; the work of Taimina  and Henderson on hyperbolic manifolds is particularly recommended \cite{Henderson:01}.  

When working in large-dimension spaces we will heed also Dantzig's remark that 
``%
one's intuition in higher dimensional space is not worth a damn!''~\cite{Albers:86}.  For purposes of quantitative analysis we will rely upon Gauss' \emph{Theorema Egregium}~\cite{Gauss:1827} to analyze the Riemannian and \Kahlerian geometric properties of gabion manifolds.  We~will prove that ``number of petals'' becomes exponentially large, relative to $\dimK$, such that the petals loosely fill the embedding Hilbert space.  In this respect our geometric analysis will parallel the informatic analysis of Nielsen and Chuang~\cite{Nielsen:00}; their~Fig.~4.18 is broadly equivalent to our Fig.~\ref{fig: QMOR geometry}(H).   

Our analysis will therefore establish two geometric properties of gabion manifolds: they are strongly curved, and they are richly endowed with straight-line rules.  We will show that these gabion properties are essential to the efficiency, robustness, and fidelity of large-scale \MOR.  Later on, in Sections~\ref{sec: error-correcting codes}--\ref{sec: design of sampling matrices}, we will establish a relation between these properties and compressive sampling (\CS) theory.

\subsubsection{\GK basis vectors are over-complete}  We will nowhere assume that the basis vectors of the underlying algebraic structure of Fig.~\ref{fig: QMOR product sum} are orthonormal; they might refer for example to the non-orthonormal gaussian basis states of quantum chemistry.  A major geometric theme of our analysis, therefore, is that the negative sectional curvature of gabion manifolds helps generically account for the observed efficiency, fidelity, and robustness of gabion-based modeling techniques in many branches of science, engineering, and mathematics.

\subsubsection{\GK manifolds allow efficient algebraic computations} 
Upon restricting our attention to the special case of gabion-\Kahler (\GK) manifolds, we will show that the existence of a ruled geodesic net allows the sectional curvature and the Riemann curvature tensors of \GK manifolds to be calculated easily and efficiently.  To anticipate, we will present data from Riemann curvature tensors having dimension up to $188$, which we believe are the largest-dimension curvature tensors yet numerically computed.  We will see that it is Kraus' ``long list of miracles'' that makes large-scale numerical curvature computations feasible, and that these same miracles are equally essential to large-scale quantum dynamical calculations.

\subsubsection{\GK manifolds support the \emph{Theorema Dilectum}}
\label{sec: TD allows compression}
One geometric idea remains that is key to our simulation recipes. For gabion manifolds to represent quantum trajectories with good fidelity, some physical mechanism must be invoked to compress quantum trajectories onto the petals of the gabion state-space.  That key mechanism is, of course, the \emph{Theorema Dilectum} that was mentioned in Section~\ref{sec: Naming and applying}.  

From a geometric perspective, the \emph{Theorema Dilectum} guarantees that noise can always be modeled as a measurement process that acts to compress trajectories onto the \GK petals.   As depicted in Fig.~\ref{fig: QMOR geometry}(I), quantum simulation can be envisioned geometrically as a process in which compression toward the \GK petals, induced by measurement processes, competes with expansion away from the petals, induced by quantum dynamical processes.   The balance of these two competing mechanisms determines the \MOR dimensionality that is required for good fidelity---the ``petalthickness.''   Algebraically this petal thickness increases in proportion to the rank of the product-sum algebraic structure of Fig.~\ref{fig: QMOR product sum}.

Thus for us, trajectory compression is not a mathematical ``trick,'' but rather is a reasonably well-understood and well-validated quantum physical mechanism, originating in the \emph{Theorema Dilectum}, that compresses quantum trajectories to within an exponentially small fraction of the Hilbert phase space.

This noise-induced trajectory compression is the loophole by which \QMOR simulations evade the no-simulation arguments of Feynman  \cite{Feynman:82}, as reviewed by Nielsen and Chuang \cite[see their Section~4.7]{Nielsen:00}. 

\subsubsection{\GK manifolds support thermal equilibria}
\label{sec: thermal reservoirs}
We will see that this covert-measurement approach encompasses numerical searches for ground states.  Specifically, by explicit construction, we will show that contact with a zero-temperature thermal reservoir can be modeled as an equivalent process of covert measurement and control, in which the role of ``temperature'' is played by the control gain, such that zero temperature is associated with optimal control.

From this \QMOR point of view, the calculation of a ground-state quantum wave function is a special kind of noisy quantum simulation, in which noise is present but masked by optimal control.  This is how \QMOR reconciles the strong arguments for the general infeasibility of \emph{ab initio} condensed-matter calculations (as reviewed by, \emph{e.g.}, Kohn~\cite{Kohn:99}) with the widespread experience that numerically computing the ground states of condensed matter systems is often, in practice, reasonably tractable \cite{Friesner:05}.  

\subsubsection{\GK manifolds support fermionic states} Readers familiar with \emph{ab initio} quantum chemistry, and in particular with density functional theory (\DFT)~\cite[see Cappele \cite{Capelle:06} for an introduction]{Kohn:99,Jones:89,Burke:05} will by now recognize that \QMOR and \DFT are conceptually parallel in numerous fundamental respects: the central role of the low-dimension \Kahler manifold of \QMOR parallels the central role of the low-dimension density functional of \DFT; the closed-loop measurement and control processes of \QMOR parallel the iterative calculation of the \DFT ground state; \QMOR's fundamental limitation of being formally applicable only to noisy quantum systems parallels \DFT's fundamental limitation of being formally applicable only to ground states; \QMOR and \DFT share a favorable computational scaling with system size.  

Yet to the best of our knowledge---and surprisingly---the geometric techniques that that this article will deploy in service of \QMOR have not yet been applied to \DFT and related techniques of quantum chemistry and condensed matter physics \cite{Friesner:05}.  A~plausible starting point is to impose an antisymmetrizing Slater determinant-type structure upon the algebraic outer products of~(\ref{fig: QMOR product sum}).  

Some analytic results that we have obtained regarding the \Kahlerian geometry of Slater determinants are summarized in Section~\ref{sec: Slater determinants}.   With further work along these lines, we believe that there are reasonable prospects of establishing a geometric/informatic interpretation, via the \emph{Theorema Egregium} and the \emph{Theorema Dilectum}, of the celebrated Hohenberg-Kohn and Kohn-Sham Theorems of \DFT \cite{Jones:89} and their time-dependent generalization the Runge-Gross Theorem \cite{Burke:05}. 

A~physical motivation for this line of research is that the \emph{Theorema Dilectum} of \QMOR and the Hohenberg-Kohn Theorem of \DFT embody essentially the same physical insight: the details of exponentially complicated details of quantum wave functions are only marginally relevant to the practical simulation of both noisy systems (\QMOR) and systems near their ground-state (\DFT).  

At present, the two formalisms differ mainly in their domain of application: \QMOR is well-suited to simulating spatially localized systems at high temperature (\emph{e.g.},~spin systems) while \DFT is particularly well-suited to simulating spatially delocalized systems (\emph{e.g.},~molecules and conduction bands) at low temperature.  In the future, as \QMOR is extended to delocalized systems while the methods of quantum chemistry are increasingly extended to dynamical systems \cite{Burke:05,Friesner:05}, opportunities will in our view arise for cross-fertilization of these two fields, both in terms of fundamental mathematics and in terms of practical applications.

\subsection{Overview of contrasts between quantum and classical simulation} In aggregate, the formal, numerical, algebraic, and geometric concepts summarized in the preceding sections and in Figs.~\ref{fig: QMOR formal}--\ref{fig: QMOR geometry} are in many respects strikingly parallel to similar concepts in the computational fluid dynamics (\CFD), solid mechanics, combustion theory, and many other engineering disciplines that entail large-scale simulation using \MOR.   

However, it is evident that quantum \MOR is distinguished from real-valued (classical) \MOR by at least four major differences, which we will now summarize.

\subsubsection{The \emph{Theorema Dilectum} is fundamental and universal} 
\label{sec: fundamental and universal}
The first difference is that the \emph{Theorema Dilectum} describes an invariance of quantum dynamics that is fundamental and universal.  Its physical meaning, as we will see, is that it enforces causality.  Nonlinear classical system do not possess any similarly universal invariance, which is in our view a major contributing reason that ``developing effective and efficient \MOR strategies for nonlinear systems remains a challenging and relatively open problem'' \cite[p.~20]{Rewienski:04}.  

Our results, both analytical and numerical, will suggest that noisy quantum systems are fundamentally no harder to simulate than nonlinear classical systems, provided that the \emph{Theorema Dilectum} is exploited to allow high-fidelity dynamical projection of quantum trajectories onto a reduced-order state-space.

\subsubsection{Quantum state-spaces are veiled} The second difference is a consequence of the first.  As discussed in Sec.~\ref{subsubsec: operations}, to fully exploit the power of the \emph{Theorema Dilectum} we are required to embrace the ultra-orthodox principle of  \emph{never looking at the quantum state space}.   Furthermore, when we examine classical state-spaces more closely, we find that they too are encumbered with ontological ambiguities that precisely mirror the ``spooky mysteries'' of quantum state-spaces.  As discussed in Sections~\ref{sec: Hilbert ontology}, this modern recognition of spooky mysteries in classical physics echoes work in the 1940s by Wheeler and Feynman  \cite{Wheeler:1945pl,Wheeler:1949is}.

\subsubsection{Noise makes quantum simulation easier} 
\label{sec: easier}
The third difference is that higher noise levels are \emph{beneficial} to \QMOR simulations, because they ensure stronger compression onto the \GK petals, which allows lower-rank, faster-running \GK state-spaces to be adopted.  Later we will discuss the interesting question of whether this principle, together with the concomitant principle ``never look directly at the quantum state-space,'' have classical analogs. We will tentatively conclude that the \emph{Theorema Dilectum} \emph{does} have classical analogs, but that the power of this theorem is much greater in quantum simulations than in classical ones.

\subsubsection{\Kahlerian manifolds are geometrically special}   
Broadly speaking, \Kahlerian geometry is to Riemannian geometry what analytic functions are to ordinary functions.  This additional structure is one of the reasons why the mathematician Shing-Tung Yau has expressed the view \cite[p.~46]{Yau:06} "The~most interesting geometric structure is the \Kahler structure.''  From this point of view, the geometry of real-valued \MOR state-spaces is mathematically interesting, and the analytic extension of this geometry to \Kahlerian \MOR state-spaces is even \emph{more} interesting. 

Let us state explicitly some of the analogies between analytic functions and \Kahler manifolds.  We recall that generically speaking, analytic functions have cuts and poles.  These cuts and poles are of course exceedingly useful to scientists and engineers, since they can be intimately linked to physical properties of modeled systems.  Similarly, the \GK manifolds that concern us have singularities, as depicted in Fig.~\ref{fig: QMOR geometry}(D).  Physically speaking, they are associated with regions of quantum state-space that locally are more nearly ``classical'' than the surrounding regions, in the sense that the local tangent vectors that generate high-order quantum correlations become degenerate. It is fair to say, however, that the deeper geometric significance of \Kahlerian \MOR singularities remains to be elucidated.

Just as contour integrals of analytic functions can be geometrically adjusted to make practical reckoning easier, we will see the \emph{Theorema Egregium} allows the trajectories arising from the drift and diffusion of noise and measurement models to be geometrically (and informatically) adjusted to match state-space geometry, and thereby improve simulation fidelity, efficiency, and robustness. 

More broadly, Yau notes \cite[p.~21]{Yau:06}: ``While we see great accomplishments for \Kahler manifolds with positive curvature, very little is known for \Kahler manifolds [having] strongly negative curvature.''   It is precisely these negatively-curved \Kahler manifolds that will concern us in this article, and we believe that their negative curvature is intimately linked to the presence of the singularities mentioned in the preceding paragraph.  We hope that further mathematical research will help us understand these connections better.

\section{The sectional curvature of gabion--\Kahler (\GK) state-spaces}
\label{sec: sectional curvature}

We will now proceed with a detailed derivation and analysis of our quantum simulation recipes.  Our analysis will ``unwind'' the preceding overview: first we analyze the geometry of Fig.~\ref{fig: QMOR geometry}, as embodying the algebraic structure of Fig.~\ref{fig: QMOR product sum}, using the numerical techniques of Fig.~\ref{fig: QMOR numerical}.  Only at the very end will we calibrate our recipes in physical terms, via the quantum physics of Fig.~\ref{fig: QMOR formal}.

\subsection{Quantum \MOR state-spaces viewed as manifolds}

To construct our initial example of a gabion state-space, we will consider the following algebraic function $\gsb{\psi}(\gsb{\mathrm{c}})$, whose domain is a four-dimensional manifold of complex coordinates $\lb{\mathrm{c}} = \{c^1,$ $c^2,$ $c^3,$ $c^4\}$ and whose range in a four-dimensional Hilbert space is the set of points that can be algebraically represented as $\{c^1c^3,$ $c^1c^4,$ $c^2c^3,$ $c^2c^4\}$.  In the notation of Fig.~\ref{fig: QMOR numerical} this function is
\begin{equation}
\label{eq: simple gabion}
	\gsb{\psi}(\gsb{\mathrm{c}}) 
	= 
	\begin{bmatrix}c^1\\c^2\end{bmatrix}\otimes\begin{bmatrix}c^3\\c^4\end{bmatrix} 
\qquad\Leftrightarrow\qquad  %
	\left\{%
		\begin{array}{@{\,}c@{\,}}\psi_1 - c^1 c^3 = 0\\ \psi_2 - c^1 c^4 = 0\\\psi_3 - c^2 c^3 = 0\\ \psi_4 - c^2 c^4 = 0\end{array}%
		,
	\right.
\end{equation}where ``$\otimes$'' is the outer product. The superscripts on the $c^i$ variables are indices rather than powers, as will be true throughout this section.  From an algebraic geometry point of view, (\ref{eq: simple gabion}) defines a \emph{projective algebraic variety}  \cite{Cox:2007} (also called a \emph{homogeneous algebraic variety}) over variables $\{\psi_i{:}\  i\in1,4\}$ that is specified above in \emph{parametric form} in terms of parameters $\{c^i{:}\  i\in1,4\}$.

By definition, our example of a gabion state-space manifold is the solution set of this algebraic variety, and thus our state space is an \emph{algebraic manifold}.  Physically speaking, $\gsb{\psi}(\gsb{\mathrm{c}})$ is the most general (unnormalized) quantum state of two spin~1/2 particles sharing no quantum entanglement.  

We~will now show that this state-space is a \Kahlerian manifold that has negative sectional curvature (under circumstances that we will describe) and that this property is beneficial for simulation purposes (for reasons that we will describe).

\subsubsubsection{Practical computational considerations}
\label{sec: practical considerations} 
We begin by remarking that the basic algebraic construct ``$\text{\texttt{arg1}} \otimes \text{\texttt{arg2}}$'' that appears in (\ref{eq: simple gabion}) can be readily implemented by the built-in functions of most scientific programming languages and libraries; for example in \MATLAB by the construct ``\texttt{reshape(\hspace{0pt}(arg1*arg2')',\hspace{0pt}[],1)}'' and in Mathematica by the construct ``\texttt{Outer[\hspace{0pt}Times,\hspace{0pt}arg1,\hspace{0pt}arg2]\hspace{0pt}//\hspace{0pt}Flatten}''.  

Similar idioms exist for the efficient evaluation of more complex product-sum structures.  Although we will not describe our computational codes in detail, they are implemented in \MATLAB and Mathematica in accord with the general ideas and principles for efficient addition, inner products, and matrix-vector multiplication that are described by Beylkin and Mohlenkamp \cite{Beylkin:05}.

\subsubsubsection{The abstract geometric point of view} 
\label{sec: abstract geometry}
From an abstract point of view, the algebraic structure (\ref{eq: simple gabion}) can be regarded as a sequence of maps
\begin{equation}
\label{eq: sequence of maps}
	\lcal{C}
\ \overset{\text{surjective}}{\to}\ 
	\lcal{K}
\ \overset{\text{injective}}{\to}\ 
	\lcal{H}
	,
\end{equation}
where $\lcal{C}$ is the manifold of complex variables $\{c^1\!,c^2\!,c^3\!,c^4\}$, the gabion manifold $\lcal{K}$ is the range of $\gsb{\psi}$ in $\lcal{H}$, and $\lcal{H}$ is the larger Hilbert space within which $\lcal{K}$ is embedded.  
To appreciate the surjective and injective nature of these maps, we notice in (\ref{eq: simple gabion}) that $\gsb{\psi}(\gsb{\mathrm{c}})$ is invariant under $\{c^1, c^2, c^3, c^4\}\to\{1, c^2/c^1, c^1c^3, c^1c^4\}$.  More generally, it is clear that one coordinate can be set to any fixed nonzero value, without altering $\gsb{\psi}(\gsb{\mathrm{c}})$, by an appropriate rescaling of the other three variables.  

In our example (\ref{eq: simple gabion}), the dimensions of the three manifolds $\lcal{C}$,  $\lcal{K}$, and $\lcal{H}$ are therefore
\begin{equation}
\label{eq: dimensions}
\dimC=2\times4=8,\qquad\dimK=2\times3=6,\qquad\dimH=2\times4=8 ,
\end{equation}where the factors of two arise because these are complex manifolds.  We see that the map $\lcal{C}\to\lcal{K}$ is surjective (because $\dimK<\dimC$), while $\lcal{K}\to\lcal{H}$ is injective (because $\lcal{K}$ is immersed in $\lcal{H}$).  

\subsubsection{Defining gabion pseudo-coordinates} We will call the variables $\{c^1, c^2 ,c^3, c^4\}$ \emph{pseudo-coor\-din\-ates}.  They are not ordinary coordinates because $\lcal{C}\to\lcal{K}$ is surjective rather than bijective, or to say it another way, open sets on $\lcal{C}$ are not charts on~$\lcal{K}$.  Whenever we require an explicit coordinate basis, we can simply designate any one $c^k$ to be some arbitrary fixed (nonzero) value, and take the remaining $\{c^i:\ i\ne k\}$ to be  coordinate functions.

In practical numerical calculations---where these algebraic structures are called ``separated representations,'' ``matrix product states,'' or ``Slater deter\-min\-ants''---pseudo\-coordinate representations are adopted almost universally. Therefore, we will sometimes simply call the $c$'s ``coordinates''; this will make it easier to link the numerical algorithm of Fig.~\ref{fig: QMOR numerical} to the geometric properties of \lcal{K}.

\subsection{Regarding gabion manifolds as real manifolds} 
\label{sec: real manifolds}
Now we will begin analyzing in detail the curvature of the gabion  manifold $\lcal{K}$.  For geometric purposes it is convenient to regard $\lcal{H}$ not as a complex vector space, but as a Euclidean space, such that $\gsb\psi$ is a vector of real numbers that in our simple example has the eight components $\{\psi^m\} = \{\Re(\psi^1),\dotsc,\Re(\psi^4),\Im(\psi^1),\dotsc,\Im(\psi^4)\}$.  Similarly, we specify real coordinates on $\lcal{C}$ via $c^k = x^k + i y^k$, and with a long-term view toward interfacing with the \Kahler geometry literature we agree to specify these real coordinates in the conventional order $\{x^1,\dotsc,x^4,y^1,\dotsc,y^4\} \equiv \{r^1,\dotsc,r^8\} = \{r^a\}$.  Thus $\{\partial/\partial r^a\}$ is a complete set of vectors on $\lcal{C}$.

\subsubsection{Constructing the metric tensor} Then the map $\gsb{\psi}:\lcal{C}\to\lcal{H}$ induces a metric tensor $g$ upon $\lcal{C}$ via the Euclidean metric of $\lcal{H}$. The components of $g$ evidently are
\begin{equation}
\label{eq: metric tensor}
g_{ab} \equiv g\left(%
\frac{\partial\hspace{1.3ex}}{\partial r^a}\,,\,\frac{\partial\hspace{1.3ex}}{\partial r^b}\right) = 
  \left[\frac{\partial \gsb\psi(\lb{\mathrm{c}}(\lb{\mathrm{r}}))}{\partial r^a}\right]\cdot%
  \left[\frac{\partial \gsb\psi(\lb{\mathrm{c}}(\lb{\mathrm{r}}))}{\partial r^b}\right]
  .
\end{equation}This also suffices to define $g$ as a metric tensor on $\lcal{K}$, provided we restrict our attention---as for \MOR purposes we always will---to functions on $\lcal{K}$ having the functional form $f(\gsb{\psi}(\lb{\mathrm{c}}(\lb{\mathrm{r}})))$, such that the tangent vectors $\{\partial/\partial r^a\}$ always act either directly or indirectly (via the chain rule) upon $\gsb{\psi}(\lb{\mathrm{c}}(\lb{\mathrm{r}}))$.  Then knowledge of $g$ allows us to compute via (\ref{eq: notation}) the velocities and path lengths of arbitrary trajectories on $\lcal{K}$, as is required of a metric on $\lcal{K}$.

\subsubsection{Raising and lowering the indices of a pseudo-coordinate basis} Considered as a covariant matrix, the indices of $g_{ab}$ range over an over-complete basis set, and therefore $g_{ab}$ is singular.  It follows that we cannot construct a contravariant matrix $g^{ab}$ in the usual manner, by taking a matrix inverse of $g_{ab}$.  To evade this difficulty we define the contravariant metric tensor to have components $g^{ab} \equiv (g_{ab})^\scriptscriptpseudoinverse$, where ``$\sp{\scriptscriptpseudoinverse}$'' is the same Moore-Penrose matrix pseudoinverse that appears in Step~B.3 of~Fig.~\ref{fig: QMOR numerical}, and that was discussed following Eq.~(\ref{eq: Dirac-Frenkel}). 

It is easy to verify that $g^{ab}$ and $g_{ab}$ act to raise, lower, and contract tensor indices in the usual manner, with a single important difference: the operation of raising followed by lowering is no longer the identity operator, but rather is a projection operator, in consequence of the general pseudoinverse identity $(X^\scriptscriptpseudoinverse X)^2 = X^\scriptscriptpseudoinverse X$.   Physically this projection annihilates tangent vectors on $\lcal{K}$ whose length is zero. 

\subsubsection{Constructing projection operators in the tangent space} From these identities it follows that at a specified point $\gsb\psi_{\lcal{K}}$ of $\lcal{K}$,  the local operator $P_{\lcal{K}}(\gsb\psi_{\lcal{K}})$ that projects vectors in $\lcal{H}$ onto the tangent space of $\lcal{K}$ at $\ket\psi$ is
\begin{equation}
\label{eq: PK projection}
\big[P_{\lcal{K}}(\gsb\psi_{\lcal{K}})\big]_{mn} = 
\sum_{a,b=1}^{\scriptscriptdimC} 
  \left[\frac{\partial \gsb\psi_{\lcal{K}}(\lb{\mathrm{c}}(\lb{\mathrm{r}}))}{\partial r^a}\right]_{m}%
  g^{ab}
  \left[\frac{\partial \gsb\psi_{\lcal{K}}(\lb{\mathrm{c}}(\lb{\mathrm{r}}))}{\partial r^b}\right]_{n}
.
\end{equation}
In the interest of compactness, we will often write $P_{\lcal{K}}$ rather than $P_{\lcal{K}}(\gsb\psi_{\lcal{K}})$.  We readily verify that the projective property $P_{\lcal{K}}P_{\lcal{K}} = P_{\lcal{K}}$ follows from the definition of $g^{ab}$ given in (\ref{eq: metric tensor}) and the general pseudo-inverse identity $X^\scriptscriptpseudoinverse XX^\scriptscriptpseudoinverse = X^\scriptscriptpseudoinverse$.  

The ability to construct the projection $P_{\lcal{K}}$ solely from tangent vectors and the local metric tensor $g$ will play a central role in our geometric analysis of $\lcal{K}$.  

\subsection{``Push-button'' strategies for curvature analysis} 
\label{sec: gauge}
At this point we can analyze $\lcal{K}$'s intrinsic geometry by either of two strategies.  The~first strategy, which can be wholly automated, is to fix in our example problem (say) $r^7=1$ and $r^8=0$ so that the remaining $\{r^1, r^2,\dotsc, r^6\}$ can be regarded as conventional coordinate functions on the six-dimensional gabion~$\lcal{K}$.  The now-restricted set of tangent vectors associated with $\{r^1, r^2,\dotsc, r^6\}$ constitutes a coordinate basis, such that (\ref{eq: metric tensor}) specifies the metric tensor for this basis.  By construction, this metric has no null vectors, and hence $g_{ab}$ is invertible.

The intrinsic geometric properties of $\lcal{K}$ can then be automatically computed by any of the many symbolic manipulation packages that are available for research in general relativity.  This automated approach allows us to ``push the button'' and discover that for our simple example the scalar Riemann curvature $R$ of $\lcal{K}$ is given by the remarkably simple expression $R=-8/(\gsb{\psi}{\cdot}\gsb{\psi})$.  

\subsubsection{The deficiencies of push-button curvature analysis} What is unsatisfying about an automated coordinate-based analysis, however, is that this simple integer result is obtained as the result of seemingly miraculous cancellations of high-order polynomials.  This method produces no insight as to \emph{why} such a simple integer result is obtained, or why the sign of the curvature is negative, or whether this simplicity is linked to $\lcal{K}$'s ruled structure.    

Another objection to coordinate-based analysis is that it forces us to ``break the symmetry'' of the coordinate manifold $\lcal{C}$ by designating arbitrary fixed values for arbitrarily selected gabion coordinates.  This is undesirable because our quantum simulation algorithms in Figs.~\ref{fig: QMOR formal}--\ref{fig: QMOR numerical} do not break this symmetry.  To do so in our geometric analysis would unnecessarily obstruct our goal of linking quantum simulation physics to the \Kahlerian geometry of $\lcal{K}$.  

We will therefore develop a Riemannian/\Kahlerian curvature analysis of the gabion manifold $\lcal{K}$ that fully respects the algebraic symmetries, not of $\lcal{K}$, but of $\lcal{C}$.  For this purpose the \emph{sectional curvature} proves to be an ideal mathematical tool.  

\subsection{The sectional curvature of gabion state-spaces} %

Because $\lcal{K}$ has a natural embedding in the Euclidean manifold $\lcal{H}$, our analysis of sectional curvature is able to follow quite closely the embedded geometric reasoning of Gauss' original derivation of the \emph{Theorema Egregium} \cite{Gauss:1827}.  This approach has the advantage of yielding immediate physical insight.  Equally important, this approach can be readily adapted to accommodate the pseudo-coordinate tangent basis that is most natural for analyzing gabion geometry.  

As depicted in Fig.~\ref{fig: QMOR geometry}(E), we choose an arbitrary point on $\lcal{K}$ and define tangent vectors $U$ and $V$ on $\lcal{K}$ to be directional derivatives
\begin{equation}
 U \equiv \sum_{a=1}^{\scriptscriptdimC} u^a\frac{\partial}{\partial r^a }
 \qquad\text{and}\qquad
 V \equiv \sum_{a=1}^{\scriptscriptdimC} v^a\frac{\partial}{\partial r^a }\,.
\end{equation}Because the map $\lcal{C}\to\lcal{K}$ is surjective, the representation of $U$ and $V$ as a sum over components $u^a$ and $v^a$ is nonunique, and we will take care to establish that our sectional curvature calculations are not thereby affected.  

\subsubsection{Remarks on gabion normal vectors} Conjugate to the tangent space of $\lcal{K}$ at a given point is the space of vectors in $\lcal{H}$ that are normal to the tangent space.  We specify $\gsbhat{n}$ to be an (arbitrarily chosen) unit vector in that normal space, \emph{i.e.}, to be a vector satisfying
\begin{equation}
\label{eq: normal relations}
\lbhat{n}\cdot\lbhat{n} = 1\
\qquad\text{and}\qquad
\gsbhat{n}\cdot \frac{\partial\gsb{\psi}(\gsb{r})}{\partial r^a} 
\equiv 
\gsbhat{n}\cdot\gsb{\psi}_{\!,a} = 0\,,
\end{equation}where we have adopted the usual notation that a comma preceding a subscript(s) indicates partial differentiation with respect to the indexed variable(s).  The sign of $\gsbhat{n}$ will not be relevant.  We remark that $\gsbhat{n}$ is not unique because the \emph{codimension} of $\lcal{K}$ (by definition $\codim \lcal{K} \equiv \dimH - \dimK$) is in general greater than unity.  Looking ahead, in some of our large-scale numerical examples $\codim \lcal{K}$ will be very large indeed, of order $2\times2^{18}\simeq512,000$.

\subsubsection{Computing the directed sectional curvature} With reference to the vectors $U$, $V$, and $\lbhat{n}$ depicted in Fig.~\ref{fig: QMOR geometry}(E), we define a scalar function $S(U,V,\lbhat{n})$, which we will call the \emph{directed sectional curvature}, to be
\begin{equation}
\label{eq: sectional curvature}
S(U,V,\lbhat{n}) = \frac{{\displaystyle \sum_{a,b,c,d = 1}^{\dimC}}
	\left|\begin{array}{cc}
		\lbhat{n}\cdot\gsb{\psi}_{\!,ac} & \lbhat{n}\cdot\gsb{\psi}_{\!,ab} \\ 
		\lbhat{n}\cdot\gsb{\psi}_{\!,cd} & \lbhat{n}\cdot\gsb{\psi}_{\!,bd} 
	\end{array}\right| u^a v^b u^c v^d
}{{\displaystyle \sum_{a,b,c,d = 1}^{\dimC}}
	\left|\begin{array}{cc}
		\gsb{\psi}_{\!,a}\cdot\gsb{\psi}_{\!,c} & \gsb{\psi}_{\!,a}\cdot\gsb{\psi}_{\!,b} \\ 
		\gsb{\psi}_{\!,c}\cdot\gsb{\psi}_{\!,d} & \gsb{\psi}_{\!,b}\cdot\gsb{\psi}_{\!,d} 
	\end{array}\right| u^a v^b u^c v^d
}\,.
\end{equation}Here $|{\ldots}|$ denotes the determinant.  

If we recall that $\gsbhat{n}\cdot\gsb{\psi}_{\!,a} = 0$, per  (\ref{eq: normal relations}), then it is straightforward to verify that $S$ is a scalar under coordinate transformations.  It further satisfies the identity \begin{equation}
	S(U,V,\lbhat{n}) = S(\alpha U + \beta V,\gamma U + \delta V,\lbhat{n})
\end{equation}for arbitrary real $\alpha$, $\beta$, $\gamma$, and $\delta$.  Thus $S(U,V,\lbhat{n})$ is a real-valued geometric invariant of the two-dimensional tangent subspace spanned by $U$ and $V$.

In the preceding paragraph we emphasize that $\alpha$, $\beta$, $\gamma$, and $\delta$ are real-valued because later on, when we admit a complex structure, it will \emph{not} be true that phase-shifting $U$ and/or $V$ leaves the sectional curvature invariant (see Section~\ref{sec: HBC}).

The denominator of (\ref{eq: sectional curvature}) has a simple physical interpretation as the geometric area of the section defined by $U$ and $V$; this quantity is often written as $|U\wedge V|^2$.   In terms of the metric function $g(U,V) \equiv \sum_{a,b} g_{ab} u^v v^b$ we have 
\begin{equation}
\label{eq: wedge}
|U\wedge V|^2 = g(U,U) g(V,V) - g(U,V)^2\,. 
\end{equation}

\subsubsection{Physical interpretation of the directed sectional curvature} For a two-dimensional surface embedded in a three dimensional space---the case considered by Gauss---the above expression reduces to the familiar expression $S(U,V,\lbhat{n}) = 1/(R_1R_2)$, where $R_1$ and $R_2$ are the principal radii of curvature of the surface. In higher dimensions $S(U,V,\lbhat{n})$ describes the Gaussian curvature of a two-dimen\-sional \emph{section} of $\lcal{K}$---a two-dimensional submanifold that is locally tangent to $U$ and $V$---that has been projected onto the three-space spanned by $\{U,V,\lbhat{n}\}$.  

For \MOR purposes, this means that whenever $S(U,V,\lbhat{n})$ is negative we are guaranteed local concavity of the state-space $\lcal{K}$ as viewed along~$\lbhat{n}$ (\emph{i.e}, as viewed ``from above'') along at least one curve that is locally tangent to some linear combination of $U$ and $V$.  The resulting physical picture is shown in Fig.~\ref{fig: QMOR geometry}(G).  For \MOR purposes our goal will be, therefore, to choose state-space manifolds such that $S(U,V,\lbhat{n})$ is negative, in the expectation that the associated local concavity of $\lcal{K}$  will improve the robustness of projective model order reduction.

\subsubsection{Definition of the intrinsic sectional curvature} Mathematicians usually prefer to describe the curvature of $\lcal{K}$ in intrinsic terms.  To accomplish this it is convenient to sum over a complete orthonormal set $\{\lbhat{n}_i\}$ of vectors tangent to $\lcal{K}$.  We use the identity $\sum_i \lbhat{n}\otimes\lbhat{n} = \sbar P_{\lcal{K}} = I-P_{\lcal{K}}$, where $I$ is the identity operator and $P_{\lcal{K}}$ is the projection matrix given in (\ref{eq: PK projection}), to obtain 
\begin{align}
S(U,V)& \equiv \sum_{i=1}^{\codim \lcal{K}} S(U,V,\lbhat{n}_i) \notag \\
	& = \sum_{a,b,c,d = 1}^{\dimC}\ \frac{
	\left[
		\gsb{\psi}_{\!,ac}\cdot\sbar P_{\lcal{K}}\cdot\gsb{\psi}_{\!,bd} -
		\gsb{\psi}_{\!,ab}\cdot\sbar P_{\lcal{K}}\cdot\gsb{\psi}_{\!,cd} 
	\right]
	u^a v^b u^c v^d
	}{|U\wedge V|^2}\,.
\label{eq: S(U,V)}
\end{align}
Now all reference to unit normals has disappeared, because we have already established in (\ref{eq: PK projection}) that $P_{\lcal{K}}$ can be described in intrinsic terms.  However, the above expression still refers to the embedding Hilbert space via $\gsb\psi$.  It will not be until later on (specifically, following (\ref{eq: Riemann components})) that we show that $S(U,V)$ is determined solely by the metric tensor and its derivatives.

\subsection{The formal definition of a gabion manifold} 
\label{sec: formal definition}
For general tangent vectors $U$ and $V$, the directed sectional curvature $S(U,V,\lbhat{n})$ and the intrinsic sectional curvature $S(U,V)$ can be either positive or negative.  We will now derive a condition on $U$ and $V$ which is sufficient for $S(U,V,\lbhat{n})$ and $S(U,V)$ to be nonpositive, and we will use this condition to motivate a formal definition of a gabion manifold.

We recall that $\lcal{K}$ is a reduced-dimension state-space manifold that is embedded in a larger-dimension Euclidean manifold $\lcal{H}$.  Thus each individual component of the state-vector $\gsb{\psi}$ of $\lcal{H}$ defines a scalar function on $\lcal{K}$.  If we wish, we may  regard the metric $g$ of (\ref{eq: metric tensor}) and the normal vector $\lbhat{n}$ of (\ref{eq: normal relations}) as intrinsically defined in terms of the scalar functions $\gsb{\psi}$; this eliminates any formal reference to the embedding manifold $\lcal{H}$.  We define a \emph{rule vector field}, or simply \emph{rule field}, to be any vector field $\lcal{V}$ on $\lcal{K}$ satisfying 
\begin{equation}
\label{eq: geodesics}
\nabla_{V}\nabla_{V}\gsb{\psi} = 0\,.
\end{equation}
The motivation for this definition is simply that the above equation is both intrinsic and geometrically covariant, and furthermore, it is manifestly satisfied by the vector field that is associated with each gabion pseudo-coordinate; these coordinates thus are canonical examples of rule fields.  We define \emph{rule lines}, or simply \emph{rules}, to be the integral curves of a rule field.  It is straightforward to show that rule lines are geodesics; this formally justifies our earlier  depiction of rules as ``straight lines'' in  Fig.~\ref{fig: QMOR geometry} and in Section~\ref{sec:geometric ideas}.  We define \emph{rule tangent vectors}, or simply \emph{rule vectors}, to be vectors that are locally tangent to a rule line.

We now formally define a gabion manifold as follows:
\begin{mydefinition}
\label{def: gabion}
A gabion manifold is a manifold endowed with rule fields whose rule tangent vectors constitute a local basis at every point of the manifold.
\end{mydefinition}
\noindent This definition is more restrictive than the usual definition of a ruled manifold \cite{Fischer:2001}, in which there is no requirement that the rule vectors provide a local basis.  Roughly speaking, therefore, a gabion manifold is a ruled manifold that is exceptionally rich in rule structure. 

Associated with a rule vector $V$, we define \emph{local rule coordinates} such that $\nabla_{V}\nabla_{V}\gsb{\psi} = 0$ takes the component form $\sum_{a,b}v^av^b \gsb{\psi}_{,ab}$ = 0.  Thus gabion pseudo-coordinates are local rule coordinates. Evaluating (\ref{eq: S(U,V)}) in local rule coordinates, we see that whenever either $U$ or $V$ is a rule vector, the first numerator term vanishes, and the remaining numerator term is nonpositive.  This proves the following theorem: 
\begin{mytheorem}
\label{thm: one}
Let $U$ be a rule vector at an arbitrary point on a manifold $\lcal{K}$, let $V$ be an arbitrary tangent vector at that same point, and let~$\lbhat{n}$ be an arbitrary unit vector normal to the tangent space.  Then the directed sectional curvature satisfies $S(U,V,\lbhat{n}) \le 0$ and therefore, the intrinsic sectional curvature satisfies $S(U,V) \le 0$.
\end{mytheorem}
\noindent In physical terms, any two-dimensional section of $\lcal{K}$ that includes a rule vector has negative sectional curvature.  Since for gabion manifolds, the local rule tangents form an over-complete local basis at each point in \lcal{K}, we see that negative sectional curvature is ubiquitously present in our gabion state-spaces.

\subsubsection{Recipes for constructing rules and rule fields} In the context of \MOR analysis, rule lines are easy to construct, and they have a clear algorithmic significance.   

Rules can  be readily constructed via the product-sum algebraic structure of Fig.~\ref{fig: QMOR product sum}, by varying any one $\{\sp{l}c^{m}_{n}\}$ while holding the others fixed.  The tangents to the rule lines then constitute an overcomplete local basis, as depicted in Fig.~\ref{fig: QMOR geometry}(C).

More generally, rule fields can be constructed by selecting an arbitrary order (column) in the product-sum algebraic structure of Fig.~\ref{fig: QMOR product sum}, selecting arbitrary basis vectors for the Hilbert subspace associated with that order (equivalent to imposing an arbitrary rotation on the basis vectors of that column's subspace), selecting an arbitrary rank (row), selecting an arbitrary element of the substate of that order and rank, choosing a coordinate system (in the strict sense of Section~\ref{sec: gauge}) such that the selected element is one of the coordinate functions, and identifying a rule field on $\lcal{K}$ with the partial derivative with respect to that coordinate.

From an algorithmic point of view, a rule vector is a direction in the state-space along which trajectories can move with great algorithmic efficiency, since only one state-space coordinate need be updated.   

\subsubsection{The set of gabion rules is geodesically complete} Whenever any two rows of the product-sum in Fig.~\ref{fig: QMOR product sum} are algebraically degenerate, the tangent space of $\lcal{K}$ will be geometrically degenerate, yet according to the construction of the geometric rules, and in particular because the underlying algebraic structure is polynomial, the rules pass through curvature singularities without disruption of their geodesic properties, as depicted in Fig.~\ref{fig: QMOR geometry}(D).  

Furthermore, it is evident that the sectional curvature (\ref{eq: S(U,V)}) diverges in the neighborhood of a rule singularity, in consequence of the divergence of the pseudo-inverse metric $g^{ab}$ that appears in the projection operator $P_{\lcal{K}}$ as given in (\ref{eq: PK projection}). Numerical experiments confirm this expectation.  These phenomena suggest that gabion manifolds might fruitfully be analyzed in terms of affine algebraic varieties.  The authors have not pursued this line of analysis.

\subsection{Gabion-\Kahler (\GK) manifolds} 
\label{sec: Kahlerian gabions}
We now specialize (\ref{eq: S(U,V)}) to complex manifolds having a \Kahlerian metric, which we will call gabion-\Kahler (\GK) manifolds.  In so doing, we will adopt certain ``tricks'' of indexing that \Kahlerian geometers use.  We begin by writing the numerator and denominator of 
(\ref{eq: S(U,V)}) in matrix notation, with all quantities still real:
\begin{subequations}
\begin{align}
\text{num} =  &  {\displaystyle \sum_{a,b,c,d = 1}^{\dimC}}
\Big[
(\lb{\psi}_{\!,ac}\cdot\lb{\psi}_{\!,bd}) - 
{\displaystyle \sum_{e,f = 1}^{\dimC}}
(\lb{\psi}_{\!,ac}\cdot\lb{\psi}_{\!,e})g^{ef}(\lb{\psi}_{\!,f}\cdot\lb{\psi}_{\!,bd})  \notag \\
& \quad - (\lb{\psi}_{\!,ab}\cdot\lb{\psi}_{\!,cd}) 
+ {\displaystyle \sum_{e,f = 1}^{\dimC}}(\lb{\psi}_{\!,ab}\cdot\lb{\psi}_{\!,e})g^{ef}(\lb{\psi}_{\!,f}\cdot\lb{\psi}_{\!,cd})
\Big] u^a v^b u^c v^d
\label{eq: S(U,V) numerator}\\
\text{den}  = & {\displaystyle \sum_{a,b,c,d = 1}^{\dimC}}
\left[
(\lb{\psi}_{\!,a}\cdot\lb{\psi}_{\!,c})(\lb{\psi}_{\!,b}\cdot\lb{\psi}_{\!,d}) - 
(\lb{\psi}_{\!,a}\cdot\lb{\psi}_{\!,c})(\lb{\psi}_{\!,c}\cdot\lb{\psi}_{\!,d})
\right] u^a v^b u^c v^d
\label{eq: S(U,V) denominator}
\end{align}
.
\end{subequations}
Now we reason as follows. $S(U,V)$ is a real number that is independent of coordinate system.  We are therefore free to analytically continue our coordinates, transforming (for example) the coordinate pair $\{x_1,y_1\}\to\{c_1,\sbar{c}_1\}$ via $c_1 = x_1+i y_1$ and $\sbar{c}_1 = x_1-i y_1$.  Since the sectional curvature is a geometric invariant, the (real) value of $S(U,V)$ will not be altered thereby, even though the coordinates themselves are now complex.

\subsubsection{\Kahlerian indexing and coordinate conventions} It is evident that analytic continuation to complex coordinates treats ${c}_1$ and $\sbar{c}_1$ as independent coordinates for symbolic manipulation purposes (such as partial differentiation), just as $x_1$ and $y_1$ are independent coordinates. It is only at the very end of a calculation, when we assign (complex) numerical values to ${c}_1$ and $\sbar{c}_1$, that they can no longer be varied independently. 

It~is helpful too to replace Latin indices with unbarred and barred Greek indices, in a convention that associates barred indices with barred coordinates (see \cite[p.~8]{Flaherty:76} or \cite{Martin:02}). Then the vector $V$ has components $\{v^1, v^2,\dotsc, \sbar{v}^{\sbar{1}}, \sbar{v}^{\sbar{2}}, \ldots\}$, for example, and is represented in terms of partial derivatives by
\begin{equation}
V = \sum_{\alpha=1}^{\dimC/2} 
	v^\alpha \frac{\partial}{\partial c^\alpha} +
	\sum_{\sbar\alpha=\sbar 1}^{\dimC} 
	\sbar v^{\sbar\alpha} \frac{\partial}{\partial \sbar{c}^{\sbar\alpha}}
	.
\end{equation}  
In this convention a barred index $\sbar k \equiv \dimC/2+k$, such that the index $\scriptstyle\sbar1$ is a shorthand for the integer $\dimC/2+1$, and  $(v^\alpha)^\star = \sbar v^{\sbar\alpha}$.

With regard to the embedding Hilbert space $\lcal{H}$, we will adopt the physics convention that the ''ket'' vector $\gsb\psi(\lb{\mathrm c})\equiv\ket{\psi(\lb{\mathrm c})}$ is a complex vector of dimension $\dimH/2$, with the ``bra'' vector $\gsbbar\psi(\lbbar{\mathrm c}) \equiv \bra{\sbar\psi(\lbbar{\mathrm c})}$ being the conjugate vector.  

Thus $\gsb\psi(\lb{\mathrm c})$ is a holomorphic function (also known as an ``analytic function'') of complex pseudo\-coordinates~$\lb{\mathrm c}$, and  $\gsbbar\psi(\lbbar{\mathrm c})$ is similarly a holomorphic function of~$\lbbar{\mathrm c}$. Defining the biholomorphic \emph{\Kahler potential function} $\kappa(\lbbar{\mathrm c},\lb{\mathrm c})$ to be 
\begin{equation}
\label{eq: Kahler potential}
\kappa(\lbbar{\mathrm c},\lb{\mathrm c}) \equiv \tfrac{1}{2}\gsbbar{\psi}(\lbbar{\mathrm c})\cdot\gsb{\psi}(\lb{\mathrm c}) = \tfrac{1}{2}\inner{\sbar\psi(\lbbar{\mathrm c})}{\psi(\lb{\mathrm c})}\,,
\end{equation}
the components of the metric tensor $g$ are given in terms of $\kappa(\lbbar{\mathrm c},\lb{\mathrm c})$ by
\begin{equation}
\label{eq: glower}
g_{\sbar\alpha\beta} = g_{\beta\sbar\alpha} = 
\frac{\partial\kappa(\lbbar{\mathrm c},\lb{\mathrm c})}{\partial \sbar{c}^{\sbar\alpha}\partial c^\beta} \quad\text{and}\quad 
g_{\alpha\beta} = g_{\sbar\alpha\sbar\beta} = 0. 
\end{equation}
It immediately follows that in any holomorphic coordinate system 
\begin{equation}
\label{eq: g upper}
g^{\alpha\sbar\beta} = g^{\sbar\beta\alpha} \quad\text{and}\quad
g^{\alpha\beta} = g^{\sbar\alpha\sbar\beta} = 0\,. 
\end{equation}

The simplest explicit example of this convention is the complex plane regarded as a two-dimensional \Kahler manifold.  Indexing its two coordinates by $\{1,\sbar 1\}$ yields coordinates  $\{c_1,\sbar{c}_{\sbar 1}\}$, which in analytic function theory are conventionally called $\{z,\sbar z\}$.  The real-valued \Kahler potential is $\kappa = c_1\sbar{c}_{\sbar 1}/2 = z\sbar z/2$, the components of the metric tensor are $g_{ab}=\left[\begin{smallmatrix}0&1\\1&0\end{smallmatrix}\right]\!/2$ and $g^{ab}=2\left[\begin{smallmatrix}0&1\\1&0\end{smallmatrix}\right]$, and the length element $ds^2$ is
 \begin{equation}
{d}s^2 = \sum_{ab}\,g_{ab}\,{d}c^a\,{d}c^b 
= 
	g_{1\sbar1}\,{d}c^1\,{d}{\sbar c}^{\sbar1} 
	+ g_{\sbar1 1}\,{d}{\sbar c}^{\sbar1}\, {d}c^1 
= \tfrac{1}{2}({d}z\,{d}\sbar z+{d}\,\sbar z{d}z)
= {d}z\,{d}\sbar z
,
\end{equation}which is the usual normalization.  Note the ubiquitous factors of~2 and~1/2, which require careful attention in practical calculations. A general scalar function on this manifold is of the form $f(\sbar z,z)$, and functions of the special form $f(z)$ are the {holomorphic} (or {analytic}) functions.  

In programming calculations on \Kahler manifolds of larger dimension, it is helpful that (\ref{eq: glower}) and (\ref{eq: g upper}) take the block-matrix forms
\begin{equation} 
\label{eq: g pseudoinverse}
\big[g_{ab}\big] = \left[
		\begin{array}{@{}c@{}c@{}}
			[\ 0\ ] & [g_{\alpha\sbar\beta}]\\{}
			[g_{\sbar\alpha\beta}] & [\ 0\ ]
		\end{array}
	\right]\quad\text{and}\quad
\big[g^{ab}\big] = \left[
		\begin{array}{@{}c@{}c@{}}
			[\ 0\ ] & [g^{\alpha\sbar\beta}]\\{}
			[g^{\sbar\alpha\beta}] & [\ 0\ ]
		\end{array}
		\right] = 
		\left[
		\begin{array}{ll}
			[\ 0\ ]\hspace*{0.7ex} & [g_{\sbar\alpha\beta}]^{\scriptpseudoinverse}\\{}
			[g_{\alpha\sbar\beta}]^{\scriptpseudoinverse}\ & [\ 0\ ]\hspace*{0.7ex}
		\end{array}
		\right]
		,
\end{equation}
where ``$[\ldots]$'' is a square matrix.  Environments like \MATLAB and Mathematica provide built-in functions for  block-matrix constructs of this type.  We further see that in consequence of $g_{ab}=g_{ba}$ and $g_{\alpha\sbar\beta}=\sbar g_{\sbar\alpha\beta}$, which follow from (\ref{eq: glower}), the individual block matrices in (\ref{eq: g pseudoinverse}) are Hermitian and semipositive; for this reason the submatrix $\big[g_{\alpha\sbar\beta}\big]$ is sometimes called the \emph{Hermitian metric} of the complex manifold.   In practical calculations it is considerably more efficient to work solely with the Hermitian metric and its pseudo-inverse, than with the larger matrix $g_{ab}$.

Numerically-minded readers who are new to the literature of \Kahler manifolds will appreciate that the above indexing conventions elegantly resolve a contradiction of our intuitions.  On the one hand, we expect that a coordinate transformation cannot change the range of an index. On the other hand, we expect on physical grounds that a manifold described by complex coordinates will need only half the number of coordinate variables as the same manifold described by real coordinates.  The resolution of this dilemma is in the block structure and symmetry properties of $g$, which ensure that in practical geometric calculations, only half the index range need be summed over.

\subsubsection{\GK sectional curvature in physics bra-ket notation} It is then a straightforward exercise to write (\ref{eq: S(U,V) numerator}--\ref{eq: S(U,V) denominator}) compactly, in the bra-ket notation of physics: 
\begin{subequations}
\begin{align}
\text{num}& = \tfrac{1}{2}
\Big[ 
\braket{\partial_{\sbar u}\partial_{\sbar u}\sbar \psi}{\sbar P_\lcal{K}}{\partial_{v}\partial_{v}\psi}
+ \braket{\partial_{\sbar v}\partial_{\sbar v}\sbar \psi}{\sbar P_\lcal{K}}{\partial_{u}\partial_{u}\psi}
-2 \braket{\partial_{\sbar u}\partial_{\sbar v}\sbar \psi}{\sbar P_\lcal{K}}{\partial_{u}\partial_{v}\psi}
\Big] 
\label{eq: S(U,V) bra-ket numerator}\\
\text{den}& =
	\Big[\inner{\partial_{\sbar u}\sbar\psi}{\partial_{u}\psi}
	\inner{\partial_{\sbar v}\sbar\psi}{\partial_{v}\psi}
	- \tfrac{1}{4}\big(
			\inner{\partial_{\sbar u}\sbar\psi}{\partial_{v}\psi}
			+ \inner{\partial_{\sbar v}\sbar\psi}{\partial_{u}\psi}
		\big)^2\Big]^2
\label{eq: S(U,V) bra-ket denominator}
,
\end{align}
where the partial derivatives and the projection operator $\sbar P_\lcal{K}\equiv I-P_\lcal{K}$ are given in terms of components by
\begin{equation}
\ket{\partial_v \psi} = \sum_{\alpha = 1}^{\dimC/2}
	v^\alpha\frac{\partial}{\partial c^\alpha}\,\ket{\psi(\lb{\mathrm c})}
\quad\text{(\emph{etc.})}
\end{equation}
\begin{equation}
\label{eq: final Kahler equation}
P_\lcal{K} = \frac{1}{2} \sum_{\alpha = 1}^{\dimC/2}\ \sum_{\sbar\beta = \sbar1}^{\dimC}
g^{\alpha\sbar\beta}
\frac{\partial^2 }{\partial c^\alpha \partial {\sbar c}^{\sbar\beta}}
\,\ket{\psi(\lb{\mathrm{c}})}\bra{\sbar\psi(\lbbar{\mathrm c})}
.
\end{equation}
\end{subequations}
The above expressions show explicitly that bra-ket notation allows the sectional curvature to be computed by summing over half-ranges of coordinate indices.  This notational compactness constitutes (from a \Kahlerian  geometry point of view) the main practical rationale for the bra-ket notation of the physics literature.

\subsubsection{Defining the Riemann curvature tensor} Now we define the \emph{Riemann curvature tensor} to be that scalar function $R(A,B,C,D)$, defined at each point on the gabion manifold $\lcal{K}$, with $A, B, C, D$ being arbitrary vectors, such that the local sectional curvature and the local Riemann curvature are related by
\begin{equation}
\label{eq: Riemann definition}
	S(U,V) = R(U,V,U,V)/ |U\wedge V|^2 = R_{abcd}\ u^a v^b u^c v^d / |U\wedge V|^2
.
\end{equation}
It is known (see \cite[Theorem~3.8]{Gallot:04} or \cite[Theorem~7.51]{Martin:02}) that the Riemann curvature so defined is unique, for both real and \Kahler manifolds, provided that the following index symmetries are imposed:
\begin{equation}
R(A,B,C,D) = -R(B,A,C,D) = -R(A,B,D,C) = R(C,D,A,B)
\label{eq: Riemann symmetries}
\end{equation}which are the conventional antisymmetries of the Riemann tensor, and provided in addition the following identity is satisfied
\begin{equation}
\label{eq: first Bianci identity}
R(A,B,C,D) + R(B,C,A,D) = R(C,A,B,D)=0\,,
\end{equation}
which is called the \emph{first Bianchi identity}.  

On real manifolds, the implicit definition (\ref{eq: Riemann definition}) of the Riemann curvature in terms of the sectional curvature is difficult to work with, in the sense that the general expression for $R_{abcd}$ as a function of $g$ turns out to be too complicated to readily derive by inspection or manipulation of (\ref{eq: S(U,V)}). 

Fortunately, on \Kahler manifolds we have the simpler definition of $S(U,V)$ given in (\ref{eq: S(U,V) bra-ket numerator}--\ref{eq: final Kahler equation}), from which the Riemann curvature can be read off as\begin{equation}
\label{eq: Riemann components}
	R_{\alpha\sbar\beta\gamma\sbar\delta} = 
		\kappa_{,\alpha\sbar\beta\gamma\sbar\delta} - 
		\kappa_{,\sbar\beta\sbar\delta\mu}\,g^{\mu\sbar\nu}\,\kappa_{,\sbar\nu\alpha\gamma}
		= g_{\alpha\sbar\beta,\gamma\sbar\delta} - 
		g_{\sbar\beta\mu,\sbar\delta}\,g^{\mu\sbar\nu}\,g_{\sbar\nu\gamma,\alpha}
,
\end{equation}
where $\kappa(\lbbar{\mathrm c},\lb{\mathrm c})$ is the biholomorphic \Kahler potential introduced in (\ref{eq: Kahler potential}) and $g^{\mu\sbar\nu}$ is the pseudo-inverse of $g_{\mu\sbar\nu}=\kappa_{,\mu\sbar\nu}$ introduced in (\ref{eq: g pseudoinverse}).  We further specify that all components of $R$ not fixed by (\ref{eq: Riemann components}) vanish, save those required by the symmetries~(\ref{eq: Riemann symmetries}).  Since the following components of $R$ \emph{cannot} be obtained from (\ref{eq: Riemann components}) by symmetry, we take them to be zero
\begin{equation}
R_{\alpha\beta c d} = R_{\sbar\alpha\sbar\beta c d} = 
R_{a b\gamma\delta} = R_{a b\sbar\gamma\sbar\delta}  = 0\,,
\end{equation}
and we see that the resulting \Kahlerian $R_{abcd}$ has a block structure similar to that of the \Kahlerian metric $g_{ab}$ in (\ref{eq: g pseudoinverse}). In consequence of this block structure, it is straightforward to verify that the Bianchi identity (\ref{eq: Riemann symmetries}) is equivalent to the following \emph{\Kahlerian Bianchi index symmetries} (which we have not found explicitly given in the literature): 
\begin{equation}
\label{eq: first Bianci index symmetries}
R_{\alpha\sbar\beta\gamma\sbar\delta}= 
R_{\gamma\sbar\beta\alpha\sbar\delta}=
R_{\alpha\sbar\delta\gamma\sbar\beta}=
R_{\gamma\sbar\beta\alpha\sbar\delta},
\end{equation}and which (\ref{eq: Riemann components}) respects. Thus the definition, symmetries, and identities (\ref{eq: Riemann definition}--\ref{eq: first Bianci identity}) all are satisfied, and we conclude that the \Kahlerian sectional curvature (\ref{eq: S(U,V) bra-ket numerator}--\ref{eq: final Kahler equation}) uniquely determines the Riemann curvature to be (\ref{eq: Riemann components}).  We remark that the elegant functional simplicity of the \Kahler-Riemann curvature tensor (\ref{eq: Riemann components}), which in many textbooks mysteriously appears only at the end of a long algebraic derivation, emerges quite simply and naturally in our Gauss-style, immersive, bra-ket derivation. 

\subsubsection{The \emph{Theorema Egregium} on \GK manifolds}
\label{sec: GK Theorema Egregium}
Because $S(U,V)$ and $R(A,B,C,D)$ depend solely on the intrinsic metric $g$, we have thus derived---solely by analysis of sectional curvature---the celebrated Gauss/Riemann \emph{Theorema Egregium} as it applies to \Kahler manifolds.  As mentioned above, we have not found in the \Kahlerian geometry literature any similar derivation of the Riemann tensor by sectional curvature analysis.  However, the \Kahler geometry literature is so vast, that we can say with confidence only that the above derivation of (\ref{eq: Riemann components}) is quite different from the usual (intrinsic) derivations in the literature (see \cite[Theorem~12.5.6]{Martin:02} and \cite[Ch.~6]{Moroianu:07}).  We remark also that minor misprints are commonplace in the mathematical literature (\emph{e.g.}, \cite[eq.~7.11 of ch.~1]{Kobayashi:87} is identical to (\ref{eq: Riemann components}) save for an incorrect index-pair contraction).  That is why we will derive some closed-form results in Section~\ref{sec: gold standards} against which large numerical codes can be compared.

To better serve our \MOR purposes, we have generalized $R_{\alpha\sbar\beta\gamma\sbar\delta}$ by allowing its indices to range over $\dimC$ rather than $\dimK$, and defined the contravariant tensor $g^{\mu\sbar\nu}$ in terms of the matrix pseudoinverse.  These extensions do not alter the functional form of (\ref{eq: Riemann definition}--\ref{eq: first Bianci index symmetries}), and they make practical numerical calculations very much easier to program, as we will see in Section~\ref{sec: Slater determinants}. 

It is apparent that the generalized metric tensor $g^{\mu\sbar\nu}$ that appears in the Riemann curvature tensor (\ref{eq: Riemann components}) is identical to the matrix $(\mypartialtwo\kappa)^{\scriptpseudoinverse}$  that appears in the simulation algorithm of Step~B.3 of Fig.~\ref{fig: QMOR numerical}.  This is the first of many links that we will establish between \Kahlerian geometry and quantum simulation.

\subsubsection{Readings in \Kahlerian geometry} Having derived the \Kahlerian Riemann curvature tensor, we will attempt a brief survey of the \Kahler geometry literature.  The literature on \Kahler geometry is comparably vast to the literature on quantum measurement and information, and so our review necessarily will be exceedingly sparse and subjective.  

Our development and indexing conventions in this section have paralleled the Riemannian conventions of Weinberg \cite{Weinberg:72}, as extended to \Kahlerian geometry by Flaherty~\cite{Flaherty:76}, as further extended to abstract notation by Martin~\cite{Martin:02} and Moroianu~\cite{Moroianu:07}.  However, our analysis has been centered upon sectional curvature, rather than Riemann curvature as in the preceding  texts.  Other suitable texts include Kobayashi~\cite{Kobayashi:87}, Frenkel~\cite{Frenkel:04}, Gallot~\cite{Gallot:04}, Flanders~\cite{Flanders:63}, Jost~\cite{Jost:97}, Hou and Ho~\cite{Hou:97}, and a lengthy review by Yau~\cite{Yau:06}.  Many more textbooks and review articles on \Kahler geometry exist, and it is largely a matter of individual taste to chose among them. 

\subsection{Remarks upon holomorphic bisectional curvature}
\label{sec: HBC}
We have seen that the sectional curvature of both real and \Kahlerian gabion manifolds is constrained by Theorem~\ref{thm: one}.  We now review additional (known) sectional curvature theorems that relate particularly to gabion-\Kahler manifolds.  These theorems concern a geometric measure called the \emph{holomorphic bisectional curvature}  \cite{Goldberg:67}.

We begin by remarking that physicists in particular are accustomed to thinking of $\gsb\psi$ and $i\gsb\psi$ as being physically the same vector.  In \Kahlerian notation, the notion of ``multiplying by $i$'' is associated with an \emph{almost complex structure} $J$, which is defined to be a linear map $J$, defined on every point in $\lcal{K}$, satisfying $J^2V=-V$ for $V$ an arbitrary tangent vector, and differentially smooth, that leaves the metric tensor invariant, \emph{i.e.}, $g(U,V)=g(JU,JV)$.  The effect of $J$ upon the components  $\{v^\alpha,\sbar v^{\sbar\alpha}\}$ of an arbitrary vector $V$ is simple:\begin{equation}\{v^\alpha,\sbar v^{\sbar\alpha}\}\overset{\scriptstyle J}{\to}\{i v^\alpha,-i \sbar v^{\sbar\alpha}\}.\end{equation}The functional form of (\ref{eq: S(U,V) bra-ket numerator}-\ref{eq: S(U,V) bra-ket denominator}) then implies the identities $S(U,V) = S(JU,JV)$ and $S(JU,V) = S(U,JV)$. 

We now consider the sum $S(U,V)+S(U,JV) = S(U,V)+S(JU,V)$, which by definition is called the \emph{holomorphic bisectional curvature}  \cite{Goldberg:67}.  The fundamental inequality of holomorphic bisectional curvature\begin{equation}
\label{eq: HCBM Theorem}
S(U,V)+S(U,JV) \le 0
\end{equation}follows immediately from  (\ref{eq: S(U,V) bra-ket numerator}-\ref{eq: S(U,V) bra-ket denominator}), because the ``multiply by $i$'' rule implies that the first term in the denominator cancels in the sum.  We will call (\ref{eq: HCBM Theorem}) the \emph{holomorphic bisectional curvature nonpositivity theorem} (\HBCN Theorem).   

The \HBCN Theorem is a long-known result \cite{Goldberg:67} that applies in general to any \Kahler manifold that is a complex submanifold of a Euclidean space.  Physically speaking, if a given \Kahlerian section $\{U,V\}$ has positive curvature, then the ``rotated'' section $\{U,JV\}$ will have negative curvature. It is readily shown~\cite{Goldberg:67} that the \HBCN Theorem implies the nonpositivity of the eigenvalues of the Ricci tensor, and therefore, the nonpositivity of the scalar curvature.  

As a technical point, the \HBCN Theorem applies only to \Kahler manifolds that have (or can be given) a complex embedding in a larger Euclidean manifold, which is the case of greatest interest in quantum \MOR applications.  The \HBCN  Theorem does \emph{not} apply to \Kahler manifolds that have no complex Euclidean embedding.  For example, \Kahler manifolds having a Fubini-Study metric can (and sometimes do) exhibit positive scalar curvature.  This is incompatible with the \HBCN Theorem, and these manifolds therefore have no complex Euclidean embedding.  

The functional form of (\ref{eq: S(U,V) bra-ket numerator}) allows us to immediately extend the \HBCN Theorem to encompass the directed sectional curvature $S(U,V,\lbhat{n})$ as follows:
\begin{mytheorem}[directed extension of the \HBCN Theorem]
\label{lemma: one}
At an arbitrary point on a \Kahler manifold $\lcal{K}$ that has a complex embedding within a Euclidean space $\lcal{H}$, let $U$ and $V$ be arbitrary tangent vectors and let $\lbhat{n}$ be an arbitrary unit vector normal to $\lcal{K}$.  Then the directed sectional curvature satisfies $S(U,V,\lbhat{n})+ S(U,JV,\lbhat{n}) \le 0$.
\end{mytheorem}

\subsubsection{Relation of Theorem~\ref{thm: one} to the \HBCN Theorem}  Our Theorem~\ref{thm: one} is a different result from both the \HBCN Theorem and its directed extension Theorem~\ref{lemma: one}.  Most obviously, Theorem~\ref{thm: one} applies to all manifolds that possess one or more rule fields, whether they are \Kahlerian or not, while both the \HBCN Theorem and our Theorem~\ref{lemma: one} apply solely to \Kahler manifolds. 

Even on \Kahler manifolds, the \HBCN Theorem and Theorem~\ref{thm: one} have substantially differing implications.  It follows from the defining equation of a rule field (\ref{eq: geodesics}) that if $\lcal{W}$ is a rule field, then so is $J\lcal{W}$, hence if $V$ is a rule vector, then so is $JV$.  It then follows from Theorem~\ref{thm: one} that $S(U,V)$ and $S(U,JV)$ are \emph{both} nonpositive, which is a~strictly stronger condition than the nonpositivity of their sum that is implied by the \HBCN Theorem.

It further follows immediately from (\ref{eq: S(U,V) numerator}--\ref{eq: S(U,V) denominator}) that if $V$ is a rule vector, then $S(U,V) = S(U,JV)$. More generally, if $U$ and $V$ are \emph{both} rule vectors, then  $S(U,V) = S(U,JV) = S(JU,V) =  S(JU,JV)$, and \emph{all} of these sectional curvatures are nonpositive.  From this ``gabionic'' point of view, we see that on gabion-\Kahler manifolds the rule vectors $U$ and $JU$ are effectively the same vector, \emph{i.e.} rule vector phases are irrelevant to sectional curvature properties, as is natural in quantum mechanical analysis. 

The ubiquity of nonpositive sectional curvature on \GK manifolds therefore can be viewed as arising from the confluence of rule structure and complex structure, both of which are associated with strong theorems that imply negative sectional curvature.

\subsubsection{Practical implications of sectional curvature theorems} From an \MOR point of view, the nonpositive sectional curvature implied by Theorem~\ref{thm: one} can be regarded as helping to ensure the robustness of \MOR on both real and complex manifolds, while the \HBCN Theorem helps makes \MOR even \emph{more} robust on complex manifolds.  This is one of two fundamental reasons why quantum \MOR can be regarded as intrinsically easier than classical \MOR (the other reason being the algorithmic resources that are provided by the \emph{Theorema Egregium}, as we will discuss in the following section).

We remark that not all sectional curvatures $S(U,V)$ are negative on \GK manifolds.  In small-dimension \GK manifolds we have constructed analytical examples of positive-curvature sections---with some trouble because neither $U$ nor $V$ can be rule vectors---and in large-dimension gabions numerical searches find them.  According to our present (limited) understanding, these positive-curvature \Kahlerian sections seem rather artificial, and so we will not discuss them further.   It is possible that their significance has eluded us.

\subsection{Analytic gold standards for \GK curvature calculations}  
\label{sec: gold standards} The Riemann tensor specifies the Gaussian curvature of every section of~$\lcal{K}$, and on large-dimension \MOR manifolds the Riemann tensor therefore carries a vast amount of information.  To compress this information (among other purposes)  it is conventional to condense the four-index Riemann tensor $R_{abcd}$ into the two-index \emph{Ricci tensor} $R_{ab}$ by
\begin{equation}
\label{eq: Ricci}
	R_{ab} = \sum_{c,d=1}^{\dimC} g^{cd}R_{cadb}
	,
\end{equation}
which in \Kahlerian index notation can be written
\begin{align}
\label{eq: Ricci Kahlerian}
R_{\alpha\sbar\beta} & = R_{\sbar\alpha\beta} \equiv 
\sum_{c,d=1}^{\dimC} g^{cd} R_{c \alpha d \sbar\beta} =
\sum_{\gamma=1}^{\dimC/2}\ 
\sum_{\sbar\delta=\sbar 1}^{\dimC} -g^{\gamma\sbar\delta} 
R_{\alpha \sbar\delta \gamma \sbar\beta}.
\end{align}Further compression is achieved by the \emph{scalar curvature} $R$ defined by
\begin{equation}
\label{eq: scalar curvature}
	R = \sum_{a,b=1}^{\dimC} g^{ab}R_{ab} = 
	\,2 \!\!\sum_{\alpha=1}^{\dimC/2}\ \sum_{\sbar\beta=\sbar 1}^{\dimC}
	g^{\alpha\sbar\beta}R_{\sbar\beta\alpha}
.
\end{equation}
\emph{Caveat:} various authors entertain diverse conventions regarding which indices should be contracted to obtain the Ricci tensor, and the factors of -1 and 2 that appear above are commonly associated with minor errors and imprecisions.

There is one sectional curvature convention, however, that is universal: the directed sectional curvature on a unit hypersphere $S^n$ (the ordinary sphere embedded in $R^3$ being $S^2$) satisfies $S(U,V,\lbhat{n}) =1$ for all linearly independent $U$ and $V$.  In a locally orthonormal basis there are $n(n-1)$ such pairs; it follows that on a unit hypersphere the eigenvalues of the Ricci tensor are all $n-1$, and the scalar curvature itself is therefore $R=n(n-1)$.  This result is a useful ``gold standard'' for testing symbolic and numerical calculations on real manifolds.

To construct a similar gold standard for testing symbolic and numerical curvature calculations on \Kahlerian manifolds, it is convenient to consider the manifold of rank-one, order-$n$ gabion states (see Fig.~\ref{fig: QMOR product sum}).  We allow the spin quantum numbers $\{j_i:\ i\in 1,n\}$ associated with successive product subspaces to vary independently.  Then by a straightforward (but not short) calculation, it can be shown that the scalar curvature of a general rank one, order-$n$ product state is 
\begin{equation}
\label{eq: analytic curvature}
R = -\frac{8}{\kappa} \sum_{k,m=1}^{n} 
\begin{cases}
j_k j_m & \text{for $m\ne k$}\\
0 & \text{otherwise}
\end{cases}
,
\end{equation}
where $\kappa$ is the \Kahler potential function of (\ref{eq: Kahler potential}).  See the following section for a summary of the assorted algebraic techniques used to obtain this result. 

To the best of our knowledge, this general analytic result has not previously appeared in the literature.  For the simple example manifold of (\ref{eq: simple gabion}), we have order $n=2$ and $j_1=j_2=1/2$, so the above yields $R=-4/\kappa=-8/{\inner{\psi}{\psi}} = -8/{(\gsb{\psi}{\cdot}\gsb{\psi})}$, in agreement with the automated  ``push-button'' analysis of Section~\ref{sec: gauge}. 

\subsection{The Riemann-\Kahler curvature of Slater determinants}
\label{sec: Slater determinants}
We now have all the tools we need to compute the scalar Riemann curvature of \emph{Slater determinant states}, which are the main state-space of \emph{quantum chemistry} \cite{Cramer:04,Slater:29,Friesner:05}.  The strategy of the calculation is straightforward, and although the details are lengthy, the final result is simple.

We begin by considering, as the simplest example having nontrivial curvature, the quantum states that are obtained by antisymmetrizing the outer products of a rank-one order-two product-sum state of spin $j=3/2$ (see Fig.~\ref{fig: QMOR product sum})
\begin{equation}
\label{eq: simple slater}
	\gsb{\psi}(\gsb{\mathrm{c}}) 
\equiv 
	\left(\begin{bmatrix}c^1_a\\[0.35ex]c^2_a\\[0.35ex]c^3_a\\[0.35ex]c^4_a\end{bmatrix}
	\otimes
	\begin{bmatrix}c^1_b\\[0.35ex]c^2_b\\[0.35ex]c^3_b\\[0.35ex]c^4_b\end{bmatrix}- 
	\begin{bmatrix}c^1_b\\[0.35ex]c^2_b\\[0.35ex]c^3_b\\[0.35ex]c^4_b\end{bmatrix}
	\otimes
	\begin{bmatrix}c^1_a\\[0.35ex]c^2_a\\[0.35ex]c^3_a\\[0.35ex]c^4_a\end{bmatrix}%
	_{\rule[-0ex]{0pt}{1.5ex}}^{\rule[-0ex]{0pt}{1.5ex}}\right).
\end{equation}%
In the language of quantum chemistry, we can equivalently regard $\gsb{\psi}(\gsb{\mathrm{c}})$  as the variational state-space of the (unnormalized) Slater determinant states of two electrons ``$a$'' and ``$b$''  occupying linear combinations of four orbitals.  

With $\gsb{\psi}(\lb{\mathrm{c}})$ given, we can compute its scalar curvature $R$ by either numerical or analytic means.  To compute $R$ numerically, we can simply set the eight pseudo-coordinates $\lb{\mathrm{c}}=\{c^i_a,c^i_b;\ i=1,4\}$ to any desired value (thereby choosing the point in state-space at which $R$ is to be evaluated) then compute first the \Kahler potential $\kappa$ from (\ref{eq: Kahler potential}),\ %
then the Riemann curvature $R$ from (\ref{eq: Riemann components}),%
\ and (\ref{eq: Ricci}--\ref{eq: scalar curvature}), 
evaluating the pseudo-inverse metric $g^{\mu\sbar\nu}$ of (\ref{eq: Riemann components}) numerically.   

Empirically we find that for our simple example (\ref{eq: simple slater}) these numerical calculations invariably yield a Riemann scalar curvature of \mbox{$R = -8/\kappa$} (to~machine precision) for all values of the pseudo-coordinates $\{c^i_a,c^i_b\}$.  

The simplicity of the numerical result motivates and guides the following analytic evaluation of $R$ in closed form.  We designate by $\gsb{\psi}_0$ the point in the state-space $\gsb{\psi}(\gsb{\mathrm{c}})$ at which the Riemann curvature is to be evaluated. We write $\gsb{\psi}_0$ in the following standard form by an appropriate choice of basis vectors\begin{equation}
\label{eq: point}
	\gsb{\psi}_0
= c^0 \left(
	\begin{bmatrix}\,1\,\\[0.35ex]0\\[0.35ex]0\\[0.35ex]0\end{bmatrix}
	\otimes
	\begin{bmatrix}0\\[0.35ex]\,1\,\\[0.35ex]0\\[0.35ex]0\end{bmatrix}- 
	\begin{bmatrix}0\\[0.35ex]\,1\,\\[0.35ex]0\\[0.35ex]0\end{bmatrix}
	\otimes
	\begin{bmatrix}\,1\,\\[0.35ex]0\\[0.35ex]0\\[0.35ex]0\end{bmatrix}%
	_{\rule[-0ex]{0pt}{1.5ex}}^{\rule[-0ex]{0pt}{1.5ex}} \right)\,.
\end{equation}An arbitrary state in a neighborhood of $\gsb{\psi}_0$ can be written as an explicit holomorphic function of five complex coordinates $\{c^0, c^3_a, c^4_a, c^3_b, c^4_b\}$ as follows:
\begin{equation}
\label{eq: coords}
	\gsb{\psi}(\gsb{\mathrm{c}}) 
= c^0 \left(
	\begin{bmatrix}1\\[0.35ex]0\\[0.35ex]c^3_a\\[0.35ex]c^4_a\end{bmatrix}
	\otimes
	\begin{bmatrix}0\\[0.35ex]1\\[0.35ex]c^3_b\\[0.35ex]c^4_b\end{bmatrix}- 
	\begin{bmatrix}0\\[0.35ex]1\\[0.35ex]c^3_b\\[0.35ex]c^4_b\end{bmatrix}
	\otimes
	\begin{bmatrix}1\\[0.35ex]0\\[0.35ex]c^3_a\\[0.35ex]c^4_a%
	\end{bmatrix}_{\rule[-0ex]{0pt}{1.5ex}}^{\rule[-0ex]{0pt}{1.5ex}} \right)\,.
\end{equation}That the above variety is in fact equivalent to (\ref{eq: simple slater}) can be demonstrated by the ``push-button'' method of computing their respective \emph{Gr\"oebner bases}  \cite{Cox:2007} and verifying that these bases generate the same \emph{homogeneous ideal} in the variables$\{\psi_i\}$.\footnote{Specifically, the Gr\"obner basis of both (\ref{eq: simple slater}) and (\ref{eq: coords}) is found to be $\{
\psi_{1},  
\psi_{6}, 
\psi_{11}, 
\psi_{16}, 
\psi_{2}  + \psi_{5},  
\psi_{3}  + \psi_{9}, 
\psi_{7}  + \psi_{10},  
\psi_{12}  + \psi_{15},  
\psi_{4}  + \psi_{13},  
\psi_{8}  + \psi_{14},  
\psi_{10}  \psi_{13}  + \psi_{5}  \psi_{15}  - \psi_{9}  \psi_{14} \}$.  The disadvantage of the Gr\"obner basis is that it obscures the symmetries of the parametric representation (\ref{eq: simple slater}); for example the rule structure of the manifold is not evident.  Furthermore, implicit representations like this one are poorly suited to sectional and Riemannian curvature calculations.}

In the transition from (\ref{eq: simple slater}) to (\ref{eq: coords}), four pseudo-coordinates $\{c^1_a, c^2_a, c^1_b, c^2_b\}$ have disappeared and been replaced by the ``0'' and ``1'' entries.  These are physically accounted as follows: $c^1_a$ and $c^2_b$ have been merged into the overall (complex) normalization~$c^0$ by a simple rescaling; $c^1_b$~and~$c^2_1$ have tangent vectors that vanish at~$\gsb{\psi_0}$, such that they do not induce coordinate charts on $\gsb{\psi}(\gsb{\mathrm{c}})$ at $\gsb{\psi_0}$ and can safely be dropped. In~numerical calculations, their tangent vectors at $\gsb{\psi_0}$ are in the null-space of $g_{\alpha\sbar\beta}$ and $g^{\alpha\sbar\beta}$.

The remainder of the analytic calculation is straightforward.  The metric tensor components $g_{\mu\sbar\nu}$ are given from (\ref{eq: coords}) by (\ref{eq: Kahler potential}--\ref{eq: glower}), and when evaluated at  $\gsb{\psi}_0$, yield a $5{\times}5$ matrix that is diagonal and nonsingular.  Computing the inverse $g^{\sbar\nu\mu}$ is therefore trivial. The Riemann components $R_{\alpha\sbar\beta\gamma\sbar\delta}$ are given by (\ref{eq: Riemann components}) and the Ricci components $R_{\mu\sbar\nu}$ are given by (\ref{eq: Ricci Kahlerian}). Finally, the scalar Riemann curvature given by (\ref{eq: scalar curvature}) is found to be $R=-8/\kappa$,  in~accord with the numerical result.

This reasoning is readily generalized.  Upon antisymmetrizing a general rank-one, order-$n$ product-sum state (see Fig.~\ref{fig: QMOR product sum}) under the exchange of all pairs of spins---thus converting it to a single $n$-particle Slater determinant---and evaluating the metric and Riemann  tensors in the diagonal basis of (\ref{eq: point}--\ref{eq: coords}), the dimensionality of the \Kahler manifold is evidently $\dim \lcal{K} = 2(1+n(n_{\text{orb}}-n))$, and the scalar curvature is found in closed analytic form to be\begin{equation}
\label{eq: analytic Slater curvature}
R = -\frac{2}{\kappa} \times 
\begin{cases}
n(n-1)(n_{\text{orb}}-n)(n_{\text{orb}}-n-1) & \text{for $n_{\text{orb}}\ge n$}\\
\text{undefined because $\gsb{\psi}(\gsb{\mathrm{c}}) = 0$}&\text{for $n_{\text{orb}} < n$}\,. 
\end{cases}
\end{equation}Here  $n_{\text{orb}}= 2j+1$ is the dimension of the individual states in the Slater product-sum, with the mnemonic ``orb'' referring to ``orbitals'' in token of the practical use of these states in quantum chemistry. Since the scalar curvature is a geometric invariant, this result holds in any basis.  

The curvature of a Slater determinant manifold having a Fubini-Study metric, \emph{i.e.}, having a \Kahler potential $
\kappa = 1/2 \log \inner{\gsbbar{\psi}}{\gsb{\psi}}$ instead of $\kappa = 1/2 \inner{\gsbbar{\psi}}{\gsb{\psi}}$, can similarly be calculated from (\ref{eq: coords}) and its multidimensional generalizations.  The result is\begin{equation}
\label{eq: Fubini-Study Slater curvature}
R =  
\begin{cases}
4 n n_{\text{orb}} (n_{\text{orb}}-n) & \text{for $n_{\text{orb}}\ge n$}\\
\text{undefined because $\gsb{\psi}(\gsb{\mathrm{c}}) = 0$}&\text{for $n_{\text{orb}} < n$}\,. 
\end{cases}
\end{equation}Physically speaking, the Fubini-Study metric describes a manifold of normalized states in which a phase rotation of $\delta\phi$ radians has a path length of zero rather than $\delta\phi$.  Furthermore, it is a straightforward (but lengthy) algebraic exercise to verify during the course of the above calculation that $g_{\alpha\sbar\beta}\propto R_{\alpha\sbar\beta}$, \emph{i.e.}, Slater determinant manifolds having a Fubini-Study metric are \emph{Ricci-Einstein manifolds} \cite{Besse:1987}. An immediate consequence is that Slater determinant manifolds are solitons under Ricci flow \cite{Chow:04b}.   We note that Ricci soliton manifolds are of central importance to the mathematical community, because they represent (in sense that can be made precise) the unique ``smoothest'' manifolds of a given topological class. In recent years this idea has been central to the \emph{Ricci flow} \cite{Chow:07} proofs of several long-standing topological conjectures. 

To the best of our knowledge, the above algebraic/geometric properties of Slater determinants have never been noted in the literature. This gap is noteworthy, in view of the central role that Slater determinants play in chemistry and condensed-matter physics, and the similarly central role of \Kahlerian algebraic geometry in numerous branches of mathematics.  It reflects what the National Research Council has called \cite{MCTCC:95}\begin{quote}\makebox[0pt][r]{``}[\ldots the anomaly that] although theoretical chemists understand sophisticated mathematics and make heavy use of the mathematical literature, they have typically not involved mathematicians directly in either the development of models or algorithms or the derivation of formal properties of equations and
solutions. In fact, theoretical chemists have become accustomed to
self-reliance in mathematics.''\end{quote}A central objective of this article's geometric approach to quantum simulation is to more closely link the formal mathematical tools of algebraic geometry to practical problems in quantum simulation, and thereby, to help ensure that the emerging discipline of quantum system engineering is not needlessly ``self-reliant in mathematics.''

\subsection{Slater determinants are Grassmannian \GK (\GGK) manifolds}
We thank our colleague Joshua Kantor for directing our attention to the facts that the Slater determinants associated with $n$ particles distributed among $n_\text{orb}$ orbitals have a natural isomorphism to the \emph{Grassmannian manifolds} classified as $G(n,n_\text{orb})$ \cite{Harris:1992}, that Grassmannian manifolds have a known presentation as projective algebraic varieties via the \emph{Pl\"ucker embedding} \cite{Fischer:2001}, and that the Pl\"ucker embedding is possessed of isometries having a Lie group structure that is known to be compatible with a K\"ahler-Einstein metric \cite{Besse:1987}.  

From this algebraic geometry point of view, the example of a Slater determinant that we present in (\ref{eq: simple slater}), having $n=2$ and $n_\text{orb}=4$, can be identified with what Harris' classic textbook \emph{Algebraic Geometry} calls \cite[p.~65]{Harris:1992} ``the first nontrivial Grassmannian---the first that is not a projective space---[namely] $G(2,4)$.''  Further investigation of this convergence of algebraic geometry with quantum chemistry is in progress.

\subsection{Practical curvature calculations for \QMOR on \GK manifolds} 
\label{sec: efflorescent curvature}
Now our appetite is whetted to ask even more difficult questions: what are the curvature properties of higher-rank \GK manifolds? And what are the implications of these curvature properties for quantum simulation?  We will turn to numerical experiments to learn more about these issues.

As we begin numerical curvature calculations on \Kahler state-spaces of increasingly large dimension, we first consider what we may expect to learn from these calculations that would have practical implications for \MOR.  

It is known \cite{Friedan:85,Mulle:99} that at each point on \lcal{K} we can construct locally Euclidean \emph{Riemann normal coordinates} such that the local curvature tensor has the expansion\begin{equation}
g_{ab} = \delta_{ab} - \sum_{c,d=1}^{\dimK}\tfrac{1}{3}\,R(0)_{cadb}y^c y^d
\end{equation}
and the local volume element $\mathrm{d}V\propto|\,{\det g}\,|^{1/2}$ therefore has the expansion
\begin{equation}
\mathrm{d}V  = \mathrm{d}V(0)\left(1 - \sum_{a,b=1}^{\dimK}\tfrac{1}{6}\,R(0)_{ab}y^ay^b \right)
.
\end{equation}
We see that the square roots of the eigenvalues of the Ricci tensor determine the length scale over which curvature effects dominate the volume of the state-space.  Physically speaking, this sets the length scale over which gabion petals are locally Euclidean.  We further see that negative curvature is associated with an exponential ``flowering'' of the state-space volume, in accord with the beautiful knitted representations of hyperbolic spaces by Taimina and Henderson \cite{Henderson:01}.  

We note that since the Ricci tensor eigenvalues are geometric invariants (meaning specifically, the eigenvalues of the mixed Ricci tensor $R^{a}_{\,b} = \sum_{c} g^{ac}R_{cb}$ are geometric invariants), we can compute them in any coordinate system we please.  We further note that our adoption of a over-complete \Kahlerian basis creates no anomalies in the Ricci eigenvalue distribution, because the extra eigenvalues of $R^{\alpha}_{\,\beta} = \sum_{\sbar\gamma} g^{\alpha\sbar\gamma}R_{\sbar\gamma\beta}$ vanish identically, due to the projective definition (\ref{eq: g pseudoinverse}) of $g^{\alpha\sbar\gamma}$.

Our first goal, therefore, will be to compute the Ricci tensor eigenvalues for high-rank \GK manifolds, expecting thereby to gain quantitative insight into the ``flowering'' gabion geometry that is depicted in Fig.~\ref{fig: QMOR geometry}(H).

As our first test case, we will consider the rank-6, order-9, spin-1/2 \GK manifold.  This manifold has (real) dimension $\dimK = 2\times9\times(6+1) = 126$ and it is embedded in a Hilbert space of (real) dimension $\dimH=2\times2^6=128$.  The gabion state-space therefore ``almost'' fills the Hilbert state-space, since only $\codim \lcal{K} = \dimH-\dimK=2$ dimensions are missing.

\begin{figure}[t]\centering
\vspace*{2ex}
\includegraphics[width=0.75\textwidth]{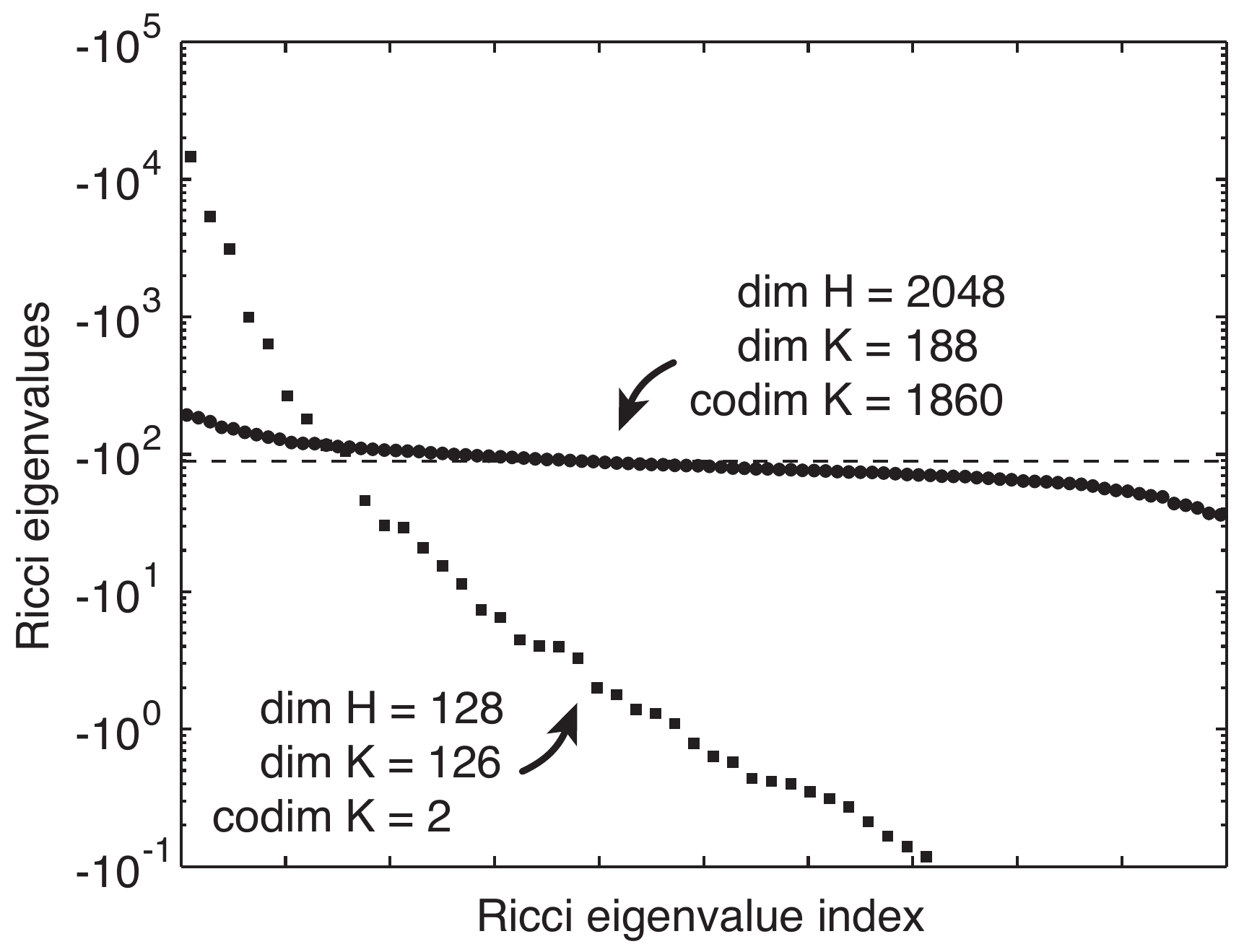}\hspace*{2em}\\
\begin{minipage}{0.8\textwidth}
\caption[Typical Ricci tensor eigenvalues for gabion-\Kahler manifolds]{%
\protect\justifying%
Typical Ricci tensor eigenvalues for gabion-\Kahler manifolds.\\[1ex]
\label{fig: QMOR Ricci eigenvalues}%
Points on the gabion-\Kahler manifold are randomly selected by first randomly generating an independent normalized product state for each gabion rank (\emph{i.e.}, each row of Fig.~\ref{fig: QMOR product sum}), then summing the ranks, then normalizing the final state.  The sparsely dotted line contains typical Ricci eigenvalues of the spin-1/2, rank-9, order-6 gabion, the densely dotted line contains the eigenvalues of the spin-1/2, rank-9, order-10 gabion.  The dashed line is an empirical \mbox{``$\text{rank}{\times}\text{order}$''} estimate of the mean Ricci eigenvalue of gabion-\Kahler manifolds having large-codimension. }
\end{minipage}
\end{figure}

Choosing a random point in $\lcal{K}$ and computing the Ricci eigenvalues numerically yields the typical eigenvalues shown in Fig.~\ref{fig: QMOR Ricci eigenvalues}.  We take the square root of the largest eigenvalue to be (roughly) the linear extent of a gabion petal; these petals evidently have an extent of order 0.01.

We emphasize that $\lcal{K}$ is large enough to contain exponentially many such petals.  See Nielsen and Chuang for a quantitative analysis \cite[Section~4.54]{Nielsen:00}, noting that the ``patches'' of Nielsen and Chuang are broadly equivalent to our petals and therefore, their Fig.~4.18 is broadly equivalent to our Fig.~\ref{fig: QMOR geometry}(H).  This mathematically justifies our physical picture of a gabion as a geometric ``flower'' having exponentially many petals.  Later on, in Sections~\ref{sec: error-correcting codes}--\ref{sec: design of sampling matrices}, we will rigorously count the number of petals using ideas from coding theory.

\subsection{Numerical results for projective \QMOR onto \GK manifolds}
\label{sec: expectation} Suppose we generate a random (normalized) quantum $\gsb{\psi}_0$, and numerically search for a highest-fidelity projection of $\gsb{\psi}_0$ onto $\lcal{K}$.  We will call this high-fidelity image point $\gsb{\psi}_{\lcal{K}}$.  Based on our geometric analysis so far, we have two strong expectations relating to quantum \MOR by projection onto $\lcal{K}$.    We expect first, that projections exist for which $|\gsb{\psi}_{\lcal{K}}-\gsb{\psi}_0|$ is no greater than $\sim 0.01$, since at greater separations the negative curvature of $\lcal{K}$ is great enough to ensure a better solution, via the mechanism of Fig.~\ref{fig: QMOR geometry}(G).   Second, we expect that the numerical search for $\gsb{\psi}_{\lcal{K}}$ will be well-conditioned. That is, it will robustly converge to a high-fidelity representation, despite the exponentially convoluted geometry of $\lcal{K}$, without becoming trapped in local minima (because every local section that contains a rule is negatively curved).

In numerical trials both expectations are fulfilled.  The achieved fidelity is in good accord with the geometric expectation of ${\sim}0.01$: the median value of $|\gsb{\psi}_{\lcal{K}}-\gsb{\psi}_0|$ in a trial of~100 projective reductions was $0.005$, and the maximal value was $0.025$.  

We computed the reduced-order \GK representation $\gsb{\psi}_{\lcal{K}}$ by a simple gradient search.  Specifically, we integrated to convergence the dynamical equation of Step~B.3 of Fig.~\ref{fig: QMOR numerical}, with the potential $\phi$ replaced by
\begin{equation}
\gsb{\phi} \to \frac{1}{2}\,\gsbbar{\psi}(\lb{\mathrm{c}})_{\!,a}\cdot
\big(\gsb{\psi}_0-\gsb{\psi}_{\lcal{K}}(\lb{\mathrm{c}})\big)
.
\end{equation}
Upon conversion to \Kahler index notation, the resulting reduction equation is manifestly geometrically  covariant:
\begin{equation}
\label{eq: trajectory}
\frac{\partial c^\alpha(t)}{\partial t} = \frac{1}{2}\,\sum_{\sbar\beta=\sbar 1}^{\dimC}
g^{\alpha\sbar\beta}(\lb{\mathrm{c}})
\left[
\frac{\gsbbar{\psi}(\lb{\mathrm{c}})}{\partial c^{\sbar\beta}}\cdot
\big(\gsb{\psi}_0-\gsb{\psi}(\lb{\mathrm{c}})\big)
\right]\,.
\end{equation}
From an algorithmic point of view, this reduction equation can be regarded as a downhill gradient search for a reduced-order representation $\gsb{\psi}_{\lcal{K}} = \lim_{t\to\infty}\gsb{\psi}(\lb{\mathrm c}(t))$ of $\gsb{\psi}_0$.  The search evidently converges when $\gsb{\psi}_{\lcal{K}}$ is directly ``underneath'' $\gsb{\psi}_0$, \emph{i.e}, when $P_{\lcal{K}}(\gsb{\psi}_{\lcal{K}})\big(\gsb{\psi}_0-\gsb{\psi}_{\lcal{K}}\big)=0$, where we recall that $P_{\lcal{K}}(\gsb{\psi}_{\lcal{K}})$ projects vectors in $\lcal{H}$ onto the tangent space of $\lcal{K}$ at $\gsb{\psi}_{\lcal{K}}$.
As an alternative to gradient search, we note that Beylkin and Mohlenkamp \cite[sec.~3.1]{Beylkin:05} describe an alternating least-squares algorithm that in their hands gives excellent results, but we have not ourselves tried this method.  In Section~\ref{sec compressible} we derive the above equation (\ref{eq: trajectory}) using the language (and nomenclature) of compressive sampling (\CS), and we describe the numerical calculations in greater detail.  

No attempt was made to optimize the efficiency of the gradient search; instead we simply reused the existing dynamical code implementing Step~B.3 of of Fig.~\ref{fig: QMOR numerical} (this code exploits the gabion algebraic structure of $\gsb{\psi}(\lb{\mathrm{c}})$ to evaluate (\ref{eq: trajectory}) efficiently).   This duplication of internal algorithmic structure again illustrates the intimate relation between trajectory calculations and geometric calculations.  

No gross failures of convergence, such as might be expected from trapping of the trajectory induced by (\ref{eq: trajectory}) on a distant gabion ``petal,'' were observed in this (or any)  of our numerical trials.  
Our geometric analysis explains this robustness as originating in the negative sectional curvature of the ruled net of the gabion state-space $\lcal{K}$, as depicted in Fig.~\ref{fig: QMOR geometry}(G).  
We now increase the order of the gabion to 10, keeping the rank at 9.  This increases the dimensionality of the Hilbert space to $2\times 2^{10} = 2048$, and the dimensionality of the \Kahlerian gabion to $\dimK = 2\times9\times(10+1)=188$.  With $\codim \lcal{K}=1048-188=1860$, our model order reduction is now discarding ${\sim}\,90\%$ of the dimensions of the larger Hilbert space.  Thus, in deliberate contrast to the previous example, the \MOR is now fairly aggressive. Typical resulting Ricci eigenvalues are shown in Fig.~\ref{fig: QMOR Ricci eigenvalues}. 

A pronounced flattening of the eigenvalue distribution is evident in this large-$\codim$ example.  The plotted straight line is simply $\text{rank}\times\text{order}$, which seems empirically to describe the average Ricci eigenvalue in this particular case, and also in many other trials that we have run.  This rule-of-thumb is simply the analytic result for rank-1 curvature (\ref{eq: analytic curvature}) multiplied by the rank. 

There is at present no analytic theory that justifies this rule-of-thumb, or explains the observed flattening of the eigenvalue distribution. Obviously, such a theory would be welcome.  We remark that to the best of our knowledge, the Ricci eigenvalues of Fig.~\ref{fig: QMOR Ricci eigenvalues}, having $\dim 
\lcal{K} = 188$, are the largest-dimension curvature eigenvalues ever numerically computed.  

We then calculated projective \MOR approximations by integrating (\ref{eq: trajectory}) as before.  We generated our targets $\gsb{\psi}_0$ by randomly selecting target points on $\lcal{K}$, and moving a distance of 0.05 along a random vector $\lbhat{n}$ perpendicular to $\lcal{K}$. 

Again, robust convergence to high-fidelity reduced-order representations was observed.  Of 100 trials, in 98 cases the separation distance was precisely $0.050$, representing convergence to the correct ``petal.''  In the remaining two cases the separation distances were $0.123$ and $0.120$, representing sporadic convergence to a wrong-but-nearby ``petal.''   Thus even wrong-petal convergence yielded a high-fidelity representation.  This robustness can again be ascribed to the negative sectional curvature of the state-space manifold's ruled net, according to the geometric mechanism depicted in Fig.~\ref{fig: QMOR geometry}(G). 

\subsection{Avenues for research in geometric quantum mechanics}  
\label{sec: geometric avenues}  
To begin, the results of the preceding section are understood only qualitatively, and rigorous bounds on projective fidelity and robustness would be very welcome.  How much of the preceding section can be understood solely as a consequence of the known sectional curvature properties of the \GK manifolds?

Although our presentation focuses on practical applications of quantum \MOR, a broad class of fundamental physics questions can be given a geometric interpretation.  We temporarily adopt the point of view of geometric quantum mechanics in which---as reviewed in Section~\ref{subsection: Overview}---the manifold $\lcal{K}$ is regarded as the ``real'' arena on which physics takes place.  

We begin by noting that that the rule-field equation (\ref{eq: geodesics}) specifies the dimensionality of Hilbert space to be the number of (linearly independent) scalar rule-fields that the (postulated) ``real'' \GK manifold of geometric quantum mechanics supports.  The question then arises, what determines the number of these embedding state-space fields?  The present article suggests that this question is best investigated from a blended informatic-algebraic-geometric point of view.

As another example, suppose $p$ and $q$ are arbitrary linear Hermitian operators in the embedding Hilbert space \lcal{H}.  At a given point of the Kahler manifold \lcal{K}, specified by a state $\ket{\psi(\lb{\mathrm c})}$, we can construct tangent vectors $V_q$ and $V_p$ whose components are
\begin{equation}
\label{eq: GQM}
v_q^\alpha = \sum_{\sbar\beta=\sbar1}^{\dimC}\,
g^{\alpha\sbar\beta}\,\frac{\partial}{\partial\sbar{c}^{\sbar\beta}} \braket{\sbar\psi(\lbbar{\mathrm c})}{q}{\psi(\lb{\mathrm c})}\qquad\text{and}\qquad {\sbar v}_q^{\sbar\alpha} = (v_q^\alpha)^\star
\end{equation}and similarly for $V_p$.  With this normalization we have $g(V_q,V_q) = \braket{\sbar\psi}{qP_{\lcal{K}}q}{\psi}$.  Physically speaking, $V_q$ defines a \Kahlerian velocity field along which the projected dynamical equation $\partial\ket{\psi(t)}/\partial t=P_{\lcal{K}}q\ket{\psi(t)}$ moves state trajectories.

The sectional curvature $S(V_p,V_q)$ then can be regarded as a fundamental property of the ``true'' physical manifold $\lcal{K}$.  The \emph{Theorema Egregium} guarantees that the sectional curvature is an intrinsic property of $\lcal{K}$, and our physical intuition suggests that it should  therefore be measurable.

A host of fundamental questions then arise quite naturally, that intimately unite physics and mathematics in the context of quantum simulation.   If the quantum sectional curvature is physically measurable---whether in reality or within projective simulations---then by what kinds of experiment?  Are these experiments practical in our real world?  Is there any experimental evidence already at-hand to indicate that quantum sectional curvature vanishes in the physical world?  If so, to what precision has this been verified?   

It is apparent that quantum chemists can apply a reverse strategy to create a geometric context for assessing the fidelity of chemical quantum simulations.  Set $p$ to be the kinetic energy of the electrons of a molecule, and $q$ to be the potential energy (including perturbations due to applied potentials), and set the gabion state-space to be a sum of Slater determinants. Then the vanishing of the sectional curvature $S(V_p,V_q)$ indicates that the state-space section generated by $\{p,q\}$ is Euclidean, as is (presumably?) desirable in chemical simulations, most particularly, density functional theory (\DFT) simulations.

And finally, to anticipate, in Section~\ref{sec: QMOR respects the Theorema Dilectum} we will discuss how the \emph{Theorema Dilectum} manifests itself upon \GK state-spaces.  This will turn out to raise thorny issues of how causality works in geometric quantum mechanics. For the present we will say no more about these difficult fundamental questions, instead referring the reader to Leggett's recent review \cite{Leggett:2002it}, and we return instead to our central topic of practical quantum spin simulations.

\subsection{Summary of the geometric analysis}  
\label{sec: summary}

We have seen that the ruled net of a \GK manifold, with its associated nonpositive sectional curvature, plays a geometric role in quantum simulation that is broadly similar to the role of polytope convexity in linear programming, namely, the role of providing geometric foundations for developing efficient and robust algorithms.  However, understanding polytope convexity has been a slow process, and this research is still being pursued along multiple mathematical lines that include algebra, geometry, information theory, and their numerous hybridizations.  We foresee that the role of algebraic geometry in classical and quantum simulations will be elucidated along similarly multidisciplinary lines, and we are conscious that the analysis presented here has barely begun this enterprise.

Even in its present early stage, geometric analysis provides grounds for confidence that quantum model order reduction and trajectory simulation can be achieved with high efficiency, fidelity, and robustness, \emph{provided} that some physical mechanism can be found to compress quantum trajectories onto the petals of gabion-type state-spaces.  That needed physical mechanism is, of course, the \emph{Theorema Dilectum}, which will be the topic of the section that follows.

\section{Designing and Implementing Large-Scale Quantum Simulations}
\label{sec: designing and implementing}

Our goal in this section is to join the geometric ideas and theorems of the preceding section with the well-known ideas and established design principles of linear quantum mechanics.  

We have reason to worry that perhaps very few principles of linear quantum mechanics will survive the transition to  reduced-order quantum \MOR mechanics. After all, we have seen that gabion-\Kahler (\GK) state-spaces are strongly curved, so that when we project quantum trajectories onto them, we (seemingly) dispense with all of the mathematical properties of Hilbert space that depend upon its linearity.  Furthermore, depending upon the degree of the model order reduction imposed, the \GK projection of \QMOR mechanics may even discard all but an exponentially small fraction of the dimensions of the embedding Hilbert space.
 
We seek, therefore, to establish that the principles of linear quantum mechanics hold true in \QMOR simulations, and in particular, to establish precisely the mechanisms by which they \emph{can} hold true.

\subsection{Organization and nomenclature of the presentation}

For convenience, whenever we derive a result that is novel or expressed in a new form, we state it as a formal (numbered) design rule that is accompanied (as needed) by formal (numbered) definitions.  On the other hand, whenever a result is unsurprising, or can be found in the literature, and is obtained from previously given equations by standard manipulations and reductions, we outline the derivation but omit details.  

We define the subject of our analysis to be \emph{\QMOR mechanics}:

\begin{mydefinition}
\emph{\QMOR mechanics} is the mechanics of a physical system simulated according to the orthodox principles of linear quantum mechanics, as modified by projection onto a lower-dimension manifold having a \Kahler geometry, for purposes of quantum model order reduction (\QMOR), as~concretely embodied in the algorithms and algebraic  structures of Figs.~\ref{fig: QMOR formal}--\ref{fig: QMOR geometry}.  
\end{mydefinition}
\noindent We seek to construct recipes by which \QMOR mechanics simulates linear quantum mechanics as closely as feasible.  To recapitulate the objectives of Section \ref{sec: Constraints upon the Analysis}, our analysis seeks to be \emph{orthodox} in its respect for linear quantum mechanics,  \emph{operational} in the traceability of its predictions to measurement processes, and \emph{reductive} in the sense that its principles are summarized by closed-form analytic design rules.  Our analysis strives also to be \emph{synoptic} in the sense that whenever we choose between equivalent analysis formalisms, we state a rationale for our choice, and we note the practical consequences of alternative choices. 

\subsection{\QMOR respects the principles of quantum mechanics} 
\label{sec: QMOR respects}

We begin by establishing that the mathematical structure of \QMOR mechanics is sufficiently rich to respect the following principles of quantum and classical mechanics:
\begin{QSEitemize} 
\item the causal invariance of the \emph{Theorema Dilectum} is respected (Sec.~\ref{sec: QMOR respects the Theorema Dilectum}), 
\item the entropy of systems in thermodynamic equilibrium is respected (Sec.~\ref{sec: QMOR respects the thermal equilibrium}),
\item the principles of classical linear control theory are respected (Sec.~\ref{sec: QMOR respects operational structure}).
\item the quantum limits to measurement noise and back-action are respected (Sec.~\ref{sec: QMOR respects the fundamental quantum limits}).
\end{QSEitemize}
If \QMOR mechanics did not respect these principles, it would scarcely be useful for simulating real-world quantum systems.  Conversely, by stating these principles in a quantitative, we gain a fairly clear picture of the path that our \QMOR analysis needs to take. 

\subsubsection{\QMOR respects the \emph{Theorema Dilectum}}
\label{sec: QMOR respects the Theorema Dilectum}

We now state the \emph{Theorema Dilectum} in an algebraic form that is well-adapted to the formal quantum \MOR algorithm of Fig{.}~\ref{fig: QMOR formal}.   Our mathematical analysis parallels that of Nielsen and Chuang \cite[see their Theorem~8.2 in Section~8.2]{Nielsen:00}, whose analysis in turn derives from a 1975 theorem of Choi~\cite{Choi:75} (see Section~\ref{sec: Naming and applying}).  

We suppose that at the end of step $n$ the \QMOR simulation algorithm of Fig{.}~\ref{fig: QMOR formal} has computed a wave-function $\ket{\psi_n}$.  For simplicity, we further suppose that precisely one measurement operator pair $\{M_{\myupsymbol},M_{\mydownsymbol}\}$ acts during the subsequent time-step.  These operators satisfy the normalization condition
\begin{equation}
\label{eq: measurement normalization}
	M_{\myupsymbol} M_{\myupsymbol}^\dagger + 
	M_{\mydownsymbol} M_{\mydownsymbol}^\dagger = I
\end{equation}
where $I$ is the identity operator.  In the absence of \GK projection the post-timestep density matrix $\rho_{n+1}$ is readily shown to be
\begin{equation}
\label{eq: linear evolution}
	\rho_{n+1} = 
	M_{\myupsymbol}\rho_n M_{\myupsymbol}^\dagger + 
	M_{\mydownsymbol}\rho_n M_{\mydownsymbol}^\dagger 
	\equiv \lcal{L}[\rho_n]
\end{equation}
where $\rho_n \equiv \ket{\psi_n}\bra{\psi_n}$.  This expression implicitly defines the well-known \emph{linear superoperator} $\lcal{L}$ to be that linear operation on (Hermitian) matrices that takes $\rho_n\rightarrow\rho_{n+1}$.  The existence and strict positivity of this linear map is one of the main defining characteristics of linear quantum mechanics (as reviewed in Section~\ref{sec: Naming and applying}).

Now we ask  ``What mathematical operations upon  $\{M_{\myupsymbol},$ $M_{\mydownsymbol}\}$ leave $\lcal{L}$ invariant?''   The \emph{Theorema Dilectum} in its algebraic form due to Choi~\cite{Choi:75} states that all such invariance operations are of the general form 
\begin{equation}
\label{eq: theorema dilectum}
\left[\begin{array}{c}
M_{\myupsymbol}\\
M_{\mydownsymbol}
\end{array}\right] \rightarrow \,\textrm{U} \left[\begin{array}{c}
M_{\myupsymbol}\\
M_{\mydownsymbol}
\end{array}\right],
\end{equation}
where $\textrm{U}$ is an arbitrary $2\times2$ unitary matrix of complex numbers (\emph{i.e.}~a matrix acting on the linear space of measurement operators, not the Hilbert space of $\ket{\psi_n}$, such that the matrix elements of $\textrm{U}$ are $c$-numbers).  This is the sole general mathematical invariance of the first two steps of the simulation algorithm of Fig.~\ref{fig: QMOR formal}, and so (from the \QMOR point of view) it is the most fundamental invariance of linear quantum mechanics.  

In Section~\ref{sec: Causality and scattering}, we will establish that $U$-transform invariance enforces physical causality.  We are thereby motivated to ask ``How is the $U$-transform invariance of the \emph{Theorema Dilectum} affected by \GK projection?''  According to the algorithm of Fig.~\ref{fig: QMOR formal}, \GK projection modifies (\ref{eq: linear evolution}) to 
\begin{eqnarray}
	\nonumber
	\lefteqn{\rho_{n+1}=%
	(P_{\scriptscriptstyle\lcal{K}})_nM_{\myupsymbol}\ket{\psi_n}\bra{\psi_n} M_{\myupsymbol}^\dagger (P_{\scriptscriptstyle\lcal{K}})_n 
	\ \displaystyle\frac{\braket{\psi_n}{M_{\myupsymbol}^\dagger M_{\myupsymbol}}{\psi_n}}
	  {\braket{\psi_n}{M_{\myupsymbol}^\dagger (P_{\scriptscriptstyle\lcal{K}})_n M_{\myupsymbol}}{\psi_n}}}\\
	  &\qquad\ +\ 
	(P_{\scriptscriptstyle\lcal{K}})_nM_{\mydownsymbol}\ket{\psi_n}\bra{\psi_n} M_{\mydownsymbol}^\dagger (P_{\scriptscriptstyle\lcal{K}})_n 
	\ \displaystyle\frac{\braket{\psi_n}{M_{\mydownsymbol}^\dagger M_{\mydownsymbol}}{\psi_n}}
	  {\braket{\psi_n}{M_{\mydownsymbol}^\dagger (P_{\scriptscriptstyle\lcal{K}})_n M_{\mydownsymbol}}{\psi_n}}
\label{eq: nonlinear evolution}
\end{eqnarray}
where we recall that $(P_{\scriptscriptstyle\lcal{K}})_n\equiv P_{\scriptscriptstyle\lcal{K}}(\ket{\psi_n})$ was defined in (\ref{eq: PK projection}) as the operator that projects onto the local tangent space of the \GK manifold at $\ket{\psi_n}$.  It is easy to check that $\tr[\rho_{n+1}]=1$ (\emph{i.e.}, probability is conserved), and that above expression reduces to the linear result (\ref{eq: linear evolution}) whenever the commutators $[(P_{\scriptscriptstyle\lcal{K}})_n,M_{\myupsymbol}]$ and $[(P_{\scriptscriptstyle\lcal{K}})_n,M_{\mydownsymbol}]$ both vanish.  

As it stands, the nonlinear projective evolution (\ref{eq: nonlinear evolution}) is not $U$-transform invariant, and hence it does \emph{not} always respect the \emph{Theorema Dilectum}.  We therefore introduce the further assumption (which our \QMOR simulations will always respect) that both $M_{\myupsymbol}$ and $M_{\mydownsymbol}$ are near to being multiples of the identity operator.  We quantify ``near''  by introducing an artificial expansion parameter $\epsilon$ such that
\begin{eqnarray}
  \label{eq: close definitions}
  M_{\myupsymbol} = c I + 
  \epsilon\,\delta M_{\myupsymbol}
  \qquad
   \delta M_{\mydownsymbol} = s I  + 
  \epsilon\,\delta M_{\mydownsymbol} 
\end{eqnarray} 
where 
\begin{eqnarray}
\label{eq: c and s}
c = \tr M_{\myupsymbol}/\dim M_{\myupsymbol}
\qquad
s = \tr M_{\myupsymbol}/\dim M_{\myupsymbol}
\end{eqnarray}
Equations (\ref{eq: close definitions}--\ref{eq: c and s}) uniquely define the products $\epsilon\,\delta M_{\myupsymbol}$ and $\epsilon\,\delta M_{\mydownsymbol}$.  Therefore we can uniquely define $\epsilon$, and thereby uniquely define $\delta M_{\myupsymbol}$ and $\delta M_{\mydownsymbol}$ too, as follows:
\begin{eqnarray}
\label{eq: epsilon definition}	
\epsilon^2 = \tr[(\epsilon\,\delta M_{\myupsymbol})^\dagger\,(\epsilon\,\delta M_{\myupsymbol})+
	(\epsilon\,\delta M_{\mydownsymbol})^\dagger\,(\epsilon\,\delta M_{\mydownsymbol})]/\dim M_{\myupsymbol}
\end{eqnarray}
Equations (\ref{eq: measurement normalization}--\ref{eq: epsilon definition}) then imply the exact normalization relations 
\begin{align}
|c|^2+|s|^2 + |\epsilon|^2 = {} & 1\\
\tr[(\delta M_{\myupsymbol})^\dagger\,(\delta M_{\myupsymbol})+
		(\delta M_{\mydownsymbol})^\dagger\,(\delta M_{\mydownsymbol})] = {} & \dim M_{\myupsymbol}\,.
\label{eq: delta M normalization}
\end{align}
and so we can regard all operator products involving $\delta M_{\myupsymbol}$ and $\delta M_{\mydownsymbol}$ to be $\Or(1)$.   

In aggregate, our definitions and normalizations ensure that $\rho_{n+1}$ as given by (\ref{eq: nonlinear evolution}) has a well-defined power series expansion in $\epsilon$; it is therefore a straightforward (but not short) calculation to verify that this expansion can be written in the form:
\begin{align}
\nonumber
	\rho_{n+1} = {}& (P_{\scriptscriptstyle\lcal{K}})_n\,\lcal{L}\big[\,\!\ket{\psi_n}\bra{\psi_n}\,\!\big]\,(P_{\scriptscriptstyle\lcal{K}})_n\ \\
	&{}+\ket{\psi_n}\bra{\psi_n}\ \,{\tr}\left[%
      \rule{0pt}{2ex}%
      (\bar P_{\scriptscriptstyle\lcal{K}})_n%
      \lcal{L}\big[\ket{\psi_n}\bra{\psi_n}\big]\,
      (\bar P_{\scriptscriptstyle\lcal{K}})_n\right]
      +{\Or}\!\left[\frac{\,\epsilon^3\!}{|c|},\frac{\,\epsilon^3\!}{|s|}\hspace{0.2em}\right] 
\label{eq: nonlinear theorema dilectum}
\end{align}
where $\bar P_{\scriptscriptstyle\lcal{K}} = I- P_{\scriptscriptstyle\lcal{K}}$.  We note that the leading terms are determined solely by $\lcal{L}$ and $P_{\scriptscriptstyle\lcal{K}}$, and therefore are  invariant under the $U$-transform (\ref{eq: theorema dilectum}) of the \emph{Theorema Dilectum}.  Furthermore, it can be shown that the value of the small parameter $\epsilon$ is itself invariant under the $U$-transform, and so is the sum $|c|^2 + |s|^2$.  And finally, we have calculated the (exact) $c$- and $s$-dependence of the $\Or(\epsilon^3)$ terms. 

We will establish in Section~\ref{sec: Physical foundations} that in the continuum limit of infinitesimally small time step intervals $\delta t$, physical quantities (for example, relaxation rates) are $\lcal{O}(\epsilon^2/\delta t)$.  Physically this means that the $\lcal{O}(\epsilon^3)$ terms in (\ref{eq: nonlinear theorema dilectum}) are negligible in the continuum limit, provided the technical conditions $|c|>0$ and $|s|>0$ are  satisfied (these technical conditions provide the rationale for calculating the $c$- and $s$-dependence of the $\Or(\epsilon^3)$ terms in (\ref{eq: nonlinear theorema dilectum})\hspace{0.1em}).  These results motivate us to adopt from Carlton Caves \cite{Caves:00} the following definition:

\begin{mydefinition}
Measurement operations satisfying $\epsilon \ll |c|$ and $\epsilon \ll |s|$ are called \emph{measurement operations of the first class} (or sometimes \emph{first-class measurements}).  
\end{mydefinition} 

\noindent The result (\ref{eq: nonlinear theorema dilectum}) then expresses the following fundamental design rule of \QMOR mechanics: 
\begin{myDesignRule} 
\label{dr: theorema dilectum}
In the continuum limit, quantum trajectory simulations of first-class measurement processes, as projected onto   state-spaces of gabion-\Kahler type (\QMOR simulations for short), respect the unitary invariance of the \emph{Theorema Dilectum}.
\end{myDesignRule}
\noindent%
In physical terms, first-class measurements are characterized in the continuum limit by stochastic drift and diffusion processes on \GK manifolds, rather than quantum jumps.  To ensure that the \emph{Theorema Dilectum} is respected, \QMOR mechanics must therefore simulate all quantum systems wholly in terms of drift and diffusion processes upon strongly-curved, low-dimension state-space manifolds of gabion-\Kahler type.  

This low-dimension curved-geometry stochastic description  of \QMOR mechanics thus contrasts sharply with the high-dimension linear-geometry ``jump-oriented'' description of quantum mechanics that is commonly given in textbooks; that is why many of our derivation methods and design rules are novel. 

Turning our attention briefly to the broader context of geometric quantum mechanics (see Sections~\ref{sec: intro to geometric} and \ref{sec: geometric avenues}), we emphasize that we do \emph{not} regard issues of causality in geometric quantum mechanics as settled by the projective generalization of the \emph{Theorema Dilectum} given in (\ref{eq: nonlinear theorema dilectum}).  We will not discuss this topic further, partly for reasons of space, but mainly because we regard it as being an exceptionally difficult and subtle topic that is intimately bound-up with the question of the existence (or nonexistence) of quantum field theory in geometric quantum mechanics.

\subsubsection{\QMOR respects thermal equilibrium}
\label{sec: QMOR respects the thermal equilibrium}
Our recipes for simulating contact with thermal reservoirs in \QMOR mechanics yield an algebraic result (not previously known) that holds exactly even in linear quantum mechanics, and that can be verified without reference to drift and diffusion equations.  We state and prove this result now, so that it can provide a well-defined mathematical target for our subsequent \QMOR analysis.

We begin with a brief summary of coherent states, referring the reader to classic textbooks, such as those by 
Gardiner~\cite{Gardiner:00}, 
Gottfried~\cite{Gottfried:66}, 
Klauder and Skagerstam~\cite{Klauder:84},   
Perelomov~\cite{Perelomov:86}, 
Rose~\cite{Rose:57}, and 
Wigner~\cite{Wigner:59}, 
for details.  We will mainly follow Gottfried's notation.

We start by identifying a spin=$j$ state having z-axis quantum number $m $ with the ket-vector $\ket{j,m}$.  Then a coherent state $\ket{\lbhat{x}}$ associated with a unit-vector spin direction $\lbhat{x}$ is by definition $\ket{\lbhat{x}} = D(\phi,\theta,0)\ket{j,j}$,
where the rotation operator $D$ that carries $\lbhat{t}=(0,0,1)$ into
$\lbhat{x}=(\sin\theta \cos\phi, \sin\theta \sin\phi,
\cos\theta)$ is 
\begin{equation}
  D(\phi,\theta,\psi)=e^{-i\phi s_{3}} e^{-i\theta s_{2}}
  e^{-\psi s_{3}}
\end{equation}
The rotation operators are well understood.  In~particular, an identity due to Wigner \cite{Gottfried:66,Rose:57} gives $\inner{j,m}{\lbhat{x}}$ in closed form as 
\begin{equation}
\label{eq: Wigner}
\quad\inner{j,m}{\lbhat{x}}
= 
D^{j}_{mj}(\phi,\theta,0) 
= 
{\binom{2j}{j+m}^{1/2}}
e^{-i m\phi}
\left(\cos\tfrac{1}{2}\theta\right)^{j+m}
\left(\sin\tfrac{1}{2}\theta\right)^{j-m}
\end{equation}
It follows that $\inner{\lbhat{x}}{\lbhat{x}} = 1$ and
$\braket{\lbhat{x}}{\lb{s}}{\lbhat{x}}= j \lbhat{x}$.   It is well known 
\cite{Perelomov:72,Perelomov:86,Gardiner:00,Radcliffe:70} (and not hard to show from (\ref{eq: Wigner})\,) that a resolution of the identity operator $I$ is
 \begin{equation}
   I = \frac{2 j+1}{4\pi}\int_{4\pi}
   \!\!\,d^{2}\lbhat{x}\,\ket{\lbhat{x}}\bra{\lbhat{x}}\,.
 \end{equation}
The \emph{$Q$-representation} and
\emph{$P$-representation} of a Hermitian operator $\rho$ are
then defined (following Perelomov's conventions \cite{Perelomov:86}) as
\begin{equation}
\label{eq:Qrep}
  Q(\lbhat{x}|\rho)
  =\braket{\lbhat{x}}{\rho}{\lbhat{x}},\\
\end{equation}
\begin{equation}
  \rho  = \frac{2j+1}{4\pi}\int_{4\pi}
  \!\!\,d^{2}\lbhat{x}\ P(\lbhat{x}|\rho)\,
  \ket{\lbhat{x}}\bra{\lbhat{x}}.
  \label{eq:Prep}
\end{equation}
Given an arbitrary Hermitian operator $\rho$, it is known that in general a $P$-representation $P(\lbhat{x}|\rho)$ can always be constructed. In brief, the construction is as follows: from the \emph{ansatz}  
$P(\lbhat{x}|\rho) = \sum_{l=0}^{\infty}\sum_{m=-l}^j a_{l,m} Y^l_m(\theta,\phi)$, with $Y^l_m(\theta,\phi)$ a spherical harmonic, a set of linear equations for the coefficients $a_{l,m}$ is obtained by substituting (\ref{eq:Prep}) into (\ref{eq:Qrep}) and expanding both sides in spherical harmonics.   Solving the resulting linear equations always yields a valid $P$-representation. However, such $P$-representation constructions are non-unique in consequence of the identity
\cite{Perelomov:72,Perelomov:86}
\begin{equation}
  \int_{4\pi}\!\!\!d^{2}\lbhat{x}\ \,
  Y^{l}_{m}(\lbhat{x})\,\ket{\lbhat{x}}\bra{\lbhat{x}}
  = 0\quad\text{for all integer $l>2j$},
   \label{eq:indeterminacy}
\end{equation}
which can be proved directly from (\ref{eq: Wigner}) as a consequence of the addition law for angular momenta.  This result shows explicitly that the set of coherent states $\ket{\lbhat{x}}$ is over-complete (as is well-known).

With the above as background, we now slow the pace of presentation. We consider the problem of finding a \emph{positive} $P$-representation for a given operator $\rho$, that is to say, a representation for which $P(\lbhat{x}|\rho)\ge0$ for all $\lbhat{x}$.  

Positive $P$-representations have the useful property (for simulation purposes) of allowing us to interpret $P(\lbhat{x}|\rho)$ as a probability distribution over spin directions $\lbhat{x}$.   But from a mathematical point of view, distressingly little is known about positive \mbox{$P$-}repre\-sen\-tations, in the sense that there is no known general method for constructing them, or even for determining whether they exist in a given case. 

We will now consider an operator that often appears in practical \QMOR simulations: the \emph{thermal operator} $\myrho[th]{j}$ defined by
\begin{equation}
\label{eq:thermal operator}
  \myrho[th]{j} = 
  \exp(-\beta \lbhat{t}\mycdot\lb{s})\,,
\end{equation}
where $\{s_1, s_2, s_3\}$ are the usual spin-$j$ operators satisfying $[s_1, s_2] = i s_3$ (and cyclic permutations), and $\lbhat{t}$ is a unit axis along which a spin is thermally polarized with inverse-temperature $\beta$. By inserting a complete set of states into \myeq{eq:Qrep} and then substituting Wigner's expression (\ref{eq: Wigner}), a well-known \cite{Radcliffe:70,Perelomov:86} closed-form expression for the $Q$-representation of $\myrho[th]{j}$ can be obtained:
\begin{align}
\nonumber
  Q(\lbhat{x}|\myrho[th]{j}) 
  &=
  \sum_{m=-j}^{j}\sum_{m'=-j}^{j} 
  \inner{\lbhat{x}}{j,m}
  \braket{j,m}{\myrho[th]{j}}{j,m'}
  \inner{j,m'}{\lbhat{x}} \\
  & =
  (\cosh\myonehalf\beta -\lbhat{x}\mycdot\lbhat{t}
  \sinh\myonehalf\beta)^{2j}\,.
\label{eq: known Q-rep}
\end{align}
We now exhibit a closed-form analytic expression for a positive $P$-representation of the thermal operator as a distribution over coherent states, which is exact for all $j$ and $\beta$:  
\begin{myDesignRule} 
\label{dr: thermal operator}
A positive $P$-representation for the spin-$j$ thermal operator $\myrho[th]{j}$ is given in terms of the $Q$-representation by \[
P(\lbhat{x}|\myrho[th]{j})   =1/Q(-\lbhat{x}|\myrho[th]{j+1})\,.
\]
\end{myDesignRule}
\noindent To our knowledge this is the first such $P$-representation given, other than the $j\to\infty$ limit of quantum optics in which $P$ and $Q$ are both simple Gaussians.

If we regard the above result solely as a mathematical theorem to be proved by the most expedient means, we can do so by treating it as an \emph{ansatz}. The resulting proof is short. Taking matrix elements of the defining relation (\ref{eq:Prep}) between states $\bra{j,m}$ and $\ket{j,m'}$, and without loss of generality setting $\lbhat{t}=(0,0,1)$, Design Rule \ref{dr: thermal operator} is equivalent to the definite integral
\begin{equation}
  e^{-\beta m}\delta_{mm'}
  = \frac{2
  j+1}{4\pi}\int_{4\pi}
  \!\!\!\!\,d^{2}\lbhat{x}\ 
  \frac{\inner{j,m}{\lbhat{x}}\inner{\lbhat{x}}{j,m'}}
  {Q(-\lbhat{x}|\myrho[th]{j+1})}\,.
  \label{eq:mmprime}
\end{equation}
Substituting the Wigner representation (\ref{eq: Wigner}) and the $Q$-representation (\ref{eq: known Q-rep}) into (\ref{eq:mmprime}), we can check by numerical integration that (\ref{eq:mmprime}) is correct for randomly chosen values of $j$, $m$, $m'$, and~$\beta$.  Thereby encouraged, we soon discover an integration strategy by which (\ref{eq:mmprime}) yields to analytic evaluation in the general case.  In brief, the $\phi$-angle integration yields the requisite $\delta_{mm'}$ factor; the $\theta$-angle integration can be transformed into an integral over rational functions in $z = \cos\theta$ via identities like $\cos^{2j} \tfrac{1}{2}\theta = (1+z)^j/2^j$; and the resulting integral is recognizably a representation of the Gauss hypergeometric series \cite[eqs.~15.1.8 {\&}~15.3.8]{Abramowitz:65} that evaluates to (\ref{eq:mmprime}).

Design Rule \ref{dr: thermal operator} is simultaneously frustrating, reassuring, and intriguing.  It is frustrating because our \QMOR analysis will answer the natural question ``Where did the \emph{ansatz} come from?'' by ``It is the solution to a Fokker-Planck equation that describes spin-systems in thermal equilibrium.'' But this answer provides no satisfying rationale for why the $P$-representation exists, or for its simple analytic form.  Furthermore, \QMOR analysis will provide no answer at all to the natural follow-on question, ``Is there a reason why the thermal operator's positive $P$-represent\-ation is simply given in terms of its $Q$-representation?''  

Design Rule \ref{dr: thermal operator} is reassuring because it tells us that \QMOR simulations \emph{can} simulate thermal equilibrium (at least in simple cases).  From a physical point of view this reassures us that the dimensional reduction associated with \QMOR mechanics preserves at least some crucial thermodynamic physics.  From a practical point of view it provides a reassuring consistency check that the (rather lengthy) chain of theorems and stochastic analysis that leads to the $P$-representation is free of algebraic errors.

And finally, Design Rule \ref{dr: thermal operator} is intriguing because it suggests that a partial answer to the open question ``Which operators have positive $P$-representations?" might be ``Only thoses operators that are proportional to a density matrix that is associated with a stationary measure\-ment-and-control process that drives input states to coherent states.''  This conjecture is intriguing because its negation is logically equivalent to ``Density matrices exist that \emph{cannot} be the result of any stationary measurement-and-control process that drives input states to coherent states.''   Such density matrices would be remarkable entities from both a mathematical and physical point of view, and in particular their existence (or nonexistence) is directly relevant to the further development of practical design rules for \QMOR simulations.  The conjectured answer is therefore intriguing whether it is true or false.

\subsubsection{\QMOR respects classical linear control theory}
\label{sec: QMOR respects operational structure}
\QMOR simulations of macroscopic objects (like \MRFM cantilevers) regard them as spin-$j$ quantum objects having very large $j$.  We will see that the resulting dynamics typically are linear.   Engineers have their own idioms for describing linear dynamical systems, which are summarized in block diagrams like this one:   
\begin{equation}
\label{eq: control block diagram}
\raisebox{-0.48\height}{\includegraphics[scale=1]{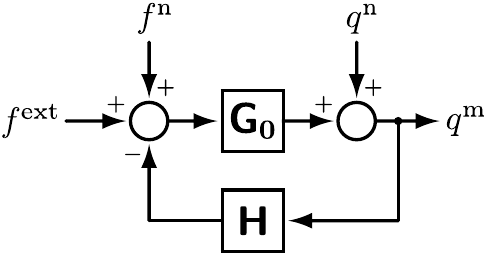}}
\end{equation}
To show that \QMOR simulations can accurately model systems like the above, we will transform the above diagram---using strictly classical methods---into an operationally equivalent form that more naturally maps onto \QMOR algorithms and geometric structures.

\begin{figure}[t]\centering
\vspace*{2ex}
\includegraphics[scale=1.1]{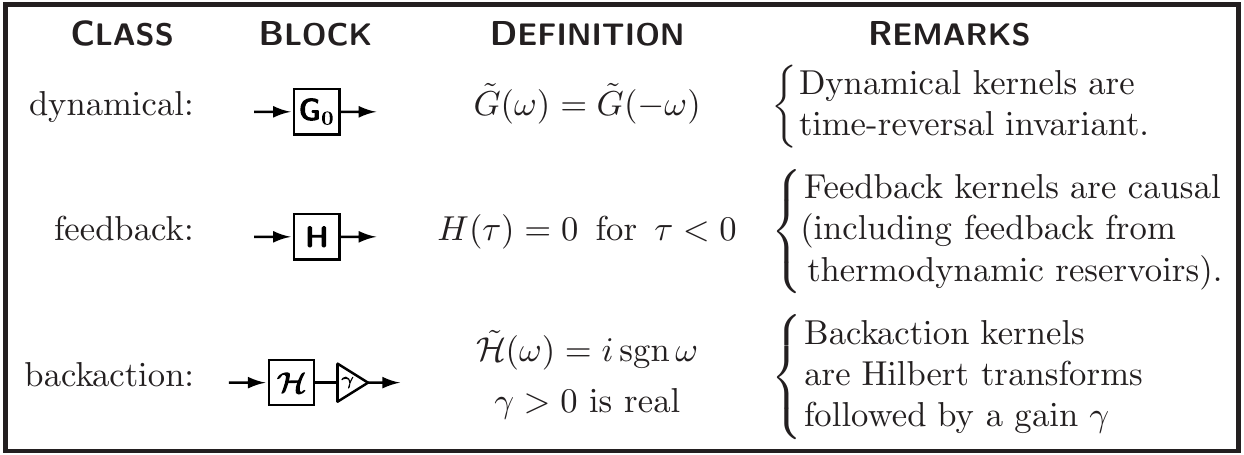}\\
\begin{minipage}{0.8\textwidth}
\caption[The three kernel classes of linear, classical \QMOR simulations]{%
\protect\justifying%
The three kernel classes of linearized \QMOR simulations.
\label{fig: QMOR kernel classes}%
}\end{minipage}
\end{figure}

\begin{figure}[p]\centering
\includegraphics[width=\textwidth]{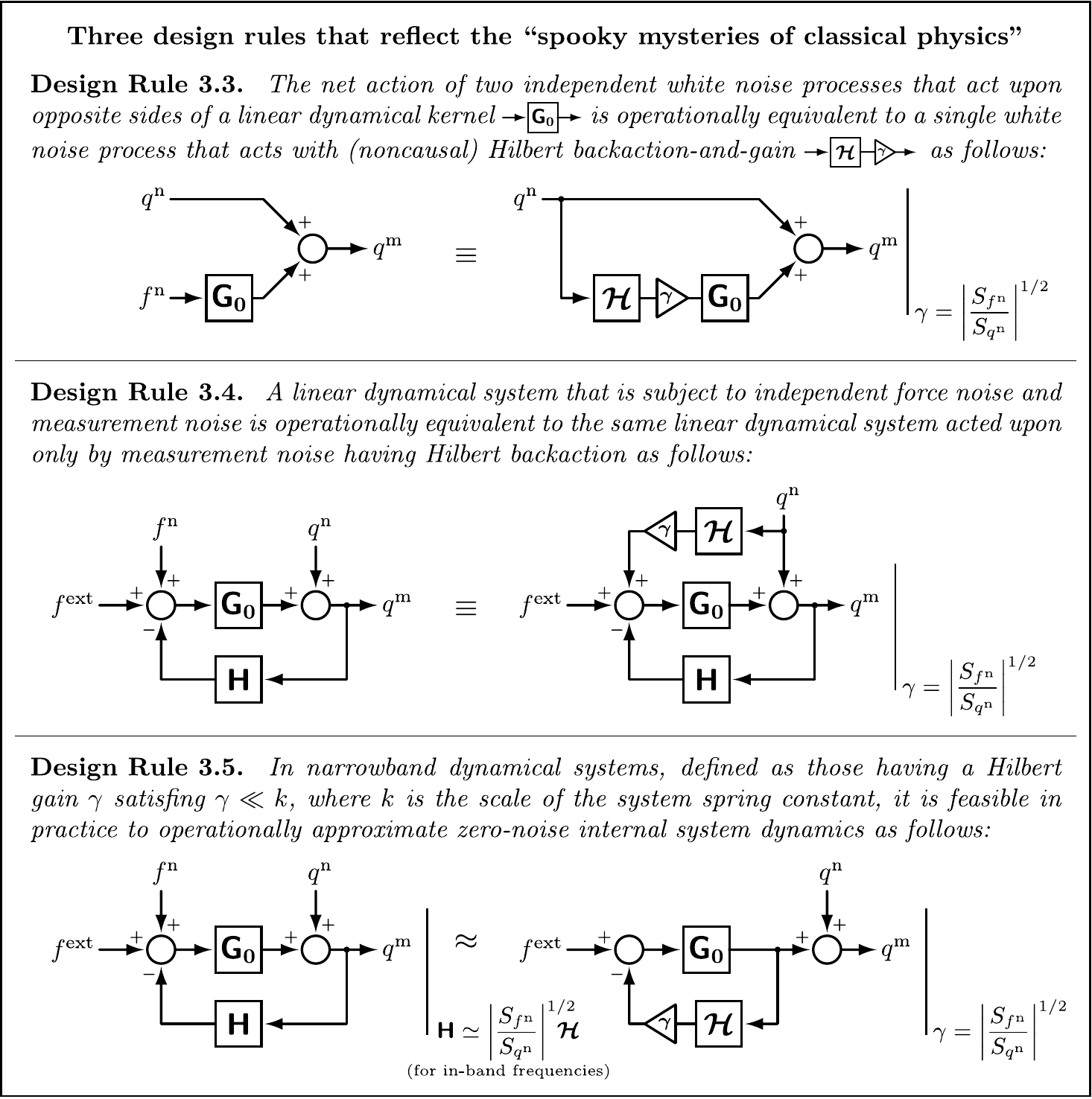}%
\caption[Three design rules that reflect ``spooky classical physics'']{%
\protect\justifying%
Three design rules that reflect ``spooky mysteries of classical physics.''\\[1ex]
\label{fig: classical linear design rules}%
The above design rules follow from elementary classical considerations as (briefly) follows.  Design Rule~3.3 follows from the time-reversal invariance of the dynamical kernel $G_0$ (see Fig.~\ref{fig: QMOR kernel classes}) plus the statistical properties of Gaussian noise.
Design Rule 3.4 is then obtained by adjoining a causal feedback kernel $H$ to the diagram of Rule~3.3.
Design Rule 3.5 then follows as a practical application of Rule~3.4, in which a causal kernel 
$H$ approximates a non-causal Hilbert backaction kernel $\lcal{H}$ within a device's finite operational bandwidth.
}
\end{figure}

For convenience, these classical equivalences are summarized as design rules in Figs.~\ref{fig: QMOR kernel classes}--\ref{fig: classical linear design rules}. Experienced researchers will recognize these equivalences as being elementary, but to the best of our knowledge, they have not previously been recognized in the literature of engineering or physics.  

As with the Feynman diagrams of physics, the block diagrams of engineering depict systems of equations.  Our diagram conventions are standard, as briefly follows.
The block diagram (\ref{eq: control block diagram}) corresponds to a set of linear relations between force noise $\myfn(t)$, measurement noise $\myqn(t)$, input external force $\myfext(t)$, and output measurement $\myqm(t)$, which we take to be classical real-valued functions.  In particular we specify that $\myfn(t)$ and $\myqn(t)$ are white noise processes having correlation functions $C$ satisfying
\begin{subequations}
\begin{align}
C(\myqn(t)\myqn(t')) &= \tfrac{1}{2}\,S_{\myqn} \delta(t-t')\\
C(\myfn(t)\myfn(t')) &= \tfrac{1}{2}\,S_{\myfn} \delta(t-t')\\
C(\myfn(t)\myqn(t')) &= 0
\end{align}
\end{subequations}
so that $S_{\myqn}$ and $S_{\myfn}$ are (one-sided) white-noise spectral densities.  A circle {\myblockdiagramAddBlock} is a node whose inputs are added and subtracted, a cross {\myblockdiagramMultBlock} is a node whose inputs are multiplied, a triangle {\myblockdiagramGainBlock} indicates a positive real scalar gain $\gamma$, and a square box {\myblockdiagramBlankKernelBlock} indicates convolution with a general real-valued stationary kernel $K$ such that
\begin{equation}
\label{eq: linear convolution}
b(t) = \int_{-\infty}^{\infty}\!\!\!dt'\ K(t-t') a(t')
\quad\underset{\text{(is depicted as)}}{\Longleftrightarrow}\quad
a\,\myblockdiagramGeneralKernelBlock\,b
\end{equation} 
Alternatively, convolution blocks can be specified in the Fourier domain.   Our Fourier transform convention is that $\tilde{a}(\omega)$ is defined to be
\begin{equation}
\tilde{a}(\omega) \equiv
\int_{-\infty}^{\infty}\!\!\!d\tau\ 
e^{-i\omega\tau} a(t)
\end{equation}
and similarly for $\tilde{b}(\omega)$, $\tilde{K}(\omega)$, \emph{etc.}  Therefore a frequency-domain description of (\ref{eq: linear convolution}) is
\begin{equation}
\label{eq: frequency convolution}
\tilde{b}(\omega) = \tilde{K}(\omega) \tilde{a}(\omega)
\quad\underset{\text{(is depicted as)}}{\Longleftrightarrow}\quad
a\,\myblockdiagramGeneralKernelBlock\,b
\end{equation}
We build our \QMOR block diagrams from three classes of linear classical kernels: dynamical kernels (\myblockdiagramDynamicKernelBlock), feedback kernels  (\myblockdiagramFeedbackKernelBlock), and backaction kernels  (\myblockdiagramHilbertKernelBlock), whose defining mathematical properties are specified in Fig.~\ref{fig: QMOR kernel classes}. 

In brief, dynamical kernels by definition are time-reversal invariant, feedback kernels by definition are causal, and the backaction kernel is the \emph{Hilbert transform} that is well-known (and much-used) by signal processing engineers.  We remark that the Hilbert transform is formally non-causal, but in practical narrow-bandwidth applications (like radio transmitters, acoustic processors, \MRFM cantilever controllers, \emph{etc{.}}) its effects can be closely approximated by a causal derivative transform.

For causal kernels, analytic continuation from the Fourier variable $\omega$ to the Laplace variable $s=i\omega$ is well-defined.  Partly because causal kernels are of central importance in control engineering, and partly by tradition, Laplace variables are more commonly adopted in the engineering literature than Fourier variables (although both are used).  However, the Laplace analytic continuation of the non-causal Hilbert transform kernel $\tilde{\lcal{H}}(\omega) = i \sgn(\omega)$ is \emph{not} well-defined.  For this reason our analysis will focus exclusively upon upon time-domain and frequency-domain (Fourier) kernel representations. 

Central to \QMOR analysis and simulation is the purely mathematical fact (which also appears in Fig.~\ref{fig: classical linear design rules} as Design Rule~3.4) that the following classical systems are operationally equivalent:
\begin{equation}
\label{eq: first equivalence}
  \myblockdiagramOne 
  \quad\equiv\quad
	\myblockdiagramTwo%
	{\hspace{0.5ex}\rule[-5ex]{0.5pt}{9ex}}_{\ %
		\gamma\,=\,\left|\frac{S_{\myfn}}{S_{\myqn}}\right|^{1/2}%
	}%
\end{equation}
By ``operationally equivalent'' we mean that the dynamical relation between the applied forces $\myfext(t)$ and the measured position $\myqm(t)$ is identical for the two sets of equations, as are the stochastic properties of $\myqm(t)$.  The internal state of the system is of course very different for the two cases, but so long as we adhere to the strict operational principle ``never observe the internal state of a system (even a classical system)'' this difference is immaterial.

Developing design rules for \QMOR mechanics is considerably easier if we habituate ourselves to the Hilbert transform that appears in all three classical design rules of Fig.~\ref{fig: classical linear design rules}. We begin by remarking that from  an abstract point of view, the Hilbert transform defines a \emph{complex structure} on the space $\lcal{R}$ of real-valued  functions $r$, that is, a map ${\lcal{H}}\!:{} \lcal{R}\to \lcal{R}$ that satisfies $\lcal{H}^2 = - I$.  This corresponds to two diagrammatic identities
\begin{equation}
\myblockdiagramFeedbackKernelSqrBlock = \myblockdiagramInverterBlock
\qquad\text{and}\qquad
\myblockdiagramFour
\end{equation}
that make it obvious (since the right-hand sides are causal) that the non-causal aspects of {\myblockdiagramHilbertKernelBlock} vanish when it is applied an even number of times. Because the even-numbered moments wholly determine the statistical properties of (zero-mean) Gaussian noise, we begin to see how it is that noncausal Hilbert transforms can appear in the equivalences of our classical design rules without inducing observable causality violations.  In the broader context of quantum simulations, causality is assured by the \emph{Theorema Dilectum}, and we will establish in a later section that the \emph{Theorema Dilectum} is directly responsible for the appearance of Hilbert transforms in the classical limit.

We saw already (in Section~\ref{sec: complex structure}) that complex structures are a defining geometric characteristic of \Kahler manifolds, so the appearance of Hilbert transforms in real-valued classical dynamics is a mathematical hint that noisy classical dynamical trajectories support a complex structure that projects naturally onto gabion-\Kahler manifolds.  This complex structure is manifest not in the (real-valued) state variables of classical systems, but in the causal properties of the response of classical systems to noise. 

\subsubsection{Remarks upon the spooky mysteries of classical physics}
\label{sec: tenets}
For teaching purposes, it is helpful (and amusing) to pretend that we live in a world in which linear control theory is taught according to a nonstandard ontology in which the Hilbert transform has a central role. This ontology was conceived as a philosophical provocation (a~\emph{heraus\-forderung} \cite{Schurrmacher:1997df}), but it has subsequently proved to be useful in teaching and a fertile source of technical inspiration.   We will call it the \emph{classical Hilbert ontology}, or sometimes just the \emph{Hilbert ontology}.  Our motivation for emphasizing the mysterious properties of classical measurement theory is similar to the motivation of Wheeler and Feynman  \cite{Wheeler:1945pl,Wheeler:1949is} in proposing their non-causal classical electrodynamic ontology proposed in the 1940s. 

The tenets of Hilbert ontology are taken to be:
\begin{QSEitemize}
\item By Occam's Razor, ontologies that invoke one noise source are preferred over ontologies that invoke two.  Therefore the Hilbert ontology regards backaction as being a  physically real phenomenon that is universally present in all noisy dynamical systems (its mathematical expression being Design Rule 3.3 of Fig.~\ref{fig: classical linear design rules}).
\item It follows (by Design Rule~3.4) that measurement noise always back-acts upon system dynamics in such a way as to ``drag'' the state of the system into agreement with the measurement.  This state-dragging Hilbert backaction has a central ontological role: it is the fundamental mechanism of nature that causes measurement processes to agree with reality.
\item As a measurement process approaches the zero-noise limit, the increasingly strong state-dragging Hilbert backaction from that measurement process dynamically  ``collapses'' the variable being measured, so as to force exact agreement with the measurement.
\item As a corollary, zero-noise measurements are unphysical, because they imply infinitely strong backaction.  That is the explanation in Hilbert ontology of why all real-world measurement processes are noisy.  
\item In narrow-band systems, it is possible to cancel the Hilbert backaction noise via causal feedback control.  This means that zero-temperature narrow-band systems can be simulated, or even realized in practice, provided that all noise processes are accessible for purposes of feedback  (as mathematically expressed by Design Rule 3.5 of Fig.~\ref{fig: classical linear design rules}).
\item Although the mathematical kernel associated with Hilbert feedback is non-causal, this noncausality cannot be exploited for purposes of communication, for the physical reason that the backaction kernel transmits only noise.  The mathematical proof is simply that Design Rules 3.3--5 can be expressed as equivalent systems having all kernels causal (as shown in Fig.~\ref{fig: classical linear design rules}).  But in Hilbert ontology this causality proof is viewed as being a purely mathematical artifice, because it postulates ``spooky'' uncorrelated forces that ``obviously'' have no physical reality (according to our Occam's Razor tenet) since no mechanism is given for them.
\end{QSEitemize}
For purposes of this article, we designate the above Hilbert ontology to be the ``true'' classical reality of the world, and we seek to provide a microscopic justification of it from orthodox quantum mechanics (recognizing of course that the Hilbert ontology has been constructed specifically to ensure that our overall \emph{heraus\-forderung} yields practical and interesting results). 

From a teaching point of view, the Hilbert ontology helps students appreciate that the mysteries and ambiguities that traditionally are taught as belonging exclusively to quantum mechanics---like ``wave functions collapsing to agree with measurement'' and ``noncausal correlations''---are manifest too in (at least one) wholly classical ontology.    

It is traditional in both the popular and the scientific literature to call these ontological mysteries and ambiguities ``spooky.'' The popularity of ``spooky'' in the scientific literature can be traced to an influential article by Mermin \cite{Mermin:1985uq}, who adopted it from an idiomatic phrase ``\emph{spukhafte Fernwirkungen}'' that Einstein uses in the Einstein-Bohr correspondence; this phrase is generally translated as ``spooky action at a distance''  \cite{Hecht:2005fk,Hardy:1998qy,Moore:2007lr}.   Given the central importance of the spookiness of quantum mechanics, it would be astonishing if this spookiness was wholly invisible at the classical level.  

Our Hilbert ontology therefore serves both to guide our calculations and to help us celebrate the ``spooky mysteries of classical physics.''  

\subsubsection{Experimental protocols for measuring the Hilbert parameters}
\label{sec: protocols}
A crucial test of an ontology is whether it motivates us to proceed to practical calculations that yield useful and/or surprising results; we now do so.  We consider a laboratory course in which students are guided by the Hilbert ontology to explore the fundamental and practical limits of low-noise sensing and amplification.   For definiteness, we assume that the students work with nanomechanical oscillators as force detectors (as in \MRFM technology), but a similar course could feasibly be organized around radio-frequency (\RF) sensing and amplification, optical sensing and amplification, or acoustic sensing and amplification.

We consider the experimental problem of measuring the two Hilbert parameters of a nanomechanical oscillator, namely the measurement noise $S_q$ and the Hilbert gain $\gamma$.  We~neglect the intrinsic damping of the oscillator, such that the dynamical kernel is ${\tilde G}_0(\omega) = 1/[m(\omega_0^2-\omega^2)]$, where $m$ is the mass of the oscillator and $\omega_0$ is the resonant frequency.

The measurement is readily accomplished by the following protocol.   Derivative feedback control is applied having a kernel $\tilde H(\omega) = i\Gamma \omega$, and the value of the controller gain $\Gamma$ is adjusted until the spectral density $S_{q^{\text{m}}}$ of the measured cantilever displacement $q^{\text{m}}(t)$ is observed to be flat in the neighborhood of the cantilever frequency $\omega_0$.   The control gain adjustment is straightforward: if the spectrum of $q^{\text{m}}(t)$ has a peak, then $\Gamma$ is too small; if the spectrum has a hole, then $\Gamma$ is too large.  

The preceding protocol is perfectly feasible in practice (\emph{i.e.}, it is not a \emph{Gedankenexperiment}).  The protocol fails only when the required dashpot gain is impractically large, such that $\Gamma\omega_0 \gtrsim k$, where $k = m \omega_0^2$ is the spring constant of the cantilever.  Such failures typically indicate that a measurement process has a noise level that is too large to be of practical interest.  

In practical cases the required control gain typically satisfies $\Gamma_0 \ll \Gamma \ll k$, where $\Gamma_0 = k/(\omega_0 Q)$ is the intrinsic damping of a cantilever having quality $Q$.  In such cases the intrinsic damping $\Gamma_0$ has negligible effect on the time-scale of the controlled response, and so we can regard the cantilever's dynamical kernel as having the time-reversal-invar\-iant dynamical form $G_0$ described in Fig.~\ref{fig: QMOR kernel classes}.  The design rules of our Hilbert classical ontology therefore can be applied without modification.   Specifically, Design Rule~3.5 determines the Hilbert backaction to be $\gamma = \Gamma\omega_0$, and determines the measurement noise to be $S_{q^{\text{n}}} = S_{q^{\text{m}}}(\omega_0)$.  

Of course, Design Rule~3.5 mathematically assures us that the adherents of traditional classical ontology can---without operational contradiction---ascribe these same measurements to a fictitious force noise $f^{\text{n}}(t)$ having spectral density $S_{f^{\text{n}}} = \gamma^2 S_{q^{\text{n}}}$, so that the vexing question of whether this force noise is  ``fictitious'' or ``real'' is operationally immaterial.

\subsubsection{\QMOR simulations respect the fundamental quantum limits}
\label{sec: QMOR respects the fundamental quantum limits}
We now apply the results of the preceding section to establish criteria that \QMOR simulations must satisfy in order to respect the known fundamental quantum limits to amplifier noise and force measurement.

We suppose that a force signal $f(t) = f_0 \cos(\omega_0 t + \phi_0)$ is applied to the oscillator.  The carrier frequency $\omega_0$ is tuned to match the oscillator resonance frequency, and the unknown magnitude $f_0$ and unknown phase $\phi_0$ of the signal are to be determined from measurement.  This is a common task in practice.

When the oscillator is configured according to the measurement protocol of the preceding section, then according to the right-hand block diagram of Design Rule~3.5 of Fig.~\ref{fig: classical linear design rules}, the mean power $p_0$ absorbed by the oscillator's (wholly classical) feedback controller during the measurement process is $p_0 = f_0^2/(2\gamma)$.  The absorbed power inferred from the (wholly classical) measurement record $q^{\text{m}}(t)$ has an equivalent (one-sided) noise \PSD $S^{\text{output}}_{p_0}$ whose expression in terms of the Hilbert parameters is readily shown to be
\begin{equation}
\label{eq: output}
    S^{\text{output}}_{p_0} = 4 p_0 \omega_0 \gamma S_{q^\text{n}}
\end{equation}
Now we connect this result with a known fundamental quantum limit on noise in power amplifiers. We adopt the definition of Caves \cite{Caves:1982gx}:
\begin{QSEquote}
An amplifier is any device that takes an input signal, carried by a collection of bosonic modes, and processes the output to produce an output signal, also carried by a (possibly different) collection of bosonic modes.  A linear amplifier is an amplifier whose output signal is linearly related to its input signal.
\end{QSEquote}
Following a line of reasoning put forth in the 1960s by Heffner  \cite{Heffner:1962fy} and by Haus \cite{Haus:1962ad}, which Caves article develops in detail \cite{Caves:1982gx}, we suppose that the input power is supplied by a bosonic mode-type device (like a resonant circuit, or an \RF wave-guide, or a single-mode optical fiber) whose power level has been independently measured with a shot-noise-limited photon counting device.  According to orthodox quantum mechanics such counting processes have Poisson statistics (we will establish in Section~\ref{sec: calibration rules} that \QMOR simulations respect this rule), and therefore the input power has a quantum-limited noise \PSD given by 
\begin{equation}
\label{eq: input}
	S^{\text{input}}_{p_0} = 2 p_0 \hbar \omega_0
\end{equation}
The measured power and phase suffice to create an arbitrary-gain replica of the input signal, and the noise-figure (\NF) of this effectively infinite-gain power amplifier is simply given from (\ref{eq: output}) and (\ref{eq: input}) by 
\begin{equation}
\label{eq: fundamental theorem}
	\NF = S^{\text{output}}_{p}/S^{\text{input}}_{p} = 2 \gamma S_{q^\text{n}}/\hbar = 2 \big(S_{f^\text{n}}S_{q^\text{n}})^{1/2}/\hbar
\end{equation}
In Caves' nomenclature, we are regarding our continuously measured oscillator as an equivalent ``{phase-insensitive linear amplifier}'' having infinite gain.  The analyses of Heffner, Haus, and Caves \cite{Heffner:1962fy,Caves:1982gx,Haus:1962ad} establish that what Caves calls the ``{fundamental theorem for phase-insensitive power amplifiers}'' is simply $\NF\ge2$ (in the infinite-gain limit), which in decibels is $10 \log_{10} 2 \simeq 3~\text{dB}$. 

We remark that the 3~dB quantum limit to power amplifier noise has been experimentally observed \cite{Helmer:1958os} and theoretically analyzed \cite{Shimoda:1957lb} since the early days of maser amplifiers in the 1950s; our review focuses upon the work of Heffner, Haus, and Caves solely because their analyses are notably rigorous, general, clearly stated, and (importantly) their predicted quantum limits are mutually consistent and consonant with subsequent experiments.  

Our result (\ref{eq: fundamental theorem}) then establishes that  the Heffner-Haus-Caves noise-figure limit finds its expression in \QMOR analysis in the following three equivalent ways:
\begin{equation}
\label{eq: fundamental equivalence}
\underset{\text{\makebox[0pt]{\rule[-0.5ex]{0pt}{2.0ex}%
\rmfamily\mdseries\scshape%
the Heffner-Haus-Caves limit%
}}}
	{\rule[-1ex]{0pt}{3.0ex}\NF\ge2}
\qquad\quad\Leftrightarrow\qquad\quad 
\underset{\text{\makebox[0pt]{\rule[-0.5ex]{0pt}{2.0ex}%
\rmfamily\mdseries\scshape%
the Braginsky-Khalili limit%
}}}
	{\rule[-1ex]{0pt}{3.0ex}S_{f^\text{n}}S_{q^\text{n}}\ge \hbar^2}
\qquad\quad\Leftrightarrow\qquad\quad\hspace{-1em} 
\underset{\text{\makebox[0pt]{\rule[-0.5ex]{0pt}{2.0ex}%
\rmfamily\mdseries\scshape%
the Hilbert limit%
}}}
	{\rule[-1ex]{0pt}{3.0ex}\gamma S_{q^\text{n}}\ge \hbar}
\end{equation}
The middle expression we recognize as the continuous-measurement version of the \emph{standard quantum limit} to force and position measurement, in precisely the quantitative form derived by Braginsky and Khalili \cite{Braginsky:1992uq}, which in turn derives from earlier seminal work by Braginsky, Vorontsov, and Thorne \cite{Braginsky:1980zm}.  Although the Braginsky-Khalili limit was derived by very different methods from the Heffner-Haus-Caves' limit, we see that the two quantum limits are equivalent.   To the best of our knowledge, this equivalence has not previously been stated in the above quantitative form.  What we have chosen to call ``the Hilbert limit'' on the right-hand side of (\ref{eq: fundamental equivalence}) has (to our knowledge) not previously been recognized anywhere in the literature, and yet from the viewpoint of Hilbert ontology it is the most fundamental of the three.

We express the above three-fold equivalence as a fundamental \QMOR design rule:
\addtocounter{mytheorem}{3}
\begin{myDesignRule}
\label{dr: fundamental quantum limits}
\QMOR simulations respect the quantum measurement limit in all its equivalent forms: the noise-figure (\NF) limit in power amplifiers ($\NF\ge2$), the standard quantum limit to the measurement of canonically conjugate variables ($S_{{f}^\text{n}}S_{q^\text{n}}\ge \hbar^2$), and the Hilbert limit that measurement noise and state-dragging Hilbert backaction cannot both be small ($\gamma S_{q^\text{n}}\ge \hbar$).
\end{myDesignRule}
\noindent We present a quantum derivation of these limits in Section~\ref{sec: calibration rules}, in the context of a more general analysis that encompasses nonlinear quantum systems.  

\subsubsection{Teaching the ontological ambiguity of classical measurement}
\label{sec: Hilbert ontology}
What should we teach students about the internal state of the system during the preceding (strictly classical) protocols for measuring the Hilbert parameters $S_{q^\text{n}}$ and $\gamma$?  

From a strictly logical point of view, this question need not be answered, since our measurement protocols and design rules are careful to make no reference to the internal state (even at the classical level).  But in practice, \emph{some} answer must be given, for the pragmatic reason that students require \emph{some} coherent story about what the systems they are measuring are ``really'' doing.   This coherence is especially necessary to those science and engineering students (a~substantial portion, in our view) whose careers will require that they extend their professional expertise from the classical to the quantum domain.

It is therefore desirable that common-sense questions from students receive common-sense answers from teachers, and it is equally desirable that these answers prepare for a classical-to-quantum educational transition that is as nearly seamless as is practicable.

Adherents of traditional classical ontology can argue cogently as follows: ``It may be operationally correct to ascribe experimental results wholly to measurement noise  and Hilbert backaction, but it is physically wrong,  because we have strong physical reasons to believe that the excitation of the oscillator is being driven by force noise from a thermal reservoir whose internal dynamics we do not observe.''  

Adherents of Hilbert classical ontology can offer an similarly cogent counter-argument, which  however requires the assertion of a definite mathematical result (in italics): ``We too believe that the observed excitation of the oscillator is in fact being driven an unobserved thermal reservoir, \emph{but the action of a unobserved thermal reservoir can be ascribed to a covert process of measurement, Hilbert backaction, and control}.''

The practical consequence of the above reasoning is that this article's embrace of Hilbert ontology has practical utility \emph{only} if we can develop well-posed \QMOR algorithms by which the action of thermal reservoirs upon a dynamical system is simulated by unobserved/covert processes of quantum measurement, Hilbert backaction, and classical control.  We develop the necessary algorithms in Section~\ref{sec: calibration rules}, using as our main mathematical tool the \emph{Theorema Dilectum} that was already given as Design Rule~\ref{dr: theorema dilectum}.  A~key mathematical result will be the positive $P$-representation that was already given as Design Rule~\ref{dr: thermal operator}; this helps us foresee the analysis path by which the \QMOR analysis program will succeed.  And of course Design Rules 3.3--5 will emerge naturally in the course of our analysis.

In summary, for students and teachers alike, a sufficient justification for embracing the Hilbert ontology is that it leads to \QMOR simulation algorithms that are computationally efficient, operationally orthodox, and mathematically novel.  A~provocative side-effect is that we learn to perceive the ``spooky mysteries of quantum physics'' as being manifest in \emph{all} noisy dynamical systems, even classical ones.

	\subsection{Physical aspects of \QMOR}
	\label{sec: Physical foundations}

We now turn our attention to the physical aspects of measurement, our goal being to establish connections between the concrete description of measurement in terms of hardware and experimental protocols on the one hand, and the measurement operator formalism of our \QMOR algorithms on the other hand.

\subsubsection{Measurement modeled as scattering}
\label{eq: scattering}
We adopt as the fundamental building block of our simulations a~single particle of spin $j$, described by a~wave function $\ket{\psi}$, and numerically encoded as a~complex vector with $\dim \ket{\psi} = 2j+1$.  We will simulate all noise and all measurement processes by scattering photons off the spin, one photon at a~time, and we associate each scattering event with a~single time step in Fig.~\ref{fig: QMOR formal}. 

We envision photon scattering as the sole mechanism by which noise is injected into our simulations, and interferometry as the sole means of measurement. We describe photon scattering as a~unitary transformation acting on the spin state
\begin{equation}
\ioptext{(before\ scattering)}\quad\ket{\psi_n}\ \to\  
\exp(i2\theta s^\ioptext{op})\,\ket{\psi_n}
\quad\ioptext{(after\ scattering)}.
\label{eq:scattering phase}
\end{equation}
A purely conventional factor of $2$ is inserted in the above to simplify our calibration rules.

We will call $s^\ioptext{op}$ a~\emph{measurement generator}.  In general $s^\ioptext{op}$ can be any Hermitian matrix, but in our simulations it will suffice to confine our attention to $s^\ioptext{op}\in\{s_1,s_2,s_3\}$, where $\{s_1,s_2,s_3\}$ are rotation matrices satisfying the commutation relation $[s_j,s_k] = i\epsilon_{jkl}s_l$.  These matrices generate the rotation group, and our discussion will assume a basic knowledge of their algebraic properties, which are discussed at length in many textbooks (\emph{e.g.}, \cite{Gottfried:66,Messiah:66,Rose:57,Wigner:59}).  

We adopt the near-universal convention of working exclusively with spin operator matrix representations that are irreducible and sparse, having dimension $2j+1$ for $j$ the spin quantum number, with~$s_3	$ a~real diagonal matrix, $s_1	$~real and bidiagonal, and $s_2$~imaginary and bidiagonal.  The scattering strength is set by the real number $\theta$, which in our simulations will always satisfy $\theta \ll 1$. 

\subsubsection{Physical and mathematical descriptions of interferometry}%
\begin{figure}[t]\centering
\vspace*{2ex}
\includegraphics[width=\textwidth]{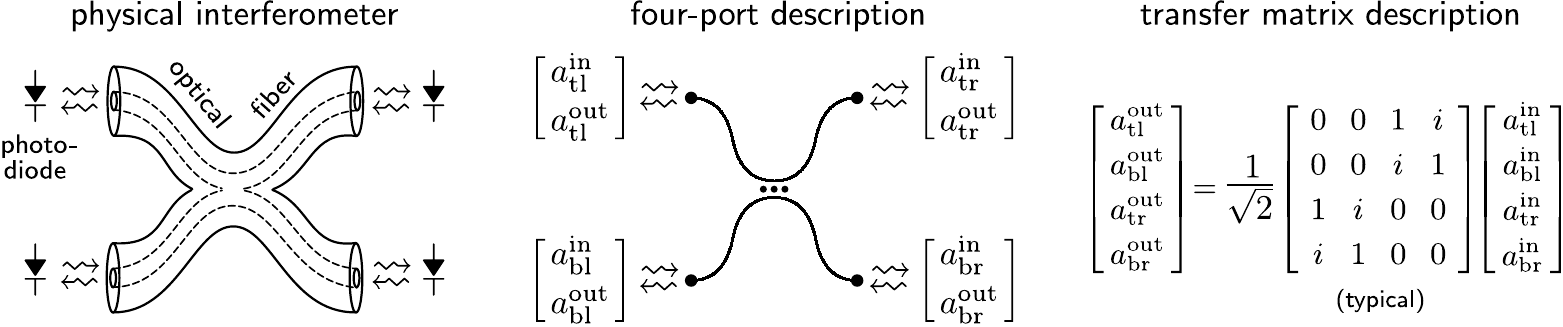}\\
\begin{minipage}{0.8\textwidth}
\caption[Three aspects of photon interferometry]{%
\protect\justifying%
Three aspects of photon interferometry.\\[1ex]
\label{fig: physical}%
Photon interferometry from the (a) physical, (b)~geometric, and (c)~algebraic points of view. See Section~\ref{sec: discussion of interferometery} for discussion.
}\end{minipage}
\end{figure}

\label{sec: discussion of interferometery}
Figure~\ref{fig: physical} provides three idealized diagrams for thinking about the physical, geometric, and algebraic aspects of photon interferometry, similar to the idealized diagrams in the Feynman Lectures \cite[chs.~5--6]{Feynman:65b} that depict the Stern-Gerlach effect. 

Figure~\ref{fig: physical}(a) depicts physical fiber-optic interferometers that  confine photons within low-loss optical fibers \cite{Rugar:89,Bruland:99}.  The region of overlap between the fibers allows photons to transfer from one fiber to another with an amplitude that is subject to engineering control.  This overlap region plays the role of the semi-silvered mirror in a~traditional Michelson interferometer.

Figure~\ref{fig: physical}(b) depicts optical couplers geometrically, as general-purpose devices for linking incoming and outgoing optical amplitudes.  Couplers thus can be braided into optical networks of essentially arbitrary topology.   

Figure~\ref{fig: physical}(c)  depicts optical couplers algebraically, as linear maps between incoming complex optical amplitudes $\lb{a}^\ioptext{in}=\{a^\ioptext{in}_\ioptext{tl},\,a^\ioptext{in}_\ioptext{tr},\,a^\ioptext{in}_\ioptext{bl},\,a^\ioptext{in}_\ioptext{br}\}$ and outgoing optical amplitudes  $\lb{a}^\ioptext{out}=\{a^\ioptext{out}_\ioptext{tl},\,a^\ioptext{out}_\ioptext{tr},\,a^\ioptext{out}_\ioptext{bl},\,a^\ioptext{out}_\ioptext{br}\}$ such that $\lb{a}^\ioptext{out}=\ioptext{S}\cdot\lb{a}^\ioptext{in}$, with $\ioptext{S}$ the \emph{optical scattering matrix}. With regard to Fig.~\ref{fig: physical}(c), the index ``tl'' is the ``top left'' port, ``br'' is the ``bottom right'' port, \emph{etc.}.  Our normalization convention is such that the probability of detection of an outgoing photon is $|a^\ioptext{out}|^2$. Optical losses in real-world couplers are small, such that $|\lb{a}^\ioptext{out}|^2 = |\lb{a}^\ioptext{in}|^2$, \emph{i.e.} the optical scattering matrix $\rm S$ is unitary.

\subsubsection{Survey of interferometric measurement methods}%

\begin{figure}[p]\centering
\includegraphics[scale=1.0]{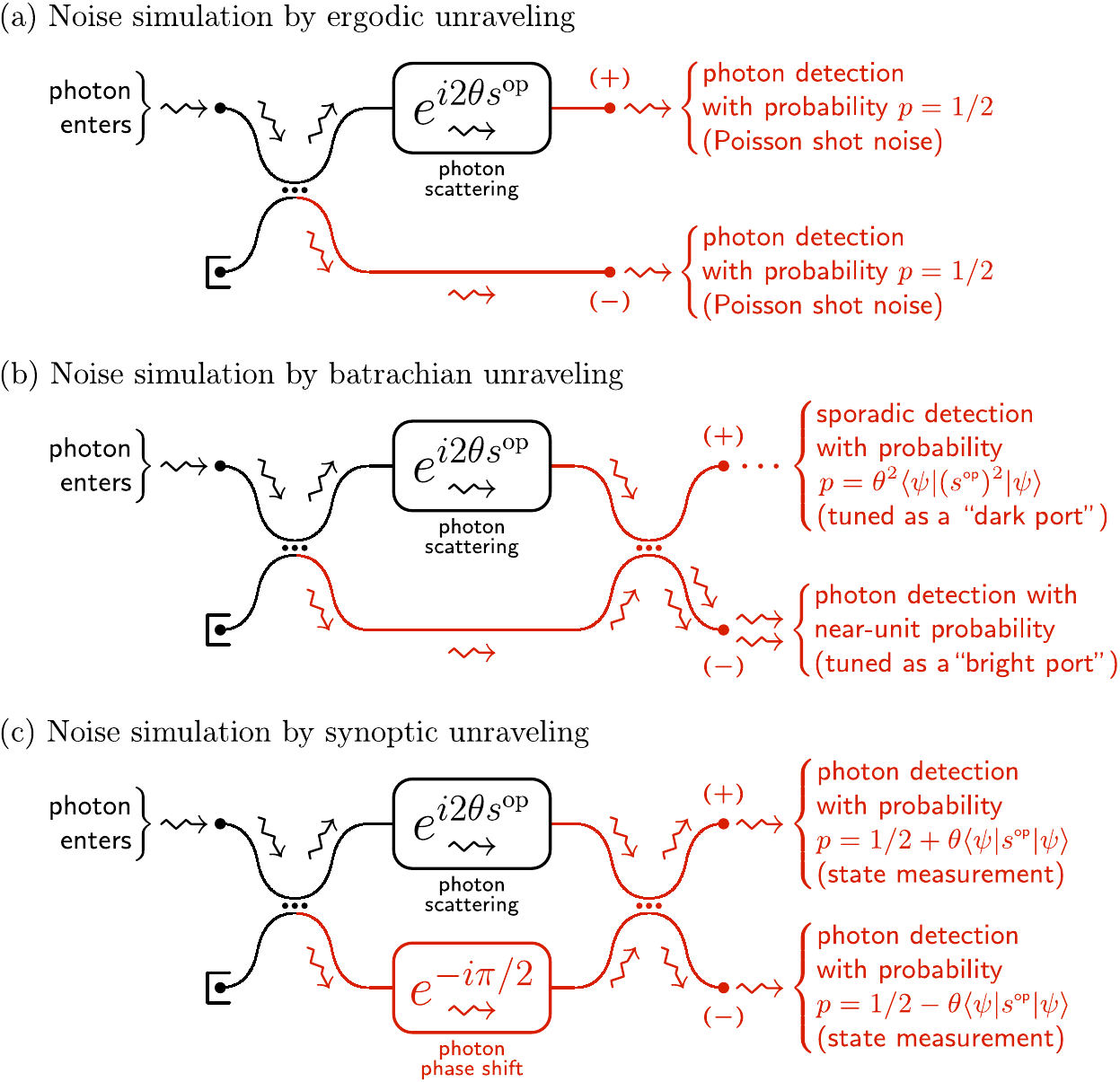}%
\caption[A physical embodiment of the \emph{Theorema Dilectum}]{%
\protect\justifying%
A physical illustration of the \emph{Theorema Dilectum}.\\[1ex]
\label{fig: Choi}%
This physical embodiment of the formal \QMOR simulation algorithm of Fig.~\ref{fig: QMOR formal}, using $2\times2$ fiber-optic couplers to scatter photons off a~spin state.   In (a) the photons are detected with equal probability, such that no information about the state is obtained.  In (b) the phase is detected via homodyne (self-interfering) interferometry, such that a~small amount of information about about the state \emph{is} obtained.  To an outside observer, the fate of the downstream photons is immaterial, hence (a) and (b) must embody physically indistinguishable noise processes, despite their differing quantum description.  This is the physical content of the \emph{Theorema Dilectum}.  From a mathematical point of view, the free choice of an arbitrary downstream unitary transform upon the photon paths is manifested in the $U$-transform invariance of (\ref{eq: theorema dilectum}).  (On-line version only: photons in red are causally downstream of the scattering, and thus can be measured in arbitrary fashion without altering the physically observable properties of the noise being simulated.) }

\end{figure}

The stochastic measurement and noise processes in our simulations can be conceived as interferometric measurements performed on each scattered photon, and this physical picture will prove very useful in designing \MOR techniques.  Such measurements require that each incoming photon be interferometrically split \emph{before} it scatters from the spin, to allow subsequent interferometric recombination and measurement. It~is convenient to conceive of this initial splitting as performed by a~$2\times2$ single-mode optical fiber coupler, as illustrated in Figs.~\ref{fig: Choi}(a--c).  The devices of this figure may be regarded as physical embodiments of the simulation algorithm of Fig.~\ref{fig: QMOR formal}. 

This physical picture embodies the idealizing assumptions that optical couplers are exactly unitary, that photon emission into the fiber takes place at equally spaced intervals $\delta t$, and that photon detectors register a~single classical ``click'' upon detection of each photon, which occurs with detection probability $|a^\ioptext{out}|^2$.  The unitarity of photon scattering and interferometric propagation then ensures that each incoming photon results in precisely one detection click. 

With respect to the algorithm of Fig.~\ref{fig: QMOR formal}, the stochastic selection of operator $M_{\myupsymbol}^k$ versus $M_{\mydownsymbol}^k$ is physically identified with these clicks, such that the sole data set resulting from a~simulation is a~set of classical binary data streams, with each stream comprising the recorded clicks for a~measurement operator pair.  Such binary streams closely resemble real-world \MRFM experimental records, in which signals corresponding to photoelectrons from an optically-monitored cantilever are low-pass filtered and recorded.  

We remark that the interferometers of Fig.~\ref{fig: Choi}(a--c) can be regarded with equal validity either as idealized abstractions or as schematic descriptions of real-world experiments.  For example, in our simulations we will regard \MRFM cantilevers as spins of large quantum number~$j$, in which case the interferometer geometry of Fig.~\ref{fig: Choi}(c) is identical (in the topology of its light path) to the real-world fiber-optic interferometers used in typical \MRFM experiments \cite{Rugar:89,Bruland:99,Rugar:04}.  

Trapped-ion experiments are examples of continuously observed small-$j$ quantum systems, since the quantum mechanics of a~two-state atom is identical to the quantum mechanics of a~spin-$\frac{1}{2}$ particle.  Such experiments presently are conducted with photons detected directly as in Fig.~\ref{fig: Choi}(b)~\cite{Powell:02,Blatt:88,Leibfried:03}. We remark that it would be theoretically interesting, and experimentally feasible, to conduct trapped ion experiments with weakly interacting photons detected interferometrically, as in Fig.~\ref{fig: Choi}(c).  

In trapped ion experiments, the observed transitions in the photon detection rates are observed to have random telegraph statistics, which are typically attributed to ``quantum jumps'' \cite{Blatt:88,Leibfried:03} or to ``instantaneous transitions between energy levels''~\cite{Powell:02}.  As was recognized by Kraus more than twenty years ago~\cite[p.~98]{Kraus:83}, ``such reasoning is unfounded,'' and we will see explicitly that observation of telegraph statistics does not imply discontinuous evolution of $\ket\psi$.   Thus, readers having a classical \MOR background need not regard discussions of quantum jumps in the physics literature as being literally true.

\subsubsection{Physical calibration of scattering amplitudes}
\label{sec: M(a)}
The calibration of our simulations will rely upon a~physical principle that is well-established, but somewhat counterintuitive.  The principle is this: measurements of the scattering phase (\ref{eq:scattering phase}) suffice to provide detailed information about the Hamiltonian that is responsible for the scattering.  Gottfried's discussion of Fredholm theory in atomic scattering~\cite[Section~49.2]{Gottfried:66} provides a~good introduction to this topic.  

As a~calibration example, we consider here the measurement process associated with the simple interferometer of Fig.~\ref{fig: Choi}(a). Using the $\rm S$-matrix of Fig.~\ref{fig: physical}(c)  to propagate the input photon in Fig.~\ref{fig: Choi}(a) through the apparatus, the measurement operators $\{M_{\myupsymbol}^\ioptext{(a)},M_{\mydownsymbol}^\ioptext{(a)}\}$ associated with detection on the {\mytextupsymbol} and  {\mytextdownsymbol} channels  are 
\begin{equation}
\label{eq: M(a)}
\left[\begin{array}{c}
M_{\myupsymbol}^{\ioptext{(a)}}(\theta,s^\ioptext{op})\\
M_{\mydownsymbol}^{\ioptext{(a)}}(\theta,s^\ioptext{op})
\end{array}\right]
= \left[\begin{array}{cc}
e^{i2\theta s^\ioptext{op}}&0\\
0&1
\end{array}\right]\frac{1}{\sqrt{2}}\left[\begin{array}{cc}
1&i\\
i&1
\end{array}\right]
\left[\begin{array}{cc}
1\\
0
\end{array}\right]
 = \frac{1}{\sqrt{2}}\left[\begin{array}{c}
e^{i2\theta s^\ioptext{op}}\\
i
\end{array}\right].
\end{equation}
The above equation follows the convention that the optical amplitudes of the top (bottom) fiber path in Fig.~\ref{fig: Choi}(a) are listed as the top (bottom) element of the above column arrays, such that successive interferometric couplings are described by successive $2\times2$~unitary matrix operations. This convention is common in optical engineering because it unites the geometric and algebraic descriptions of Fig.~\ref{fig: physical}(b--c).  In the context of our quantum simulations, each element of the above arrays is formally an operator on the wave function $\ket{\psi}$, but for  $c$-number (complex number) array elements like ``0'', ``1'' and ``$i$'' an implicit identity operator is omitted for compactness of notation. These identity operators physically correspond to events that are not dynamically coupled to the spin state, like photon emission, propagation through interferometers, and subsequent detection.  

The resulting overall measurement operators $\{M_{\myupsymbol}^{\ioptext{(a)}},M_{\mydownsymbol}^{\ioptext{(a)}}\}$ describe the state-dependence of the scattered phase of the detected photon, and it is evident that they satisfy the measurement operator completeness relation~(\ref{eq: measurement normalization}).  

\subsubsection{Noise-induced Stark shifts and renormalization}
We now consider an optical scattering effect known as the \ACacro Stark shift, as induced by the scattering process of Fig.~\ref{fig: Choi}(a).   Applying the simulation algorithm, we find that during a~time $\Delta t \gg \delta t$ the final state of the simulation accumulates a~state-dependent phase shift such that 
\begin{equation}
\ket{\psi(\Delta t)} = \exp(i2  n_\myupsymbol \theta s^\ioptext{op} )\,\ket{\psi(0)}, \end{equation}
where $n_\myupsymbol$ is the number of photons detected on the \mytextupsymbol-channel.  It is easy to show that $n_\myupsymbol$ has mean $\mu_\myupsymbol	= \Delta t/(2 \delta t)$ and standard deviation $\sigma_\myupsymbol	=  \sqrt{\mu_\myupsymbol/2}$.  

We see that the average effect of the photon scattering is equivalent to a~dynamical Hamiltonian $H^\ioptext{op}$, such that
\begin{equation}
\label{eq: Stark Hamiltonian}
H^\ioptext{op} = -\theta s^\ioptext{op} \hbar/\delta t,
\end{equation}which we identify as the effective Hamiltonian of a~Stark shift.   The Stark shift fluctuates due to statistical fluctuations in the number of photons detected on the \mytextupsymbol-channel, such that in the continuum limit the photon detection rate $r(t)$ is a~stochastic process having mean $\mu_r = 1/(2\delta t)$ and white-noise spectral density $S_{r} = 1/(4\delta t)$ (in a~two-sided spectral density convention).  This implies that the Stark shift fluctuations have an operator-valued power spectral density (\PSD)
\begin{equation}
\label{eq: Stark PSD}
S_{H^\ioptext{op}} =  
\theta^2 (s^\ioptext{op})^2 \hbar^2/\delta t = 
(H^\ioptext{op})^2 \delta t.
\end{equation} 
This result calibrates the externally-observable Stark shift parameters $\{H^\ioptext{op},S_{H^\ioptext{op}}\}$ in terms of the internal simulation parameters $\{\theta s^\ioptext{op},\delta t\}$ and \emph{vice versa}. 

The preceding two equations (\ref{eq: Stark Hamiltonian}--\ref{eq: Stark PSD}) reflect the well-known phenomenon in physics that interaction with a measurement process or thermal reservoir \emph{renormalizes} the physical properties of a system.  But as presented here, these same equations exhibit a known pathology in the limit $\delta t \to 0$: if we take $\theta \propto \sqrt{\delta t}$ such that  the Stark noise (\ref{eq: Stark PSD}) has a finite limit, then the magnitude of the Stark shift Hamiltonian (\ref{eq: Stark Hamiltonian}) diverges.  The origin of this (unphysical) infinite energy shift is our  (equally unphysical) assumption that measurement ``clicks'' occur at infinitely short intervals, such that the \PSD of the measurement noise is white; such white noise processes inherently are associated with infinite energy densities. 

In the present article we will simply repair this white-noise pathology by simply adding a counter-term to the measurement process, such that the Stark shift is zero in all our measurement processes.  In the language of renormalization theory, we redefine all of our measurement operators so that they refer to the ``dressed'' states of the system.   See (\emph{e.g.}) \cite{Sidles:03} and \cite{Leggett:2004uo} for further discussion of this delicate point, whose detailed analysis is beyond the scope of the present article. 

\subsubsection{Causality and the \emph{Theorema Dilectum}} 
\label{sec: Causality and scattering}
In Section \ref{sec: QMOR respects the Theorema Dilectum} we discussed the \emph{Theorema Dilectum} from a mathematical point of view that defined it terms of the (unitary) $U$-transformation of (\ref{eq: theorema dilectum}).  In Fig.~\ref{fig: Choi}(a--c) we associate the $U$-transformation with the physical choice of what to do ``downstream'' with photons that have scattered off the spin.  If we regard this downstream choice as being made made by Alice, and we assume that Bob is independently monitoring the same spin as Alice, then the physical content of the \emph{Theorema Dilectum} is that Alice's downstream measurement choices can have no observable consequences for Bob, and in particular, Alice cannot establish a communication channel with Bob via her measurement choices.

We therefore physically identify $\textrm{U}$ in (\ref{eq: theorema dilectum}) with the downstream optical couplers in Fig.~\ref{fig: Choi}(a--c), such that the algebraic freedom to specify an arbitrary unitary transform $\textrm{U}$ is identified with the physical freedom to ``tune'' the interferometer by adjusting its coupling ratio and fiber lengths (which adjust the phase of the output amplitudes relative to the input amplitudes).  Because experimentalists are familiar with this process and with the phrase ``interferometer tuning'' to describe it, we adopt the word ``tuning'' to describe the process of adapting $\textrm{U}$ to optimize $\{M_{\myupsymbol},M_{\mydownsymbol}\}$ for our simulation purposes.

\subsubsection{The \emph{Theorema Dilectum} in the literature}%
\label{sec: dilectum literature}

The invariance associated with $\textrm{U}$ has received most of its attention from the physics community only recently; it is not mentioned in most older quantum physics texts.  Both Preskill's class notes~\cite{Preskill:06} and Nielsen and Chuang's text~\cite{Nielsen:00} give physics-oriented proofs of the \emph{Theorema Dilectum}. The crux of all such proofs is that the choice of $\textrm{U}$ does not alter the density matrix associated with the ensemble average of all possible trajectories, such that the choice has no observable consequences. 

Carmichael \cite[p.~122]{Carmichael:93} seems to have been among the first to use the now-widely-used term \emph{unraveling} in describing quantum trajectory simulations, and he explicitly recognized that this unraveling is not unique:%
\begin{quote}We will refer to the quantum trajectories as an \emph{unravelling of the source dynamics} since it is an unraveling of the many tangled paths that the master equation evolves forward in time as a~single package. It is clear [\ldots] that the unraveling is not unique.\end{quote}%
(emphasis in the original). Diverse points of view regarding the ambiguity of trajectory unraveling can be found in the physics literature.  Rigo and Gisin~\cite{Rigo:96} argue that it is central to our understanding of the emergence of the classical world, and they make their case by presenting four different unravelings of a~single physical process; we will adopt a~similar multiple-unraveling approach in analyzing the \IBM single-spin experiment (see Section~\ref{sec: Single-spin MRFM simulations}).  Percival's text adopts the equally valid but sharply contrasting view that~\cite[p.\,46]{Percival:98}: ``the ambiguity [\ldots] is a~nuisance, so it is helpful to adopt a~convention which reduces this choice.''  Breuer and Petruccione~\cite{Breuer:02} simply state a~result equivalent to the \emph{Theorema Dilectum}, without further comment or attribution.  

Preskill's course notes and Nielsen and Chuang's text both take a~middle point of view. They  briefly describe the \emph{Theorema Dilectum} as a~``surprising ambiguity'' \cite[sec.~3.3]{Preskill:06} that is ``surprisingly useful''~\cite[sec.~8.2.4]{Nielsen:00}.  The word ``surprising'' invites readers to think for themselves about unravelling, and the word ``ambiguity'' suggests (correctly) that the implications of this invariance are not~fully understood.   Nielsen and Chuang's text notes that up to the present time, the main practical application of the \emph{Theorema Dilectum} has come in the theory of quantum error correction, where the freedom to choose unravelings that facilitate the design of error correction algorithms has been ``crucial to a~good understanding of quantum error correction''~\cite[sec.~8.2.4]{Nielsen:00}.  

In the context of quantum computing theory, Buhrman \emph{et al.} \cite{Buhrman:06} have exploited the informatic invariance of the \emph{Theorema Dilectum} to show that certain logical gates that are essential to universal quantum computing, when made noisy, can be indistinguishably replaced by randomly-selected gates from a restricted set of gates that can be simulated classically.  Their proofs build upon earlier work by many authors \cite{Aharonov:96,Harrow:03,Bravyi:05} and in particular upon the Gottesman-Knill theorem \cite{Gottesman:97}.  To our knowledge, this is the first formal quantum informatic proof that additional noise makes systems easier to simulate.

\subsection{Designs for spinometers}
\label{sec: spinometers}
\label{sec: calibration rules}

We now present some basic designs for measurement operators constructed from the fundamental set of spin operators $\{s_1,s_2,s_3\}$ for particles of arbitrary spin~$j$.  

We call this family of measurement operators \emph{spinometers}, and we will characterize their properties in such sufficient detail that in Section~\ref{sec: Single-spin MRFM simulations} we will be able to simulate the single-spin \MRFM experiment both numerically and by closed-form analysis.  Most of these results have not appeared previously in the literature.

		\subsubsection{Spinometer tuning options: ergodic, synoptic, and batrachian} 

Starting with the fundamental spinometer pair $\{M_{\myupsymbol}^\ioptext{(a)},M_{\mydownsymbol}^\ioptext{(a)}\}$ that was defined in (\ref{eq: M(a)}) and physically illustrated in Fig.~\ref{fig: Choi}(a), \emph{ergodic spinometers} are constructed by applying the following tuning 
\begin{eqnarray}
\nonumber
\left[\begin{array}{c}
M_{\myupsymbol}^{\ioptext{erg}}(\theta, s^\ioptext{op})\\
M_{\mydownsymbol}^{\ioptext{erg}}(\theta, s^\ioptext{op})
\end{array}\right]
& = \left[\begin{array}{cc}
e^{-i\theta s^\ioptext{op}}&0\\
0&e^{-i\theta s^\ioptext{op}}
\end{array}\right]\left[\begin{array}{cc}
1&0\\
0&e^{-i\pi/2}
\end{array}\right]
\left[\begin{array}{c}
M_{\myupsymbol}^{\ioptext{(a)}}\\
M_{\mydownsymbol}^{\ioptext{(a)}}
\end{array}\right],\\
 & = \frac{1}{\sqrt{2}}\left[\begin{array}{c}
e^{i\theta s^\ioptext{op}}\\
e^{-i\theta s^\ioptext{op}}
\end{array}\right].
\label{eq: ergodic}
\end{eqnarray}
Note that immediately following photon detection, and independent of the channel on which the photon is detected, a compensating Hamiltonian $\theta s^\ioptext{op}$ is applied to cancel the mean Stark shift that was noted in Section~\ref{sec: M(a)}.  The fluctuating portion of the Stark shift is \emph{not} thereby cancelled, and it follows that ergodic spinometers are well-suited for simulating physically realistic noise, \emph{e.g.}, random magnetic fields that decohere spin states.

We construct \emph{batrachian spinometers} from ergodic spinometers by adding a downstream coupler, as shown in  Fig.~\ref{fig: Choi}(b).  Our phase and tuning conventions are:
\begin{eqnarray}
\fl\qquad
\nonumber 
\left[\begin{array}{c}
M_{\myupsymbol}^{\ioptext{bat}}(\theta, s^\ioptext{op})\\
M_{\mydownsymbol}^{\ioptext{bat}}(\theta, s^\ioptext{op})
\end{array}\right]
& =
\left[\begin{array}{cc}
e^{-i\pi/2}&0\\
0&e^{-i\pi/2}
\end{array}\right]
\frac{1}{\sqrt{2}}
\left[\begin{array}{cc}
1&i\\
i&1
\end{array}\right]
 \left[\begin{array}{cc}
1&0\\
0&e^{i\pi/2}
\end{array}\right]
\left[\begin{array}{c}
M_{\myupsymbol}^{\ioptext{erg}}\\
M_{\mydownsymbol}^{\ioptext{erg}}
\end{array}\right], \\
& = \left[\begin{array}{c}
\sin\theta s^\ioptext{op}\\
\cos\theta s^\ioptext{op}
\end{array}\right].
\label{eq: batrachian}
\end{eqnarray}
The upper-right output port is tuned to be as dark as possible, such that detection clicks occur only sporadically; in experimental interferometery this is called \emph{dark port tuning}.   Each click that a dark port records is accompanied by a discrete jump in the wave function, hence the name ``batrachian'' for these measurement operators.  We will see that batrachian tuning is well-suited to the analysis of data statistics.  

We construct \emph{synoptic spinometers} similarly, but with a different phase tuning, as shown in Fig.~\ref{fig: Choi}(c).  Algebraically our tuning convention is 
\begin{eqnarray}
\nonumber 
\left[\begin{array}{c}
M_{\myupsymbol}^{\ioptext{syn}}(\theta, s^\ioptext{op})\\
M_{\mydownsymbol}^{\ioptext{syn}}(\theta, s^\ioptext{op})
\end{array}\right]
& = \left[\begin{array}{cc}
e^{-i\pi/4}&0\\
0&e^{-i\pi/4}
\end{array}\right]
\frac{1}{\sqrt{2}}
\left[\begin{array}{cc}
1&i\\
i&1
\end{array}\right]
\left[\begin{array}{c}
M_{\myupsymbol}^{\ioptext{erg}}\\
M_{\mydownsymbol}^{\ioptext{erg}}
\end{array}\right], \\
& = \frac{1}{\sqrt{2}}\left[\begin{array}{c}
\cos\theta s^\ioptext{op}+\sin\theta s^\ioptext{op}\\
\cos\theta s^\ioptext{op}-\sin\theta s^\ioptext{op}
\end{array}\right].
\label{eq: synoptic}
\end{eqnarray}
We will see that synoptic spinometers \emph{do} provide information about the quantum state---hence the name ``synoptic''---and also that they compress quantum trajectories.

We can now discern the general strategy of \QMOR analysis: we model physical noise in terms of ergodic operators, we predict data statistics by the analysis of batrachian operators, and we compress simulated trajectories by applying synoptic operators.

\subsubsection{Spinometers as agents of trajectory compression}

The following derivations assume a knowledge of basic quantum mechanics at the level of Chapters~2 and~8 of Nielsen and Chuang~\cite{Nielsen:00} (an alternative text is Griffiths~\cite{Griffiths:05}), knowledge of coherent spin states at the level of Perelomov~\cite[eqs.~4.3.21--45]{Perelomov:86} (alternatively see Klauder~\cite{Klauder:84} or del Castillo~\cite{Castillo:04}), and knowledge of stochastic differential equations at the level of Gardiner~\cite[sec.~4.3]{Gardiner:85} (alternatively see Rogers~\cite{Rogers:00}). 

Generally speaking, the design rules of this section were first heuristically suggested by the Hilbert ontology of Section~\ref{sec: QMOR respects}, then confirmed by numerical experiments, and finally proved by analysis.  Some of the lengthier proofs would have been difficult to discover otherwise; this shows the utility of the Hilbert ontology backed-up by numerical exploration.

Nowhere in the derivations of this section will we make any assumption about the dimensionality of the Hilbert space in which the trajectory $\{\ket{\psi_n}\}$ resides.  Therefore we are free to regard $\ket{\psi_n}$ as describing a spin-$j$ particle that is embedded in a larger multi-spin Hilbert space.  Thus all the theorems and calibrations that we will derive will be applicable both to the single-spin \MRFM Hilbert space of Section~\ref{sec: Single-spin MRFM simulations} and to the large-dimension ``spin-dust'' spaces that we will discuss in Section~\ref{sec: Spin-dust simulations}.

\subsubsection{Spinometers that einselect eigenstates}
\label{sec: compress}

We define a \emph{uniaxial spinometer} to be a measurement process associated with a single pair of measurement operators having generator $s^\ioptext{op}$.  We can regard $s^\ioptext{op}$ as an arbitrary Hermitian matrix, since in a uniaxial measurement there are no other operators for it to commute with. We consider ergodic, synoptic, and batrachian tunings as defined in (\ref{eq: ergodic}--\ref{eq: batrachian}).  Without loss of generality we assume $\tr(s^\ioptext{op})^2=\dim \ket\psi$, \emph{i.e.}, the mean-square eigenvalues of $s^\ioptext{op}$ are unity, which sets the scale of the coupling~$\theta$.

For a general state $\ket{\psi_n}$ and general Hermitian operator $s^\ioptext{op}$, we define the operator variance $\Delta_n(s^\ioptext{op})$ to be
\begin{equation}
  \label{eq:variance}
  \Delta_n(s^\ioptext{op}) = \braket{\psi_n}{\left(s^\ioptext{op}-\braket{\psi_n}{s^\ioptext{op}}{\psi_n}\right)^{2}}{\psi_n}\end{equation}
remarking that in a finite-$j$ Hilbert space
\begin{equation}
\label{eq: uniaxial inequality}
\Delta_n(s^\ioptext{op})
\left\{\begin{array}{ll}
  =0&\mbox{if $\ket{\psi_n}$ is an eigenstate of $s^\ioptext{op}$,}\\
  >0&\mbox{otherwise}.
\end{array}\right.
\end{equation}
Physically speaking, the smaller the variance $\Delta_n(s^\ioptext{op})$, the smaller the quantum fluctuations in the expectation value $\braket{\psi_n}{s^\ioptext{op}}{\psi_n}$.  We will now calculate the rate at which measurement operators act to minimize this variance.

Considering an ensemble of simulation trajectories, we define the ensemble-averaged variance at the $n$-th simulation step to be $E[\Delta_n(s^\ioptext{op})]$.  The algorithm of Fig.~\ref{fig: QMOR formal} evolves this mean variance according to 
\begin{equation}
\label{eq: pseudocode}
\fl\ E[\Delta_{n+1}(s^\ioptext{op})]=\sum_{k=1}^m\,
\sum_{j\in\{\myupsymbol,\mydownsymbol\}} 
\hspace{-1.3ex}E\!\left[\braket{\psi_n}{(M^k_j)^\dagger(s^\ioptext{op})^2M^k_j}{\psi_n}
-\frac{\braket{\psi_n}{(M^k_j)^\dagger s^\ioptext{op}M^k_j}{\psi_n}^2}
{\braket{\psi_n}{(M^k_j)^\dagger M^k_j}{\psi_n}}\right]
\end{equation}
For compactness we write the increment of the variance as $\delta\Delta_n(s^\ioptext{op}) \equiv \Delta_{n+1}(s^\ioptext{op})]-\Delta_n(s^\ioptext{op})$.  Then for ergodic, synoptic, and bratrachian tunings the mean increment is 
\begin{equation}
\label{eq: uniaxial enselection}
E[\delta\Delta_n(s^\ioptext{op})] = \left\{
\begin{array}{l@{\hspace{1.5em}}l}
  0&\mbox{\textit{ergodic tuning}},\\
  -4\theta^{2} E[\Delta_n^2(s^\ioptext{op})]&
  \mbox{\textit{synoptic tuning}},\\
  -\theta^{2} E[\lcal{F}_n(s^\ioptext{op})]&
  \mbox{\textit{batrachian tuning}}.
  \end{array}\right.\hspace{-2ex}
\end{equation}
These results are obtained by substituting in \myeq{eq: pseudocode} the spinometer tunings of (\ref{eq: ergodic}--\ref{eq: batrachian}), then expanding in $\theta$ to second order.  Here $\lcal{F}$ is the non-negative function
\begin{equation}
\lcal{F}_n(s^\ioptext{op})=\frac{\big(\,
\braket{\psi_n}{(s^\ioptext{op})^3}{\psi_n}-\braket{\psi_n}{s^\ioptext{op}}{\psi_n}\braket{\psi_n}{(s^\ioptext{op})^2}{\psi_n}\,\big)^2}{\braket{\psi_n}{(s^\ioptext{op})^2}{\psi_n}}.
\end{equation}
Each term in the sequence $\{E[\Delta_1],E[\Delta_2],\ldots\}$ is nonnegative by \myeq{eq:variance}, and yet for synoptic and batrachian tuning the successive terms in the sequence are non-increasing (because in \myeq{eq: uniaxial enselection} the quantities  $\Delta_n^2(s^\ioptext{op})$ and $\lcal{F}_n(s^\ioptext{op})$ are nonnegative and there is an overall minus sign acting on them); the sequence therefore has a limit.  For synoptic tuning the limiting states are evidently such that $\Delta_n(s^\ioptext{op})\to 0$, while for batrachian tunings $\lcal{F}_n(s^\ioptext{op})\to 0$, which in both cases implies that the limiting states are eigenstates of $s^\ioptext{op}$. This proves
\begin{myDesignRule}
\label{th:uniaxial}
Uniaxial spinometers with synoptic or batrachian tunings, but not ergodic tunings, 
asymptotically einselect eigenstates of the measurement generator.
\end{myDesignRule}
		\subsubsection{Convergence bounds for the einselection of eigenstates}
We now prove a bound on the convergence rate of Design Rule~\ref{th:uniaxial}.  For \QMOR purposes, this bound provides an important practical assurance that an ensemble of uniaxially observed spins never becomes trapped in a ``dead zone'' of state space. 

To prove the convergence bound, we notice that in the continuum limit $\theta\ll1$ the increment~\myeq{eq: uniaxial enselection} can be regarded as a differential equation in simulation time $t\equiv n\,\delta t$.  For synoptic tuning the inequality $E[\Delta_n^2(s^\ioptext{op})]\ge E[\Delta_n(s^\ioptext{op})]^2$ then allows us to derive---by integration of the continuum-limit equation---the power-law inequality
\begin{equation}
\label{eq: uniaxial bound}
E[\Delta_n(s^\ioptext{op})] \le E[\Delta_0(s^\ioptext{op})]/(1+4n\theta^2).
\end{equation}This implies that the large-$n$ variance is $\Or(n^{-1})$. This proof nowhere assumes that the initial ensemble is randomly chosen; therefore the above bound applies to \emph{all} ensembles, even those whose initial quantum states are chosen to exhibit the slowest possible einselection.  We conclude that for synoptic tuning the approach to the zero-variance limit is never pathologically slow.  We have not been able to prove a similar bound for batrachian tuning, but numerical experiments suggest that both tunings require a time $t \sim \delta t/\theta^2$ to achieve einselection.  Proofs of stronger bounds would be valuable for the design of large-scale \QMOR simulations.

\subsubsection{Triaxial spinometers}

We now consider \emph{triaxial spinometers}, in which three pairs of synoptic measurement operators (\ref{eq: synoptic}) are applied, having as generators the spin operators $\{s_x, s_y, s_z\}$, applied with couplings $\{\theta_x, \theta_y, \theta_z\}$.  

\subsubsection{The Bloch equations for general triaxial spinometers}
In the general case we take $\theta_1 \ne \theta_2 \ne \theta_3$.  We define $\lb{x}_n = \{x_n, y_n, z_n\} = j \braket{\psi_n}{\gsb s}{\psi_n}$ to be the polarization vector at the $n$'th simulation step.  This vector is normalized such that $|\lb{x}_n|\le 1$, with $|\lb{x}_n| = 1$ if and only if $\ket{\psi_n}$ is a coherent state.  We further define $\delta\lb{x}_n=\lb{x}_n-\lb{x}_{n-1}$.  Taking as before $E[\ldots]$ to be an ensemble average over simulations, such that the density matrix of the ensemble is $\rho_n = E[\ket{\psi_n}\bra{\psi_n}]$, and therefore $E[\lb{x}_n] = j \tr \lb{s}\rho_n$, we readily calculate that the Bloch equation that describes the average polarization of the ensemble of simulations is
\begin{equation}
\label{eq: infinite temperature Bloch equations}
\left[\begin{array}{c}
\rule[-1ex]{0pt}{3.25ex}
E[\delta x_{n}]\\
E[\delta y_{n}]\\
E[\delta z_{n}]
\rule[-1ex]{0pt}{2.5ex}
\end{array}\right] =
-\frac{1}{2}\left[\begin{array}{ccc}
\rule[-1ex]{0pt}{3.25ex}
\theta_y^2+\theta_z^2 & 0 & 0 \\
0 &\theta_x^2+\theta_z^2 & 0 \\
0 & 0 &\theta_x^2+\theta_y^2 
\rule[-1ex]{0pt}{2.5ex}
\end{array}\right]
\left[\begin{array}{c}
\rule[-1ex]{0pt}{3.25ex}
E[x_{n-1}]\\
E[y_{n-1}]\\
E[z_{n-1}]
\rule[-1ex]{0pt}{2.5ex}
\end{array}\right]
\end{equation}
Since it depends only linearly upon $\rho_n$, the above expression is invariant under the $U$-transform of the \emph{Theorema Dilectum}.  We are free, therefore, to regard our spinometers as being ergodically tuned (\ref{eq: ergodic}), such that the simulation can be equivalently regarded, not as three competing axial measurement processes, but as independent random rotations being applied along the $x$-axis, $y$-axis, and $z$-axis.  The above Bloch equation therefore has the functional form that we expect upon  purely classical grounds.

\subsubsection{The einselection of coherent states}
Now we confine our attention to balanced triaxial spinometers, \emph{i.e.}, those having with $\theta_1 = \theta_1 = \theta_1 \equiv \theta$, such that no one axis dominates the measurement process.  Numerical simulations suggest that for synoptically  tuned measurement processes, in the absence of entangling Hamiltonian interactions, simulated quantum trajectories  swiftly converge to coherent state trajectories, regardless of the starting quantum state.  We adopt Zurek's (exceedingly useful)  concept of \emph{einselection} \cite{Zurek:03} to describe this process.  We now prove that synoptic spinometric observation processes always induces einselection by calculating a rigorous lower bound upon the rate at which einselection occurs.

Given an arbitrary state $\ket\psi$, we define a \emph{spin covariance} matrix $\Lambda_n$ to be the following $3\times3$ Hermitian matrix (of $c$-numbers): \begin{equation}
\label{eq: lambda}
(\Lambda_n)_{kl} \equiv 
\braket{\psi_n}{s_{k}s_{l}}{\psi_n}-
\braket{\psi_n}{s_{k}}{\psi_n}
  \braket{\psi_n}{s_{l}}{\psi_n}.
\end{equation}
This matrix covariance is a natural generalization of the scalar variance $\Delta_n(s^\ioptext{op})$ (\ref{eq:variance}), and in particular it satisfies a trace relation that is similar to \myeq{eq: uniaxial inequality}
\begin{equation}
\label{eq:triaxial inequality}
\tr\Lambda_n
\left\{\begin{array}{ll}
  =j&\mbox{if $\ket{\psi_n}$ is a coherent spin state,}\\
  >j&\mbox{otherwise}.
\end{array}\right.
\end{equation}
Here a \emph{coherent spin state} is any spin-$j$ state $\ket{\lbhat{x}}$, conventionally labeled by a unit vector~$\lbhat{x}$, such that $\braket{\lbhat{x}}{\gsb{s}}{\lbhat{x}} = j\lbhat{x}$  (see, \emph{e.g.}, Perelomov~\cite[eq.~4.3.35]{Perelomov:86}).   The algorithm of Fig.~\ref{fig: QMOR formal} evolves the mean spin covariance according to 
\begin{align}
\label{eq: pseudocode II}
\nonumber
(E[\Lambda_{n+1}])_{lm}=
\sum_{k=1}^m\,
\sum_{j\in\{\myupsymbol,\mydownsymbol\}} 
\hspace{-1.3ex}
E\Bigg[&\braket{\psi_n}{(M^k_j)^\dagger s_l s_m M^k_j}{\psi_n}\\
&\displaystyle -\frac{\braket{\psi_n}{(M^k_j)^\dagger s_l M^k_j}{\psi_n}\braket{\psi_n}{(M^k_j)^\dagger s_m M^k_j}{\psi_n}}
{\braket{\psi_n}{(M^k_j)^\dagger M^k_j}{\psi_n}}\Bigg]
\end{align}
For compactness we define the $\Lambda$-increment $\delta\Lambda_n\equiv\Lambda_{n+1}-\Lambda_n$.  Then by a series expansion of \myeq{eq: pseudocode II} similar to that which led from \myeq{eq: pseudocode} to (\ref{eq: uniaxial enselection})---but with more indices---we find that for ergodic, synoptic, and batrachian tuning the mean increment is
\begin{equation}
  \label{eq:triaxial increment}
\tr E[\delta\Lambda_n] = \left\{
\begin{array}{l@{\hspace{1.5em}}l}
  0&\mbox{\textit{ergodic tuning}},\\
  -4\theta^{2}\ \tr E[\Lambda_n\mycdot\Lambda^{\star}_n]&
  \mbox{\textit{synoptic tuning}},\\
  \mbox{(see text)}&
  \mbox{\textit{batrachian tuning}}.
  \end{array}\right.\hspace{-2ex}
\end{equation}
The ``see text'' for batrachian tuning indicates that we have found no closed-form expression simpler than several dozen terms; numerical experiments show that for this tuning the covariance exhibits random jump-type fluctuations that seemingly have no simple limiting behavior. In contrast, synoptic tuning's increment has a strikingly simple analytic form, which was guessed as an \emph{ansatz} and subsequently verified by machine algebra.  

Proceeding as in the proof of Theorem~1, and temporarily omitting the subscript~$n$ for compactness, we now prove that for $\Lambda$ computed from $\ket\psi$ by \myeq{eq: lambda}, the scalar quantity $\tr\Lambda\mycdot\Lambda^{\star}$ is non-negative for all $\ket\psi$, and vanishes if and only if $\ket{\psi}$ is a coherent state.  We remark that this proof is nontrivial because $\tr\Lambda\mycdot\Lambda^{\star}$ is by no means  a positive-definite quantity for general Hermitian $\Lambda$; an example is $\Lambda = {\big[\begin{smallmatrix}\hfill0&+i\\-i&\hfill0\end{smallmatrix}\big]}$, for which $\tr\Lambda\mycdot\Lambda^{\star}=\tr{\big[\begin{smallmatrix}-1&\hfill0\\\hfill0&-1\end{smallmatrix}\big]}=-2$.  

Because $\Lambda$ is a Hermitian $3\times3$ matrix, it can be decomposed uniquely into a real symmetric matrix $\sbar{\Lambda}$ and a real vector $\lb{v}=\braket{\psi}{\gsb{s}}{\psi}/j$ by  $\Lambda_{ik}=\sbar{\Lambda}_{ik}+i/2\,\sum_{l=1}^3 \epsilon_{ikl}v_l$.   Because the increment \myeq{eq:triaxial increment} is a scalar under rotations, without loss of generality we can choose a reference frame having basis vectors $\{\lbhat{x},\lbhat{y},\lbhat{z}\}$ such that $\{\lb{v}{\cdot}\lbhat{x},\lb{v}{\cdot}\lbhat{y},\lb{v}{\cdot}\lbhat{z}\}=\{0,0,z\}$ and $z\ge0$.  In this reference frame, the following decomposition is valid for any Hermitian matrix~$\Lambda$ (\emph{i.e.},~it~holds for $\sbar\Lambda$ an arbitrary symmetric matrix and $z$ an arbitrary real number):
\begin{align}
\nonumber\tr\Lambda\mycdot(\Lambda^{\star})
= {}& j^2(1-z^2)/2+
j^4(1-z^2)^2/4 \\ 
&{}+
           2 p_{\ioptext{a}}+
\tfrac{1}{2} p_{\ioptext{b}}+
\tfrac{1}{2} p_{\ioptext{c}}+
\tfrac{1}{4} p_{\ioptext{d}}+
\tfrac{1}{2} p_{\ioptext{e}}+
           j p_{\ioptext{f}},
\label{eq:asymmetric decomposition}
\end{align}
where the residual terms $p_\ioptext{a},p_\ioptext{b},\ldots,p_\ioptext{f}$ are
\begin{equation}
\label{eq: probs}
\mbox{\begin{tabular}[b]{@{}r@{\ =\ }l@{\hspace{2.5em}}r@{ = }l}
  $p_{\ioptext{a}}$&$\sbar{\Lambda}_{12}^{2}+ \sbar{\Lambda}_{13}^{2}+\sbar{\Lambda}_{23}^{2}$,&
  $p_{\ioptext{d}}$&$(j^2(1-z^2)-2\sbar\Lambda_{33})^{2}$\\
  $p_{\ioptext{b}}$&$(\sbar{\Lambda}_{11}-\sbar{\Lambda}_{22})^{2}$,&
  $p_{\ioptext{e}}$&$(\sbar{\Lambda}_{11}+\sbar{\Lambda}_{22})^{2}-(j(j+1)-
  (\sbar{\Lambda}_{33}+j^2 z^2))^{2}$\\
  $p_{\ioptext{c}}$&$(\sbar\Lambda_{33})^{2}$,&
  $p_{\ioptext{f}}$&$j^{2}(1-z^2)-\sbar{\Lambda}_{33}$
\end{tabular}}
\end{equation}
We now prove that each term in this decomposition is non-negative.  The terms $p_{\ioptext{a}}$, $p_{\ioptext{b}}$, $p_{\ioptext{c}}$, and $p_{\ioptext{d}}$ are non-negative \emph{prima facie}.  The term $p_{\ioptext{e}}$ vanishes for arbitrary $\ket{\psi}$ in consequence of the spin operator identity $\sbar{\Lambda}_{11}+\sbar{\Lambda}_{22}+\sbar{\Lambda}_{33} + j^2z^2 =\braket{\psi}{s_1^2+s_2^2+s_3^2}{\psi}=j(j+1)$.   That the remaining terms are non-negative in general follows from the spin operator inequalities $-j\le\braket{\psi}{s_3}{\psi}\le j$ and $0\le\braket{\psi}{s_3^2}{\psi}\le j^2$, which together with our reference-frame convention imply the inequalities $-1\le z\le1$ and $\sbar\Lambda_{33}<j^2 (1-z^2)$.  

Next, we show that the sum of terms \myeq{eq:asymmetric decomposition} vanishes if and only if $\ket\psi$ is coherent, \emph{i.e.}, if and only if $\ket\psi=\ket{\lbhat{z}}$.  It is a straightforward exercise in spin operator algebra to show that $\ket\psi = \ket{\lbhat{z}}$ if and only if all of the following are true: $z=1$, $\sbar\Lambda_{33}=j^2$, $\sbar\Lambda_{11}=\sbar\Lambda_{22}$ and $\sbar{\Lambda}_{12}=\sbar{\Lambda}_{23}=\sbar{\Lambda}_{13}=0$; it follows that  \myeq{eq:asymmetric decomposition} vanishes if and only if $\ket\psi=\ket{\lbhat{z}}$.  By reasoning similar to Theorem~1, we conclude: 
\begin{myDesignRule}
\label{th:triaxial}
Triaxial spinometers with synoptic tunings 
asymptotically einselect coherent spin states.
\end{myDesignRule}

\subsubsection{Convergence bounds for the einselection of coherent states} 
\label{sec: coherent bounds}
We now exploit the identity~(\ref{eq:asymmetric decomposition}) to prove a bound on the convergence of Design Rule~\ref{th:triaxial}.  Our strategy is similar to our previous proof of the bound for Design Rule~\ref{th:uniaxial}. 
Substituting the identity $j^2(1-z^2)= \tr\Lambda_n-j$ in the first two terms of (\ref{eq:asymmetric decomposition}), and taking into account the non-negativity of the remaining terms $p_\ioptext{a}$, $p_\ioptext{b}$, $p_\ioptext{c}$, $p_\ioptext{d}$, and $p_\ioptext{e}$, we obtain the following quadratic inequality in $(\tr\Lambda_n-j)$:
\begin{equation}
\tr\Lambda_n\mycdot\Lambda^{\star}_n\ \ge\ 
\tfrac{1}{2}(\tr\Lambda_n-j)+\tfrac{1}{4}(\tr\Lambda_n-j)^2\ge 0.
\end{equation}
As an aside, our starting identity (\ref{eq:asymmetric decomposition}) was devised so as to imply a general inequality having the above quadratic functional form, in service of the proof that follows, but we have not been able to prove that the above coefficients $\{\tfrac{1}{2},\tfrac{1}{4}\}$ are the largest possible.

Upon taking an ensemble average of the above inequality, followed by substitution in \myeq{eq:triaxial increment}, followed by a continuum-limit integration, we obtain the following convergence bound for the ensemble-averaged trace covariance:
\begin{equation}
\tr E[\Lambda_n]-j\ \le\ \frac{2 (\tr E[\Lambda_0]-j)}%
{(\tr E[\Lambda_0]-j)(\exp({2n\theta^2})-1)+2\exp(2n\theta^2)}.
\end{equation}
It is instructive to restate this bound in terms of simulation time $t=n\,\delta t$.  Taking the continuum limit $\Lambda_n\to\Lambda(t)$, noting that the timescale $T_1=\delta t/\theta^2$ is the conventional $T_1$ that appears in the Bloch equations (\ref{eq: infinite temperature Bloch equations}), defining for compactness of notation the initial trace covariance to be $\kappa_0 \equiv \tr E[\Lambda(0)]-j$, and assuming for the sake of discussion that $\kappa_0\gg 1$ (\emph{i.e}, we assume that the initial ensemble is far-from-classical) the functional form of the above bound exhibits three asymptotic intervals, whose $t$-dependence is respectively $\lcal{O}(1)$, $\lcal{O}(1/t)$, and $\lcal{O}(\exp(-2t/T_1))$:
\begin{equation}
\fl\hfill(\tr E[\Lambda(t)]-j)\ \lesssim
\left\{
\begin{array}{l@{\hspace{1em}}l}
\kappa_0 (1-\kappa_0 t/T_1 ) &\mbox{for}\ \ 0\le t/T_1 \lesssim 1/\kappa_0,\\[0.5ex]
T_1/t &\mbox{for}\ \ 1/\kappa_0\lesssim t/T_1 \lesssim 1\\[0.5ex]
2 \exp(-2t/T_1)&\mbox{for}\ \ 1 \lesssim t/T_1  \\
\end{array}
\right.
\hfill
\label{eq: coherent bound}
\end{equation}
The $\lcal{O}(1)$ and $\lcal{O}(1/t)$ behavior is functionally similar to the convergence bound established in \myeq{eq: uniaxial bound} for the eigenvalue variance  of Theorem~1, namely, an initial linear decrease, followed by an $\lcal{O}(1/t)$ fall-off.   Unique to triaxial spinometry (as far as the authors know) is the final exponentially rapid convergence to a coherent state.  

We note that convergence is complete within a time ${\sim}\,T_1$ that is independent of both the spin quantum number~$j$ and the overall dimensionality of the Hilbert space in which the spin is embedded. %
As with Theorem~1, this is a worst-case bound that applies to \emph{all} ensembles, including (for example) exotic ensembles initialized with ``Schr\"{o}edinger's cat'' states.  More particularly, it applies to large-dimensional ensembles in which each of~$n$~spins in a Hilbert space of overall dimension~$(2j+1)^n$ is synoptically observed. 

\subsubsection{Implications of einselection bounds for quantum simulations} 
We now begin to have a quantitative appreciation of the geometric assertion of Fig.~\ref{fig: QMOR geometry}(i), that quantum simulations can be regarded as theaters in~which the trajectory compression of synoptic observation opposes the creation of entanglement by Hamiltonian dynamics, with the balance between compression and expansion determining the dimensionality of the \QMOR state space required for accurate simulation.  

Even stronger convergence bounds than those we have proved would be valuable in designing \QMOR simulations.  Especially useful would be more tunings in which noise is realized as an \emph{entangled measurement}.  Physically speaking, an entangled measurement is performed by interferometrically splitting a photon along $n$ paths, scattering the photon from a different spin along each path, then recombining and measuring the photon by freely choosing among any of an exponentially large set of braidings and interferometric couplings of the downstream optical fibers.  

The analysis of such noise-equivalent tunings would require mathematical methods considerably more sophisticated than those we have deployed in this article.  A known consequence of the Holevo-Schumacher-Westmoreland (\HSW) theorem (which is the quantum analog of the Shannon channel capacity theorem) is that entangled measurements are necessary to maximize the information capacity of quantum channels~\cite[sec{.}~12.3.2]{Nielsen:00}.  

If~we hypothesize that quantum trajectory compression is in some sense proportional to information extracted by measurement, then the \HSW theorem tells us that entangled measures will be more effective for \QMOR purposes than the single-spin measures that we consider in this article.  It is likely, therefore, that the search for more efficient \QMOR techniques will benefit considerably from continued progress in  quantum information theory.

\subsubsection{Positive $P$-representations of the thermal density matrix}
		
Now we focus upon control and
thermodynamics. For $\lbhat{t}$ the thermal axis defined
in (\ref{eq:thermal operator}), we modify the synoptic spinometer
matrices such that
\begin{eqnarray}
  \label{eq:control}
    M^{k}_{\myupsymbol} &= 
    e^{-i \alpha \theta (\lbhat{t}\times\lb{s})_{k}}
    [\cos (\theta s_{k}) + \sin (\theta s_{k})]/\sqrt{2}\,,\\
    M^{k}_{\mydownsymbol} &= 
    e^{+i \alpha \theta (\lbhat{t}\times\lb{s})_{k}}
    [\cos (\theta s_{k}) - \sin (\theta s_{k})]/\sqrt{2}\,,
\end{eqnarray}
where $\alpha$ is the control gain. We will call this a
\emph{closed-loop triaxial spinometer} with \emph{unitary
feedback}, because (as we will see) the unitary operators $\exp (\pm i
\alpha\theta\,\lbhat{t}\times\lb{s})$ act cumulatively to align
the spin axis with $\lbhat{t}$.

Closing the control loop does not alter the coherent
einselection because the sole effect of a \emph{post hoc}
unitary operator on $\gsb\sigma_{n}$ is a spatial
rotation. Since
$\tr\gsb{\sigma}_{n}\mycdot\gsb{\sigma}_{n}^{\star}$
is a rotational scalar, \myeq{eq:triaxial increment} still
holds. Thus we have
\begin{myDesignRule}
\label{dr: closed-loop einselection}
Closed-loop triaxial spinometers with unitary feedback 
asymptotically einselect coherent states.
\end{myDesignRule}
\noindent The density matrix $\rho$ of an ensemble of closed-loop triaxial spinometer simulations is described by sequence $\{\rho_{1},\rho_{2}, \ldots\}$ whose
increment is
\begin{equation}
  \label{eq:thermal increment}
  \delta\rho_{n} =\sum_{k=1}^{3} \left(
  M^{k}_{\myupsymbol}\rho_{n}M^{\dagger k}_{\myupsymbol} +
  M^{k}_{\mydownsymbol}\rho_{n}M^{\dagger k}_{\mydownsymbol}
  -\rho_{n}\right)\,,
\end{equation}
By a straightforward (but not short) calculation we find that $\delta\rho_{n}$
vanishes for $\rho_{n}=\rho^{\ioptext{th}}$ if and only if the
closed-loop gain $\alpha$ satisfies
\begin{equation}
  \label{eq:alpha beta}
  \alpha = -\tanh \tfrac{1}{4}\beta
  \quad\ioptext{or}\quad
  1/\alpha = -\tanh \tfrac{1}{4}\beta\,.
\end{equation}
The following two trigonometric identities hold for either choice, and will be used in Section~\ref{sec: clicks} to establish that the choice is immaterial in practical numerical simulations.
\begin{equation}
  \label{eq:alpha identity}
 1/\alpha + \alpha = -2 \coth \tfrac{1}{2}\beta
 \quad\ioptext{and}\quad
 1/\alpha- \alpha  = -2 \csch \tfrac{1}{2}\beta 
\end{equation}
Defining as usual the dimensionless temperature $T=1/\beta$, we see that an optimal control gain $\alpha\to\pm 1$ establishes a temperature $T\to\mp0$, while a control gain $|\alpha|\ne1$  establishes a finite temperature. We will establish  later on that $\rho^{\ioptext{th}}$ solves $\delta\rho_{n}=0$ uniquely, because the Fokker-Planck equation for $\rho$ has a unique stationary solution (thus the approach of the density matrix $\rho$ to thermodynamic equilibrium never ``stalls'' or becomes trapped at false solutions). These results prove
\begin{myDesignRule}
\label{th:thermal theorem}
The density matrix of an ensemble of closed-loop triaxial spinometer simulations is asymptotically thermal.
\end{myDesignRule}
\noindent 
To connect \myeq{eq:thermal increment} with the
thermodynamic literature, we set $\lbhat{t}=(0,0,1)$ and
expand to order $\theta^{2}$. The result is equivalent to
a thermal model given by Perelomov (eq.~23.2.1
of~\cite{Perelomov:86}). Gardiner gives a similar model
(eq.~10.4.2 of \cite{Gardiner:85}). In Lindblad form
we find
\begin{equation}
  \label{eq:Lindblad}
  \begin{array}[b]{r@{\,}l}
  \delta\rho_{n} = 
  -\tfrac{1}{2} \gamma(\nu+1)&(s_{+}s_{-}\rho-2s_{-}\rho s_{+}+\rho s_{+}s_{-})\\
  -\ \tfrac{1}{2}\gamma\nu&(s_{-}s_{+}\rho-2s_{+}\rho s_{-}+\rho s_{-}s_{+})\\
  -\ \theta^{2}&(s_{3}s_{3}\rho-2s_{3}\rho s_{3}+s_{3}s_{3})\,,
  \end{array}
\end{equation}
where $s_{+}= (s_{1}+ i s_{2})/\sqrt{2}$ and $s_{-}=
(s_{1} - i s_{2})/\sqrt{2}$ are raising and lowering
operators, and we have adopted Perelomov's variables
$\gamma = -4\alpha\theta^{2}$ and $\nu=
-1/2-(\alpha + 1/\alpha)/4$.

\subsubsection{The spin-1/2 thermal equilibrium Bloch equations} 
The special case of spin-1/2 particles in thermal equilibrium often arises in practice.  Setting the polarization axis $\lbhat{t}=\lbhat{z}$, and allowing independent spinometric couplings $\{\theta_x, \theta_y, \theta_z\}$ as in (\ref{eq: infinite temperature Bloch equations}), we find that the finite-temperature synoptic measurement operators (\ref{eq:control}) imply the following asymmetric Bloch equations (valid for $j=1/2$ only):
\begin{equation}
\label{eq: finite temperature Bloch equations}
\left[\begin{array}{c}
\rule[-1ex]{0pt}{3.25ex}%
E[\delta x_{n}]\\[0.5ex]
E[\delta y_{n}]\\[0.5ex]
E[\delta z_{n}]%
\rule[-1ex]{0pt}{2.5ex}%
\end{array}\right] =
-\frac{1}{2}\left[\begin{array}{r@{}l}%
\rule[-1ex]{0pt}{3.25ex}%
(\alpha^2\theta_x^2 + \theta_y^2 + \theta_z^2)\,&E[x_{n-1}]\\[0.5ex]
(\theta_x^2 + \alpha^2\theta_y^2 + \theta_z^2)\,&E[y_{n-1}] \\[0.5ex]
(1+\alpha^2) (\theta_x^2 + \theta_y^2)\big(&E[z_{n-1}]+\tanh\tfrac{1}{2}\beta\big)%
\rule[-1ex]{0pt}{2.5ex}
\end{array}\right]
\end{equation}As expected on thermodynamic grounds, we see that the equilibrium polarization is $E[z]=-\tanh\frac{1}{2}\beta$.  These equations are a generalization of the usual Bloch equations, in the sense that the relaxation rates along the $x$-, $y$-, and $z$-axes can differ independently.  We remark that for $j>1/2$ the thermal Bloch equations do not have a closed analytic form; that is why this more general case is not considered here.

\subsubsection{The spinometric \Ito and Fokker-Planck equations}
\label{sec: ito and fokker-planck}
Now we focus on \Ito and Fokker-Planck equations, aiming by our analysis to obtain both the already-validated positive $P$-representation of Design Rule~\ref{dr: thermal operator} and (in the large-$j$ limit) both the linear~Design Rules 3.3--5 and the fundamental quantum limits of Design Rule~\ref{dr: fundamental quantum limits}.

We define a binary data three-vector $\lb{d}_{n} = (d^1_n, d^2_n, d^3_n)$ by
\begin{equation}
\label{eq: data stream}
  d^k_n=
\left\{\begin{array}{ll}
  +1&\ioptext{for\ }\ket{\psi_{n+1}} \propto  
  M^{k}_{\myupsymbol}\,\ket{\psi_{n}}\,,\\
  -1&\ioptext{for\ }\ket{\psi_{n+1}} \propto  
  M^{k}_{\mydownsymbol}\,\ket{\psi_{n}}\,,\\
\end{array}\right.
\end{equation}
Then $\{\lb{d}_1,\lb{d}_2,\ldots\}$ is a
binary data record with calibration
$E[\lb{d}_{n}] =  g_{\ioptext{s}}\,E[\lb{x}_{n}]$,
where  \begin{equation}
g_{\ioptext{s}} =
2\theta j
\end{equation}
is the spinometer gain.
We define a zero-mean stochastic variable
$\lb{W}_{n}$ by
\begin{equation}
  \lb{d}_{n} =  g_{\ioptext{s}}\,\lb{x}_{n} + \lb{W}_{n}\,,
\end{equation}
such that (to leading order in $\theta$) $\lb{W}_{n}$ has
the second-order stochastic properties of a discrete
Wiener increment:
\begin{equation}
E[(\lb{W}_{n})_{k}(\lb{W}_{n'})_{k'}] =
\delta_{nn'}\delta_{kk'}\,.
\end{equation}
Then via an identity valid for $\ket{\psi_{n}}$ a coherent
state,
\begin{equation}
  \label{eq:coherent identity}
  \braket{\psi_{n}}{s_{k}s_{l}}{\psi_{n}} = 
  \myonehalf j\delta_{kl} + j(j-\myonehalf) x_{k}x_{j}
  +\myonehalf ij\,\epsilon_{klm}x_{m}\,,
\end{equation}
the spinometer increments (\ref{eq:control}--b) are
equivalent to an \Ito increment
\begin{equation}
  \label{eq: Ito}
  \delta\lb{x}_{n} = \lb{x}_{n+1}-\lb{x}_{n} 
  = g_{\ioptext{s}}^{2}\lb{a}(\lb{x}_{n})  + g_{\ioptext{s}}\,\lb{b}(\lb{x}_{n}) 
  \mycdot \lb{W}_{n}\,.
\end{equation}
For the drift vector $\lb{a}$ and diffusion matrix
$\lb{b}$ we find
\begin{subequations}
\begin{eqnarray}
  \label{eq:a}
  \lb{a}(\lb{x}) =\ &-\frac{1}{4j^{2}}\lb{x}\, 
  [\alpha(2 j -1)\lb{x}\mycdot\lbhat{t}+
  (1+\myonehalf\alpha^{2})]\,\nonumber\\
  &+\,\frac{1}{4j^{2}}\lbhat{t}\,[\,\alpha (2j+1)-
  \myonehalf\alpha^{2}\,\lb{x}\mycdot\lbhat{t}\ ]\,,\\
  \lb{b}(\lb{x}) =\  & \frac{1}{2 j}[\lb{I}-\lb{x}\otimes\lb{x}+
  \alpha (\lbhat{t}\otimes\lb{x}-\lb{x}\mycdot\lbhat{t}\,\lb{I})]\,.
  \label{eq:b}
\end{eqnarray}
\end{subequations}
Design Rule~\ref{dr: closed-loop einselection}, asserts that the \Ito increment
\myeq{eq: Ito} confines the trajectory of $\lb{x}_{n}$
to the unit sphere.  The mean increment of the $m$'th
radial moment $|\lb{x}|^{m}$ must therefore vanish when
$|\lb{x}|=1$. We check this by direct calculation,
finding
\begin{align}
\nonumber
  \delta E_{n}[|\lb{x}|^{m}] 
  \propto{}&
  \myonehalf m(m-2)E[\lb{x}_{n}\mycdot\lb{b}
  (\lb{x}_{n})\mycdot\lb{b}^{\dagger}(\lb{x}_{n})\mycdot\lb{x}_{n}]\,\\
  & + m
  E[|\lb{x}_{n}¥|^{2}
  (\lb{x}_{n}\mycdot\lb{a}(\lb{x}_{n})
  +\myonehalf\tr\lb{b}(\lb{x}_{n})\mycdot\lb{b}^{\dagger}(\lb{x}_{n})
  )]\,,
  \label{eq:on the sphere}
\end{align}
which indeed vanishes for the $\lb{a}$ and $\lb{b}$ of
(\ref{eq:a}--b). By well-known methods \cite{Gardiner:85}, the \Ito
increment (\ref{eq: Ito}) immediately yields a
Fokker-Planck equation for the \PDF $p_{\!j}(\lbhat{x})$. Setting
$z=\lbhat{x}\mycdot\lbhat{t}$ we obtain the stationary state equation 
\begin{align}
\nonumber 
  0 ={} & - \frac{\partial}{\partial z}
  [\alpha (1+z^{2}) + 2 j \alpha (1-z^{2}) -z(1+\alpha^{2})] p_{\!j}(z)\\
  & +\frac{1}{2}\frac{\partial^{2}}{\partial z^{2}} 
  [(1-z^{2})(1-2\alpha z+\alpha^{2})]p_{\!j}(z)\,,
\end{align}
which (when properly normalized) has a unique solution
\begin{align}
  p_{\!j}(\lbhat{x}) = {}& 
  (\alpha + 1/\alpha - 2 z)^{-2j-2} \\
  & \times (2j+1)\big[(\alpha +  1/\alpha - 2)^{-2j-1}-(\alpha +  1/\alpha + 2)^{-2j-1} \big]^{-1}/\pi
\end{align}
in which we see that the symmetry $\alpha\to1/\alpha$ is indeed respected. Substituting $\alpha + 1/\alpha = -2 \coth \frac{1}{2}\beta$ per \myeq{eq:alpha identity}, and adjusting the normalization to match the $P$-representation convention (\ref{eq:Prep}) yields Design Rule~\ref{dr: thermal operator}.

\subsubsection{The standard quantum limits to linear measurement}
\label{eq: the standard quantum limits}
To connect these results to Design Rules 3.2--\ref{dr: fundamental quantum limits}, we  first write the \Ito equation
\myeq{eq: Ito} in Langevin form by substituting
\begin{eqnarray}
\delta\lb{x}_{n} & \to \textstyle{\int_{t}^{t+\delta t}\!\!dt'}\,
\lbdot{x}(t')\\
\lb{a}(\lb{x}_{n}) & \to r \textstyle{\int_{t}^{t+\delta t}\!\!dt'}\,
\lb{a}\lb(\lb{x}(t')\lb)\,,\\
\lb{b}(\lb{x}_{n})\mycdot\lb{W}_{n} & \to rg_{\ioptext{s}} 
\textstyle{\int_{t}^{t+\delta t}\!\!dt'}\,
\lb{b}\lb{(}\lb{x}(t')\lb{)}\mycdot\lb{x}^{\ioptext{N}}(t')\,,
\end{eqnarray}
where $r=1/\delta t$ is the rate at which increments occur, and
$\lb{x}^{\ioptext{N}}(t)$ is white noise with cross-correlation
\begin{equation}
  \label{eq:white noise}
  E[x^{\ioptext{N}}_{k}(t)x^{\ioptext{N}}_{k'}(t')] =
\delta_{kk'}\delta(t-t')/(rg_{\ioptext{s}}^{2} =
\delta_{kk'}\delta(t-t')/(4rj^2\theta^2)\,.
\end{equation}
Then \myeq{eq: Ito} becomes the integral of the Langevin
equation
 \begin{subequations}
\begin{equation}
  \label{eq:Langevin}
  \lbdot{x} = rg_{\ioptext{s}}^{2}
  [\lb{a}(\lb{x})+ \lb{b}(\lb{x})\mycdot(\lb{x}^{\ioptext{M}}-\lb{x})]\,,
\end{equation}
where $\lb{x}^{\ioptext{M}}(t) =
\lb{x}(t)+\lb{x}^{\ioptext{N}}(t)$ is the measured spin axis.

We see that $\lb{x}(t)$ is dynamically attracted toward
the measured axis $\lb{x}^\ioptext{M}(t)$. Even open-loop
spinometers exhibit this attraction, since for $\alpha=0$
we find
\begin{equation}
  \label{eq:Langevin II}
  \lbdot{x}|_{\alpha=0} = rg_{\ioptext{s}}^{2}
  [-\frac{1}{4j^{2}¥} \lb{x}+\frac{1}{2j}(\lb{I}-\lb{x}\otimes\lb{x})
  \mycdot(\lb{x}^{\ioptext{M}}-\lb{x})]\,.
\end{equation}
\end{subequations}
\noindent We remark that in uniaxial spinometry we saw that a similar
einselection-by-attraction generates the ``collapse'' of
$\ket{\psi_{n}}$ to an eigenstate, as described by 
Design Rule~\ref{th:uniaxial}.  This attraction is of course
a fundamental tenet of the Hilbert ontology of Section~\ref{sec: Hilbert ontology}.

We now transform \myref{eq:Langevin II} to the
second-order Newtonian equation of an oscillator. To do
this, we introduce a spring $k$ and frequency $\omega_{0}$
by defining the operators
\begin{subequations}
\begin{eqnarray}
q^{\ioptext{op}} &= (\hbar \omega_0/j k)^{1/2}\, 
\big({+}s_{1}\cos\omega_{0}t-s_{2}\sin\omega_{0}t\,\big)\,,\\ 
p^{\ioptext{op}} &= (k\hbar/j\omega_0)^{1/2}\,
\big({-}s_{1}\sin\omega_{0}t-s_{2}\cos\omega_{0}t\,\big)\,.
\end{eqnarray}
\end{subequations}We confine our attention to those coherent states that have $z\simeq -1$, which with our sign conventions means systems having positive inverse temperature $\beta$, negative Hilbert feedback gain $\alpha$, and oppositely directed spin $\lbhat{x}$  and polarization $\lbhat{t}$, such that $\lbhat{x}\cdot\lbhat{t}\simeq -1$.  For these states the canonical
commutator $[q^{\ioptext{op}},p^{\ioptext{op}}] = -i\hbar s_{3}/j
\simeq i\hbar$ holds in the large-$j$ limit. Defining the coherent oscillator
coordinate $q(t)$ to be
\begin{equation}
\label{eq: q definition}
  q(t) = (j\hbar\omega_0/ k)^{1/2}\, 
  \big(x(t)\cos\omega_{0}t-y(t)\sin\omega_{0}t\big)\,,
\end{equation}
we find that \myeq{eq:Langevin II} takes the linearized Newtonian
form
\begin{subequations} %
\begin{align}
  \label{eq:Newtonian one}
  m\sddot{q}&{}= -k\,q+f^{\ioptext{n}}\,,\\
    q^{\ioptext{m}} &{}=q+q^{\ioptext{n}} \,,
\end{align}
\end{subequations} %
Here the spring $k$, mass $m=k/\omega_0^2$, and coordinate $q$ are to be understood in a generalized sense in which the system energy is $\tfrac{1}{2}m\sdot{q}^2+\tfrac{1}{2}kq^2$.  The measurement noise $q^{\ioptext{n}}(t)$ 
is given from (\ref{eq: q definition}) in terms of the spinometer noises $x^{\ioptext{N}}(t)$ and $y^{\ioptext{N}}(t)$ of (\ref{eq:white noise}) by
\begin{equation}
  q^{\text{n}}(t) = (j\hbar\omega_0/ k)^{1/2}\, 
  \big(x^{\text{N}}(t)\cos\omega_{0}t-y^{\text{N}}(t)\sin\omega_{0}t\big)\,,
\end{equation}
and we find from (\ref{eq:white noise}) and (\ref{eq: q definition}) that the measurement noise $q^{\text{n}}(t)$ has a \PSD $S_{q^{\text{n}}}$ of 
\begin{equation}
\label{eq: Sqn definition}
  \left.S_{q^{\text{n}}}(\omega)\right|_{\omega\simeq\omega_0} = \hbar\omega_0/(4 k r j \theta^2)
\end{equation}
The force noise $f^{\text{n}}(t)$ is then determined from (\ref{eq:Langevin II}) to be $f^{\text{n}}(t) = \gamma \lcal{H}[q^{\text{n}}(t)]$, where $\lcal{H}$ is the Hilbert transform and the Hilbert gain $\gamma$ is found to be
\begin{equation}
\gamma =4 k r j \theta^2/\omega_0
\end{equation}
\subsubsection{Multiple expressions of the quantum noise limit}
\label{sec: multiple expressions}
It follows from the preceding results that the \PSD of the spinometric force noise $f^{\text{n}}(t)$ can be expressed in multiple equivalent forms:
\begin{subequations}
\begin{align}
\nonumber 
	&\hspace{1.5em}\text{\bfseries expression} 
	&&\hspace{1.25em}\text{\bfseries physical interpretation}\\
\label{eq: force forms}
  \left.S_{f^{\text{n}}}(\omega)\right|_{\omega\simeq\omega_0} 
  & = \gamma^2 \left.S_{q^{\text{n}}}(\omega)\right|_{\omega\simeq\omega_0} 
  &&\text{force noise $\propto$ measurement noise }
  \\ 
\label{eq: force forms B}  
	& = \hbar^2/{\left.S_{q^{\text{n}}}(\omega)\right|_{\omega\simeq\omega_0}} 
  &&\text{force noise $\propto$ 1/(measurement noise)}
  \\ 
  \label{eq: force forms C} 
  & = \hbar \gamma 
  &&\text{force noise $\propto$ Hilbert backaction gain}
  \\
  \label{eq: force forms D} 
  & = 4 k \hbar r j \theta^2/\omega_0 
  &&\text{raw spinometer parameters}
\end{align}
\end{subequations}
Each of the above relations has a plausible claim to expressing the ``most natural'' or ``most fundamental'' relation between measurement noise and force noise \ldots\ despite the fact that no two physical interpretations are the same, and even though the interpretations given (\ref{eq: force forms}) and (\ref{eq: force forms B}) seem contradictory.  We further see by Design Rule~\ref{dr: fundamental quantum limits} that these spinometric relations saturate the Hilbert noise limit ($\gamma S_{q^{\text n}} = \hbar$), the Braginsky-Khalili limit ($S_{q^{\text n}} S_{f^{\text n}} = \hbar^2$), and the Hefner-Haus-Caves limit ($\NF = 2$); thus in some sense all of these fundamental quantum limits are embodied in the above family of spinometric relations.  

Acknowledging the self-consistency of this diversity, and appreciating its mathematical origin in the diversity of equivalent noise models that are supported by the \emph{Theorema Dilectum}, helps us appreciate how the quantum noise literature can be so immensely large, and support so many different notations, physical arguments, and conclusions, and yet maintain its internal consistency.

In a teaching environment, it is not practical to sustain a dispassionately anarchical equality among physical interpretations (\ref{eq: force forms}-d).  This article's Hilbert ontology (Section \ref{sec: tenets}) designates (\ref{eq: force forms}) to be the fundamental relation, because it embodies the central Hilbert tenet that ``measurement noise always back-acts upon system dynamics in such a way as to bring the state of the system into agreement with the measurement.'' This choice is justified solely because it yields useful guiding principles for efficient quantum simulations.

\subsection{Summary of the design rules} 
In summary, we have established by Design Rules \ref{th:uniaxial}, \ref{th:triaxial} and \ref{dr: closed-loop einselection} the quantum mechanism by which synoptic noise processes compress simulated quantum trajectories onto lower-dimension \GK manifolds (as was promised in Section~\ref{sec: TD allows compression}).  We have established by Design Rule \ref{th:thermal theorem} that the effects of thermal reservoirs can be modeled as equivalent processes of covert measurement and control (as was promised in Section~\ref{sec: thermal reservoirs}).  And we have established by Design Rule \ref{dr: fundamental quantum limits} that the Hefner-Haus-Caves, Braginsky-Khalili, and Hilbert quantum noise limits are respected by \QMOR simulations (as was promised in Section~\ref{sec: QMOR respects the fundamental quantum limits}).  

The focus of the remainder of this article is to show, by explicit examples, that these design rules are sufficient to ``enable the reader to proceed to the design and implementation of practical quantum simulations, guided by an appreciation of the geometric and informatic principles that are responsible for the overall simulation accuracy, robustness, and efficiency'' (as was promised in the Introduction).

\section{Examples of quantum simulation}
\label{sec: simulations} 
Now we turn our attention toward applying the preceding results in implementing practical quantum simulations.  

\subsection{Calibrating practical simulations}
\label{sec: clicks}

Our simulations provide data via the binary stream of defined in (\ref{eq: data stream}), which is low-pass filtered to produce a classical data record.  We now work through, in detail, the process of computing and calibrating this data stream, and ensuring that it is numerically well-conditioned.

We begin by considering the problem of determining, from physical system parameters, the measurement operation parameters $\{\theta,\alpha\}$ in (\ref{eq:control})
and the clock rate $r=1/\delta t$.  In essence this calibration process requires that we invert systems of equations that include the Bloch equations (\ref{eq: infinite temperature Bloch equations}) and 
(\ref{eq: finite temperature Bloch equations}), the Langevin equation (\ref{eq:Langevin II}), and the mapping of spinometer parameters onto oscillator parameters (ref{eq: q definition}).  

Whenever our simulations include a projective step, we must also ensure that the $\epsilon$-parameter of the projected \emph{Theorema Dilectum} (\ref{eq: nonlinear theorema dilectum}) satisfies $\epsilon\ll 1$.  Physically speaking, imposing the small-$\epsilon$ condition ensures that the simulated trajectories evolve by drift and diffusion, rather than by ``quantum jumps'' that may be projectively ill-conditioned.   Equally importantly, it ensures that they respect the informatic causality that is guaranteed by the \emph{Theorema Dilectum}.

\subsubsection{Calibrating the Bloch equations}
A common system to be simulated is a spin $j=\tfrac{1}{2}$ in contact with a thermal reservoir.  We desire that the three thermal relaxation rates along the $x$-, $y$-, and $z$- axes be $\{\Gamma_x, \Gamma_y, \Gamma_z\} = \{1/T_x, 1/T_y, 1/T_z\}$ and that the equilibrium thermal density matrix be $\rho^{\text{th}}\propto\exp(-\beta s_z)$, such that the equilibrium spin polarization $p_0$ is $-\tanh\tfrac{1}{2}\beta$ as in (\ref{eq: finite temperature Bloch equations}).  

We thus have four physical parameters $\{\Gamma_x, \Gamma_y, \Gamma_z, \beta\}$ with which to determine five raw spinometer parameters $\{\theta_x, \theta_y, \theta_z, \alpha, r\}$.  Needing one more physical parameter, we note that the $\epsilon$-parameters of the \GK-projected \emph{Theorema Dilectum} (\ref{eq: nonlinear theorema dilectum}) is given for $j=1/2$ spinometers from (\ref{eq: finite temperature Bloch equations}) by  $\epsilon^2 = \theta^2 (1+\alpha^2)/4$, and so we impose as our fifth condition $\epsilon \lesssim 0.1$ for all three spinometers (the cut-off 0.1 yielding in our experience well-converged numerical results). The equations below then follow from (\ref{eq:thermal increment}), (\ref{eq:alpha identity}), and (\ref{eq: finite temperature Bloch equations}): 
\begin{subequations}
\begin{align}
\label{eq: Bloch A}
	\alpha = {}& -\tanh\tfrac{1}{4}\beta\quad\text{or}\quad{-1}/\tanh\tfrac{1}{4}\beta
	\quad\text{(freely chosen)}\\
\label{eq: Bloch B}
 \epsilon_x^2 ={}&
 \left[\,\Gamma_z
	-\sgn(1-\alpha^2)(\Gamma_x-\Gamma_y)\cosh\tfrac{1}{2}\beta\,\right]/(4r)\\
\label{eq: Bloch C}
 \epsilon_y^2 ={}&
 \left[\,\Gamma_z
	+\sgn(1-\alpha^2)(\Gamma_x-\Gamma_y)\cosh\tfrac{1}{2}\beta\,\right]/(4r)\\
\label{eq: Bloch D}
 \epsilon_z^2={}&\left[\,\Gamma_x + \Gamma_y - \Gamma_z\,\right]/(4 r)\\
\label{eq: Bloch E}
 \theta^2_i = {}&\,4\epsilon^2_i/(1+\alpha^2)\ \text{for}\ i\in\{x,y,z\}
\end{align}
\end{subequations}
Bloch-parameter calibration proceeds as follows.  We first determine the spinometer gain $\alpha$ from (\ref{eq: Bloch A}), and we will see that the choice ``gain too big'' versus ``gain too small'' is immaterial.  The value of the spinometer click-rate $r$ is then set from (\ref{eq: Bloch B}-d) by requiring that $\min\{\epsilon_x, \epsilon_y, \epsilon_z\} \lesssim 0.1$ for the reasons noted above.  The values of the three spinometer phases  $\{\theta_x, \theta_y, \theta_z\}$ are then determined from (\ref{eq: Bloch E}).

We remark upon three features. First, we see that insofar as simulation efficiency and numerical conditioning are concerned, the choice between the two options for the feedback gain $\alpha$ in (\ref{eq: Bloch A}) is immaterial, since according to the above construction the spinometer click-rate $r$ and the $\epsilon$-parameters are unaffected.  Second, for the special case $T_z=T_1$ and $T_x=T_y=T_2$ the above results reduce to the usual Bloch equations.  Third, the positivity of the $\epsilon$-parameters in (\ref{eq: Bloch B}-d) requires that the Bloch relaxation rates satisfy the inequality 
\begin{equation}
|\Gamma_x-\Gamma_y| \cosh\tfrac{1}{2}\beta \le \Gamma_z \le \big(\Gamma_x+\Gamma_y\big)
\end{equation}
The authors suspect that the above Bloch inequality is tight, in the sense that no spin-1/2 Lindblad-form master equation can violate it, but we have not proved this.   We remark that the above triaxially asymmetric Bloch equations and their associated relaxation rate inequality have (to the best of our knowledge) not appeared in the literature before.  

\subsubsection{Calibrating test-mass dynamics in practical simulations}
Now we consider test-masses (\emph{e.g.}, \MRFM cantilevers) in contact with thermal reservoirs.  Calibration proceeds by a line of reasoning similar to the above.  We take the test-mass to be described by two physical parameters: the (dimensionless) temperature $\beta = \hbar\omega_0/(k_{\text{B}}T)$  and the (dimensionless) quality $Q$ of the ring-down wave-form $q(t)\propto \cos(\omega_0 t) \exp(\omega_0 t/(2 Q))$.  The four spinometer parameters to be determined are $\{j,\theta,\alpha,r\}$.  The spin number $j$ we take to satisfy $j\gg 1$, and we will find that the precise value chosen for $j$ is immaterial.  For coherent states with $z\simeq -1$ as discussed in Section~\ref{sec: ito and fokker-planck}, the $\epsilon$-parameter is found to be $\epsilon \simeq j \theta^2 (1+\alpha^2)$.  Calibration proceeds as follows:
\begin{subequations}
\begin{align}
\label{eq: test mass A}
	\alpha = {}& -\tanh\tfrac{1}{4}\beta\quad\text{or}\quad{-1}/\tanh\tfrac{1}{4}\beta
	\quad\text{(freely chosen)}\\
\label{eq: test mass B}
\epsilon^2={}&\big(\omega_0/(2Qr)\big)\,\coth\tfrac{1}{2}\beta\\
\label{eq: test mass C}
j\theta^2 ={}& \epsilon^2/(1+\alpha^2)
\end{align}
\end{subequations}
We first determine the spinometer gain $\alpha$ from (\ref{eq: test mass A}).  The value of the spinometer click-rate $r$ is then set from (\ref{eq: test mass B}) by requiring that $\epsilon \lesssim 0.1$ as in the Bloch equation case, and the spinometer phase $\theta$ is determined from (\ref{eq: test mass C}).  We remark again that insofar as simulation efficiency is concerned, the choice between the two options for the feedback gain $\alpha$ in (\ref{eq: test mass A}) is immaterial, since the spinometer click-rate $r$ is unaffected.  We see also that for fixed quality $Q$, the simulation rate $r$ is $\lcal{O}(\coth\tfrac{1}{2}\beta)$, which physically means that hot cantilevers are computationally more expensive to simulate than cold ones, as is reasonable.

\subsubsection{Calibrating purely observation processes}
It can happen that we wish to directly observe a spin-1/2 particle along a single axis, nominally the $z$-axis, in an observation process in which the measured $z$-axis polarization $z^{\text{m}}(t) = z(t)+z^{\text{n}}(t)$ has a specified (one-sided) noise \PSD $S_{z^{\text{n}}}$.  No thermodynamical feedback is applied.  Then by reasoning similar to the preceding cases, we find from (\ref{eq:white noise}) that the calibration relations are 
\begin{subequations}
\begin{align}
\epsilon^2 & =  1/(2 r S_{z^{\text{n}}}) \\
\theta^2 &= 4 \epsilon^2
\end{align}
\end{subequations}
and as before, the spinometer click rate $r$ is determined by requiring $\epsilon \lesssim 0.1$.

Similarly, if we wish to simulate the continuous observation of a test-mass coordinate $q(t)$, with no thermodynamical feedback, the required calibration equations are given from (\ref{eq:white noise}) and (\ref{eq: q definition}) in terms of the measurement noise \PSD $S_{q^\text{n}}$ by
\begin{subequations}
\begin{align}
\epsilon^2 & =  \hbar\omega_0/(4 r kS_{q^\text{n}}) \\
j\theta^2 &= \epsilon^2
\end{align}
\end{subequations}
where again the spinometer click rate $r$ is determined by requiring $\epsilon \lesssim 0.1$.

We remark that in all of the above cases the raw binary data stream (\ref{eq: data stream}) of the simulation must be low-pass filtered in order to obtain a (noisy) data record that will have the above statistical properties within the filter passband having.  This filtering closely models the way that real experimental signals (for example, a continuously measured cantilever motion) are displayed upon oscilloscopes.

\subsection{Three single-spin \MRFM simulations}
\label{sec: Single-spin MRFM simulations}

\begin{figure}[t]\centering
\vspace*{2ex}
\includegraphics[width=1.0\textwidth]{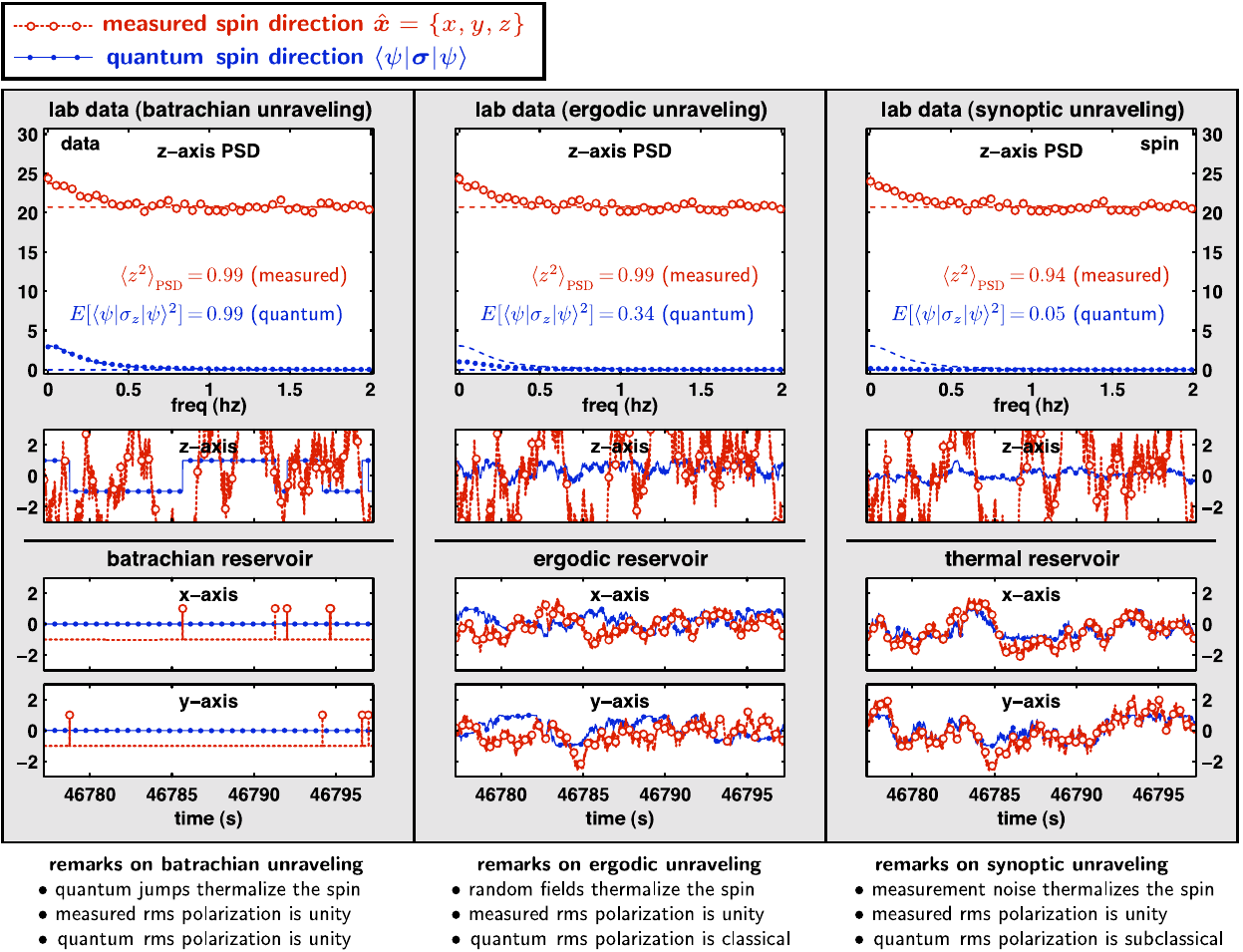}\\
\begin{minipage}{0.8\textwidth}%
\caption[Simulation of single electron moment detection by \MRFM]{%
\protect\justifying%
Simulation of single electron moment detection by \MRFM.
\label{fig: single spin simulation}
}\end{minipage}
\end{figure}

With reference to Fig.~\ref{fig: single spin simulation}, we now turn our attention to the simulation of the \IBM single-spin \MRFM experiment  \cite{Rugar:04} . We will initially present the simplest possible class of simulations that reproduce the data of that experiment, postponing a discussion of more detailed simulations until Section~\ref{sec: realistic cantilever models}.  Our goal is to illuminate the central role of the \emph{Theorema Dilectum} in answering the question that was raised in Section~\ref{sec: how does the S-G effect work}: ``How does the Stern-Gerlach experiment work?''

All three columns of Fig.~\ref{fig: single spin simulation} show a simulated thirteen-hour experiment (the length of the \IBM experiment).  The time-spacing $\delta t= 1/r$ between spinometric clicks is set to 7.1~ms; thus approximately $6.6\,\times\,10^6$ time-steps were simulated.  In each column, the experimental data are simulated as arising from three competing spinometric processes.  Spin relaxation was simulated by $x$-axis and $y$-axis spinometers having $\theta_x = \theta_y = 0.093$.  The consequent spin relaxation time from (\ref{eq: finite temperature Bloch equations}) is $T_z = 2/\big(r(\theta_x^2+\theta_y^2)\big) \simeq 0.76\text{\ s}$ as was observed in the \IBM experiment.  Measurement effects were simulated by a $z$-axis spinometer having $\theta_z = 0.026$, and the consequent measurement noise \PSD from (\ref{eq:white noise}) is $ S_{z^{\text{n}}} = 2/(r\theta_z^2)$, which numerically corresponds to a noise level of 11.5~Bohr magnetons of noise in one root-Hertz of bandwidth, as was observed in the \IBM experiment.  

For visualization purposes only, all time-domain data streams shown in Fig.~\ref{fig: single spin simulation} were low-pass filtered with a time constant $\tau=T_z=0.76\text{\ s}$.

The following discussion is insensitive to the above experimental details, and applies to all experiments of Stern-Gerlach type in which the signal-to-noise ratio (\SNR) is low, such that continuous monitoring over extended periods of time is required to observe the effect.

In the next three sections, we simulate the spin relaxation of the \IBM single-spin \MRFM experiment by three different unravelings: batrachian, ergodic, and synoptic.  We will see that the three unravelings lead to three very different classes of quantum trajectories, and hence, three very different answers to the question ``How does the Stern-Gerlach experiment work?''  Nonetheless, as guaranteed by the \emph{Theorema Dilectum}, we will find that the simulated experimental data are identical for all three unravelings.  We now work through the mathematical and physical details of how this comes about.

\subsubsection{A batrachian single-spin unraveling}
\label{sec: batrachian unravelling}

The left-hand column of Fig.~\ref{fig: single spin simulation} shows a simulation in which thermal noise is unravelled as a batrachian process, whose measurement operations are given algebraically in (\ref{eq: batrachian}) and which are depicted in hardware-equivalent form in Figure~\ref{fig: Choi}(b).  This is by far the easiest simulation to analyze in closed form: the  spin polarization jumps randomly between $\pm1$, driven by the batrachian jumps of the thermal reservoir, while being continuously measured by the (noisy) cantilever.  

The simulated data stream is therefore a random telegraph signal with added white noise, such that the mean-square quantum spin polarization inferred from the data is unity.  We conclude that from the batrachian point of view, the Stern-Gerlach effect (meaning, that the mean-square spin polarization is measured to be unity) comes about because noise is a quantized jump process, such that the  mean-square spin polarization always \emph{is} unity.

\subsubsection{An ergodic single-spin unravelling}

The middle column of Fig.~\ref{fig: single spin simulation}(b) shows a simulation in which thermal noise is unravelled as an ergodic process, whose measurement operations are given algebraically in (\ref{eq: ergodic}) and which are physically depicted in Figure~\ref{fig: Choi}(a).  Physically speaking, the spin polarization is driven by random magnetic fields, such that the mean-square quantum polarization is 1/3.   

Now a subtle effect comes into play.  The $z$-axis measurement process back-acts upon the spin state, such that whenever an ``up'' fluctuation in the data is observed, the spin state is ``dragged'' toward a positive polarization.  This effect is evident in the simulated data.  The consequence of state-dragging back-action is that the measured mean-square polarization is larger than the mean-square polarization of the underlying quantum state.  

It would be quite a complicated task to calculate the resulting data statistics from (\emph{e.g.}) the appropriate \Ito, Langevin, and Fokker-Planck equations.  Fortunately, the \emph{Theorema Dilectum} does this mathematical work for us: the data statistics are guaranteed to be \emph{exactly} the same random telegraph statistics as in the Batrachian case. 

We conclude that from the ergodic point of view, the Stern-Gerlach effect (meaning, that the mean-square spin polarization is measured to be unity) comes about because measurement is a Hilbert process (meaning, it accords with the state-dragging Hilbert back-action ontology of Section \ref{sec: tenets}).

\subsubsection{A synoptic single-spin unravelling}

The right-hand column of Fig.~\ref{fig: single spin simulation}(c) shows a simulation in which thermal noise is unravelled as an synoptic process, whose measurement operations are given algebraically in (\ref{eq: synoptic}) and which are physically  depicted in Figure~\ref{fig: Choi}(c).  

In synoptic unravelling, all processes are measurement processes, and each process seeks to align the spin polarization along its own axis.  In our simulation, the $x$-axis and  $y$-axis measurement processes are considerably stronger than the $z$-axis process.  In consequence of the Hilbert state-dragging effect, the spin polarization now points predominantly in the equatorial direction, such that the mean-square quantum polarization is only $\sim0.05$. 

Again it would be quite a complicated task to calculate the resulting data statistic from  \Ito equations, \emph{etc.}, and again the \emph{Theorema Dilectum} does this mathematical work for us: as in the preceding two cases, the data statistics are random telegraph statistics with added white noise. We conclude that from the synoptic point of view, the Stern-Gerlach effect is not associated with ``wave function collapse,'' but rather comes about (as in the ergodic case) because measurement is a Hilbert process.

\subsection{So how does the Stern-Gerlach effect \emph{really} work?}
We are now in a position to answer more completely the question ``How does the Stern-Gerlach effect \emph{really} work?''  We answer as follows: ``Nothing definite can be said about the internal state of noisy systems, either at the classical or at the quantum level.  It is best to pick an ontology that facilitates rapid calculations and suggests interesting mathematics.  For purposes of large-scale quantum simulation, a particularly useful ontology is one in which all noise processes are conceived as equivalent covert measurement processes.  In this ontology, the Stern-Gerlach effect works because  competing measurement processes exert a Hilbert back-action mechanism that `drags' quantum states into agreement with measurement.  In consequence of these competing Hilbert measurements, experimental data having the statistics of random telegraph signals are obtained even when no quantum jumps are present.'' 

Of course, we saw in the three simulations of Fig.~\ref{fig: single spin simulation}a--c  that other explanations are perfectly reasonable, and that is why in Section~\ref{sec: Constraints upon the Analysis} we embraced Peter Shor's maxim: ``Interpretations of quantum mechanics, unlike Gods, are not jealous, and thus it is safe to believe in more than one at the same time,'' to which we now append the caveat ``provided that all interpretations respect the fundamental mathematical and physical invariance of the \emph{Theorema Dilectum}.''

\subsection{Was the \IBM cantilever a macroscopic quantum object?}
\label{sec: realistic cantilever models}
The least realistic element of the proceeding simulations is the modeling of the cantilever as a single $z$-axis spinometer having quantum number $j=1/2$.  A more realistic model would have treated the cantilever as a large-$j$ quantum object subject both to thermal noise processes and to experimental measurement processes.  However, we can appeal to the \emph{Theorema Dilectum} to show that these refinements would not change the simulated data at all.  The reason is that both the cantilever thermal reservoir and the experimental (interferometric) cantilever measurement process can be modeled as synoptic processes that compress the cantilever's quantum state to a coherent state.  Then modeling the spin relaxation as a batrachian process, the output of the resulting (effectively semi-classical) batrachian simulation will be precisely the random telegraph signal that was obtained in the simpler batrachian simulation of Section~\ref{sec: batrachian unravelling} above.  

It follows by the \emph{Theorema Dilectum} that \emph{all} quantum simulations of the cantilever, even elaborate large-$j$ simulations in which a non-coherent quantum  cantilever state is entangled with the quantum spin state, will simulate the same random telegraph data statistics as the simpler simulations already given, and in particular, will yield an observed mean-square polarization of unity.

This leads to an interesting question: what was the ``real'' quantum state of the \IBM cantilever?  We have seen that this question has a well-posed answer only insofar as there is agreement upon the ``real'' noise and observation processes acting upon the cantilever, such that the tuning ambiguity of the \emph{Theorema Dilectum} does not come into play.  

If we stipulate that the ``real'' cantilever thermal noise and the ``real'' spin relaxation are due to ergodic physical processes, then the \IBM experiment can only be ``really'' described in terms of a spin-state that is quantum-entangled with the cantilever state, in which the observed mean-square polarization of unity is ``really'' due to the state-dragging Hilbert back-action associated with the cantilever measurement process.  In other words, the \IBM experiment ``really'' observed the cantilever to be a macroscopic quantum object.

As quantum objects go, the \IBM cantilever was exceptionally large \cite{Rugar:04}: its resonant frequency was $\omega_0/(2\pi) = 5.5\text{\ kHz}$, its spring constant was $k= 0.011\text{\ mN/m}$, and its motional mass was $m = k/\omega_0^2 = 9.1\text{\ pg}$.  The preceding paragraphs are an argument for regarding this cantilever to be among the stiffest, slowest, most massive dynamical systems whose quantum nature has been experimentally confirmed.  Such measurements are significant from a fundamental physics point of view, in probing the limits of quantum descriptions of macroscopic objects, as reviewed by Leggett \cite{Leggett:2002ek,Leggett:2004uo,Leggett:2002ph,Leggett:2002it}.

Any line of reasoning that is as brief as the preceding one, about a subject that is as subtle as macroscopic  quantum mechanics, is sure to have loopholes in it.  A major loophole is our modeling of decoherent noise as a Markovian process.  As reviewed by Leggett \emph{et al.}\ \cite{Leggett:1987pi}, spin decoherence in real experiments is (of course) due to non-Markovian quantum-entangling interactions. We now turn our attention to the algorithmic and numerical challenges of simulating such systems.

\subsection{The fidelity of projective \QMOR in spin-dust simulations}
\label{sec: Spin-dust simulations}
\label{sec: spin-dust}

As test cases, we computed what we will call \emph{spin-dust} simulations.  Spin-dusts are quantum systems that are deliberately constructed so as to have no symmetries or spatial ordering.  Their sole purpose is to provide a well-defined test-bed for numerical and analytic studies of the fidelity of projective quantum model order reduction.

Spin-dusts couple pairs of spin-1/2 particles $\{j,k\}$ via a dipole-dipole interaction Hamiltonian $H_{jk}$ that is given by
\begin{equation}
H_{jk} = \begin{cases}
\lb{s}_j\cdot\left[I-3\lb{n}_{jk}\otimes\lb{n}_{jk}\right] \cdot\lb{s}_k\ & \text{for $j\ne k$}\\
\lb{s}_j\cdot\lb{n}_{jk}& \text{for $j=k$}
\end{cases}
\end{equation}
The unit vectors $\lb{n}_{jk}$ are chosen randomly and independently for each $\{j,k\}$, and we note that self-coupling is allowed.  Physically we can think of spin-dusts as broadly analogous to---but less structured than---systems such as the interacting spins in a protein molecule.

In our simulations each spin is randomly coupled to four other spins, in addition to its self-interaction.  Then it is easy to show that $\tr H = 0$ and $\tr H^2/ \dim \lcal{H} = n_\text{spin}$, which is to say, the per-spin energy of our spin-dusts has zero mean and unit variance. The time-scale of the spin dynamics of the system is therefore unity.  We further stipulate that each spin is subject to a triaxial spinometric observation process having relaxation time $T_x = T_y = T_z = 10$.  Thus the time-scale of decoherent observation is ten-fold longer than the dynamical time-scale.   

Simulations were computed with a time-step $\delta t = 0.1$ and spinometric couplings $\theta_x = \theta_y = \theta_z = 0.1\hspace{0.1ex}$, using the sparse matrix routines of \emph{Mathematica}.  The numerical result was an ``exact'' (meaning, full Hilbert space) quantum trajectory $\ket{\psi_0(t)}$.  These trajectories were then projected on \GK manifolds of various order and rank by the numerical methods of Section~\ref{sec: expectation}.   The main focus of our numerical investigations was the fidelity of the projected states $\ket{\psi_\lcal{K}(t)}$ relative to the exact states $\ket{\psi_0(t)}$.  

\subsubsection{The fidelity of quantum state projection onto \GK manifolds}
\label{sec: projection fidelity}
\begin{figure}[t]\centering
\vspace*{2ex}
\includegraphics[width=1.0\textwidth]{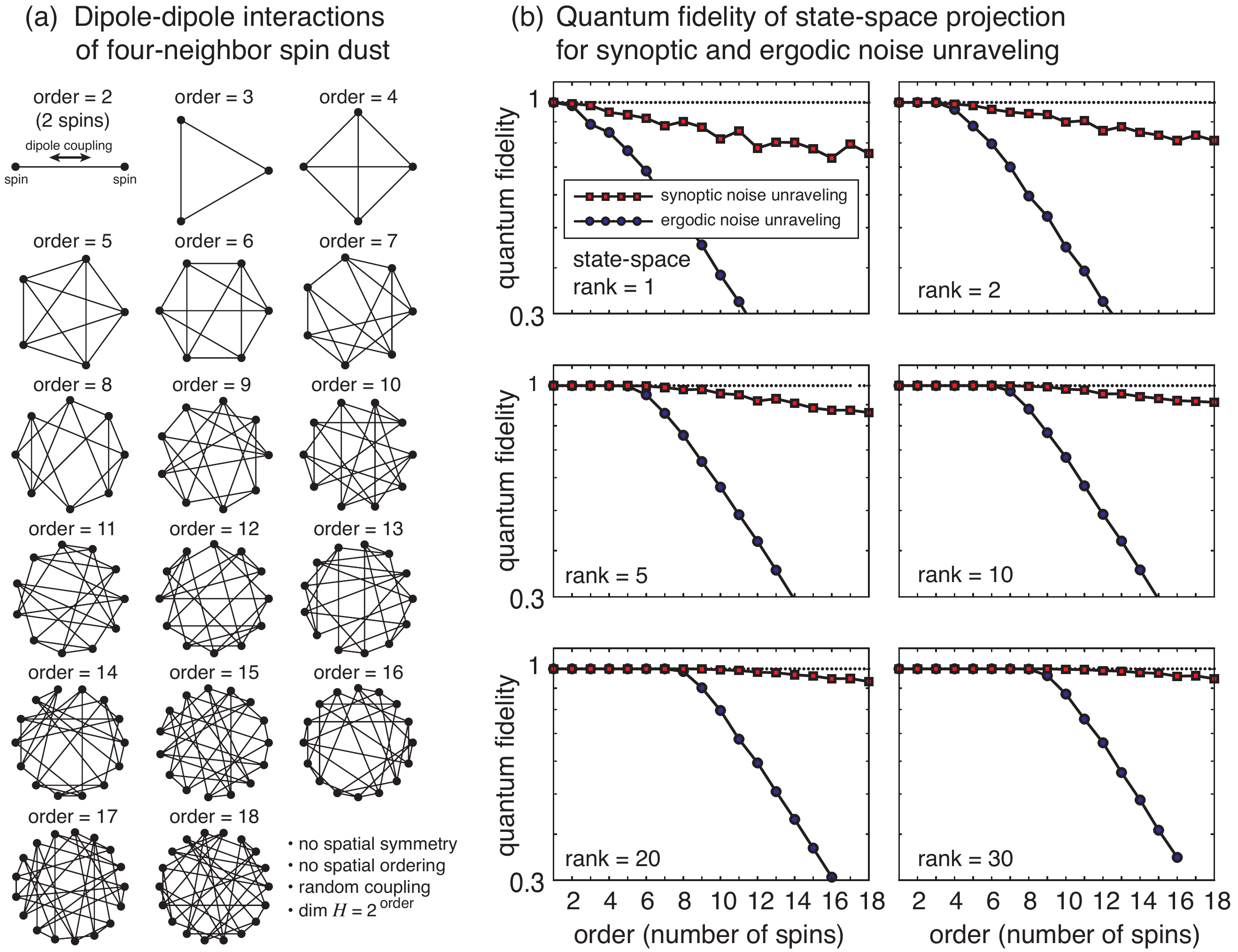}\\
\begin{minipage}{0.8\textwidth}
\caption[The dependence of \QMOR fidelity upon \GK order and rank]{%
\protect\justifying%
\label{fig: gabion-Kahler order and rank}%
The dependence of \QMOR fidelity upon \GK order and rank.%
}\end{minipage}
\end{figure}

With reference to Fig.~\ref{fig: gabion-Kahler order and rank}(a), simulations were conducted with numbers of spins $n \in 1,18$, having random dipole coupling links as depicted.  The median quantum fidelity was then computed, as a function of $n$, for \GK rank $r\in\{1, 2, 5, 10, 20, 30\}$ (see Fig.~\ref{fig: QMOR product sum} for the definition of \GK rank).  Both synoptic and ergodic unravelings were simulated.  Typically $\ket{\psi_0}(t)$ was projected at thirty different time-points along each simulated trajectory, always at times $t>100$ to ensure that memory of the (randomly chosen) initial state was lost.  We remark that numbers of spins $n>18$ could not feasibly be simulated on our modest computer (an Apple G5). 

The quantum fidelity of a projected state $\ket{\psi_{\lcal{K}}}$ was defined to be \cite[Section~9.2.2]{Nielsen:00}
\begin{equation}
f = |\inner{\psi_\lcal{K}}{\psi_0}|/\big[\inner{\psi_\lcal{K}}{\psi_\lcal{K}}\inner{\psi_0}{\psi_0}\big]^{1/2} .
\end{equation}
As shown in Fig.~\ref{fig: gabion-Kahler order and rank}(b), for ergodic unravellings large-$n$ quantum fidelity fell-off exponentially, while for synoptic unravelings large-$n$ fidelity remained high.    

No mathematical explanation for the observed exponential fall-off in ergodic unravelling fidelity is known.  The asymptotic large-$n$ behavior of the synoptic fidelity also is unknown.  In particular, for systems of hundreds or thousands os spins, would the empirical rule-of-thumb ``\GK rank fifty yields high fidelity for spin-dust systems'' still hold true?  These are important topics for further investigation.

The achieved high-fidelity algorithmic compression was large: an 18-spin exact quantum state $\ket{\psi_0(t)}$ is described by $2^{18}$ independent complex numbers, while an order-18 rank 30 \GK state $\ket{\psi_\lcal{K}(t)}$---as seen at lower right in Fig.~\ref{fig: gabion-Kahler order and rank}(b)---is described by $30\times(18+1)=570$ independent complex variables.  The dimensional reduction is therefore 460-to-1.

\subsubsection{The fidelity of spin polarization in projective \QMOR}
\label{sec: local measures}
\begin{figure}[t]\centering
\vspace*{2ex}
\includegraphics[width=1.0\textwidth]{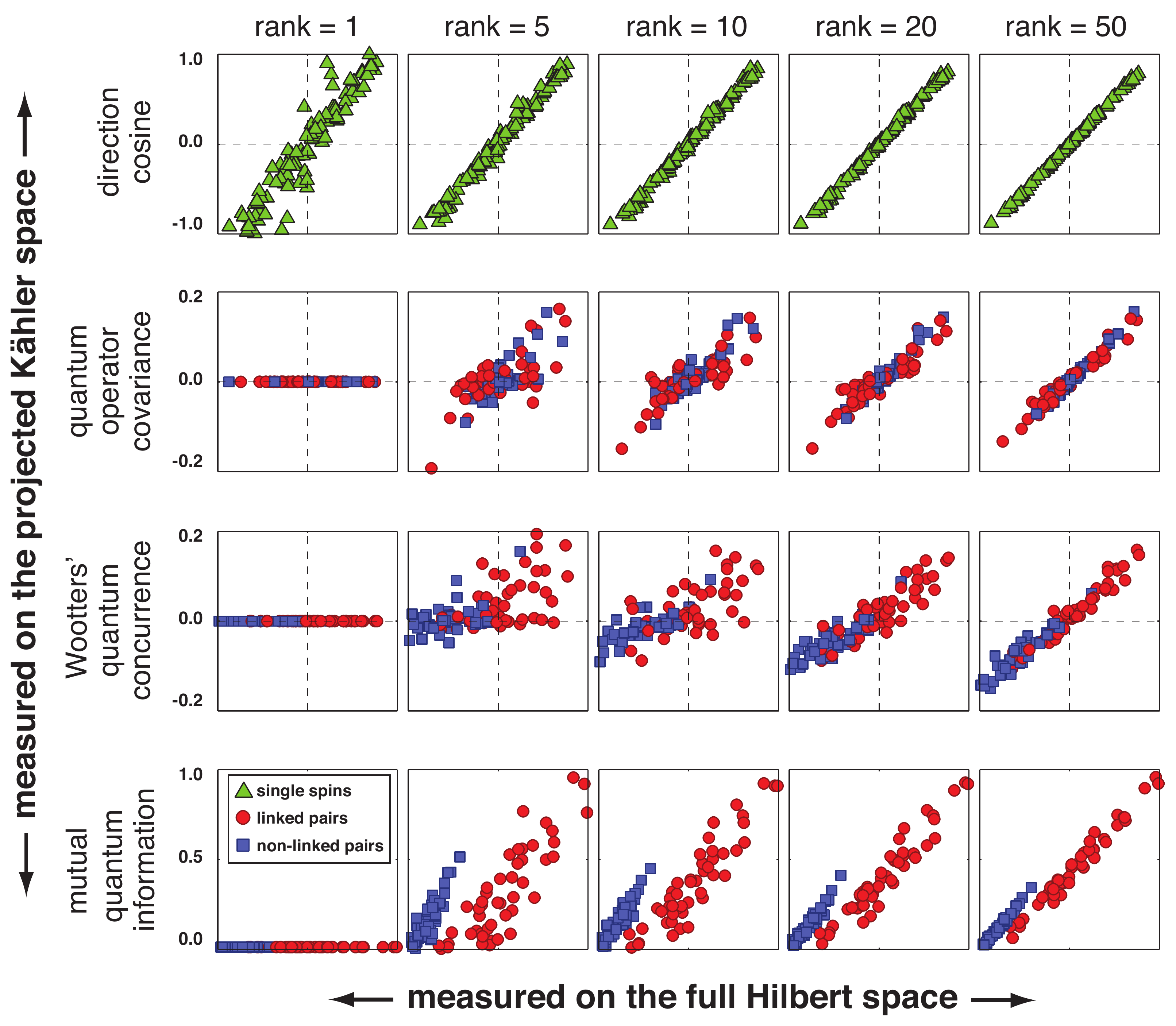}\\
\begin{minipage}{0.8\textwidth}
\caption[Measures of projective fidelity.]{%
\protect\justifying%
\label{fig: quantum measures of QMOR fidelity}
Measures of projective fidelity for $n=15$ spin-dust.%
}\end{minipage}
\end{figure}

We next turned our attention to measures of local quantum fidelity, as depicted in Fig.~\ref{fig: quantum measures of QMOR fidelity}.  All simulated trajectories in this figure were for $n=15$ spin-dust.   The first such measure we consider are the direction cosines $\braket{\psi}{\lb{s}\cdot\lbhat{m}}{\psi}/j$ for randomly spins, randomly chosen trajectory points, and randomly chosen unit vectors $\lbhat{m}$.  One hundred randomly chosen data points are shown. We observe that the rank-one \GK manifold does an excellent job of representing the spin direction cosines, which can be regarded as (essentially) classical quantities.

\subsubsection{The fidelity of operator covariance in projective \QMOR}\
As a measure of pair-wise quantum correlation, we examined the spin operator covariance.  With reference to (\ref{eq: lambda}), this quantity is given by 
\begin{equation}
\Sigma_{jk} = 4 \Lambda_{kl}=
\braket{\psi}{\sigma_{j}\sigma_{k}}{\psi}-
\braket{\psi}{\sigma_{j}}{\psi}\braket{\psi}{\sigma_{k}}{\psi}.
\end{equation}which vanishes for rank-1 (product states).  

The second row of  Fig.~\ref{fig: quantum measures of QMOR fidelity} plots $\Sigma_{kl}$ for one hundred randomly chosen trajectory points, and randomly chosen spins, having randomly chosen indices $j$ and $k$.  We observe that \GK ranks in the range 20--50 are necessary for projection to preserve pair-wise quantum correlation with good accuracy.

\subsubsection{The fidelity of quantum concurrence in projective \QMOR}
As a measure of pairwise quantum entanglement, we examined Wooters' quantum concurrence \cite{Wootters:1998oq}.  The concurrence is computed as follows.  Let $\rho_\subab$ be the reduced density matrix associated with spins {\small\sffamily A} and {\small\sffamily B}.  Let  $\lambda_i$ be eigenvalues of the non-Hermitian matrix $\rho_\subab{\stilde\rho}_\subab$, in decreasing order, where $\stilde{\rho}_\subab = (\sigma_y\otimes\sigma_y)\rho^\star_\subab(\sigma_y\otimes\sigma_y)$.  
Then the concurrence $c$ is defined to be\begin{equation}
c=\sqrt{\lambda_1}-\sqrt{\lambda_2}-\sqrt{\lambda_3}-\sqrt{\lambda_4}
\end{equation}
It can be shown that the concurrence vanishes for product states, and that spins are pairwise entangled if and only if $c>0$.  

In the third row of Fig.~\ref{fig: quantum measures of QMOR fidelity}, we observe that coupled spin-pairs are far more likely to  be quantum-entangled than non-coupled spin-pairs, as expected on physical grounds.  We further observe that \GK ranks in the range 20--50 are necessary for projection to preserve concurrence with good accuracy.

\subsubsection{The fidelity of mutual information in projective \QMOR}

As a measure of pair-wise quantum information, we examined von Neumann's mutual information \cite[Section 11.3]{Nielsen:00}, which is computed as follows. For a general density matrix $\rho$ we define the von Neummann entropy $S(\rho)=-\tr \rho\log_2 \rho$.  Then for spins {\small\sffamily A} and {\small\sffamily B} the mutual information is given by \begin{equation}
S(\rho_\suba)+S(\rho_\subb)-S(\rho_\subab)
\end{equation}
It is known that the mutual information vanishes for product states, and that this quantity is otherwise positive in general.  

In the fourth row of Fig.~\ref{fig: quantum measures of QMOR fidelity}, we observe that coupled spin-pairs share more mutual quantum information than non-coupled pairs, as expected on physical grounds.  We further observe that \GK ranks in the range 20--50 are necessary for projection to preserve mutual quantum information with good accuracy.

As a quantitative summary of this observation, the 15-spin simulations of Fig.~\ref{fig: quantum measures of QMOR fidelity} predict 15 single-spin density matrices $\rho_\suba$ and 105 pairwise reduced density matrices $\rho_\subab$ (in addition to higher-order correlations).   Each of the single-spin density matrices has 3 (real) degrees of freedom, and each pairwise density matrix introduces 9 more (real) independent degrees of freedom, for a total of 990 independent degrees of freedom associated with the one-spin and two-spin reduced density matrices.   In comparison, the rank-50 \GK manifold onto which the quantum states are projected is described by \Kahlerian coordinates having (it can be shown) 1600 locally independent coordinates.  Using 1600 state-space coordinates to encode 990 physical degrees of freedom represents a level of \MOR fidelity that (obviously) cannot be improved by more than another factor of two or so.  The mathematical origin of this empirical algorithmic efficiency is not known.

\subsection{Quantum state reconstruction from sparse random projections}
\label{sec: sparse random projections}

We will conclude our survey of spin-dust simulations with some concrete calculations that are motivated by recent advances in the theory of compressive sampling (\CS) and sparse reconstruction.  It will become apparent that synoptic simulations of quantum trajectories mesh very naturally with \CS methods and ideas.  To the best of our knowledge, this is the first description of \CS methods applied to quantum state-spaces.  

Our analysis will mainly draw upon the ideas and methods of Donoho \cite{Donoho:2006cr} and of Cand\`{e}s and Tao \cite{Candes:2007ve}, and our discussion will assume a basic familiarity with these and similar \CS articles \cite{Candes:2007vn,Candes:2006ys,Candes:2006ly,Candes:2006uq,Candes:2006zr}, especially a recent series of articles and commentaries on the Dantzig selector \cite{Meinshausen:2007rw,Bickel:2007yr,Efron:2007jc,Cai:2007ci,Ritov:2007ev,Candes:2007ls,Friedlander:2007ul}.  Our analysis can alternatively be viewed as an extension to the quantum domain of the approach of Baraniuk, Hegde, and Wakin \cite{Chinmay-Hegde:2007fv,Wakin:2006ty,Baraniuk:2008fj} to manifold learning \cite{Tenenbaum:2000rm} from sparse random projections.  

Our objectives in this section are:\begin{QSEitemize}\item establish that synoptically simulated wave functions $\gsb{\psi}_0$ are \emph{compressible objects} in~the sense of Cand\`{e}s and Tao \cite{Candes:2007ve}, 
\item establish that high-fidelity quantum state reconstruction from sparse random projections is algorithmically tractable,
\item describe how nonlinear \GK projection can be described as an embedding within a larger linear state-space of a convex optimization problem, and thereby 
\item specify algorithms for optimization over quantum states in terms of the Dantzig selector (a linear convex optimization algorithm) of Cand\`{e}s and Tao \cite{Candes:2007ve}.
\end{QSEitemize}
At the time of writing, the general field of compressive sensing, sampling and simulation is evolving rapidly---``Nowadays, novel exciting results seem to come out at a furious pace, and this testifies to the vitality and intensity of the field''~\cite{Candes:2007ls}---and our overall goal is to provide mathematical recipes by which researchers in \emph{quantum} sensing, sampling and simulation can participate in this enterprise.

\subsubsection{Establishing that quantum states are compressible objects} 
\label{sec compressible}
To establish that $\gsb{\psi}_0$ is compressible, it suffices to solve the following sparse reconstruction problem.  We begin by specifying what Donoho and Stodden \cite{Donoho:2006yu} call the \emph{model matrix} and what Cand\`{e}s and Tao \cite{Candes:2007ve} call the \emph{design matrix} to be an $n\times p$ matrix $\mathrm{X}$.  The projected state $\gsb{\phi}_0 = \mathrm{X}\gsb{\psi}_0$ is given, and our reconstruction task is to estimate the $\gsb{\psi}_0$ (the ``model'') from the $\gsb{\phi}_0$ (the ``sensor data'').  In general $n\le p$ and we particularly focus upon the case $n\ll p$.  

We initialize the elements of the design matrix $\mathrm{X}$ with i.i.d.\ zero-mean unit-norm (complex) Gaussian random variables.  Then the rows and columns of $\mathrm{X}$ are \emph{approximately} pairwise orthogonal, such that $\mathrm{X}$ satisfies the approximate orthogonality relation $\mathrm{X}\mathrm{X}^\dagger \simeq p\,\mathrm{I}$ and therefore satisfies the approximate projective relation $(\mathrm{X}^\dagger\mathrm{X})^2 \sim p \mathrm{X}^\dagger\mathrm{X}$.   As a remark, if we adjust $\mathrm{X}$ to make these orthogonality and projective relations exact instead of approximate---for example by setting all $n$ nonzero singular values of $\mathrm{X}$ to unity---our sparse reconstructions are qualitatively unaltered.  

In \CS language, we have specified random design matrices $\mathrm{X}$ that satisfy the uniform uncertainty principle (\UUP) \cite{Candes:2007ve,Candes:2006ys,Candes:2005ve}, meaning (loosely) that the columns of $\mathrm{X}$ are approximately pairwise orthogonal.  See \cite{Candes:2007ve} for a definition of \UUP design matrices that is more rigorous and general.

From a geometric point of view, this means we can regard $\mathrm{X}^\dagger\mathrm{X}$---which will turn out to be the mathematical object of interest---as a projection operator from our (large) $p$-dimensional quantum state-space onto a (much smaller) $n$-dimensional subspace.    

We have already seen in Sections~\ref{sec: expectation} and \ref{sec: Spin-dust simulations} that the following minimization problem can be tractably solved by steepest-descent methods:
\begin{equation}
\min_{\lb{\mathrm c}} 
	\big\|
			{\gsb\psi}_0
			-{\gsb\psi}_{\kappa}(\lb{\mathrm c}) 
	\|^2_{l_2}.
\label{eq: l2 norm}
\end{equation}
where we have adopted the \CS literature's practice of specifying the ${l_2}$ norm explicitly.  Here ${\gsb\psi}_{\kappa}(\lb{\mathrm c})$ is a vector of multilinear gabion-\Kahler (\GK) polynomials as defined in Section~\ref{sec: sectional curvature} and depicted in Fig.~\ref{fig: QMOR product sum}. Inspired by the \CS literature, we investigate the following \CS generalization of (\ref{eq: l2 norm}): 
\begin{equation}
\min_{\lb{\mathrm c}} \big\|\mathrm{X}
\left({\gsb\psi}_{\kappa}(\lb{\mathrm c}) - {\gsb\psi}_0\right)
\big\|^2_{l_2} = 
\min_{\lb{\mathrm c}} \big\|
{\gsb\phi}_0-\mathrm{X}{\gsb\psi}_{\kappa}(\lb{\mathrm c}) 
\big\|^2_{l_2}
\label{eq: nonlinear generalization}
\end{equation}
Now we are minimizing not on the full Hilbert space, but on the $n$-dimensional subspace projected onto by $\mathrm{X}$.  We recognize the right-hand expression as a nonlinear \Kahlerian generalization of a standard minimization problem (it is discussed \emph{e.g.}\ by Donoho and Stodden \cite[eq.~3]{Donoho:2006yu} and by Cand\`{e}s and Tao \cite[eq.~1.15]{Candes:2007ve}).  To make this parallelism more readily apparent, we can write the above minimization problem in the form
\begin{equation}
\min_{\beta} \big\|
{y-\mathrm{X}\beta} 
\big\|^2_{l_2}\quad\text{s.t.}\quad\beta={\gsb\psi}_{\kappa}(\lb{\mathrm c})
\label{eq: second nonlinear generalization}
\end{equation}
for some choice of $\lb{\mathrm c}$, where we have substituted ${\gsb\phi}_0\to y$ and introduced $\beta$ as an auxiliary variable.  Comparing the above to the well-known \LASSO minimization problem \cite{Donoho:2006yu,Friedlander:2007ul}
\begin{equation}
\min_{\beta} \big\|
{y-\mathrm{X}\beta} 
\big\|^2_{l_2}\quad\text{s.t.}\quad\big\|\beta\big\|_{l_1}\le t
\label{eq: LASSO nonlinear generalization}
\end{equation}
for some $t$, we see that the sole change is that the \LASSO problem's $l_1$ sparsity constraint $\big\|\beta\big\|_{l_1}\le t$ has been replaced with the \GK~representability constraint $\beta={\gsb\psi}_{\kappa}(\lb{\mathrm c})$.   We remark upon the parallelism that both constraints are highly nonlinear in $\beta$.

But this parallelism in itself does not give us much reason to expect that the minimization (\ref{eq: second nonlinear generalization}) is tractable, since we saw in Section~\ref{sec: sectional curvature} that the space of feasible solutions ${\gsb\psi}_{\kappa}(\lb{\mathrm c})$ is (floridly) nonconvex.  Consequently, unless some ``\GK magic'' of comparable algorithmic power to the well-known ``$l_1$ magic'' of \CS theory \cite{Candes:2005qf} should come our rescue, there seems to be little prospect of computing the minimum (\ref{eq: second nonlinear generalization}) in practice.

Persisting nonetheless, we compute successive approximations $\{\lb{\mathrm{c}}_1,\lb{\mathrm{c}}_2, \ldots, \lb{\mathrm{c}}_i\}$ by a projective generalization of the same steepest-descent method that produced the results of Figs.~\ref{fig: gabion-Kahler order and rank} and \ref{fig: quantum measures of QMOR fidelity}.  Specifically,  we expand the \GK coordinates via $\lb{\mathrm c}_{i+1} = \lb{\mathrm c}_i + \delta\lb{\mathrm c}_i$ and iterate the resulting linearized equations in $\delta\lb{\mathrm c}_i$
\begin{equation}
\label{eq: projected Kahler}
\delta\mathrm{\lb{c}}_i = 
-\big(
	\mathrm{A}^\dagger\mathrm{X}^\dagger\mathrm{X}\mathrm{A}
\big)^{\!\scriptpseudoinverse}\,
	\mathrm{A}^\dagger\mathrm{X}^\dagger\mathrm{X}
	\big(
		{\gsb\psi}_{\kappa}(\lb{\mathrm c}_i)-\gsb{\psi}_0
	\big).
\end{equation}
Empirically, good minima are obtained from $\lcal{O}(\dimK)$ iterations of this equation from randomly-chosen starting-points.  This benign behavior is surprising, given that our objective function (\ref{eq: nonlinear generalization})  is a polynomial in $\lcal{O}(\dimK)$ variables having $\lcal{O}\big((\dimH)^2\big)$ independent terms, because generically speaking, finding minima of large polynomials is computationally infeasible.

According to the geometric analysis of Section~\ref{sec: formal definition}, the existence of feasibly computed minima is explained by the rule structure of \GK state-space, which ensures that almost all state-space points at which the increment (\ref{eq: projected Kahler}) vanishes are saddle points rather than local minima, in consequence of the nonpositive directed sectional curvature that is guaranteed by Theorem~\ref{thm: one}.

We now discuss \GK geometry from the alternative viewpoint of \CS theory, further developing the idea that \GK rule structure provides the underlying geometric reason why \CS ``works'' on \GK state-spaces.

\subsubsection{Randomly projected \GK manifolds are \GK manifolds} With reference to the algorithm of Fig.~\ref{fig: QMOR numerical}, we immediately identify $(\mathrm{A}^\dagger\mathrm{X}^\dagger\mathrm{X}\mathrm{A})^{\scriptpseudoinverse}$ in (\ref{eq: projected Kahler}) as the \Kahlerian metric of a \GK manifold having an algebraic \Kahler potential (see (\ref{eq: Kahler potential})) that is simply
\begin{equation}
\kappa(\lbbar{\mathrm c},\lb{\mathrm c}) = \tfrac{1}{2}\gsbbar{\psi}(\lbbar{\mathrm c})\mathrm{X}^\dagger\mathrm{X}\gsb{\psi}(\lb{\mathrm c}).
\end{equation}  Since $\mathrm{X}$ is constant, we see that the projected \Kahler potential is a biholomorphic polynomial in the same variables and of the same order as the original \Kahler potential.  It follows that a projected \GK manifold is itself a \GK manifold, and in particular the \GK rule structure is (of course) preserved under projection, and this means that all of the sectional curvature theorems of Section~\ref{sec: sectional curvature} apply immediately to \QMORCS on \GK state-spaces.

This \GK inheritance property is mathematically reminiscent of the inheritance properties of convex sets and convex functions, and it suggests that a calculus of \GK polynomials and manifolds might be developed along lines broadly similar in both logical structure and practical motivation to the calculus of convex sets and convex functions that is presented in the standard textbooks of \CS \cite{Boyd:2004kx} (we discuss this further in Section~\ref{sec: why?}).  

\subsubsection{Donoho-Stoddard breakdown at the Cand\`{e}s-Tao bound} 
\begin{figure}[p]%
\centering%
\vspace*{2ex}
\hspace*{-1em}\includegraphics[width=0.925\textwidth]{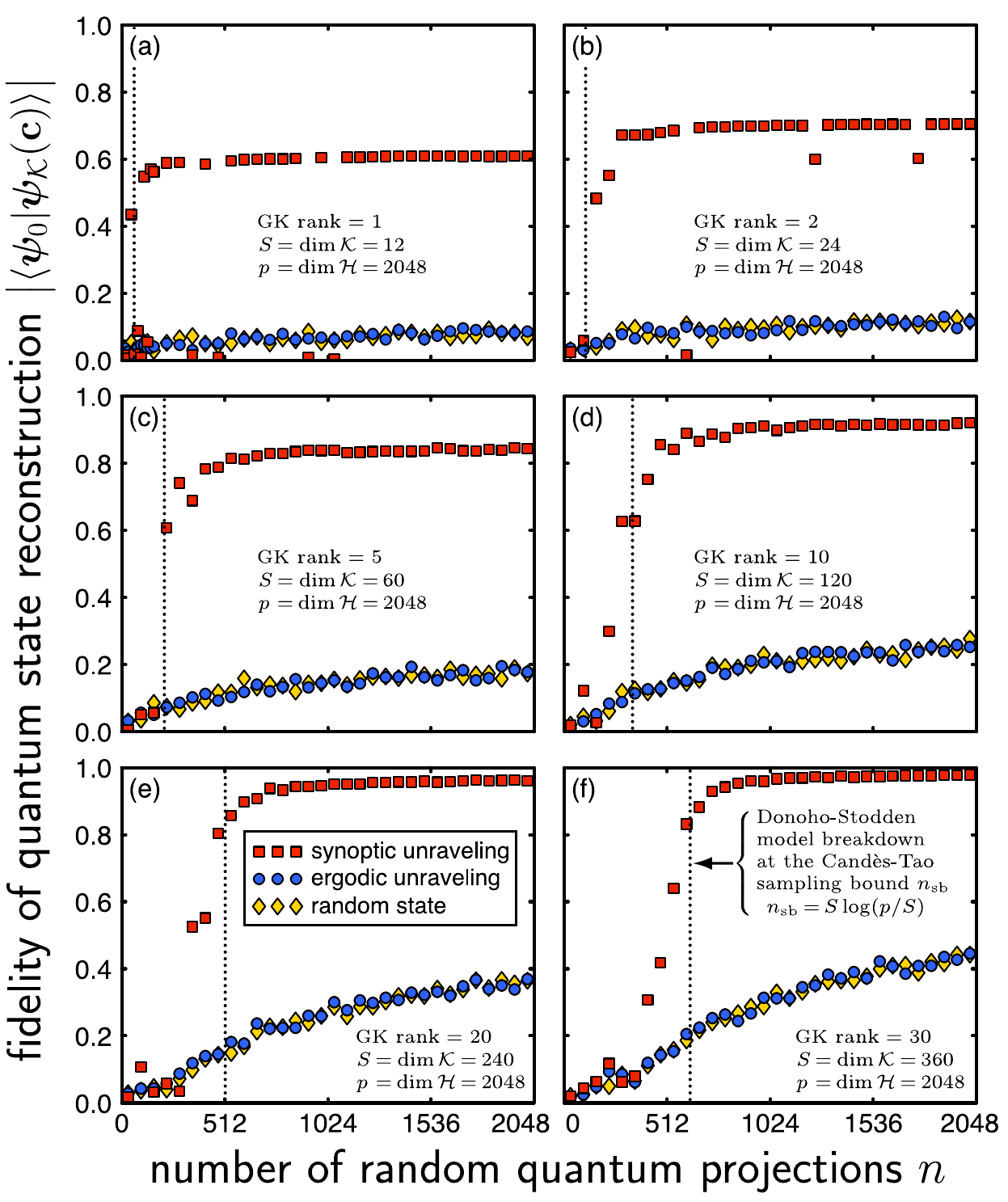}\\
\begin{minipage}{1\textwidth}%
\caption[Quantum state reconstruction from sparse random projections]{%
\protect\justifying%
\label{fig: sparse random projections}%
Quantum state reconstruction from sparse random projections.  A (typical) state
from an 11-spin trajectory was reconstructed from sparse projections
onto random subspaces (horizontal axis), and the resulting quantum fidelity was evaluated (vertical axis).  Each point represents a single minimization of (\ref{eq: projected Kahler}), by iteration of (\ref{eq: projected Kahler}) with a conjugate gradient correction, from a random starting point chosen independently for each minimization.  Convergence to ``false'' local minima was sporadically encountered for low-rank \GK projections (graphs~a--b, \GK ranks 1 and 2) but not for higher-rank \GK projections (graphs~c--f, \GK ranks 5, 10, 20, and 30).  The onset of Donoho-Stodden breakdown was observed to occur near the Candes-Tao bound (\ref{eq: candes-tao limit})---plotted as a dotted vertical line---for all \GK ranks tested.  The ergodic spin-dust simulation yielded states whose reconstruction properties were indistinguishable from random states, as expected.
}\end{minipage}
\end{figure}

Putting these ideas to numerical test, using the same spin-dust model as in previous sections, we find that random compressive sampling \emph{does} allow high-fidelity quantum state reconstruction, \emph{provided} that the state trajectory to be reconstructed has been synoptically unraveled (Fig.~\ref{fig: sparse random projections}).

These numerical results vividly illustrate what Donoho and Stodden \cite{Donoho:2006yu} have called ``the breakdown point of model selection''  and we note that Cand\`{e}s and Tao have described similar breakdown effects in the context of error-correcting codes \cite{Candes:2005ve}.  Surprisingly, it does not appear to have been recognized that a similar breakdown occurs in quantum modeling whenever too many wave function coefficients are reconstructed from too few projections.

For discussion, we direct our attention to the rank-30 block of Fig.~\ref{fig: sparse random projections}(f), where (as labeled) we reconstruct the $p = \dimH = 2^{n_\text{spin}} = 2048$ (complex) components of $\gsb{\psi}_0$ from $n$ random projections onto a \GK manifold having $S = \dimK = (\text{\GK rank}) \times(1+\log_2(\dimH)) = 30\times12 = 360$ (complex) dimensions.  All six blocks (a-f) of the figure are similarly labeled with $p$ and $S$.  

Given our example with $p=2048$ and $S=360$, what does \CS theory predict for the minimum number $n$ of random projections required for accurate reconstruction?  According to Cand\`{e}s and Tao \cite{Candes:2007ve}
\begin{QSEquote}
With overwhelming probability, the condition [for sparse reconstruction] holds 
for $S = \lcal{O}\big(n/\log(p/n) \big)$. In other words, this setup only requires $\lcal{O} \big(\log(p/n) \big)$ observations per non\-zero parameter value; for example, when $n$ is a nonnegligible fraction of $p$, one only needs a handful of observations per nonzero coefficient. In practice, this number is quite small, as few as 5 or 6 observations per unknown generally suffice (over a large range of the ratio $p/n$).
\end{QSEquote}
We naively adapt the above Cand\`{e}s-Tao big-$\lcal{O}$ sampling bound to the case at hand by recalling that $S = n/\log(p/n)$ implies $n \simeq S \log(p/S)$ for $p \gg n$ \cite{Corless:1996ma}.  We therefore expect to observe Donoho-Stodden breakdown at a (complex) projective dimension $n_{\text{sb}}(p,S)$ (which we will call the \emph{sampling bound}) that to leading order in $S/p$ is
\begin{equation}
n_{\text{sb}}(p,S)  \simeq S\log(p/S)%
\big|_{\scriptstyle\begin{array}[t]{@{}r@{\,=\,}l}
S&\dimK\\
p&\dimH
\end{array}}%
\label{eq: candes-tao limit}
\end{equation}
Here we recall that $\dimH$ is the (complex) dimension of the Hilbert space within which the efflorescent \GK state-space manifold of (complex) dimension $\dimK$ is embedded (see Sections~\ref{sec: GK Theorema Egregium} and~\ref{sec: efflorescent curvature}, and also (\ref{eq: dimensions}), for discussion of how to calculate $\dimK$).  

The above sparsity bound accords remarkably well with the numerical results of Fig.~\ref{fig: sparse random projections}.  This empirical agreement suggests that \QMOR and \CS may be intimately related, but on the other hand, there are the following countervailing reasons to regard the agreement as being possibly fortuitous: 
\begin{QSEenumerate}
\item the Cand\`{e}s-Tao bound applies to state-spaces that are globally linear, whereas we are minimizing on a \GK state-space that is only locally linear, and 
\item the onset of Donoho-Stodden breakdown in Fig.~\ref{fig: sparse random projections} is (experimentally) accompanied by the onset of multiple local minima of (\ref{eq: nonlinear generalization}), which are not present in the convex objective function of Cand\`{e}s and Tao, and 
\item high-accuracy numerical agreement with a ``big-$\lcal{O}$'' estimate is fortuitous; the agreement seen in Fig.~\ref{fig: sparse random projections} is better than we have a reason to expect.
\end{QSEenumerate}
So although the Cand\`{e}s-Tao bound seems empirically to be the right answer to ``when does Donoho-Stodden breakdown occur in quantum model order reduction?'' the numerical calculations do not explain \emph{why} it is the right answer.  

We now present some partial results---which however are rigorous and deterministic insofar as they go---that begin to provide a nontrivial explanation of why the Cand\`{e}s-Tao bound applies in the sparse reconstruction of quantum states.  The basic idea is to embed the nonlinear minimization (\ref{eq: l2 norm}) within a larger-dimension problem that is formally convex.  We will show that this larger-dimension optimization problem can be written explicitly as a Dantzig selection.  

The main mathematical tool that we will need to develop is sampling matrices $X$ whose (small) row dimension is $n=\dim H$, and whose (large) column dimension $p$ is a power of $\dim H$.  These matrices are too large to be evaluated explicitly---they are what Cai and Lv call ``ultrahigh-dimensional''~\cite{Cai:2007ci}.  A novel aspect of our analysis is that we construct these matrices deterministically, such that their analytic form allows the efficient evaluation of matrix products.

\subsubsection{Wedge products are Hamming metrics on \GK manifolds} 
\label{sec: error-correcting codes}
Let us consider how the efflorescent \GK geometry that we described in Sections~\ref{sec: efflorescing geometry} and \ref{sec: efflorescent curvature} can be made the basis of a deterministic algorithm for constructing good sampling matrices.  

Our basic approach is to construct a deterministic lattice of points on \GK manifolds, together with a labeling of the lattice for which the Hamming distance between two labels is a monotonic function solely of the wedge product between that pair of points, such that the larger the Hamming distance between two points, the closer they approach to mutual orthogonality.   The problem of constructing good sampling matrices then becomes equivalent to the problem of constructing good error correction codes.

We now construct the desired \GK lattice.  We consider sampling matrices $\mathrm{X}$ whose columns are not random vectors, but rather are constrained to satisfy $\gsb\psi = {\gsb\psi}_{\kappa}(\lb{\mathrm c})$ for some \GK state-space ${\gsb\psi}_{\kappa}(\lb{\mathrm c})$.  We wish these vectors to be approximately orthogonal.  To construct these vectors (and simultaneously assign each vector a unique code-word), we specify an \emph{alphabet} of four \emph{characters} $\{a,b,d,e\}$, we identify the four characters with the four vertices of a tetrahedron having unit vectors $\{\lbhat{n}_a,\lbhat{n}_b,\lbhat{n}_d,\lbhat{n}_e\}$, and we identify the unit vectors with the four spin-$j$ coherent states $\{\ket{\lbhat{n}_a},$ $\ket{\lbhat{n}_b},$ $\ket{\lbhat{n}_d},$ $\ket{\lbhat{n}_e}\}$, such that for $\gsb{s}=\{s_x,s_y,s_z\}$ the usual spin operators, the tetrahedron vertices are $\lbhat{n}_a = \braket{\lbhat{n}_a}{\gsb{s}}{\lbhat{n}_a}/j$, \emph{etc}.  Soon it will become apparent that the vertices of any polytope, not only a tetrahedron, suffice for this construction, and that the vertices of Platonic solids are a particularly good choice.  

We recall from our study of \GK geometry that a wedge product (\ref{eq: wedge}) can be associated to each letter-pair $\{a,b\}$ as follows
\begin{equation}
|a\wedge b|^2 \equiv 
  \inner{\lbhat{n}_a}{\lbhat{n}_a}\inner{\lbhat{n}_b}{\lbhat{n}_b} -
  \inner{\lbhat{n}_a}{\lbhat{n}_b}\inner{\lbhat{n}_b}{\lbhat{n}_a} 
\end{equation}
From Wigner's identity (\ref{eq: Wigner}) we have $|\inner{\lbhat{n}_a}{\lbhat{n}_b}|^2 = |D^j_{jj}(0,\theta_{ab},0)|^2 = \cos(\theta_{ab}/2)^{4j}$ where $\cos(\theta_{ab}) = \lbhat{n}_a\cdot\lbhat{n}_b$, so the spin-$j$ wedge product is easily evaluated in closed form as 
\begin{equation}
|a\wedge b|^2_j = 1 - \cos(\theta_{ab}/2)^{4j}
\label{eq: first of four}
\end{equation}
which for our tetrahedral alphabet is simply
\begin{equation}
|a\wedge b|^2_j = \begin{cases}
0 & \text{for} \quad a = b \\
1 - 9^{-j} & \text{for}\quad a \ne b 
\end{cases}
\label{eq: tetrahedral alphabet}
\end{equation}
Here and henceforth we have added a subscript $j$ to all wedge products for which an analytic form is given that depends explicitly on the total spin $j$.

Now we specify a \emph{dictionary} to be a set of $n$-character \emph{words} $\{w^k\}$  with each word associated with an ordered set of tetrahedral characters $w^k = \{c^k_1,$ $c^k_2,$ $\ldots\,,$ $c^k_n\}$.  We further associate with each word a \emph{petal-vector} $\ket{w^k}$ 
\begin{equation}
\ket{w^k} = \ket{\lbhat{n}_{c^k_1}} \otimes \ket{\lbhat{n}_{c^k_2}} \ldots \otimes \ket{\lbhat{n}_{c^k_n}}
\label{eq: third of four}
\end{equation}
Given two words $\{w^i,w^k\}$, their mutual \emph{Hamming distance} $h(w^i,w^k)$ is defined to be the number of symbols that differ between $w^i$ and $w^k$.  We also can associate with any two petal-vectors their mutual wedge product $|w^i \wedge w^k|$ defined by 
\begin{equation}
|w^i \wedge w^k|^2 = 
  \inner{w^i}{w^i}\inner{w^k}{w^k} -
  \inner{w^i}{w^k}\inner{w^k}{w^i} 
\label{eq: fourth of four}
\end{equation}
The Hamming distance $h(w^i,w^k)$ is a \emph{Hamming metric} on our codeword dictionary, and it is easy to show that this code metric is related to the petal-vector wedge product $|w^i \wedge w^k|$ by the simple expressions
\begin{align}
	|w^i \wedge w^k|^2&  = 1 -
	\prod_{m=1}^{n_\text{spin}} |\inner{\lbhat{n}_{c^i_m}}{\lbhat{n}_{c^k_m}}|^2 
	& \text{in general, from (\ref{eq: third of four}--\ref{eq: fourth of four})}\\
	|w^i \wedge w^k|^2_j & = 1 - 9^{-j\,h_j(w^i,w^k)}
	& \text{\hspace{-3.5em}tetrahedral dictionary, from (\ref{eq: first of four}--\ref{eq: tetrahedral alphabet})}
\label{eq: tetrahedral Hamming distance}
\end{align}
or equivalently for a tetrahedral petal-vector dictionary
\begin{align}
h_j(w^i,w^k) = -\log_9(1-|w^i \wedge w^k|^2_j)/j
\label{eq: main result}
\end{align}
The main result of this section is the above monotonic functional relation between a Hamming distance and a wedge product.  We are not aware of previous \CS work establishing such a relation.

The practical consequence is that given a dictionary of $n$-character words  $\{w^k\}$ having mutually large Hamming distances (\emph{i.e.}, a good error correcting code), this construction deterministically specifies a set of nearly-orthogonal petal-vectors $\{\ket{w^k}\}$ (\emph{i.e.}, good vectors for sparse random sampling).  Conversely, the general problem of constructing a deterministic set of nearly-orthogonal petal-vectors is seen to be precisely as difficult as deterministically constructing a good error correcting code.

\begin{table}[t]\centering
\vspace*{1ex}
\includegraphics[width=\textwidth]{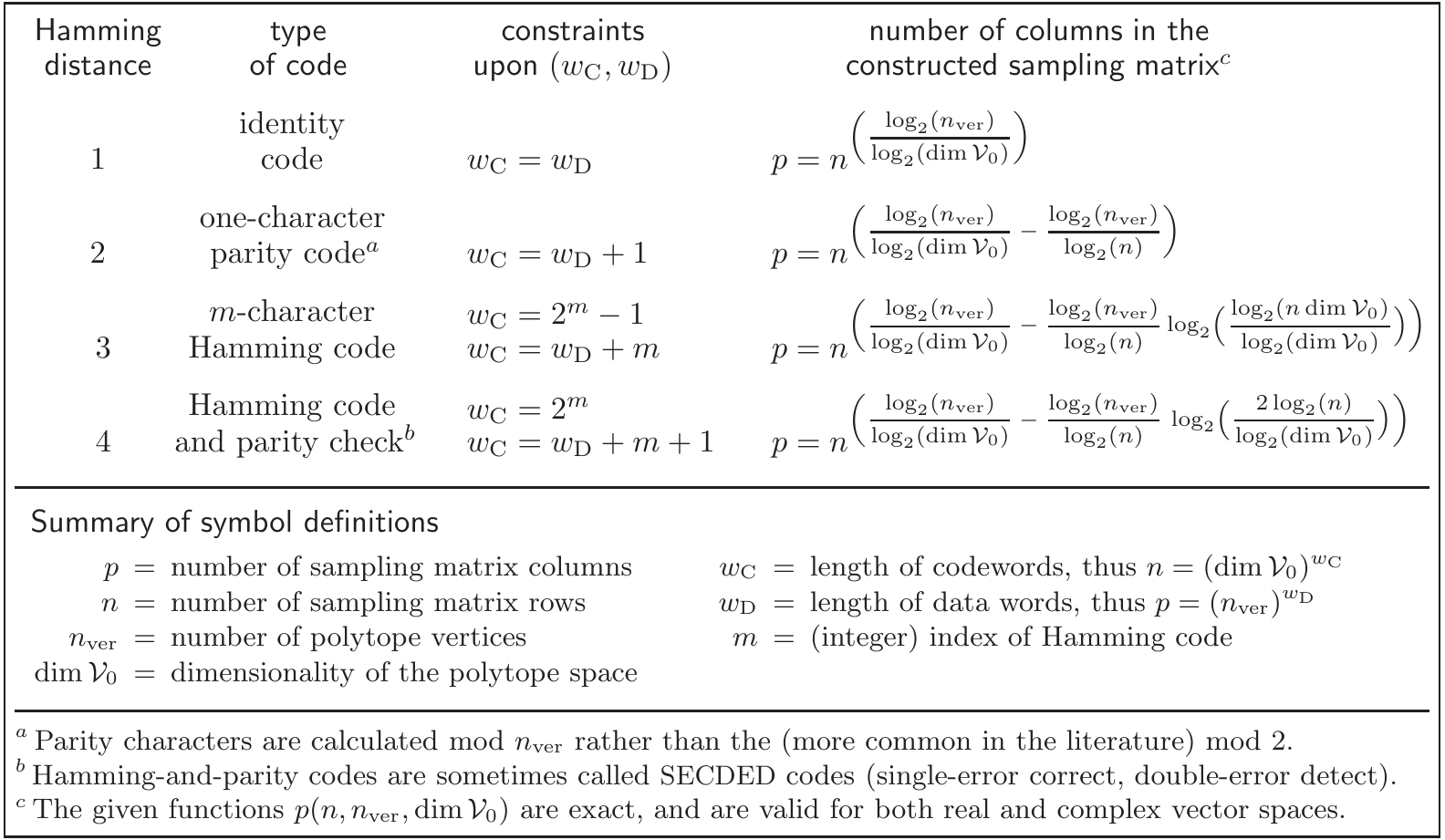}\\
\caption[Recipes for deterministic construction of sampling matrices]{%
\protect\justifying%
\label{table: sampling matrices}%
Recipes for deterministically constructing sampling matrices by the methods of Section~\ref{sec: error-correcting codes}.  The primary design variables are taken to be the dimensionality of polytope space $\dim \lcal{V}_0$, the number of polytope vertices $n_\text{ver}$, and the desired number of sampling matrix rows $n$.  All expressions are exact, and the results apply to both real and complex  vector spaces.  The expressions  are organized so as to make manifest that increased minimal Hamming distance is associated with decreased column-length $p$; this is the central design trade-off.  
}
\end{table}%

\subsubsection{The $n$ and $p$ dimensions of deterministic sampling matrices}
The column dimension $p$ of the petal-vector sampling matrices thus constructed is given in Table~\ref{table: sampling matrices} for Hamming distances 1--4 as a function of the row dimension $n$, the number of polytope vertices $n_\text{ver}$, and the dimensionality of the polytope space $\dim\lcal{V}_0$ (\emph{e.g.}, for our tetrahedral construction  $n_\text{ver}=4$ and $\dim\lcal{V}_0 = 2$).  We see that for fixed row dimension $n$, larger Hamming distances are associated with smaller column dimension $p$, as is intuitively reasonable: the more stringent the pairwise orthogonality constraint, the smaller the maximal dictionary of sampling vectors that meet this constraint.

The construction has a further dimensional constraint as follows: it is straightforward for values of $n$ that are powers of two (because the tetrahedral construction can be used), more complicated when $n$ has factors other than $2$ (because larger-dimension polytope vertices must be specified), and infeasible when $n$ is large and prime.  These constraints are reminiscent of similar constraints that act upon the fast Fourier transform, and arise for basically the same number-theoretic reason.

\subsubsection{Petal-counting in \GK geometry via coding theory}
\label{sec: design of sampling matrices}
These sampling theory results have a direct quantitative relation to the efflorescent \GK geometry that we discussed in Sections~\ref{sec: efflorescing geometry} and \ref{sec: efflorescent curvature}.  Specifically, we are now able to construct a petal-vector description of \GK manifolds, and verify that they indeed have exponentially many petals.

We consider a spin-$\tfrac{1}{2}$ rank-1 \GK manifold having $n_\text{spin}$ spins.  The preceding tetrahedral \GK construction deterministically generates a dictionary of petal-words $\{w^k: k\in 1,4^{n_\text{spin}}\}$ in one-to-one correspondence with petal-vector states $\{\ket{w^k}: k\in 1,4^{n_\text{spin}}\}$.  This dictionary of states of course exponentially over-complete, since its number of words is $4^{n_\text{spin}} =  2^{n_\text{spin}}\ \dimH$.  Yet we also know that random pairs of petal-vectors in our tetrahedral dictionary are pairwise orthogonal to an excellent approximation, because their median Hamming distance is $3{n_\text{spin}}/4$, and consequently from (\ref{eq: tetrahedral Hamming distance}) their median pairwise wedge product is $|w^i \wedge w^k|^2 = 1 - 3^{-3 n_\text{spin}/4}$.  

As a concrete exercise in petal-counting, we consider a system of $n_\text{spin} = 16$ spin-$\tfrac{1}{2}$ particles.  The tetrahedral construction generates a dictionary of $4^{16} = 2^{32}$ petal-vectors for this system, each word of which labels a petal whose state-vector has a wedge separation of $|w^i \wedge w^k|^2 \ge 2/3$ from the state-vector of all other petals.  A subset of that petal dictionary having minimal Hamming distance 4 is specified by the \SECDED code of Table~\ref{table: sampling matrices}.  This \SECDED subset has Hamming parameter $m=4$, and hence $m+1=5$ characters out of 16 in each word are devoted to error-correcting.  The~resulting (smaller) error-corrected dictionary has $4^{16-5} = 4^{11} = 2^{22}$ petal-vectors, and the sampling matrix whose columns are the petal-vectors of this dictionary therefore has $n=2^{16}$ rows and $p=2^{22}$ columns, whose column wedge products satisfy the (exact) pairwise inequality
\begin{equation}
|w^i \wedge w^k|^2   \ge  1 - 3^{-4}\quad\text{for all}\quad i\ne k
\end{equation} 
These calculations confirm our previous conclusion from Riemann curvature analysis, that even a rank-one \GK manifold contains exponentially many petals.  They also illustrate that the deterministic construction of high-quality sampling matrices involves sophisticated trade-offs in error-correcting codes.   

\subsubsection{Constructing a Dantzig selector for quantum states} 
\label{sec: quantum Dantzig selector}
We now have all the ingredients we need to establish the following remarkable principle: any quantum state that can be written as a (sparse) sum of petal-vectors can be recovered from sparse random projections by convex programming methods.  The method is as follows.  We approximate the minimization problem (\ref{eq: l2 norm}) that began this section in the form 
\begin{equation}
\min_{\lb{\mathrm c}} 
	\big\|
			\gsb{\psi}_0
			-\gsb{\psi}_{\kappa}({\mathrm c}) 
	\|^2_{l_2} 
\simeq
\min_{\tilde w} 
	\big\|
			\gsb{\psi}_0
			-X \tilde w
	\big\|^2_{l_2}
\label{eq:step one of transformation}
\end{equation}
Here the columns of $X$ are the petal-states of our dictionary  (the preceding tetrahedral dictionary will do) and $\tilde w$ is a column vector of \emph{petal coefficients} (one coefficient for every vector in our dictionary).

We then further approximate the above minimization problem in any of several standard forms \cite{Friedlander:2007ul}, see also \cite{Candes:2005qf,Efron:2007jc,Meinshausen:2007rw}.  These forms include (\emph{e.g.}):
\begin{subequations}
\begin{align}
\min_{\tilde w} 
	\big\|
			\tilde w
	\big\|_{l_1}	
	\quad\text{s.t.}\quad&%
\big\|
			X^{\dagger}(\gsb{\psi}_0
			-X \tilde w)
	\big\|_{l_\infty}{\,\le\,}\lambda
	&\text{Dantzig selector}
\label{eq: quantum Candes-Tao}\\
\min_{\tilde w} \big\|
{\gsb\psi_0-X \tilde w} 
\big\|^2_{l_2}\quad\text{s.t.}\quad&%
\big\|\tilde w \big\|_{l_1}\le \lambda
&\text{\LASSO}
\label{eq: quantum Lasso}\\
\min_{\tilde w} 
\big\|\tilde w\big\|_{l_1}
+ & \lambda \big\|
{\gsb\psi_0-X \tilde w} 
\big\|^2_{l_2} 
&\text{basis pursuit}
\label{eq: basis pursuit}
\end{align}
\end{subequations}
Here $\lambda$ is a parameter that is adjusted on a per-problem basis.  Although the relative merits of the above optimizations are the subject of lively debate, for many practical problems they all work well.  The Dantzig selector optimization (\ref{eq: quantum Candes-Tao}) in particular can be posed as an explicitly convex optimization problem that can be solved by a Dantzig-type simplex algorithm \cite{Candes:2007ve} (among other methods).

To recapitulate, the key physical idea behind the first step (\ref{eq:step one of transformation}) of the above two-step transformation is to represent a general state as a sparse superposition of petal-states.   The key mathematical idea behind the second step (\ref{eq: quantum Candes-Tao}--c) is to approximate the resulting sparse minimization problem as any of several forms that can be efficiently solved by numerical means.  

A second key mathematical idea is that the column dimensions of $X$ can be very large---much larger than the Hilbert space dimension $\dimH$---provided that efficient algorithms exist for calculating the product $X\tilde w$ without calculating either $X$ or $\tilde w$ explicitly (as was discussed earlier in Section~\ref{sec: practical considerations}).  This is why deterministic methods for constructing $X$ are essential to the feasibility of quantum optimization by Dantzig selection and related methods.  

This construction provides a non-trivial mathematical explanation of why the numerical optimizations of this article are well-behaved: the early coarse-grained, non-linear stages can be regarded as implicitly solving a convex optimization problem over petal-states, and the later fine-grained stages are solving a problem which is linear to a reasonable approximation.  

Boyd and Vandenberghe's textbook ambitiously asserts \cite[in the Preface]{Boyd:2004kx} 
\begin{QSEquote}
With only a bit of exaggeration, we can say that, if you formulate a practical problem as a convex optimization problem, then you have solved the original problem.
\end{QSEquote}
But this assertion must be regarded with caution when it comes to convex optimization over quantum state-spaces, because the matrices and vectors involved are of enormously larger dimension than is usually the case in convex optimization.  Figueiredo, Nowak, and Wright \cite{Figueiredo:2007ya} and also Cai and Lv \cite{Cai:2007ci} discuss this domain, and it is clear that Cai and Lv's conclusion  ``Clearly, there is much work ahead of us'' applies especially to compressive quantum sensing, sampling, and simulation.

\begin{table}[t]\centering
\vspace*{1ex}
\includegraphics[width=\textwidth]{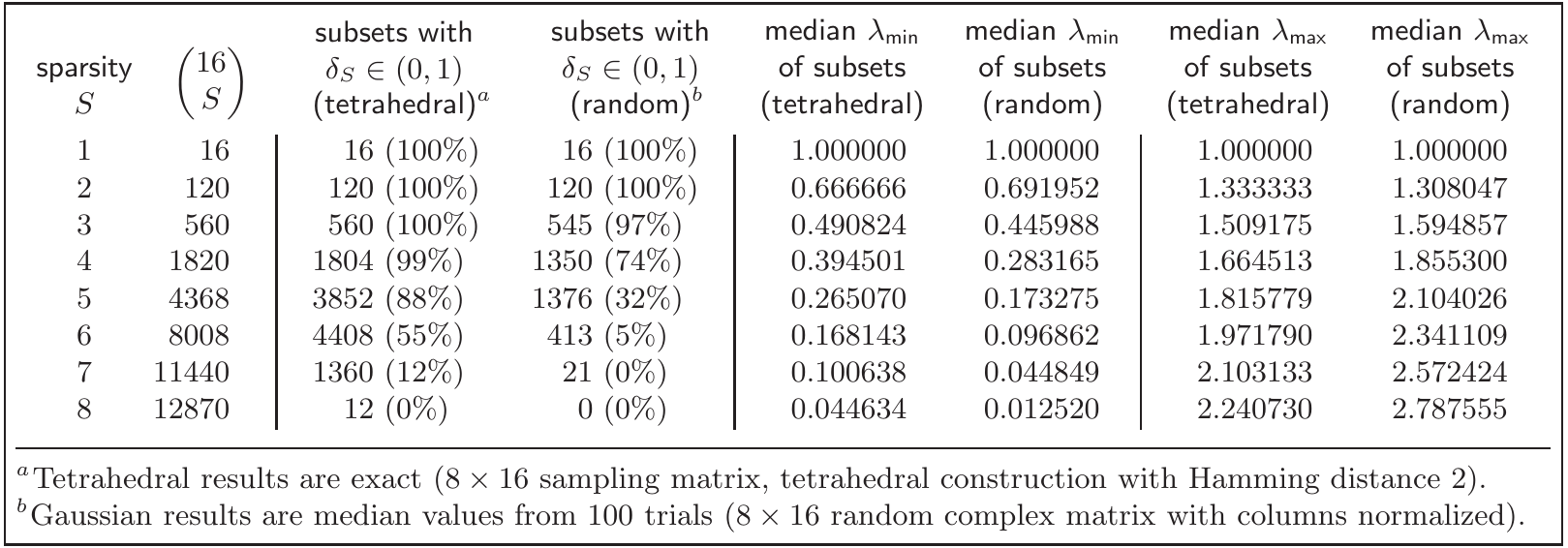}\\
\caption[\RIP properties of deterministic versus random $8\times 16$ sampling matrices]{%
\protect\justifying%
\label{table: eightsixteen sampling matrices}%
\RIP properties of $8\times 16$ sampling matrices created via a deterministic tetrahedral construction, contrasted with same-size random Gaussian sampling matrices.
By definition, matrices for which 100\%\  of subsects have $\delta_S \in (0,1)$ are \RIP in order $S$.  For all sparsities the tetrahedral construction yields \RIP properties that are superior to random constructions.}
\end{table}%

\subsubsection{\RIP properties of deterministic versus random sampling matrices} It is clear from the preceding discussion that over-complete dictionaries of word-states $\{\ket{w^i}\}$ having the approximate orthogonality property $\inner{w^i}{w^j} = (XX^\dagger)_{ij} \simeq \delta_{ij}$ are desirable both for simulation purposes and for sampling purposes.  Stimulated by the work of Candes and Tao \cite{Candes:2007ve}, an extensive and rapidly growing body of work  characterizes such matrices in terms of the \emph{restricted isometry property} (\RIP).  We now briefly discuss the \RIP of tetrahedral sampling matrices, mainly following the notation and discussion of Baraniuk \emph{et al{.}}~\cite{Baraniuk:2008oa}.  

We regard the word indices $i$ and $j$ in $\inner{w^i}{w^j}$ as the rows index and column index of a Hermitian matrix.  We specify a subset $T$ of word indices, and we define the \emph{sparsity} $S$ of that subset to be  $S=\#T$.  Then $\inner{w^i}{w^j}_T \equiv \inner{w^i}{w^j}:\,i,j \in T$ is an $S\times S$ Hermitian matrix, which we take to have minimal (maximal) eigenvalues $\lambda_\text{min}$ ($\lambda_\text{max}$).  Then the \emph{isometry constant} $\delta_S$ of Candes and Tao is by definition
\begin{equation}
 \delta_S = \max (1-\lambda_\text{min},\lambda_\text{max}-1)
\end{equation}
Our word-state dictionary is said to have the \emph{restricted isometry property for order $S$} iff $\delta_S \in (0,1)$ for all subsets $T$ having sparsity $S$.  Physically speaking, a dictionary of $p$ word-states having the \RIP property in order $S$ has the property that any set of $S$ words is (approximately) mutually orthogonal.

Testing for the \RIP property is computationally inefficient, since (at present) no known algorithm is significantly faster than directly evaluating $\lambda_\text{min}$ and $\lambda_\text{max}$ for all $\binom{p}{S}$ distinct subsets $T$.  Referring to Table~\ref{table: sampling matrices}, we see that a spin-$\tfrac{1}{2}$ tetrahedral dictionary of three-letter words, one of which is a parity-check character, such that the minimal Hamming distance is two, yields a sampling matrix having $p = 4^2 = 16$ columns and  with $n=2^3 = 8$ rows.   To calculated the \RIP properties of this dictionary, the maximum sparsity we need to investigate is $S=n=8$, for which $\binom{p}{n} = \binom{16}{8} = 12870$ subsets must be evaluated, which is a feasible number.   As summarized in Table~\ref{table: eightsixteen sampling matrices}, the tetrahedral construction yielded sampling matrices having the \RIP property for sparsity $S=1,2,3$, while for higher values of $S$ the fraction of subsets having  $\delta_S \in (0,1)$ dropped sharply.

For purposes of comparison, we computed also the median \RIP properties of $8\times 16$ random Gaussian matrices.  We found for all values of sparsity, the \RIP properties of Gaussian random matrices were inferior to those of the deterministic tetrahedral construction.   We are not aware of any previous such random-versus-deterministic comparisons in the literature.  Since is is known that the Gaussian random matrices are \RIP in the large=$p$ limit, we were surprised to find that their \RIP properties are unimpressive for moderate values of $p$.

In preliminary studies of larger matrices, we found that known asymptotic expressions for the extremal singular values of Gaussian random sampling matrices---due to Mar{\v{c}}enko and Pastur \cite{Marcenko:1967pi}, Geman \cite{Geman:1980hb}, and Silverstein \cite{Silverstein:1985yf}, as summarized for \CS purposes by Cand\`{e}s and Tao \cite[see their Sec.~III]{Candes:2005ve}---were empirically accurate for tetrahedral sampling matrices too, for all values of the row dimension $n\le 256$ and all values of the sampling parameter $S\le n$.

We emphasize however that although the average-case performance of these petal-vector sampling matrices is empirically comparable to Gaussian sampling matrices, their worst-case performance is presently unknown, and in particular such key parameters as their worst-case isometry constants are not known.   

As Baraniuk \emph{et al{.}}~\cite{Baraniuk:2008oa} note: ``the question now before us is how can we construct matrices that satisfy the \RIP for the largest possible range of $S$.''   It is clear that answering this question, in the context of the deterministic geometric construction given here, comprises a challenging problem in coding theory, packing theory, and spectral theory, involving sophisticated trade-offs among the competing goals of determinate construction, large (and adjustable) $p/n$ ratio in the sampling matrix, and small isometry constants for all values of the sparsity parameter $S\le n$.

\subsubsection{Why do \CS principles work in \QMOR simulations?} 
\label{sec: why?}
Guided by the preceding analysis, we now try to appreciate more broadly why \CS principles ``work' in \QMOR simulations by systematically noting mathematical parallels between the two disciplines.  We will see that these  parallels amount to an outline for extending the mathematical foundations of \CS to provide foundations for \QMORCS.

As our first parallel, we remark that what Cand\`{e}s and Tao call \cite{Candes:2007ve} \emph{compressible objects} are ubiquitous in both the classical and quantum worlds.  This ubiquity is not easily explained classically, and so almost always it is simply accepted as a fact of nature; for example almost all visual fields of interest to human beings are compressible images.  In contrast, we have seen that the ubiquity of \emph{quantum} compressible objects has a reasonably simple explanation: most real-world quantum systems are noisy, and noisy systems can be modeled as synoptic measurement processes that compress state trajectories; working through the mathematical details of this synoptic compression was of course our main concern in Section~\ref{sec: designing and implementing}.  From this quantum  informatic point of view, it is a fundamental law of nature that any quantum system that has been in contact with a thermal reservoir (or equivalently, a measurement-and-control system) is a compressible object.

The second parallel is the availability of what the \CS field calls \emph{dictionaries} \cite{Berg:2007uo} of the natural elements onto which both classical and quantum compressible objects are projected.  For example, wavelet dictionaries are well-suited to image reconstruction.  In the \QMOR formalism of this article, the parallel quantum dictionary is (of course) the class of multilinear biholomorphic \GK polynomials that define the \Kahlerian geometry of \QMOR state-spaces (Section~\ref{sec: sectional curvature}).  This is not a linear dictionary of the type generally discussed in the \CS literature, but rather is an algebraic generalization of such dictionaries.  In the language of Donoho \cite{Donoho:2006cr}, open quantum systems exhibit a generalized \emph{transform sparsity} whose working definition is the existence of high-fidelity projections  onto \GK manifolds.

The third parallel is the existence of robust, numerically efficient methods for projection and reconstruction.  It is here that the mathematical challenges of aligning \QMOR with \CS are greatest.  In our own research we have tried non-\CS/non-\QMOR  optimization techniques---like  regarding $\gsb\psi_0 - \gsb\psi(\lb{\mathrm{c}})=0$ as the definition of an algebraic variety, and decomposing it into a Gro\"{e}bner basis---but in our hands these methods perform poorly.  Turning this observation around, it is possible that the efficient methods of \QMOR-\CS might find application in the calculation of (specialized algebraic forms of) Gro\"{e}bner bases.

Although a substantial body of literature exists \cite{Udriste:1994eu} for minimizing functions that are convex along geodesic paths on Riemannian state-spaces---which generalizes the notion of convexity on Euclidean spaces---there does not seem to be any similar body of literature on the convexity properties of holomorphic functions on \Kahlerian state-spaces.  

We have previously quoted Shing Tung Yau's remark \cite[p.~21]{Yau:06}: ``While we see great accomplishments for \Kahler manifolds with positive curvature, very little is known for \Kahler manifolds [having] strongly negative curvature.''   By the preceding construction, we now appreciate that (negatively curved) \GK manifolds have embedded within them lattices that display all the intricate mathematical structure of coding theory---so that it is not surprising that the geometric properties of these manifolds resists easy analysis.  It seems that the ultrahigh-dimensional model selection of Cai and Lv \cite{Cai:2007ci} can be described---with more-or-less equal mathematical justification---in terms of the differential geometry of ruled manifolds, or alternatively in terms of coding theory, or alternatively in terms of optimization theory.   

There is also the as-yet unexplored practical issue of whether quantum optimization of $l_2$-type functions over \GK polynomials like (\ref{eq: l2 norm}) is more efficient, less efficient, or comparably efficient to \CS-type optimization over petal-words of $l_1$-type functions like (\ref{eq: quantum Candes-Tao}--c).   This question is analogous to the long-standing issue in \CS of whether interior-point methods are superior to edge-and-vertex polytope methods ... the answer after five decades of \CS research being "yes, sometimes."

Efficient numerical means for evaluating the Penrose pseudo-inverse of (\ref{eq: projected Kahler}) are needed, as this inversion is the most computationally costly step of our sparse reconstruction codes as they are presently implemented.  Preconditioned conjugate gradient techniques are one attractive possibility \cite{Ciegis:2005qq,Gravvanis:2005xe,Gupta:1995rw}, because these techniques lend themselves well to the large-scale parallel processing.  The algebraic structure of the \GK metric tensor creates additional algorithmic challenges and opportunities that (so far as the authors are aware) have not been addressed in the computing literature.

Finally, suites of test problems and open-source software tools have contributed greatly to the rapid development of \CS theory and practice \cite{Berg:2007uo}, and it would be valuable to have a similar suite of problems and tools for the simulation of open quantum systems.

\section{Conclusions}

As Terence Tao has remarked \cite{Tao:2007fr}
\begin{QSEquote}
The field of [partial differential equations] has proven to be the type of mathematics where progress generally starts in the concrete and then flows to the abstract, rather than \emph{vice versa}. 
\end{QSEquote}
The recipes in this article have resulted from a flow in the opposite direction, from abstract to concrete, in which abstract ideas from quantum information theory, algebraic geometry, quantum physics, and compressive sensing have found find concrete embodiment in practical recipes for large-scale quantum simulation.   

\subsubsection*{Key abstract ideas from quantum information theory (\QIT)} 
Our recipes have adopted from quantum information theory the key idea that noise processes can be modeled as covert measurement processes.  This leads naturally to the idea that quantum states that have been in contact with a thermal reservoir (or equivalently, a measurement and control process) are compressible objects.

\subsubsection*{Key ideas from algebraic geometry}
Our recipes have adopted  from algebraic geometry the key idea that reduced-order quantum state-spaces can be described as geometric objects, using the language and methods of algebraic and differential geometry.  In particular, quantum trajectories can be described in terms of drift and diffusion processes upon state-space manifolds, just as in classical modeling and simulation theory.

\subsubsection*{Key ideas from quantum physics theory}
Our recipes have adopted  from theoretical quantum physics the key idea that quantum states have multiple unravelings, and that the efficiency of a calculation can be optimized by choosing an appropriate unravelling. 

\subsubsection*{Key ideas from quantum physics experiments}
Our recipes have adopted  from experimental quantum physics the key idea that mathematical ingredients of quantum simulation map one-to-one onto familiar physical systems such as measuring devices, and also the operational principle that the main deliverable of a quantum simulation is accurate prediction of the results of physical measurements.

\subsubsection*{Key ideas from compressive sensing, sampling, and simulation (\CS)}
Our recipes have adopted  from \CS the idea that optimization problems involving compressible objects (like quantum states) can often be transformed into convex optimization problems.  This can lead to both faster algorithms and improved physical insight.

\subsection{Concrete applications of large-scale quantum simulation}

By combining the preceding abstract ideas, the objective that began this article \begin{QSEquote}
\ldots\ \theObjective.
\end{QSEquote}
has now been achieved in a preliminary sense, albeit there is much further work to be done.  Now we consider some practical applications.

\subsubsection{The goal of atomic-resolution biomicroscopy}
The goal of atomic-resolution microscopy was the main motivation for developing the simulation algorithms described in this article. 
This goal was proposed as early as 1946 by Linus Pauling who envisioned ``If it were possible to make visible the individual molecules of the serum proteins and other proteins of similar molecular weight, all the uncertainty which now exists regarding the shapes of these molecules would be dispelled'' \cite{Pauling:46}.  
Later that same year, John von Neumann (possibly speaking as a reviewer of Pauling's proposal \cite{Kay:1998th}) wrote a letter to Norbert Wiener \cite{Neumann:1997qw} that embraced and extended Pauling's vision.  The letter expressed a strikingly modern vision of atomic-level structural and systems biology:
\begin{QSEquote}
There is no telling what really advanced electron-microscopic techniques will do. \ldots\ A ``true'' understanding of [viral-scale] organisms may be the first step forward and possibly the greatest step that may at all be required. I would, however, put on ``true'' understanding the most stringent interpretation possible: That is, understanding the organism in the exacting sense in which one may want to understand a detailed drawing of a machine, \emph{i.e.} finding out where every individual nut and bolt is located.
\end{QSEquote}
It was not until 1959 that Richard Feynman---who spent a sabbatical year working as a biochemist \cite{Edgar:1962nx}---issued his famous challenge: ``Is there no way to make the electron microscope more powerful? \ldots\ Make the microscope one hundred times more powerful, and many problems of biology would be made very much easier''   \cite{Feynman:59}.

Unfortunately, the problem of electron-beam radiation damage to fragile biological molecules proved intractable \cite{Henderson:95}, and so the Pauling-von Neumann-Feynman challenge of achieving atomic-resolution biomicroscopy remained unanswered for several decades.  

New ideas were needed, and three key ideas that emerged in ensuing decades were magnetic resonance imaging, nanotechnology, and quantum measurement theory.  The early stages of development of each of these new fields was slow, not because fundamentally new concepts of mathematics or physics were required---the key concepts were reasonably familiar to Pauling, von Neumann, and Feynman's generation---but because each field sought to push  familiar concepts to extreme limits.  Magnetic resonance was a familiar concept; exploiting the tiny magnetic resonance signals for \THREED imaging purposes was novel.  Making devices smaller was a familiar concept; fabricating micron-scale and nanometer-scale devices was novel.  Quantum measurement was a familiar concept; studying in detail the quantum evolution of small, continuously observed systems was novel.  

These research fields are united in magnetic resonance force microscopy (\MRFM), which was conceived explicitly as a means of meeting the Pauling-von Neumann-Feynman challenge of atomic-resolution biomicroscopy \cite{Sidles:91,Sidles:92,Sidles:92b,Sidles:95}. 

\subsection{The acceleration of classical and quantum simulation capability}
\label{sec: embracing}
Simulation technologies, both classical and quantum, began an immense surge of progress during the Pauling-von Neumann-Feynman era, and this surge has continued to the present day.  This is true especially in the classical domain, where simulation tools have become essential to system-level engineering \cite{Johnson:05,Ramo:84}.  In an era in which a new aircraft, a new processor chip, or a new drug can readily incur system development costs in excess of one billion dollars, engineering processes that maximize confidence, reliability, and economy have become a practical necessity.  

Feynman, in his seminal 1982 article \emph{Simulating physics with computers} \cite{Feynman:82}, argued that generic quantum physics problems are exponentially hard to simulate on a classical computer. But Feynman's analysis was vague about what constitutes generic quantum physics. 
The viewpoint we have developed in this article is that any quantum system that has been in contact with a thermal reservoir is---in principle at least---a~compressible object (in the language of \CS theory) and thus is amenable to simulation with classical resources.

The present status of quantum simulation algorithms has, in our experience, striking parallels to the status of linear programming algorithms during 1947--2004 \cite{Dantzig:63}.  It was clear for many decades that linear programming methods were exceedingly useful for solving practical problems, but their mathematical foundations in convex set theory were established only very slowly (see Spielman and Teng's 2004 article \cite{Spielman:2004ad} for a review and a reasonably definitive solution).   

This slow-but-steady progress illustrates Dantzig's principle \cite{Albers:86}:\begin{QSEquote}
In brief, one's intuition in higher dimensional space is not worth a damn! Only now, almost forty years after the time the simplex method was first proposed, are people beginning to get some insight into why it works as well as it does.\end{QSEquote}and also Feynman's words  \cite{Feynman:66}
\begin{QSEquote}
I think the problem is not to find the best or most efficient method to proceed to a discovery, but to find any method at all. \ldots\ [That is why] it is useful to have a wide range of physical viewpoints and mathematical expressions of the same theory.
\end{QSEquote}

\subsection{The practical realities of quantum system engineering in \MRFM}
\label{sec: emerging}
The practical experience of operating an \MRFM devices is very much like operating a small satellite that is distant from the experimenter not in space, but in scale.  In particular, the state-space of the \MRFM device includes the quantum state-space of spins in the sample.  The \QMOR analysis methods of this article were conceived specifically to allow the efficient \emph{a priori} modeling of this quantum state-space, by methods that have the potential to substantially extend present capabilities in modeling large-scale spin-systems \cite{Al-Hassanieh:2006qf}.  To the extent that future progress in \QMOR analysis allows this goal to be achieved, then the optimization of \MRFM technology can proceed partially \emph{in silico}, which will help retire technical risk, speed development, and build alliances among sponsors, researchers, enterprises, and customers.

\subsection{Future roles for large-scale quantum simulation}
\label{sec: challenges}
In our view, the single most important role for quantum system engineering (\QSE) will be to help sustain the exponentially cumulative technological progress that characterized the 20st century.  In the context of computing this exponentiation is popularly known as ``Moore's Law,'' but it is fair to say that similar exponentially cumulative progress is evident in fields such as nanotechnology, information theory, and (especially) biology.

A recent theme issue of \emph{\IBM Journal of Research and Development} describes  large-scale simulation codes running on ``Blue Gene'' hardware that is approaching peta\-flop-scale computation speeds \cite{IBM-Blue-Gene-Team:2008xy}.  Both classical \cite{Kumar:2008sf,Fitch:2008xy,Raman:2008rw,Fisher:2008yq,Ethier:2008nx} and quantum \cite{Vranas:2008qf,Gygi:2008kq,Bohm:2008nr} simulations are reviewed, and it is fair to say that the boundary between these two kinds of simulations is becoming indistinct, in particular when it comes to computing inter-atomic potentials that are both numerically efficient (classical) and accurate (quantum).   This continues a sixty-year record of mutually supportive progress in hardware, software, and algorithm development \cite{Dongarra:2008xy}.

From a geometric point of view, modern multi-processor computer architectures are exceptionally well-suited to the efficient computation of fundamental geometric objects such as polytopes, metrics, drift vectors, gradients, and diffusion tensors. These fundamental geometric objects are the raw building blocks---both conceptually and as software libraries---for broad classes of system simulations.  In particular, the recipes of this article demonstrate that both classical and quantum systems can be simulated using a shared set of abstract mathematical ideas and concrete software tools that are well-matched to distributed computing architectures.

In summary, twenty-first century technologies seek to maximize the pace, coordination, and reliability of technology development, to create products that press against the quantum quantum and thermodynamic limits of device speed, sensitivity, size and power.  For all such technologies, the quantum simulation recipes of this article promise to be useful.

\subsection*{Acknowledgements}

Author J.~A.~Sidles gratefully acknowledges the insights gained from discussions with Al Matsen on quantum chemistry, Rick Matsen on regenerative medicine, and Eric Matsen on informatic biology.

This work was supported by the Army Research Office (\ARO) MURI program under contract {\#}W911NF-05-1-0403, and by the National Institutes of Health \NIH/\NIBIB under grant {\#}EB02018.  Prior support was provided by the \ARO under contact {\#}DAAD19-02-1-0344, by \IBM under subcontract {\#}A0550287 through \ARO contract {\#}W911NF-04-C-0052 from \DARPA, and by the \NSF/\ENG/\ECS under grant {\#}0097544.

\newpage 
\bibliographystyle{jphysicsB}%

\end{document}